# A Survey of Skyline Query Processing

Christos Kalyvas and Theodoros Tzouramanis

*Department of Information and Communication Systems Engineering, University of the Aegean*
*Karlovassi, Samos, 83200, Greece*
*{chkalyvas, ttzouram}@aegean.gr*

**Abstract**

Living in the Information Age allows almost everyone have access to a large amount of information and options to choose from in order to fulfill their needs. In many cases, the amount of information available and the rate of change may hide the optimal and truly desired solution. This reveals the need of a mechanism that will highlight the best options to choose among every possible scenario. Based on this the skyline query was proposed which is a decision support mechanism, that retrieves the value-for-money options of a dataset by identifying the objects that present the optimal combination of the characteristics of the dataset. This paper surveys the state-of-the-art techniques for skyline query processing, the numerous variations of the initial algorithm that were proposed to solve similar problems and the application-specific approaches that were developed to provide a solution efficiently in each case. Aditionally in each section a taxonomy is outlined along with the key aspects of each algorithm and its relation to previous studies.

# Table of Contents

# 1. Introduction

The rapid growth of decision support systems and the increasing size of multidimensional data lead researchers to seek for new efficient methods for data processing in order to retrieve useful insights. The *operational research* science is related with the support of decision making by using various advanced analytical methods such as mathematical models, statistical analysis and data mining. Some of these analytical methods may be rank-aware approaches that contain scoring functions, such as those used in Top-K queries. Although, in many cases may not be desired to define a cumulative scoring function in order to retrieve the best results of a dataset since this will reduce the potential multi-dimensional comparisons of data to a single scalar value.

Taking this into account skyline queries deflect from the strict ranking approach of top-k queries and directed to an approach that is more understandable by hummans. Opposed to top-k queries where specific ranking functions and criteria are used, skyline queries assume that every user has a series of preferences over the attributes of data. Those preferences indicate what user's likes and dislikes (e.g. "I like the sea more than the mountains" or "I prefer to go vacations on an island rather than on a mountain). All the preferences are considered equivalent and will help to discard the items of the dataset that will not be preferred by anyone. This results in a small subset that contains the most interesting and preferred items based on all the preferences of all users. This set will be the skyline set or pareto optimal set.

In recent years, skyline query processing has become an important issue in database research for extracting interesting objects from multi-dimensional datasets. The skyline query processing is applicable in many applications that require multi-criteria decision making without using cumulative functions in order to define the best results but based on user's preferences. The skyline operator filters out a set of interesting points based on a set of evaluation criteria from a potentially large dataset of points. A point is considered as interesting, if there is not any other point better than that in all the evaluation criteria. The popularity of the skyline operator is mainly due the paradigm's simplicity and its applicability on multi-criterion decision support with respect to user preferences.

To be more precise, consider a typical skyline query example for a house purchase. In this problem, we suppose that a house might be interesting for somebody if no other house is both cheaper and closer to a metro-station. It is considered that as the distance of a house from a point of highly (general) interest is decreased (in this case a metro-station), the objective value (price) of the house is increased. So the user tries to find the best money-to-value ratio that satisfies his/her own preferences.

Table 1 represents a set of eight houses that a user found to be sold in the vicinity of a particular metro station. Each row in the table contains information, which can be used to identify the most interesting houses. To make the example simple there exist only two numeric attributes (dimension) for the houses. One attribute will be price and the other will be distance from the metro-station. In this case first evaluation criterion is minimizing the distance from the metro-station and the second one is minimizing price. Every evaluation criterion is considered as a single dimension in the d-dimensional space.

| House | price (in thousand €) | Distance (m) |
|---|---|---|
| H1 | 100 | 1500 |
| H2 | 1400 | 500 |
| H3 | 700 | 600 |
| H4 | 1300 | 1000 |
| H5 | 900 | 1300 |

| H6 | 1600 | 100 |
| H7 | 400 | 300 |
| H8 | 200 | 1200 |
| H9 | 1000 | 200 |
| H10 | 500 | 1400 |
| H11 | 500 | 900 |

**Table 1 Dataset of houses**

**Figure 1** illustrates the skyline of the existing set of houses. Houses H2, H3, H4, H5, H10 and H11 do not belong on the skyline as they are no one's top choice because for each one of them exists at least one house which is better in terms of price or distance. Houses H1, H6, H7, H8 and H9 are the most interesting ones and so belong to the skyline. All the skyline points are connected by a line. The skyline is essentially the boundaries of the union of dominance area of all skyline point. To make it easily understood, the dominance area of a 2-dimensional point is the North-East quadrant of the space that occurs by imaginably drawing a x-y axis system with origin point the point of interest that is examined. The dominance area of a point will be inside the dominance area of a second point, noted as the first point dominates the second one, only if the first point is as good or better in all dimensions and better in at least one dimension based on the evaluation criteria. The skyline would refer to those points that are not dominated by any other point. In the house-metro station example "better" is minimizing the values.

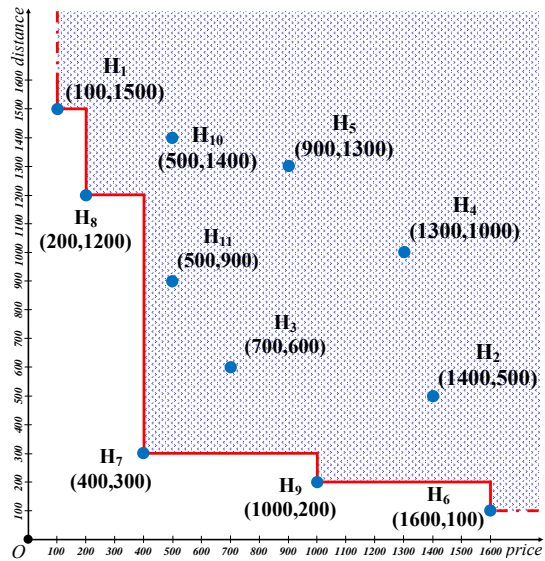

**Figure 1: Skyline of a set of houses**

Skyline queries can also involve more than two dimensions. For instance, a buyer could be interested in houses that are near to a metro-station, are cheap, have high square footage and low communal costs. The main idea of the skyline operator is to give the user the overall view of all interesting results and then let him/her to make a decision.

## 1.1. Applications

Some indicative applications areas for which skyline queries **[19]** are useful are **customer information services**, **decision support** and **decision making systems**. For instance, a skyline query can be used by travel agencies to find a reasonable priced hotel near the sea or to find good salespersons which have low salary **[19]** Additionally reverse skyline queries **[44]** can assist in **market research** applications in order to find if a specific product is appealing to consumers or to identify the best location for a new branch store. Also can be applied also in **economics [108]** where can support microeconomic data mining or even in **continuous data stream environments [117]** such as stock exchange systems. Additionally it can be used on **location based systems (LBS)** in order to identify the shortest root to a destination or the closest point of interest among many **[75], [98]**. Another application is **distributed query optimization.** This can be particularly useful in cloud architectures where data are scattered among servers or in the case where **Quality of (web) services [3]** is the primary goal. Skyline queries can also be used to focus on a **subspace of attributes [196]** in order to identify the skyline on a small subset of the dimensions of the dataset that are defined. Skyline queries have also applications in **computer security** and especially on problems concerning **privacy [28]** and **authentication [113]**. Skyline computation in metric space **[30]** can assist the *DNA searching problem* in **bioinformatics**. Finally skyline queries are applicable in a wide **variety of data types** such as partial ordered **[24]**, and incomplete **[91]** or uncertain data **[138, 87]** .

## 1.2. Related Problems

Many similar problems and operators related with skyline queries have been studied in the literature. For example, the Top-*K* query **[81]** retrieves the best *K* objects that minimize a specific preference function. The difference from skyline query is that the output changes according to a user-specified input function and the retrieved points are not necessary part of the skyline. The k-nearest neighbor (k-NN) query **[135]**, in another example, requires the existence of a query point *p* and outputs the k objects closest to *p,* in increasing order of their distance. In this case the difference from the skyline query is that k-NN query retrieves answers according to the proximity of a given point and not based on domination to other points. Finally, convex hull **[18, 144]** contains the points that are enclosed by the polygon that is defined from the minimum and maximum skyline (i.e. minimizing and maximizing values based on the evaluation criteria) of the given set of points. The main difference from a skyline is that it defines an area of interest rather than a line with individual interesting points.

## 1.3. Contribution

This survey gives an overview of skyline query computation techniques that have been proposed so far. It reports on the source of skyline problem and gives a detailed coverage of state-of-the-art skyline query processing techniques. It also introduces taxonomy to classify skyline queries based on their general characteristics and it presents in a categorized manner the applications of skyline queries. To the best of our knowledge, this is the first study that gives a complete and comprehensive overview of the research that has been done on skyline query computation.

## 1.4. Outline

The rest of the survey is organized as follows. **Section 2** reports on the preliminaries notions and on most popular algorithms for general skyline query computation. **Section 3** reports on the skyline queries family. **Section 4** reports the applications and problems that the skyline operator can solve. Finally **section 5** reports on the cardinality estimation of the resulted skyline set and on efficient methods for handling updates of the dataset.

**Figure 2** illustrates the number of citations by year that where received in the broad research area of skyline computation.

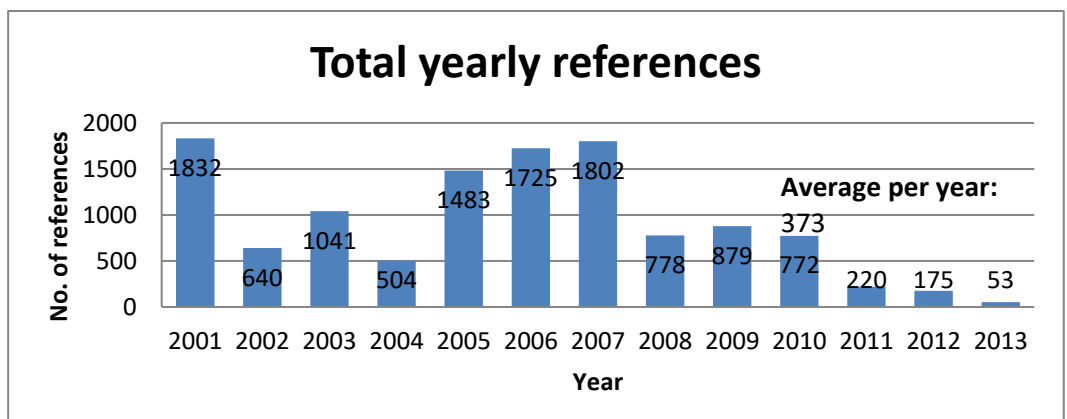

**Figure 2: Yearly references.**

**Figure 3** illustrates approximately the average number of references that an article received in each year.

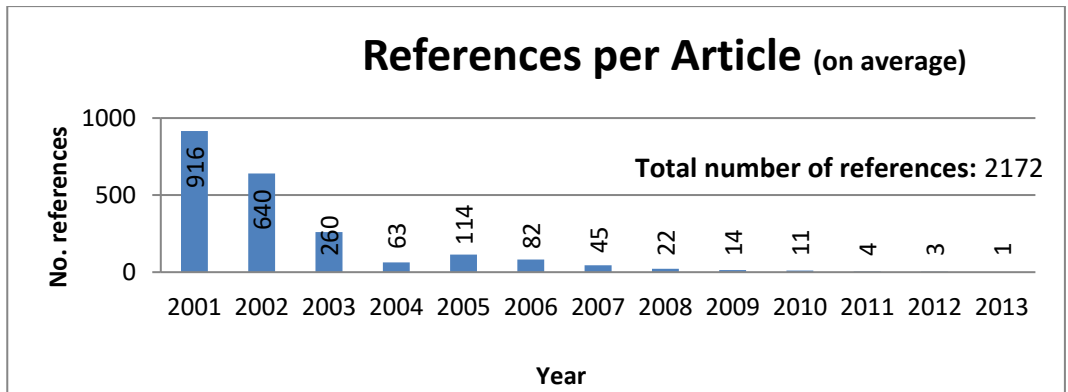

Figure 3: References per article.

**Figure 4** illustrates the number of articles appeared in each year.

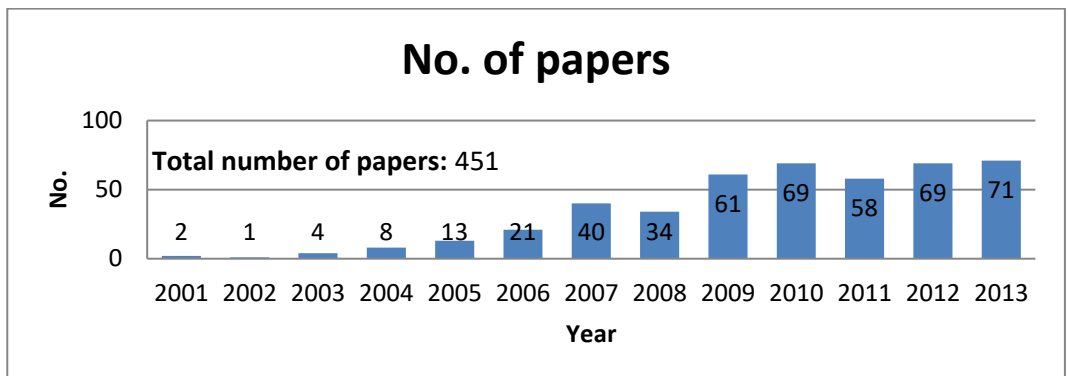

Figure 4: Number of articles per year.

## 2. Skyline

Before going into details on skyline processing methods, this section will provide some preliminary definitions and will present an overview of the fundamental skyline queries computation algorithms.

The computation of the skyline in database research is equivalent to determining the *maximal vector problem* in *computational geometry* **[144],** or equivalently the *pareto* optimal *set* **[99, 144]** problem in *operations research*. The *maximal vector problem* is to find the subset of a set of vectors such that each one of them is not *dominated* by any other vector from that set. Considering that those vectors are *points* in a *k*-dimensional space, then the maximal vectors **[15]** can also be called *admissible points* **[8]** and the maximal set of vectors as *Pareto set*. This class of problems was extensively studied by the mathematical community in the decade of 60s.

As mentioned in **[19]** the skyline problem considers that the dataset cannot fit completely in the main memory (RAM) in order to be processed. This is more likely to be the case in modern database systems where the dataset is retrieved from an external memory such as disks. Methods that do not rely on external memory are DD&C **[99],** LD&C **[15]** and FLET **[14]**.

Authors in **[59]** proved that the initial algorithms, proposed for maximals **[99, 144]** which are based on the divide-and-conquer approach **[13]** (which divides the initial problem in equal sub-problems and then tries to solve each one separately, combining the results in the last step of the process) have quite bad performance with respect to the dimensionality of the initial problem. Additionally these algorithms assume that the whole dataset fits into memory and they do not take into account memory limitations and thus cannot be directly applied in a database scenario. Such kind of approaches suffers for the "curse of dimensionality" [16] which was first used by Bellman [12] and is often used to indicate that high dimensionality causes problems in resolving due to increased computational cost. This problem was observed and solved with the introduction of the skyline operator [19] which proposes a *divide and conquer* algorithm suitable for external memory and shows how it can be integrated into a database system. An overview of the foundamentals skyline algorithms in [134], [169].

## 2.1. Skyline problem & properties

Skyline queries are a popular and powerful paradigm for incorporating user preferences into relational queries and extracting interesting points from a set of points. The main difference from the previous described approaches is that instead of finding vectors or points, a skyline queries finds the maximals over a set of *tuples or the so-called* set of Pareto-optimal tuples. Those tuples are those that are not *dominated* by any other tuple in the same relation. One of the nice properties of the Skyline of a given set Ds of points is that any set of evaluation criteria that arise from user's preferences can be modeled in the form of a monotone scoring function $f: D_i \to R$, like L1 norm $f(x,y) = x + y$ or Euclidean norm $f(x,y) = \sqrt{x^2 + y^2}$. If p ∈ Ds and minimizes (or maximizes) the scoring function, then p is in the Skyline. That means, regardless how a user weights his/her preferences towards price and distance of houses, s/he will find a house that matches his/her preferences in the Skyline. In this example for simplicity, is assumed that skylines are computed with respect to minimum (min) conditions (minimizing the scoring function) on all dimensions. In detail, using the *min* condition, a point *p dominates* another point *r* if the coordinate of *p* on at least one axe is smaller than the corresponding coordinate of *r*, and no larger on any of the remaining axis. This implies that *p* is preferable to *r* according to any *preference* (*scoring*) function which is monotone on all attributes. Furthermore, for every point **p** in the Skyline, there exists a monotone scoring function *f* such that **p** minimizes (or maximizes) that scoring function. This ensures that the skyline will contain all the preferable houses no matter how users weight their preferences. More formally, given a d-dimensional space D={$d_1$ ,..., $d_d$} and a set *Ds* of points that belongs in D, a point p ∈ Ds can be represented as P = {$p._{d1}$,...,$p._{dj}$}, 1<=j<=d, where p.dj is the value of the *jth*-dimension of the point. Assume that the dataset Ds contains the points Ds={$p_1$ ,..., $p_n$}. The notation pi.dj ≥ 0, with $1 \leq j \leq d$ and $1 \leq I \leq n$, is used to denote the *j*-th dimensional value of the *pi point.* Assume that for each dimension $d_j$ there exists a total ordering relation, denoted by '<' or '>' according to the user's preferences. Without loss of generality in our examples we will use the '<' relation.

**Definition 1:** Dominate
Given points p, r ∈ Ds, p dominates r, denoted as p ≺ r , if and only if ∃ j∈[1,d] such that p.dj<r.dj and ∀ i ∈[1,d]-{j}: p.di ≤ r.di ∎

Dominance has the property of a transitive relation. That is if p dominates r and r dominates t, then p also dominates t **(Figure 5)**. This is given more formally in the next proposition.

**Proposition 1:** Transitivity
Given points p, r, t∈ Ds, if p≺r and r≺t, then p≺t. ∎
Transitivity can be used to eliminate from further consideration a single point or a group of points that are dominated by a point p which in its turn is dominated by a new point r.

Through previous analysis, the domination between two points was explained. In contradiction If two points *p, r ∈ Ds* do not dominate (denoted with ≺>) each other simultaneously (that is p≺>r and r≺>p) **(Figure 6)** are considered as incomparable in *Ds*, and denoted with *p ~Ds r or simply p ~ r*. More formally:

**Proposition 2:** Incomparability
Given two points p, r ∈ Ds, if p≺>r and simultaneously r <> p, then p and r are incomparable on Ds (i.e. p~r). This property helps in determining if one or more points can be skyline points. A point in the skyline set must be incomparable to all other points of the set.

For example consider the partitions on **Figure 6** that could be derived from the original dataset using a Divide & Conqueror approach. If we first examine partition 1 and identify at least a single skyline point (in any location inside of it) then partition 4 could be completely pruned. Furthermore since partitions 3 and 2 are incomparable the skyline points in one partition do not affect the skyline points in the other partition and there is no case that points from partition 3 dominate points in partition 2 and vice versa.

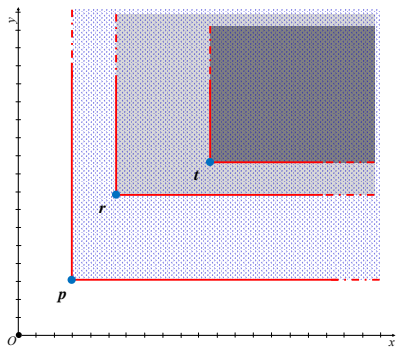

**Figure 5: Transitivity dominance.**

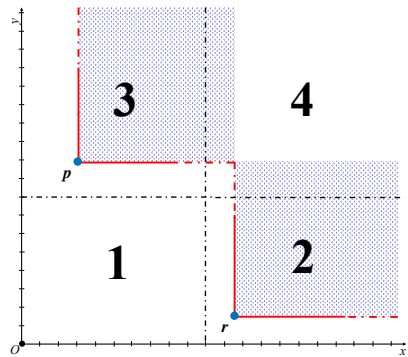

**Figure 6: Incomparable points.**

The skyline of a dataset of n points refers to those points that are not dominated (are incomparable) by any other point. That is, a data point p is a skyline point if there does not exist any point on the dataset that dominates p.

**Definition 2:** Skyline
A data point p ∈ Ds is a skyline point iff ∄ r∈ Ds such that r≺p. ∎

Notice that in order a point to be a skyline point it is not needed necessary to dominate another point in the dataset. Additionally the skyline set of a dataset is unique.

The math notations that will be used in the subsequent discussion are summarized in **Table 2.**

| Notation | Definition |
|---|---|
| D | d-dimensional space |
| Ds | Input dataset for skyline computation (set of points) |
| d | Number of dimensions of DS |
| $d_i$ | one dimension ($1 \leq i \leq d$) |
| p, r, t | Data points |
| s | Skyline point |
| $p.d_i$ | i-th dimension of the point p |
| q | query |
| $S_{DS}$ | Set of skyline points of DS |
| f | Monotone function |
| p ≺ r | p dominates r |
| $p \prec^q r$ | p dominates r with regard to q |
| $p \prec_{Ds} r$ | p dominates r in the Dataset Ds |
| $p \prec^q_{Ds} q$ | p dominates r in the Dataset Ds with regard to q |
| $p \;^\varepsilon\!\prec r$ | p ε-dominates r |
| p≺≻r | p does not dominates r |
| p ~ r | p and r are incomparable |
| $S_{DS}^q$ | Skyline set S of dataset Ds with regard to the query point q |

**Table 2: Math Notations**

Apart from the formal definition of the skyline there are some additional related interesting features. A skyline query tries to find an optimal solution for a user, based on **multiple, and sometimes conflicting, goals**. For example, a user may be interested in buying an economic house in Athens that is also close to a metro-station. In general case, houses that are near to a metro-station are expected to be more expensive (because they are preferred by the majority of buyers), therefore his/her preference for an economic house contradicts his preference for a house close to the metro station. Additionally there may be **no single optimal answer** (or answer set) that satisfies exactly the preferences of the user, but rather there could be numerous answers that are close in satisfaction of his/her preferences. In the same example, it is unlikely that there exists a house that is the cheapest among all houses and is at the same location with the metro-station, (because houses near the metro-station are preferable by most buyers and a house in a distance will try to attract buyers with a lower price). Instead, one can expect to find in the skyline, among others, a list of economic houses such that those nearer to the metro-station to be slightly more expensive. Thus, users are typically **looking for satisficing answers** (decision making support). For the same query, different users with similar personal preferences, which are not exactly satisfied by a single optimal answer, may finally find different answers appealing. A person may be willing to pay a little more to be closer to the metro-station and another may be contented with a cheaper house as long as it is convenient to go by foot. In conclusion it is important to present all interesting answers that may fulfill a user's need.

## 2.2. SQL Extension

The skyline operator was first introduced in [19] where authors proposed an SQL syntax which extends SQL's SELECT statement by an optional SKYLINE OF clause which is syntactically similar to an ORDER BY clause. This way a user can find all the interesting points by specifying the skyline dimensions and the criteria that will be used, using one of *MIN*, *MAX* and *DIFF* directives. The SKYLINE OF clause is executed after *scan, join,* and *group-by* operators and before a final *sort* operator such as an ORDER BY clause (some exceptions are discussed in [19]). According to the syntax a skyline query would be expressed as:

> SELECT <attributes> FROM <relations> WHERE <conditions> GROUP BY <attributes> HAVING <conditions> SKYLINE OF d1 [MIN| MAX| DIFF], . . ., dn [MIN| MAX| DIFF]

A SKYLINE OF clause contains columns $d_1,...,d_n$ which represents a list of attributes (which can be noted as skyline dimensions) over which a user can apply his/her preferences in order to rank the dataset. Note that it does not matter the order in which dimensions are specified, but it is important for each dimension to have a natural total ordering (in the case of general skylines queries) as integers, floats or dates. The directives MIN, MAX and DIFF, can specify whether the value in that dimension should be minimized, maximized, or distinct from all the others. It is important to mention that a one dimensional Skyline is equivalent to a MIN or MAX SQL query without a SKYLINE OF clause.

In the general *house-metro station* example, a user who is looking for the most interesting houses, which are cheap and near to the metro station, would prefer both price and distance dimensions to be minimized. In that case the query could be written in SQL syntax as follows:

> SELECT * FROM Houses WHERE city = 'Athens' SKYLINE OF price MIN, distance MIN

The simplest way to implement a skyline query on top of a relational database system is by translating the Skyline query into a naive-nested SQL query which will compare every tuple (point) with every other tuple. The above skyline query is equivalent to the following standard SQL query:

```
SELECT * FROM Houses h WHERE h.city = 'Athens' AND NOT EXISTS (
    SELECT * FROM Houses h1 WHERE h1.city = h.city AND
        h1.distance <= h.distance AND h1.price<=h.price AND
            (h1.distance <h.distance OR h1.price < h.price));
```

Thought the problem can be very well expressed in **SQL,** this approach shows very poor performance. Particularly it involves a self-join over a table which is essentially a $\vartheta$-join rather than an equality-join.

This way a ϑ-join [128] will evaluate all the combinations of the tuples that involves and satisfy the relation θ, causing a significant computational overhead.

## 2.3. Fundamental skyline algorithms

Existing skyline computation methods can be classified into two categories, depending on whether or not rely on pre-computed indexes on data. Index-based methods have better performance, since they avoid accessing the entire data collection, but have limited applicability due to the necessity of an indexed dataset. Additionally multi-dimensional indexes like R-trees have their own limitations as they suffer from the well-known curse of dimensionality. Not index-based methods are more *generic*, in the sense that they do not require any specialized access structure to compute the skyline.

Authors in [19] introduces a Block Nested Loop (BNL) algorithm which like the naive nested-loop algorithm repeatedly reads the set of tuples and eliminates points by finding other points in the dataset that dominate them. BNL allocates a buffer (window) in main memory that contains a number of points in order to sequentially track the dominance between them **(Figure 7)**. The algorithm reads the input data and each point is retrieved and compared against the points in the buffer. In the first run of the algorithm no point will exist in the buffer so it's trivial to insert the first point in the buffer. For the next runs if the point retrieved is dominated by at least one point in the buffer there is no need to continue the comparison with the others points that maybe exist in it and the point is discarded. Otherwise if the point is incomparable or dominates one or more points in the buffer, those points that are dominated are removed from the buffer and the new point is inserted. **Figure 7** illustrates the algorithm in its fifth iteration in which has processed houses H1, H2, H3, H4, H5 from the input dataset. As seen points H4 and H5 are dominated by one or more points in the buffer and so are discarded from further processing. For further considerations suppose that the buffer has size 3, meaning that can store up to three entries.

If in any stage the buffer becomes full, a different approach is followed. Once this happens, the rest of the input is processed differently and a temporary overflow disk file is used **(Figure 8)** to store the points that where compared and characterized as incomparable or dominated existing points in the list and cannot be further placed in the window. Such a point is house H6, which is incomparable with the houses already existing in the buffer and thus is placed on the temporary file. Nevertheless the dominated points in the window are still discarded as before right after each dominance comparison. After the dataset has been read now the temporary file is used as input for the next passes of the algorithm. After the first run all the points of the input will be either inside the window or in the temporary file. Points that inserted in the window before any other point was inserted in the temporary file are guaranteed to be skyline points. This can be checked by assigning a timestamp to each point that exists in the window and the temporary file.

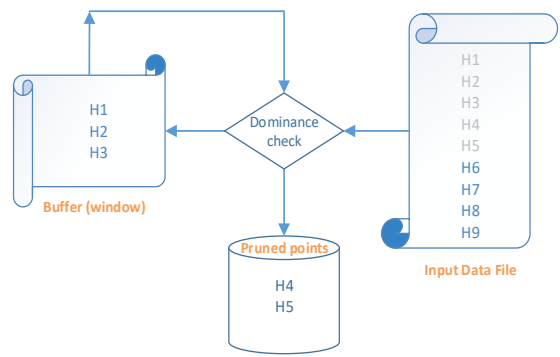 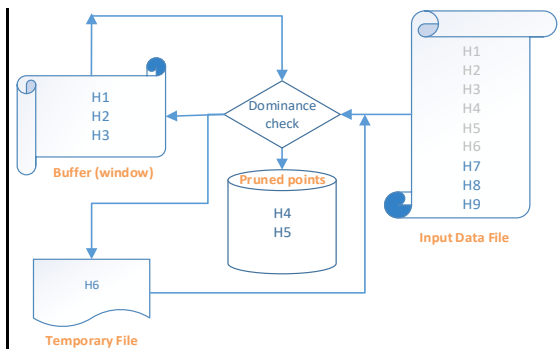

Figure 7: BNL without temporary file. | Figure 8: BNL with temporary file.

The algorithm may require a large number of passes until the complete skyline is computed and eventually terminate as at the end of each pass the size of the temporary file will be decreased. BNL works well if the size of the resulted skyline is small and in best case fits into the window which will result in the termination of the algorithm in one iteration. BNL algorithm cannot compute skyline points progressively. Its performance is very sensitive to the number of dimensions and to the underlying data distribution. Especially, it is good for up to five dimensions for a uniform distribution but its performance degrades if the distribution tends towards an anti-correlated distribution.

***Complexity:*** The complexity for the best-case is $O(n)$ and in the worst-case is $O(n^2)$, where n is the size of the dataset. Complexity concerns the case where the entire set of candidate skyline points at any time, fit in the main memory buffer. The worst-case complexity concerns the case where every point must be compared with every other point, which can happen in the case of a buffer window with size 1.

### 2.3.1. D&C

The D&C (Divide-and-conqueror) algorithm proposed in **[19]** is an extension of the two-way partitioning divide-and-conqueror algorithms proposed in **[99, 144].** These earlier proposed algorithms do not scale well for large datasets, since they do not take into account main memory limitations. The D&C algorithm recursively divides the input dataset in m partitions $\{P_1,....P_m\}$ (m-way partitioning), in order each one of them to fit in the main memory **Figure 9.** The partitions boundaries are determined by computing the q-quintiles of the dataset which results in the division of the dataset into q-1 equal subsets. Then a local (partial) skyline $S_i$ is computed for each partition $P_i$ with $1 \leq i \leq m$. Finally the algorithm computes the global skyline by progressively merging the local ones based on a bushy merge tree **Figure 10** and **Figure 11**. This way points that belong to one partition and are dominated by points of another partition can be removed. The points that left from the merging process are the skyline points and the algorithm terminates returning the resulted set.

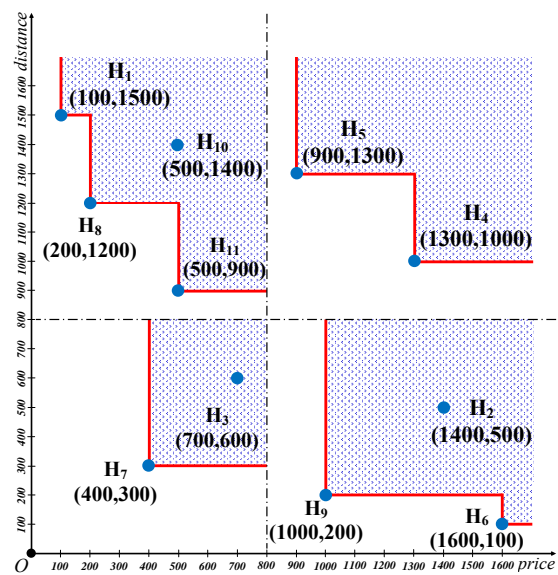

**Figure 9: Divide and conqueror**

As BNL algorithm, so D&C cannot produce skyline points progressively since the first skyline point can be generated only when the entire dataset has been scanned. Moreover, as the main memory size increases it performs better as it requires the partitions to be in-memory. D&C is less sensitive than the BNL to the number of dimensions and correlations in the database.

***Complexity:*** The algorithm can find the skyline points in $O(n)$ in the best case which essentially is the merging cost of the divided partitions and in $O(n^2)$ in the worst case which is the cost of identifying the skyline points in each sub-partition and the cost of merging the results.

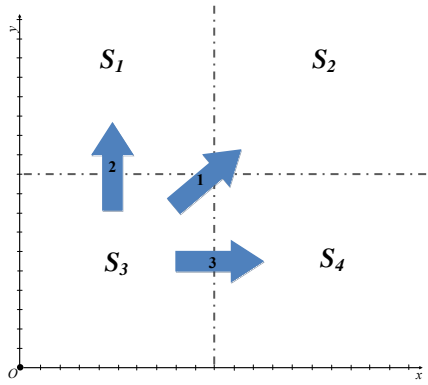

Figure 10: Merging process

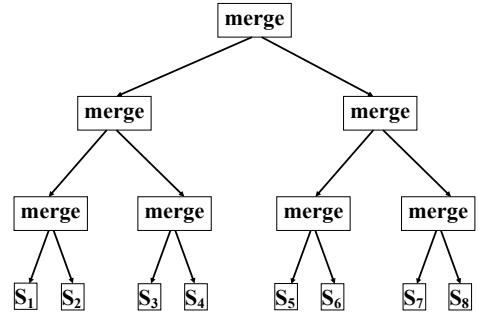

Figure 11: Bushy merge tree

### 2.3.2. Bitmap

To resolve the problem of progressive skyline computation [164] proposed the index-based Bitmap algorithm which encodes all data into a bitmap structure in order to identify the skyline points by exploiting the speed of a bitwise & operation. Bitmap is a progressive algorithm which means that it does not need to scan the complete dataset in order to return results and is based on a bitmap structure which encodes all the required information in order to determine if a point belongs in the skyline.

---

**ALGORITHM 1:** Bitmap [164]

**Input:** A Dataset D.
**Output:** The Set of skyline points of dataset D.

1. **for** each point $x = (x_1, x_2, ..., x_d)$ in the Dataset D
2.     let $x_i$ be the $q_i$th distinct value in dimension i
3.     $A \leftarrow BitSlice(q_1, 1)$
4.     **for** $i = 2$ to d **do**
5.         $A \leftarrow A\ \&\ BitSlice(q_i, i)$
6.     $B \leftarrow BitSlice(q_1 - 1, 1)$
7.     **for** $i = 2$ to d **do**
8.         $B \leftarrow B\ |\ BitSlice(q_i - 1, i)$
9.     $C \leftarrow A\ \&\ B$
10.    **if** $C == 0$ **then**
11.        output x

---

At this point was outlined the pseudocode of Bitmap algorithm. The $BitSlice(q_i,i)$ function returns the bitslice for the $q_i^{th}$ distinct value of the $i^{th}$ dimension. If the bitslice does not exists, then $BitSlice(q_i,i)$ will be equal to 0. In the case where C equals to 0 (line 10) for a given point x, then no point dominates x and thus point x is a skyline point.

In order to describe the algorithm assume that a point $p=\{p_{.d1}, ...., P_{.dj}\}$, $1 \leq j \leq d$ in a d-dimensional space is represented by an m-bit vector. From those m-bits each p.di is represented by a number of ki bits. Each ki has as many bits as the number of distinct coordinate values of all the points of the dataset in that dimension and thus $m = \sum_{i=1}^{d} k_i$.

To incorporate the house-metro station example, there are 10 distinct values for dimension price and 11 distinct values for dimension distance. That means k1+k2=10+11=21 and thus m=21. Considering the *min* annotation and assuming that $p.d_m$ is the j-th smallest number on the i-th dimension it can be represented by the $k_i$ bits setting the ($k_i$-$j_i$+1) most significant bits to 1 and the rest to 0. In detail, value 400 is the third largest value among the 10 in the first dimension so in its bit-representation the first (10-3) +1=8 most significant bits will be assigned to 1 and all the other to 0. The results of the mapping process are shown in **Table 3.** Next the algorithm needs to determine if a point is a part of the skyline or not.

In our case will check points H7, which from previous example is a skyline point and H4 which is an ordinary point. In order the algorithm to compare the points obtains the array of bit-vectors of all points and transposes it in an m-length array of bit-slices. Each bit-slice $V_i$ corresponds to the sum of the i-th bit-value of the dimension, of all points. The bit-length of bit-slices depends on the number of points. The two bit-slices of House H7 for the two dimensions are shown in bold in the **Table 3**.

After the construction of the bit-slices the algorithm performs 3 bitwise operations among 2 sets of bit-slices. The first set contains the bit-slices $V_x$, $V_y$ (one for each dimension) where resides the last bit of the point which is equal with one .The second set contains the next in order bit-slices $V_{x+1}$ , $V_{y+1}$ of those that selected in the previous set. In the case that the bit-slices of the previous step is the last in order then is used the zero bit-slice (all bits zero). The first bitwise operation A will be an AND operation between $V_x$ and $V_y$. The second bitwise operation B will be an OR operation between $V_{x+1}$ and $V_{y+1}$ . The third bitwise operator C would be also an AND operation between the results of the two previous operations. If the result of the final operation is zero then the tested point is a skyline point.

For Point H7 A=$V_x$ AND $V_y$ ={ 10111011111 AND 00000110100 } = 00000010000, which indicates that the points that have values in each dimensions that are greater or equal to this point is only the point H7. The second operations B= $V_{x+1}$ OR $V_{y+1}$= { 10000001000 OR 00000100100 } = 10000101100, which shows that points which have some of it's dimension better than H7 are the points H1, H6, H8 and H9. The last operation C= A AND B = {00000010000 AND 10000101100} = 00000000000 which shows that their is no house that dominates H7.

In the case of House H4 A=$V_x$ AND $V_y$ ={ 10111011111 AND 01110110101 }= 00110010101 which indicates that houses H3, H4, H7, H9, H11 are equal or better in each dimension. Operation $V_{x+1}$ OR $V_{y+1}$= {10101011111 OR 01100110101}= 11101111111 indicates that points {H1-H3}, and {H5-H11} are better in at least one dimension from H4. The final operation C=A AND B = {00110010101 AND 11101111111} =00100010101 indicates that points H3, H7, H9, H11 dominate point H4.

| House | coordinates | Bitmap representation |
|---|---|---|
| H1 | ( 100 , 1500 ) | ( 1111111**1**11 , 10000000**0**00 ) |
| H2 | ( 1400 , 500 ) | ( 1100000**0**00 , 11111111**0**00 ) |
| H3 | ( 700 , 600 ) | ( 1111110**0**00 , 11111110**0**00 ) |
| H4 | ( 1300 , 1000 ) | ( 1110000**0**00 , 11111000**0**00 ) |
| H5 | ( 900 , 1300 ) | ( 1111100**0**00 , 11100000**0**00 ) |
| H6 | ( 1600 , 100 ) | ( 1000000**0**00 , 11111111**1**11 ) |
| **H7** | **( 400 , 300 )** | ( 1111111**1**00 , 11111111**1**00 ) |
| H8 | ( 200 , 1200 ) | ( 1111111**1**10 , 11110000**0**00 ) |
| H9 | (1000 ,200 ) | ( 1111000**0**00 , 11111111**1**10 ) |
| H10 | ( 500 , 1400 ) | ( 1111111**0**00 , 11000000**0**00 ) |
| H11 | ( 500 , 900 ) | ( 1111111**0**00 , 11111100**0**00 ) |

**Table 3 Bitmapped dataset**

Even if Bitmap is a progressive algorithm it must consider all points of the dataset in order to compute the full Skyline which tends to be an expensive operation because, for each point inspected must retrieved the bitmaps of all points. Additionally the algorithm does not allow the user to give preferences in which order the results are produced but rather points are returned depending on the clustering of the data. Finally bitmaps perform well when the number of distinct values per dimension is small.

***Complexity:*** -

### 2.3.3. INDEX

Among the Bitmap algorithm **[164]** authors additionally proposed the *Index* algorithm, inspired from the rank aggregation algorithm proposed in **[52]**, which partitions the entire d-dimensional dataset into d ordered lists. It uses a specialized B-tree to index each point by a transformation mechanism that maps high-dimensional points into single dimension point. Note that it can use any single dimension index structure and not only a b-tree. The data points are mapped to y and ordered as

y=$d_{min}$ + $x_{min}$. A point $p = (p.d_1, p.d_2, \ldots, p.d_d)$ of the dataset belongs to the i-th list  (1 ≤ i, j ≤ d) if it's p.$d_i$ value is minimum among all p.dj values, that is p.$d_i$≤p.$d_j$ for all i ≠ j. In each list points are organized in batches and sorted in an ascending (or non-ascending) order of their distinct minimum (or maximum) value in that dimension. Each batch is identified by the minimum value of the point that represents. Points with the same minimum value in each list are organized in the same batch. Each batch is processed according to its ascending index value and the algorithm tries to determine if it belongs to the skyline. If a batch has more than one point a local skyline is computed which is then checked if it can be merged to a global one.

---

**ALGORITHM 2:** Index **[164]**

**Input:**   A Dataset D. (B-tree)
**Output:** The Set of skyline points of dataset D.

1. **for** i = 1 to d **do**
2.    $f_i$ ← True
3.    $t_i$ ← traverseTreeMax(root, i)
4.    $max_i$ ← maxValue($t_i$)
5.    $min_i$ ← minValue($t_i$)
6. mn ← $\max_{i=1}^{d}$ $min_i$
7. mx ← $\max_{i=1}^{d}$ $max_i$
8. **for** i = 1 to d **do**
9.    **if** mn > $max_i$ **then**
10.      $f_i$ ← False
11. j ← 1
12. S ← 0
13. **while** there are some partitions to be searched **do**
14.    **for** i = 1 to d **do**
15.      **if** $max_i$ == mx **then**
16.        $P_j$ ← $t_i$
17.        $S_j$ ← 0
18.        $t_i$ ← getNextLeftElement($t_i$)
19.        **while** (maxValue($t_i$) == mx) **do**
20.          mn ← max(mn; minValue($t_i$))
21.          $P_j$ ← $P_j$ ∪ $t_i$
22.          $t_i$ ← getNextLeftElement($t_i$)
23.          $max_i$ ← maxValue($t_i$)
24.        $S_j$ ← computePartitionSkyline($P_j$)
25.        S ← S ∪ computeNewSkyline($S_j$ , S)
26.        j ← j + 1
27.    mx ← $\max_{i=1}^{d}$ $max_i$
28.    **for** i = 1 to d **do**
29.      **if** mn > $max_i$ **then**
30.        $f_i$ ← False

---

At this point the pseudocode of Index algorithm was outlined. The $f_i$ flag indicates whether or not the algorithm must continue to search on the $i^{th}$ dimension. When the flag $f_i$ is false, it means that all the remaining points are dominated and thus the partition does not need to be further searched. Routine maxvalue(t) returns the maximum value of tuple t among all dimensions and respectively minValue(t) returns the minimum value. The traverseTreeMax(root,i) routine traverses the B-tree in order to obtain the tuple with the largest value in dimension i. Routine getNextLeftElement(t) returns the left element of t, computePartitionSkyline(P) computes the skyline of a set of points P and computeNewSkyline ($S_i$ , S) computes the new skyline set derived from $S_i$ taking into account the already found skyline set S.

In the case of the hotel metro-station example the houses that belong to the first list and have their first coordinate minimum among the two are H1, H5, H8, H10, H11  and houses that have their second coordinate minimum and belong to the second list are  Houses H2, H3, H4, H6, H7 and H9 **(Table 4)**. Houses H10 and H11 have the same minimum value so they belong to the same batch. When the algorithm starts loads the first batch from each list. The two first batches have minimum value 100, so the algorithm process with the batch from the first list. Point H1 is added to the skyline list because is a single point and the skyline list is empty. The next batch the algorithm handles is H6 which was considered previously. The point is incomparable so is added to the list. The next batches from each list which are loaded are H8 and H9 that again have the same Min value which is 200, so the algorithm continues with the one on the first list. H8 is incomparable so it is added to the list. The next point in the first list has Min value 500 so the algorithm continues with the previous considered H9 which is added to the list. Algorithm continues by loading the batches {H10, H11} and H7 from each list respectively. The batch with the smallest minimum value is H7 which is added to the list because it's not dominated by any point in it. At this step the algorithm terminates because both the coordinates of H7 are smaller than or equal to the minimum value of the next batch {H11, H10}, H2 on the two lists. In this case the algorithm does not need to proceed further because all the remaining point will be dominated by H7 and thus algorithm terminates returning the set of skyline point.

For clarity reasons is explained what should happened in the case {H11, H10} was processed. This batch has two points so in that case the algorithm would calculate the local skyline of the batch. The resulted point (or points) would be checked if they could be a part of the skyline. Both points do not dominate each other so both of them belong to the local skyline. In this case algorithm will check both points if can be added to the skyline list.

| Min$_1$ | Dimension 1 | Dimension 2 | Min$_2$ |
|---|---|---|---|
| Min$_1$= 100 | H1 ( 100 , 1500 ) | H6 ( 1600 , 100 ) | Min$_2$= 100 |
| Min$_1$= 200 | H8 (200 , 1200) | H9 (1000 , 200 ) | Min$_2$= 200 |
| Min$_1$= 500 | { H11 (500 , 900 ) , H10 (500 , 1400) } | H7 ( 400 , 300 ) | Min$_2$= 300 |
| Min$_1$= 900 | H5 ( 900 , 1300) | H2 ( 1400 , 500 ) | Min$_2$= 500 |
|  |  | H3 ( 700 , 600 ) | Min$_2$= 600 |
|  |  | H4 ( 1300 , 1000 ) | Min$_2$= 1000 |

Table 4 Index approach

The Index algorithm can quickly return skyline points in bursts (since it examines collection of points together) but does not support user-defined preferences since the order of the skyline points that are returned is fixed and depends on the value distribution of the data.

*Complexity:* -

### 2.3.4. NN

The Nearest Neighbor (NN) algorithm [97] is the first algorithm that uses the widespreaded R*-tree [11, 65] index structure in order to massively eliminate points by avoiding redundant dominance checks. The algorithm recursively applies the NN search, using an existing algorithm such as [149, 70] which is based on any monotone distance function (i.e. $L_1$-norm or Euclidean norm($L_2$-norm)). At the beginning applies an NN search to find the point with the minimum distance (mindist) from the beginning of the axes (when the problem is to be minimized) and inserts the resulted point into the skyline. The resulted NN point partitions the space in four partitions. One partition contains only points that are dominated by this point and thus can be removed. A second partition contains no point according to NN search and the other two partitions will be processed recursively through a to-do list in order to output the skyline result. If a region is empty is not sub divided any further and is removed from the To-Do list. The algorithm terminates when the To-Do list is empty.

**ALGORITHM 3:** NN [97]

**Input:** A Dataset D (r-tree).
Distance function $f$ (e.g., Euclidean distance)
**Output:** The Set of skyline points of dataset D.

1. T = {(∞,∞)}
2. **while** (T≠∅) **do**
3.    ($m_x$ , $m_y$) = takeElement (T)
4.    **if** (∃ boundedNNSearch (O, D, ($m_x$ , $m_y$), $f$) **then**
5.       ($n_x$, $n_y$) = boundedNNSearch (O, D, ($m_x$ , $m_y$), $f$)
6.       T = T ∪ {($n_x$ , $m_y$) , ($m_x$ , $n_y$)}
7.       output n
8.    End if
9. End while

At this point the pseudocode of NN algorithm was outlined. The list T represents the regions to be processed. Each time a new nearest neighbor is identified the region T is subdivided into smaller regions. The boundedNNSearch(O, D, ($m_x$ , $m_y$), $f$) function takes as input a point O, a dataset D, a region ($m_x$ , $m_y$), a distance function $f$ and identifies the nearest neighbor ($n_x$, $n_y$) taking into account that must be inside the region ($m_x$ , $m_y$).

Algorithm starts by searching for the nearest neighbor from the origin point defined (in this case the start of axes). The nearest point to the origin is H7 (400,300) with mindist 700, based on the $L_1$, and is guaranteed to be a skyline point. This point partitions the data space in four regions **Figure 12**. Region 1 contains no points according to the definition and properties of nearest neighbor. Region 4 contains all the points that have greater coordinate values than those of the nearest neighbor point. Thus the points that belong to this region are dominated by the NN point and so they can be pruned massively (this could efficiently done with the r-tree implementation). Region 2 contains the points [0, 400) [300, ∞) and region 3 that contains all the points that belong to [400, ∞) [0, 300). The set of partitions resulted after the discovery of a skyline point must be inserted in a *to-do* list so the algorithm removes the initial region and inserts in their position regions 2 and 3, which are needed to be investigated. The algorithm recursively calls itself on Region 2 and Region 3.

In the recursion of Region 2, which is the first region of the To-Do list, the algorithm makes again an NN search in order to find the next skyline point. The NN point that is retrieved on R2 {[0,*400*) [300, ∞)} is H8 (200, 1200) with mindist 1400 which is inserted in the skyline list. Due to the discovery of the NN point Region 2 is divided in 4 partitions **(Figure 13)**. Region 2.1 will be [0,200) [300, 1200), region 2.2 [0,200)[1200, ∞), region 2.3 [200,400) [300, 1200), and region 2.4 [200,400)[1200,∞). As mentioned before regions 2.1 and 2.4 are not needed to be considered. As a final step the algorithm removes region 2 from the To-Do list and inserts regions 2.2 and 2.3. Next the algorithm will recursively call itself on the first region on the To-Do list which is region 2.2. An NN query in this region will return point H1 (100, 1500) with mindist 1600 that is added to the skyline list. Due to the discovery of the NN point, region 2.3 is divided in 4 partitions which are region 2.2.1 [0,100)[1200,1500), region 2.2.2 [0,100)[1500,∞), region 2.2.3 [100,200)[1200,1500) and region 2.2.4 [100,200)[1500,∞). As previously regions 2.2.2 and 2.2.3 are inserted in the To-Do list and processed recursively.

Algorithm will perform the next NN query starting with the first region on the To-Do list which is 2.2.2. The region is empty so the algorithm discards it and process the next one. Region 2.2.2 is also empty so it's discarded and so the region 2.3. The only region remaining is Region 3. As with Region 2 the algorithm will recursively call itself until the To-Do list is empty, where in that case algorithm terminates and returns the final skyline list. The skyline points returned by processing Region 3 are H9 (1000, 200) with mindist 1200 and H6 (1600, 100) with mindist 1700.

The algorithm improves the divide-and-conquer algorithm by applying the D&C framework on datasets indexed by R*-trees. NN is the first algorithm that gives the user control over the process by tendentiously selecting on-demand the preferred region to be processed and allows him/her to give preferences by altering the scoring function on-the-fly. On the downside, the algorithm has large I/O overhead, especially in high dimensional spaces, due to the recurrent access of the R*-tree. Additionally the To-do list size may exceed the size of the dataset for as low as 3 dimensions **[133]**. Finally is mentioned that in the general case of *d*>2, regions overlap in such a way that the same Skyline point can be found more than once. For that reason authors proposed some additional elimination methods for datasets with d>2.

***Complexity:*** -

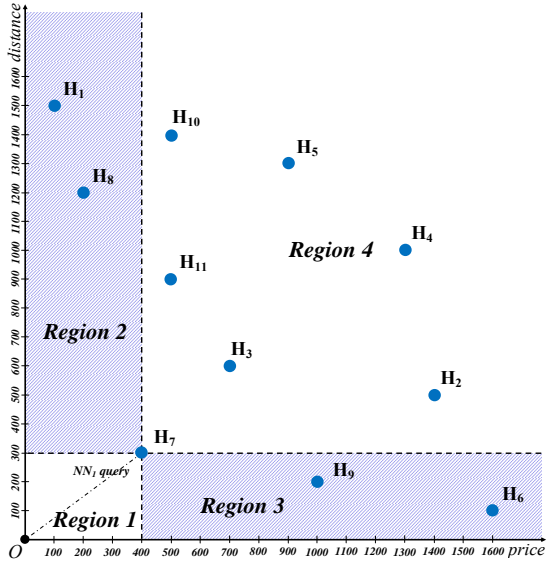
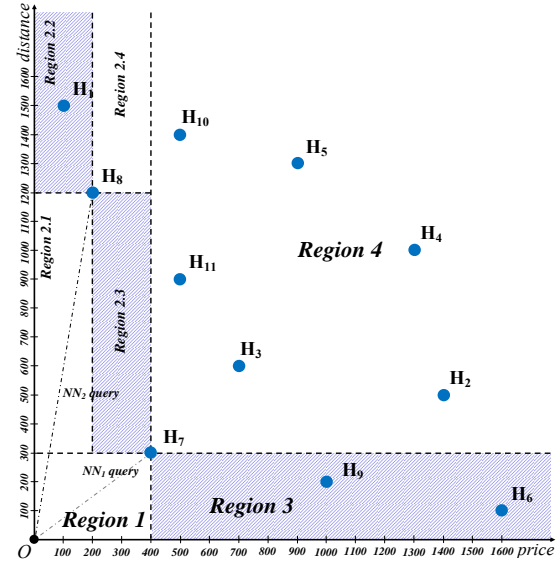

Figure 12: Regions after 1st NN query.  Figure 13 Regions after 2nd NN query.

| # of NN query | To-do List space partitions | Skyline Points |
|---|---|---|
| 0 | $[0,\infty),[0,\infty)$ | ∅ |
| 1st | R2{ [0,*400*) [300, ∞) } and R3{ [400,∞) [0,*300*) } | H7 |
| 2nd | R2.2{ [0,200)[1200, ∞) }, R2.3{ [200,400)[300,1200) } and R3{ [400,∞) [0,*300*) } | H7, H8 |
| 3rd | R2.2.2{ [0,100)[1500,∞) }, R2.2.3 [100,200)[1500,∞ ) }, R2.3{ [200,400)[300,1200) } and R3{ [400,∞) [0,*300*) } | H7, H8, H1 |
| 4th | ~~R2.2.2{ [0,100)[1500,∞) }, R2.2.3{ [100,200)[1500,∞ ) }, R2.3{ [200,400)[300,1200) }~~ and R3.2{ [400,1000)[200,300) }, R3.3{ [1000,∞)[0,200) } | H7, H8, H1, H9 |
| 5th | ~~R3.2{ [400,1000)[200,300) }~~, R3.3.2 { [100,200)[100,1600) }, R3.3.3{ [1600,∞)[0,100) } | H7, H8, H1, H9,H6 |
| 6th | Empty | H7, H8, H1, H9,H6 |

Table 5 To-Do list based on NN query.

### 2.3.5. BBS

Both NN and BBS **[133, 134]** apply nearest neighbor search techniques mentioned previously to progressively output skyline points from datasets that are indexed by *R\**-trees in order to massively eliminate points from being checked for dominance. BBS algorithm is an improvement of NN algorithm. In contradiction with NN that searches R*-tree many times, BBS traverses the R*-tree once. **Table 6** illustrates the indexed dataset. Data points are organized in the R*-tree, in which each internal R-tree node can hold up to three entries, and that each leaf node can hold also up to three entries. In the example, an intermediate entry $e_i$ of the R-tree of **Figure 14** corresponds to the minimum bounding rectangle (MBR) of a node $N_i$ of the R-tree, while a leaf entry corresponds to a data point $H_i$ **(Figure 15)**. As in the NN algorithm mindist denotes the minimum distance of a point or an MBR from an origin point. The *mindist* of a point is computed according to the L1 norm as the sum of its coordinates and the *mindist* of a MBR as the *distance* of its lower-left corner from the origin point. The algorithm uses the best-first search paradigm to traverse the R-tree, in such order that it always evaluates and expands, among all un-visited nodes, the tree node closest to the origin. All the candidate entries are kept in a *heap* until they are no longer useful. Entries in the heap are sorted in ascending order of their mindist. Skyline points are generated iteratively and stored in a list *in* the main memory, for dominance validation. Next the pseudocode of BBS algorithm is presented.

| ALGORITHM 4: BBS [133, 134] |
|---|
| **Input:** A Dataset D (r-tree). |
| **Output:** The Set of skyline points of dataset D. |
| |
| 1. $S=\emptyset$ // list of skyline points |
| 2. insert all entries of the root $D$ in the heap |
| 3. **while** heap not empty **do** |
| 4. remove top entry $e$ |
| 5.    **if** $e$ is dominated by some point in $S$ **do** discard $e$ |
| 6.    **else** // $e$ is not dominated |
| 7.       **if** $e$ is an intermediate entry **then** |
| 8.         **for** each child $e_i$ of $e$ **do** |
| 9.           **if** $e_i$ is not dominated by some point in $S$ **then** |
| 10.            insert $e_i$ into heap |
| 11.         **else** // $e$ is a data point |
| 12.           insert $e_i$ into $S$ |
| 13. **end while** |

Initially, the root of the *R*-tree is inserted in the heap. At each step, the top heap entry with the smaller *mindist* is removed. If it is a R*-tree node, its children, which are not dominated by any current skyline point, are inserted into the heap. If it is a point (leaf node), it is tested for dominance with the skyline points found so far by issuing an enclosure query. If the examined point (or region) is entirely enclosed by any skyline candidate's dominance region, then the point (or the entire region) is dominated. Notice that every entry is checked twice for dominance because an entry in the heap may become dominated by skyline points discovered after its insertion. In the end all the points, except of those where one of its ancestor nodes has been pruned, will be examined. In order to efficiently examine the dominance relationship, is maintained an in-memory *R*-tree that contains the skyline points found so far. When the heap is empty the algorithm terminates. Initially, BBS inserts all the child entries of the root of the *R*-tree into the heap.

The algorithm begins with region $e_1$ in its heap. As it proceeds it iteratively processes the (leaf/intermediate) entry which has the minimum *mindist* value and if it's an intermediate entry $e_i$ is expanded and its non-dominated children are inserted to the heap, ordered by their *mindist*. After the expansion of $e_3$ the first entry of the heap is a leaf node. The list of skyline points is empty so $H_7$ is inserted in the list. Next $e_8$ is expanded and $H_9$ is inserted to the list since it is not dominated by $H_7$. In the next step $e_2$ is expanded. Region $e_4$ is inserted in the heap, but region $e_5$ is dominated by the found skyline point H7 so the region is discarded. Next region to be expanded is $e_4$. The points on the heap are sequentially checked if they are dominated by any so-far found skyline point and if not are inserted to the list. From this comparison points $H_8$, $H_1$, $H_6$ are inserted to the list and point $H_8$ is discarded. Now the only region left in the heap is e7 which is not expanded because is dominated by the skyline point $H_7$ and $H_9$.

| Action | Heap contents | Skyline points |
|---|---|---|
| Initial state | ($e_1$,200) | ∅ |
| Expand $e_1$ | ($e_3$,500), ($e_2$,1300) | ∅ |
| Expand $e_3$ | ($e_6$,700), ($e_8$,1100), ($e_2$,1300), ($e_7$,1800) | ∅ |
| Expand $e_6$ | ~~($H_7$,700)~~, ($e_8$,1100), ($e_2$,1300), ($H_3$,1300), ($H_{11}$,1400), ($e_7$,1800) | $H_7$ |
| Expand $e_8$ | ~~($H_9$,1200)~~, ($e_2$,1300), ($H_3$,1300), ($H_{11}$,1400), ($H_6$,1700), ($e_7$,1800) | $H_7$, $H_9$ |
| Expand $e_2$ | ($e_4$,1300), ($H_3$,1300), ($H_{11}$,1400), ($H_6$,1700), ~~($e_5$,1800)~~, ($e_7$,1800) | $H_7$, $H_9$ |
| Expand $e_4$ | ~~($H_3$,1300)~~, ~~($H_8$,1400)~~, ~~($H_{11}$,1400)~~, ~~($H_1$,1600)~~, ~~($H_6$,1700)~~, ($e_7$,1800) | $H_7$, $H_9$, $H_8$, $H_1$, $H_6$ |
| ~~Expand $e_7$~~ | empty | $H_7$, $H_9$, $H_8$, $H_1$, $H_6$ |

Table 6: Heap contents of BBS

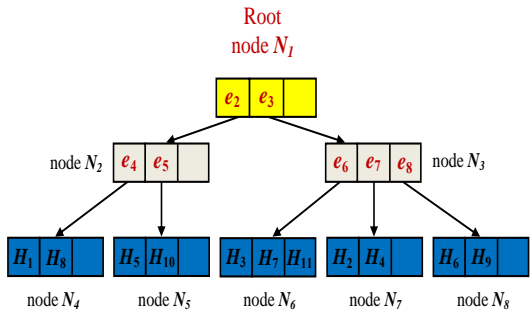
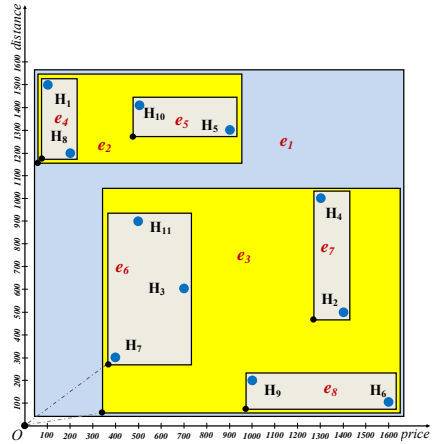

Figure 14: Dataset indexed by the R-tree

Figure 15: Minimum bounding rectangles (MBRs)

One of the most important properties of BBS is that it guarantees the minimum I/O costs and equivalently R-tree page accesses. Additionally, along with the NN algorithm can incorporate user preferences in general skyline computation. However its performance can deteriorate due to many unnecessary dominance checks and due to high dimensionality is falling in the curse of dimensionality [16]. We mention that an R-tree is efficient for up to 5 dimensions.

***Complexity:*** -

### 2.3.6. SFS

The sort-filter-skyline (SFS) algorithm [35] improves BNL performance by presorting the input dataset in an ascending order according to a monotone preference function f, such as the sum of coordinates of a point on all dimensions, or optimized as entropy (assuming in both cases that values have been normalized in (0,1) non-inclusive). Presorting enforce that a point *p* dominating another point *q* will be visited before *q*. This ensures the progressive behavior of SFS and the reduction of the number of pairwise comparisons between points. The algorithm examines the data points by the ascending order of their scores and keeps an in-memory buffer that has the till now found skyline candidate points in a similar way as that on BNL. At beginning the buffer is initially empty. A point is read from the sorted dataset and if it is not dominated by a skyline point in the buffer is inserted into it. The dominance tests in SFS are performed by an exhaustive search on the existing skyline points.

Authors have found that the entropy scoring function $E_D(p) = \sum_{i=1}^{d} \ln(p'.d_i + 1)$, where p'.$d_i$ is the normalize value of *p.di* in (0,1) non-inclusive, yields the most effective discarding during the skyline computation. Intuitively, the smaller entropy value a point has, the less likely is to be dominated.

**Value Normalization:** There are several ways to normalize the values of a dataset. One case is to divide all the values of the dataset with the maximum value found over this. That is $f(p.d_i) = (p.d_i/max)$, where max is the maximum value observed in the dataset. In this case the dataset would be normalized in the [0,1] inclusive which uses efficiently all the range of [0,1]. But in this example this is not the case because the values are needed to be normalized in (0, 1) non-inclusive. A case to achieve this is by dividing all the values of the dataset with a higher value than the maximum value of the dataset. For memorization reasons and simplicity of numbers and computations we choose to divide all values by 10.000. That is $f(p.d_i) = (p.d_i/10000)$. We note that this is not considered as a user preference since in another case we could assume that the distance between the two furthest locations of the town is 2000 meters and divide with this number.

| ALGORITHM 5: SFS [35] | |
|---|---|
| **Input:** | A Sorted Dataset D (Heap). |
| **Output:** | The Set of skyline points of dataset D. |

```
1.   unfinished = True
2.   while (unfinished) do
3.       T=open_cursor(Heap)
4.       unfinished = False
5.       while (next_tuple(T,t)) do
6.           if ("T is not dominated") then
7.               if ("window is full") then
8.                   unfinished = True
9.                   break
10.              else
11.                  "Add t to window."
12.      if (unfinished) then
13.          S=open_new_file_(second_pass)
14.          write(S,t)
15.          while(next_tuple(T,t)) do
16.              If ("t is not dominated") then
17.                  write (S , t)
18.      free(Heap)
19.      close(S)
20.      Heap=SecondPass
21.  "Write window tuples to output."
22.  "Clear window."
```

At this point the pseudocode of SFS algorithm was outlined. The algorithm takes as input a sorted dataset. If the buffer (window) is large enough to process all the tuples it finishes in one pass otherwise multiple passes are necessary.

To demonstrate the algorithm the entropy scoring function was used. Points of the dataset of the house metro-station example will be normalized and sorted as in **Table 7** by an ascending order of their score and will be processed in this order. The first point that is inserted in the buffer is H7 since the buffer is empty. The second point is H9 which is incomparable to H7 so it's inserted to the buffer. H3 is dominated by H7 so it's discarded. H8 is incomparable with H7 and H9 that are already in the buffer so it's added. Point H11 is dominated by H7 so it's discarded. Point H1 is incomparable to the points that belong in the buffer so it's added to the buffer. The same hold for point H6. The rest four points are dominated by a point in the buffer (H2 by H7 and H9, H10 by H7 and H8, H5 by H7 and H8, H4 by H7 and H9) so they are discarded. It is observed that the killer-dominant points are first in the presorted dataset which ensures maximum discarding with minimum comparisons. An indication for this is that the algorithm processed seven points in order to find the five skyline points out of the total eleven points.

| House ($h_i$) | Price | Distance | $E_D(h_i)$ | # points that dominate |
|---|---|---|---|---|
| **H7** | 0,04 | 0,03 | 0,068779515 | 6 |
| **H9** | 0,1 | 0,02 | 0,115112807 | 2 |
| H3 | 0,07 | 0,06 | 0,125927557 | - |
| **H8** | 0,02 | 0,12 | 0,133131313 | 2 |
| H11 | 0,05 | 0,09 | 0,13496786 | - |
| **H1** | 0,01 | 0,15 | 0,149712273 | 0 |
| **H6** | 0,16 | 0,01 | 0,158370336 | 0 |
| H2 | 0,14 | 0,05 | 0,179818427 | - |
| H10 | 0,05 | 0,14 | 0,179818427 | - |
| H5 | 0,09 | 0,13 | 0,208395329 | - |
| H4 | 0,13 | 0,1 | 0,217527813 | - |

**Table 7 : pre-sorted Dataset**

The main drawback of SFS is that it cannot adapt to different user preferences and has to scan the entire dataset to return a complete skyline, as with BNL. Nevertheless it can be stopped early returning some of the skyline points. The significant advantage over BNL is that reduces the number of comparisons needed.

**Complexity:** Algorithm's complexity for best case is $O(dn + nlogn)$ and in the worst case $O(dn^2)$, where d is the number of dimensions and n the size of the dataset. This runtime involves the sorting phase which sorts points by their volume.

### 2.3.7. LESS

Skyline algorithm LESS (linear elimination sort for skyline) **[59]**, is an optimized version of SFS, which achieves a better average performance. As with SFS it sorts the dataset based on the entropy scoring function which has the advantage of pushing the killer-dominant points in the beginning of the sorted dataset. The algorithm implements two optimizations.

The first optimization in the first pass of the external sorting process makes use of a buffer called elimination-filter (EF), which keeps a small set of points (copies of them) that have the best entropy scores seen so far. This set will be used in order to prune efficiently and as early as possible the dominated tuples of the dataset. The input dataset is divided in b blocks and each block of points is read in order to be sorted (i.e. using quicksort). During sorting the algorithm compares the points of the block with those of the EF. If the point from the block is dominated by a point in the EF, it is discarded. Otherwise if the point is incomparable or dominates other points in the EF it is inserted (a copy of it) in the EF and the points of EF that are dominated are discarded. It is noted that points of the EF buffer are not guaranteed to be maximals.

The second optimization combines the final pass of the external sorting process (last merge step of the b blocks) with the first pass of the skyline-filter (SF) process (i.e. first pass of the BNL component of SFS), which eliminates the remaining dominated tuples in order to get the final skyline. As in SFS and BNL, may be required multiple passes of the SF component in order to compute the final skyline. If the SF buffer becomes full, then an overflow file will be created. In general the EF filter reduces effectively the size of the input dataset that will be processed by the SF process and additionally the combination of the final pass of the EF process with the first pass of the SF process saves always one pass from the computation of the skyline. LESS is not be applicable in scenarios in which one has no direct control on the algorithm used to sort tuples. Additionally as in SFS all points on the dataset should be scanned at least once after sorting.

**Complexity:** It's complexity for the best case is O(kn) and in the worst case O(kn$^2$), where n is the number of points and k the number of dimensions.

### 2.3.8. SaLSa

SaLSa (Sort and limit skyline algorithm) algorithm **[9]** is an improvement of SFS and LESS which strives to avoid scanning the complete sorted dataset as opposed with the two previous algorithms. As SFS and LESS, it does not have an index structure and is the first algorithm that exploits the values of a monotone scoring (limiting) function to sort the dataset and effectively limit the number of point to be read and compared by using a threshold value.

Author's suggestion is an optimal sorting function, which orders the points according to the value $fmin(p) = (\min_{i \in [1,d]} p.d_i, sum(p))$, which is the minimum coordinate value of a point among all dimensions and $sum(p) = \sum_{i=1}^{d} p.d_i$ is the second sorting element that works as a tie-breaking rule. Letting S be the current set of skyline points, for each point $p_i \in S$ let $\underline{p_i} = \max_j \{p_i.d_j\}$, which is the maximum coordinate value of a point. The threshold value that is used during the filter-scan process to check whether all points in the rest of the sorted dataset are dominated, in order for the algorithm to stop, is set as $p_{stop} = \arg\min_{i \in S} \{\underline{p_i}\}$. That is P$_{stop}$ equals with the minimum $\underline{p_i}$ value calculated so far based on the existing skyline points. The computation of P$_{stop}$ can be done incrementally by simply updating the value at each skyline point insertion in O(1) time.

The algorithm during the filter-scan process reads and examines the points one at a time. Each time a new point is read, is compared against the current skyline list. If it's dominated by any point is discarded, otherwise is inserted in the skyline list and algorithm checks it's termination trigger. If the current threshold P$_{stop}$ is smaller or equal than the point's f$_{min}$ value, then the algorithm terminates and returns the set of skyline points. This termination condition guarantees that all later examined data points should not be part of the skyline list, avoiding this way scanning the entire dataset.

| ALGORITHM 6: | SaLSa |
|---|---|
| **Input:** | A Dataset U. |
| | A monotone sorting function *f*. |
| **Output:** | The Set of skyline points of dataset U. |

1. $S \leftarrow \emptyset$, $U \leftarrow r$, *stop* $\leftarrow$ *false*, *pstop* $\leftarrow$ undefined
2. sort $U$ according to *f*
3. **while** not *stop* ∧ $U\_= \emptyset$ **do**
4.     $p \leftarrow$ get next point from $U$, $U \leftarrow U\setminus\{p\}$
5.     **if** $S \not\succ p$ **then** $S \leftarrow S \cup \{p\}$, update $p_{stop}$ **then**
6.     **if** $p_{stop} \succ U$ **then** *stop* $\leftarrow$ *true* **then**
7. return $S$

At this point the pseudocode of SaLSa algorithm was outlined. Initially is defined the set U which contains the unread tuples. The $p_{stop}$ point is used to terminate the reading of tuples. S contains all the skyline points.

In order the algorithm to compute the skyline, the values of the dataset is needed to be normalized in the range of [0,1] inclusive. Since this is not applicable in many cases author's suggest as a solution to normalize the values of the dataset as $f(p.d_i) = (p.d_i - min_i)/(max_i - min_i)$, where $min_i$ is the minimum coordinate value on the i-th dimension and $max_i$ is the maximum coordinate value. To demonstrate the algorithm the house-metro station dataset is normalized as sawn in **Table 8.** The first point of the sorted dataset is H1. At this point, before point H1 is read, $p_i$ and $P_{stop}$ are undefined. Since the set of skyline points is empty H1 is inserted in it. Values $p_1, p_5$ and $P_{stop}$ are calculated since a skyline point was found. The new $p_1$ value is 1, which is the largest value among the two coordinate values of the point and $p_{stop} = p_1$ since it's the only $p_i$ value due to the only one skyline point. Next point in the dataset is H6 which is a skyline point because it is not dominated by H1. Because of the insertion of H6 in the skyline list $p_2$ is set to 1 since it's the largest coordinate among the two of the point H6. $P_{stop}$ remains 1 since the new $p_i$ value is not smaller than the old one. Next point in the dataset is H8. It is not dominated and thus is a skyline point which triggers the computation of the $p_3$ value that equals with 0,785714. The $P_{stop}$ value is now set to 0,785714 also, since the value $p_3$ was smaller than the current $P_{stop}$ value. Next point in the dataset is H9 with a $p_4$ value equals with 0,6. Since $p_4$ is smaller than the current $P_{stop}$ value, $P_{stop}$ is set to 0,6. Next point is H7. It's $p_5$ value is 0,2 since is the biggest value among the two coordinate values of the point, which also triggers the altering of the $P_{stop}$ value to 0,2 since the new value is smaller. Next point is H11. It's $f_{min}(H11)$ value is bigger than the current $P_{stop}$ value which terminates the algorithm and returns the list with the skyline points. It is observed that were processed only points that were actually skyline points and the rest were discarded saving unnecessary computations. On the downsides of the algorithm is that it's performance is affected by data distribution and high dimensionality, since the pruning power of the *stop object* is limited. Additionally because the dataset is based on a fixed ordering for each attribute, the algorithm cannot be used for arbitrary preference specifications. The advantage of the algorithm is that it can stop efficiently before the complete dataset is readied.

*Complexity:* -

| House | Price | Distance | $f_{min}(h)$ | Sum(h) | $p_i$ | $P_{stop}$ |
|---|---|---|---|---|---|---|
| H1 | 0 | 1 | 0 | 1 | 1 | 1 |
| H6 | 1 | 0 | 0 | 1 | 1 | 1 |
| H8 | 0,067 | 0,786 | 0,067 | 0,853 | 0,786 | 0,786 |
| H9 | 0,600 | 0,071 | 0,071 | 0,671 | 0,600 | 0,600 |
| H7 | 0,200 | 0,143 | 0,143 | 0,343 | 0,200 | 0,200 |
| H11 | 0,267 | 0,571 | 0,267 | 0,838 | - | Stop! $F_{min}(H_{11}) \geq P_{stop}$ |
| H10 | 0,267 | 0,929 | 0,267 | 1,196 | - | |
| H2 | 0,867 | 0,286 | 0,286 | 1,153 | - | |
| H3 | 0,400 | 0,357 | 0,357 | 0,757 | - | |
| H5 | 0,530 | 0,857 | 0,533 | 1,387 | - | |
| H4 | 0,800 | 0,643 | 0,643 | 1,443 | - | |

Table 8: pre-sorted Dataset

## 2.4. Summary

### 2.4.1. Criteria for online/progressive algorithms

In general a batch-oriented algorithm will return the complete skyline faster than an online algorithm. In contradiction an online algorithm will return faster than the batch-oriented algorithm a part of the skyline but it will take much longer to compute the complete skyline. Authors in **[69, 97]** suggested a set of criteria in order to evaluate the behavior and applicability of a progressive algorithm.

1. **Progressiveness:** A part of the final set of skyline points should returned instantly and the remaining skyline points gradually.
2. **Absence of false negative**: The algorithm, given enough reasonable time, should eventually produce the complete set of skyline points.
3. **Absence of false positives:** The points that the algorithm returns should be guaranteed to be skyline points and not a temporary skyline points that will be discarded later.
4. **Fairness:** The algorithm should not favors points that are particularly good in one dimension.
5. **Incorporation of preferences:** User should be able to make preferences on the order that the skyline points are returned, while the algorithm is running.
6. **Universality:** The algorithm should be easily integrated into an existing database system and be applicable to any dataset distribution and dimensionality making use of standardized technology.

In **Table 9** the algorithms are classified based on those criteria.

| Algorithm | Progressiveness | Absence of false misses / Absence of temporary false hits | Absence of false hits | Fairness | Incorporation of preferences | Universality |
|---|---|---|---|---|---|---|
| D&C | × | √ | √ | √ | × | √ |
| Bitmap | √ | √ | √ | √ | × | √ |
| Index | √ | √ | √ | × | × | × |
| NN | √ | √ | √ | √ | √ | √ |
| BBS | √ | √ | √ | √ | √ | √ |
| BNL | × | √ | × | √ | × | √ |
| SFS | √ | √ | √ | √ | × | √ |
| LESS | √ | √ | √ | √ | × | √ |
| SaLSa | √ | √ | √ | √ | × | √ |

Table 9: Classification of progressive algorithms.

## 2.4.2. Algorithm Classification

The state of the art index based skyline algorithm is BBS. On the other hand the state of the art algorithm that does not require indexing or preprocessing is SaLSa. **Table 10** summarizes some basic properties of all algorithms.

| Algorithm | Based-on | Index | D&C | Pre-processing | Sorted data | Main problem |
|---|---|---|---|---|---|---|
| Nested-loop join | Θ-joins [128] | × | × | × | × | Join cost |
| Bitmap [164] | - | Bit mapping | × | bitmaps | × | Lack of user interaction and bitmapping |
| Index [164] | - | Specialized B-tree | √ | Index - based | Scoring function | Lack of user interaction |
| NN [97] | NN search and D&C scheme | Multi-dimensional index ( R*-tree) | √ | Index - based | Minimum distance from origin point | I/O accesses |
| BBS [133] | NN | Multi-dimensional index ( R*-tree) | √ | Index - based | Minimum distance from origin point | many dominance checks / R-tree dimensionality |
| D&C [19] | maximal vector computation [99, 144] | × | √ | × (Partial skylines can be assumed) | × | Not online/ curse of dimensionality |
| BNL [19] | Naive Nested-loop | × | × | × | × | Not online |
| SFS [35] | BNL | × | × | Sort - based | Entropy scoring function | reads all dataset / Lack of user interaction |
| LESS [59] | SFS | × | × | Sort - based | Entropy scoring function | Sorting / reads all the dataset |
| Salsa [9] | SFS | × | × | Sort - based | Min/Max Scoring function | Sorting / reads all the dataset |

**Table 10 Classification of skyline query algorithms.**

**Figure 16** illustrates the skyline algorithms via a tree structure in chronological order. The entries [99, 144] concern the maximal vector computation. Black lines indicate that the algorithm heavily depends or improves a previous algorithm and red dashed line indicates that the algorithm shares some general main ideas in order to compute the skyline.

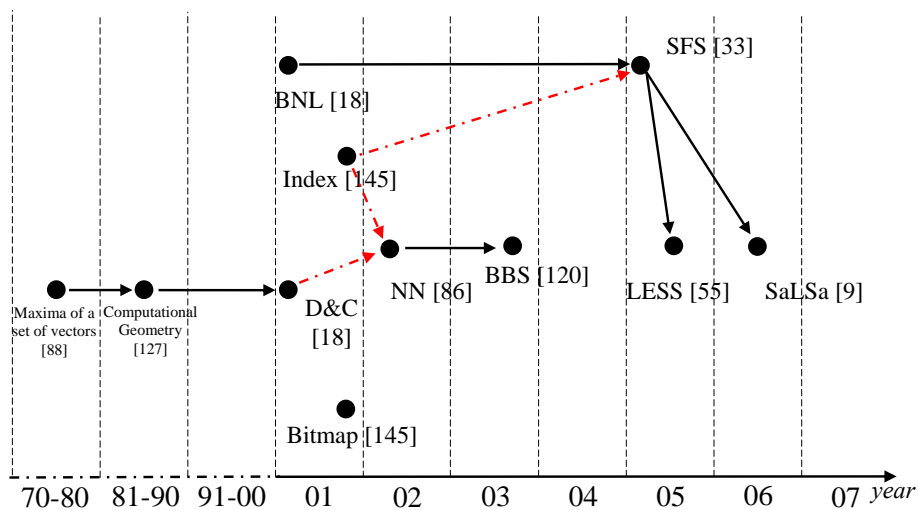

**Figure 16: Chronological order of fundamental skyline algorithms.**

# 3. Skyline Family

This section will reason about the variations of skyline queries. The main idea and notion of skyline query is maintained. Each outlined variation is applicable and can solve different aspects of a problem.

## 3.1. Constrained Skyline Queries

There are cases where a skyline query may return too many objects. This can happen if the dimensionality of the dataset is large or the dataset is anti-correlated. Additionally users may be interested to investigate a particular subspace than the whole data space. For the previous reasons user may specify constrains on some dimensions to express those restrictions. Each constraint is typically expressed as a range along a dimension of the dataset. The constrained skyline queries are very usefull in skyline maintenance in the presence of point deletions or insertions as will be shown in **section 5.2**.

For this type of problems, a general variant of the skyline queries are the *constrained skyline queries* **[133, 134]** In this type of queries users are interesting in finding the skyline points of a subset of the original dataset, which satisfies one or more constraints. Given a set of constraints, a *constrained skyline query*, will return the most "interesting" points of the dataset defined by these constraints. For example the user may be only interested for "interesting" houses in the distance range from the metro-station of 400 to 1250 and price range from 100 to 1500. For the house-metro example the *constrained skyline query* will return points H8, H11, H3, H2 **[Figure 17]** that are enclosed in the shaded region and are skyline points in that region. Point H4 which also belongs in the region will be discarded since it is dominated by H11 and H3. A *constrained skyline query* can be well expressed using the SKYLINE OF and WHERE clause. The query of the previous example would be:

SELECT * FROM Houses WHERE ((distance ≥ 400 AND distance ≤ 1250) AND (price ≥ 100 AND price ≤ 1500)) SKYLINE OF price MIN, distance MIN

**Definition 3:** Constrained region
Given a d-dimensional dataset Ds a constrained region $C=\{c_1,c_2,...c_d\}$ is determined by *d* sub-constraints $c_i$ where each one expresses a range along each dimension of the dataset. That is $c_i=\{c_i min, c_i max\}$ where $c_i min$ and $c_i max$ are the minimum and maximum range restriction values on the i-th dimension. ∎

**Definition 4:** Constrained Skyline
Given a dataset Ds and a constrained region C⊆Ds, a constrained skyline will contain all the points p∈C (p.di∈ $c_i$, ∀ i ∈[1,d]) where ∀p,r∈C, ∃ j∈[1,d] such that p.dj<r.dj and ∀ i ∈[1,d]-{j}:p.di ≤ r.di ∎

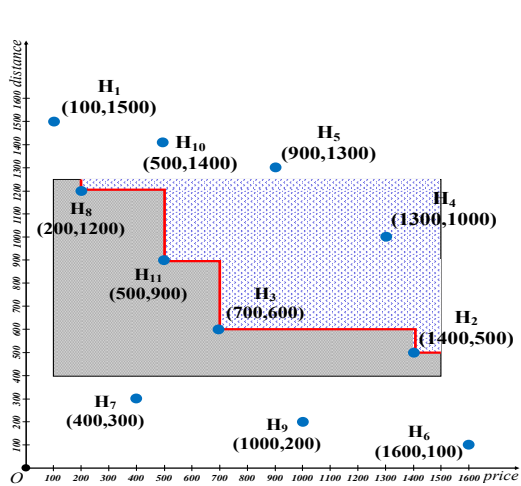

Figure 17: Constrained Skyline.

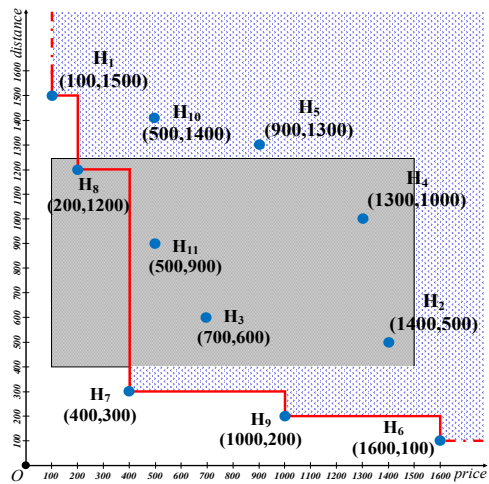

Figure 18: Skyline with constrains.

Another type of queries with similar name that might confuse the reader, are *skyline queries with constraints* **[197].** This type of queries, given a set of constraints, returns the computed skyline set of the whole dataset restricted by the constraints that were placed. For the house-metro example and the constraints mentioned above, the *skyline queries with constraints* computes the skyline of the whole dataset and then applies the constraints to the retrieved skyline set and returns only the point H8 **[Figure 18]**. In this case the SQL query would be:

SELECT * FROM (SELECT * FROM Houses SKYLINE OF price MIN, distance MIN) skyline_result

WHERE ((distance ≥ 400 AND distance ≤ 1250) AND (price ≥ 100 AND price ≤ 1500))

In general a *constrained skyline query* is computed over the restricted dataset by the constraints that were placed, while the *skyline query with constraints* is computed over the whole dataset and then the resulted set is restricted by the constraints. Thus the results of both types of queries will be different (in the majority of cases) for the same dataset.

## 3.2. Dynamic skyline queries (DSQ)

A Dynamic skyline query is a variation of the original skyline query which was first introduced in **[133, 134].** In this type of queries the *dynamic coordinates* of each point are given by a set of *distance (dynamic) functions* that are based on the distance between a given query/reference point q and a point p of the original dataset. The term *original space/dataset* refers to the original d-dimensional space/dataset and equivalently the term *original coordinates* to the coordinates of a point in the original space. The produced data space that occurs from the distance functions and the query point will be called *dynamic space* and the coordinates of a point in it, *dynamic coordinates*.

A *dynamic* skyline query of a d-dimensional data space DS specifies a new *d'*-dimensional data space DS' based on the original space and depicted as an inner coordinate system. To achieve this transformation specifies m (m≤d) dimension functions *f*. Each function takes as parameters one or more original coordinates of each point and maps them in a new single dynamic coordinate. That is, each point p of the original d-dimensional data space is mapped to a new d'-dimensional point $p' = (f_1(p), \ldots , f_{d'}(p))$ where each $f_i$ is referred as a distance function. Then the dynamic skyline applied on DS with respect of functions $f_i$ specified by a query point q returns the original skyline of the new transformed d'-dimensional space DS'.

To simplify the definition of the dynamic skyline it is assumed, without loss of generality, that DS and DS' have the same dimensionality (d=d'). Additionally for a given query point q each distance function is defined as the obsolete distance, of the i-th dimension's value of point p of the dataset DS from the i-th dimension's value of query point q, $f_i(p)=|q.d_i-p.d_i|$.

Note that dynamic skylines can have a more general class of distance functions such as Euclidian distance. In addition they can be employed in conjunction with constrained and ranked queries (by placing weights on dimensions). An example, is the case where the absolute distance functions can receive different weights and the result of distance functions is constrained by a threshold value i.e. find the top-3 houses within 1km given that the price is twice as important as the distance, where k is specified by user.

**Definition 5:** Dynamic dominance.
Given a dataset Ds, a query-reference point q in the workspace and two points p, r ∈ Ds, point p dynamically dominates point r with regard to the query point q, denoted as $p \prec^q r$ if and only if $\exists$ j∈[1,d] such that $|q.d_i - p.d_j|<|q.d_i - r.d_j|$ and $\forall$ i ∈[1,d]-{j}: $|q.d_i - p.d_j|\leq|q.d_i - r.d_j|$ ∎

**Definition 6:** Dynamic skyline.
Given a query-reference point q in the workspace, the dynamic skyline set of Ds with regard to the query point q, denoted as $S_{DS}^q$, consists of the points of the dataset that are not dynamically dominated by any other point. That is, $S_{DS}^q =\{p\in DS|\nexists r\in DS:r \prec^q p\}$ ∎

| House | price (in thousand €) | Coordinate X | Coordinate Y |
|---|---|---|---|
| H1 | 100 | +900 | +1200 |
| H2 | 1400 | +300 | +400 |
| H3 | 700 | -360 | +480 |
| H4 | 1300 | +600 | -800 |
| H5 | 900 | +500 | -1200 |
| H6 | 1600 | +60 | -80 |
| H7 | 400 | +240 | +180 |
| H8 | 200 | -960 | +720 |
| H9 | 1000 | -192 | +56 |
| H10 | 500 | -1120 | -840 |
| H11 | 500 | -720 | -540 |

Table 11: 3-dimensional dataset of the house-metro station example.

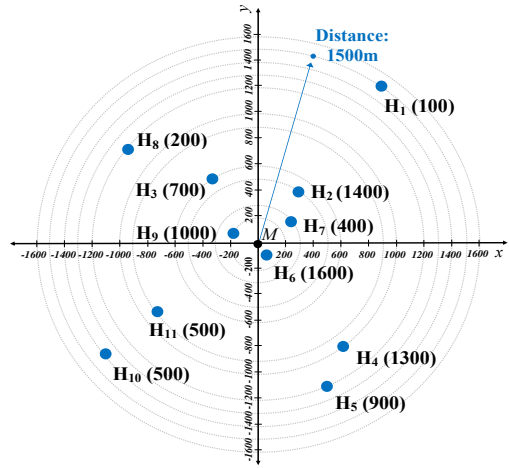

Figure 19: Initial position of houses and their prices in a coordinate system with origin point the metro station.

In a dedicated example assume that the 2-dimensional dataset of the house- metro station example [Table 1] was calculated dynamically from a previous 3-dimensional dataset [Table 11] that had as attributes the price of each house and it's position (X,Y) in a 2-dimensional map with origin point O(0,0) the metro station's position Figure 19. In the case of this example and its needs, the 3-dimensional dataset is projected in a 2-dimensional one by using as a query point the position of the metro station and as distance functions the functions $f_1(H)=(H.d1)$ and $f_2(H) = \left(\sqrt{(q.d_2 - H.d_2)^2 + (q.d_3 - H.d_3)^2}\right)$. This way the relative coordinated position of a house with respect the metro station is converted to the Euclidean distance of the house from the metro station (with price attribute intact) as shown in table [Table 1].

As with the original skyline, in order for BBS to compute a dynamic skyline query, it processes the (leaf/intermediate) R-tree entries in ascending order of their *mindist*. In this case the *mindist* of a point (leaf entry) from the query point q is computed as, $f(H) = \left(\sqrt{(H.d_2 - q.d_2)^2 + (H.d_3 - q.d_3)^2} + H.d_1\right)$. The mindist of an MBR with range ($[e.d_1min,e.d_1max][e.d_2min,e.d_2max][e.d_3min,e.d_3max]$), from the query point q, is computed as the mindist($[e.d_1min,e.d_1max][e.d_2min,e.d_2max],(q.d_1,q.d_2)$) + $e.d_3min$ where the first term is the mindist between the query point and the lower-left corner of the 2D rectangle [$e.d_1min,e.d_1max$] [$e.d_2min,e.d_2max$].

In a more general example it is assumed that the user needs to find the dynamic skyline of the house-metro station dataset DS. The dynamic functions that will be used are the obsolete distances of points in the dataset from the specified query point and thus points of DS are mapped in the new space as shown in [Figure 20], with the same dimensionality as the original space (d=d'). In detail points H1, H2, H3, H6, H7, H8, H9, H10, H11 are projected to points H1', H2', H3', H6', H7', H8', H9', H10', H11' respectively [Table 12] with regard the query point q and the dimension functions $f_1(H)=|q.d1 - H.d1|$ and $f_2(H)=|q.d2 - H.d2|$. The dynamic skyline for the selected query point contains houses H3', H11' which are essentially points H3, H11.

| House | price (in thousand €) | Distance (m) | Dynamic price (in thousand €) | Dynamic Distance (m) |
|---|---|---|---|---|
| H1 | 100 | 1500 | 1500 | 1500 |
| H2 | 1400 | 500 | 1400 | 1100 |
| H3 | 700 | 600 | 900 | 1000 |
| H4 | 1300 | 1000 | 1300 | 1000 |
| H5 | 900 | 1300 | 900 | 1300 |
| H6 | 1600 | 100 | 1600 | 1500 |
| H7 | 400 | 300 | 1200 | 1300 |
| H8 | 200 | 1200 | 1400 | 1200 |
| H9 | 1000 | 200 | 1000 | 1400 |
| H10 | 500 | 1400 | 1100 | 1400 |
| H11 | 500 | 900 | 1100 | 900 |

**Table 12: Original and dynamic dataset.**

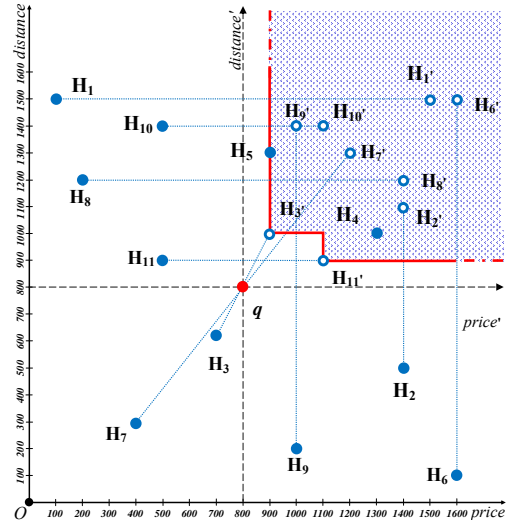

**Figure 20: Dynamic skyline.**

A DSQ query can be seen as a query from the buyer's perspective, by identifying the houses that are most interesting to him.

### 3.2.1. Spatial Skyline queries (SSQ)

The spatial skyline query (SSQ) [155, 156] can be considered as a more restricted special case of the dynamic skyline queries. It considers multiple query points at the same time and relies on the existence of a multi-dimensional Euclidean space to derive geometric bounding structures, such as convex hull and Voronoi diagram in order to reduce the search space. Given a dataset DS and a set of query points Q, a Spatial Skyline Query retrieves those points of DS which are not spatially dominated by any other point in DS with respect to Q. Specifically a point p∈DS spatially dominates a point r∈DS with respect to Q, if and only if p is closer to at least one query point q∈Q as compared to r and has in the best case the same distance as r to the rest of the query points, i.e., no other object is closer to all the given query points simultaneously.

**Geometric notations**

**Convex Set:** A set S of points, that exist on a plane over ℝ, is called convex set if and only if for any two points p,r∈S, the segment (line) that connects them resides entirely in S (i.e. all the points of a circle or a hexagon)∎

**Convex Hull:** The convex hull of a set S of points over ℝ, is the intersection of all the convex sets containing S ∎

A counter example of a convex hull would involve the dashed line in **Figure 21**. If the segment of the red line which belongs between houses H4 and H6 was replaced be the dashed line which contains house H2, the set S would not be a convex set since the line that connects houses H4 and H6 would not reside in S.

**Voronoi diagram:** Given a set S of n points over ℝ that exist on a plane, the Voronoi diagram of S, is the subdivision of the plane in n cells, where each cell contains only one point of S, called *generator*. The important property is that any point (except the generator) in a particular cell will be always closer to the point that generates this cell **(Figure 22)**∎

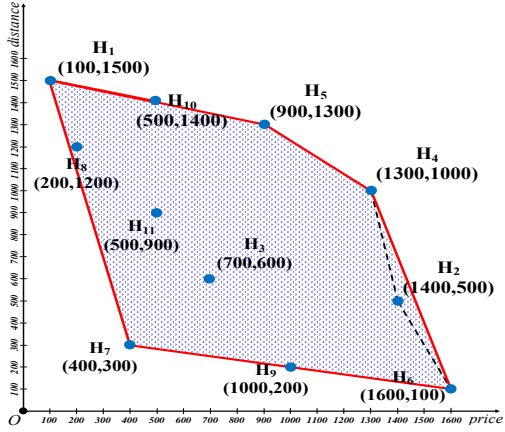

Figure 21: Convex Hull of the house-metro station dataset.

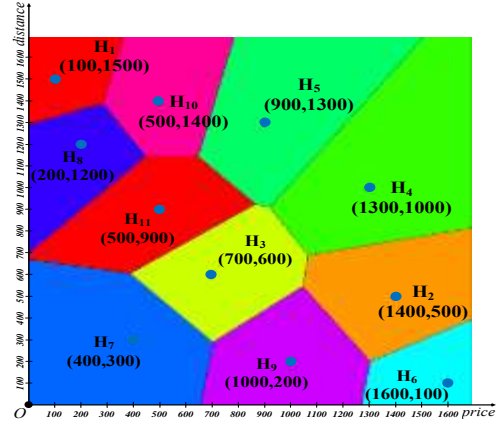

Figure 22: Voronoi diagram of the house-metro station dataset.

In order to reduce the search space authors give two important theorems, that is, the spatial skyline points are those points either within the convex hull [42] of query points or having their own Voronoi cells [42] intersect with boundaries of the convex hull of query points. Also they proposed the R-tree-based $B^2S^2$ algorithm and the Voronoi-based $VS^2$ algorithm for the *spatial skyline queries*. Both algorithms are efficient in cases where only Euclidean distances are considered as dimension functions, but their search structures are inefficient in high-dimensional and metric spaces. Additionally proposed a Voronoi-based continuous $VCS^2$ algorithm in order efficiently update a spatial skyline considering that the location of query point q changes and extended their work in [156] by considering spatial skyline queries in the metric space of Spatial Network Databases (SNDB) such as road networks where spatial objects are restricted in predefined locations/routes.

Following is presented the pseudocode of $B^2S^2$ algorithm. The algorithm has as an input a set of query points Q and a dataset R indexed by an R-tree. The SR(p,Q) refers to the union of all circles $C(q_i,p)$ which have as center a query point $q_i \in Q$ and their radious is equal to the distance from p. By *CH(Q)* and $CH_U(Q)$ refers to the convex hull of the set of points Q and the set of its vertices, respectively.

| **ALGORITHM 7:** $B^2S^2$ (set Q) [155, 156] |
|---|
| **Input:** A set of query points Q. <br> A dataset R (R-tree). |
| **Output:** The set of spatial skyline points of dataset R. |
| 1. compute the convex hull *CH(Q)*; |
| 2. set $S(Q) = \{\}$; |
| 3. box $B = MBR(R)$; |
| 4. minheap $H = \{(R, 0)\}$; |
| 5. **while** H is not empty **do** |
| 6.    remove first entry e from H; |
| 7.    **if** e does not intersect with B **then** discard e; |
| 8.    **if** e is inside *CH(Q)* or e is not dominated by any point in *S(Q)* **then** |
| 9.      **if** e is a data point p **then** |
| 10.        add p to *S(Q)*; |
| 11.        $B = B \cap MBR(SR(p,Q))$; |
| 12.      **else** // e is an intermediate node |
| 13.        **for** each child node e' of e |
| 14.          **if** e' does not intersect with B **then** |
| 15.            discard e'; |
| 16.          **if** e' is inside *CH(Q)* or e' is not dominated by any point in *S(Q)* **then** |
| 17.            add (e', mindist(e', $CH_V(Q)$)) to H; |
| 18. **return** *S(Q)*; |

In order to demonstrate the use of spatial skyline queries consider for example that from a set of home heating oil delivery stations (data points P) the user wants to identify a candidate subset in order to dispatched delivery trucks to multiple houses (query points Q). This candidate subset includes those stations that are not dominated by any other station with respect to all the houses, and hence they are the spatial skylines.

In [46] is proposed the Multi-Source Skyline Query (MuSSQ) in road networks where the network distance between two locations needs to be computed on-the-fly and the attributes are defined to be the shortest path length from data points to query points. This type of query is described in more detail in **section 4.5.** In [207] is proposed the *Location-Dependent Skyline Query (LDSQ)* for multi-objective distance optimization, considering a continuous changing user location (query point). In **[94]** authors consider spatial skyline computation with user preference information in addition to distances. Also they extend the query processing algorithm in order to return at least *k* good objects (where *k* is a user specified number) even when the original skyline contains fewer than *k* items. In **[64]** is proposed the *Direction-based Spatial Skyline Query* (*DSSQ),* which finds the best objects by comparing them in terms of distance from a mobile user and also by considering the direction that the user moves, rather than only distance as in traditional spatial skyline queries. In **[161, 160]** authors propose Manhattan Spatial Skyline Queries (MaSSQ) and develop an efficient algorithm for spatial skyline queries using the L1 norm, also known as Manhattan distance. Readers must not associate Manhattan Spatial Skyline Queries (MaSSQ) with Multi-Source Skyline Queries (MuSSQ).

### 3.3. Reverse skyline queries (RSQ)

A Reverse Skyline Query **[44]** retrieves these points in the database whose dynamic skylines contain a given query point. This type of query as opposed with the DSQ can be seen as a query from the real estate company perspective. For example given the ideal preferences of potential house buyers, as points in a two-dimensional space, the reverse skyline query can answer the question if it make sense to offer a house q (as a query point) to one of the potential buyers. The house q (becoming an origin point) will be interesting for a buyer, if it will be part of the dynamic skyline of his preferences (that represent the dataset points). Another example would be the selection of a new store's location. A reverse skyline query on a customer database, with respect to a query point q that represents the new location of the store would return those customers who are potentially interested in the new store. Then the strategy is to select the location that maximizes the number of customers.

**Definition 7:** Reverse skyline.
Given a dataset *Ds* in a *d*-dimensional space *D* and a query point *q* ($q_1, q_2, ..., q_d$) ∈ D, the reverse skyline query of *Ds* with regards to *q* retrieves the set of points $RSL_q(Ds)$ ⊆ *Ds* for which *q* is a dynamic skyline point of *Ds* with regards to all points in $RSL_q(Ds)$, that is, $RSL_q(Ds)$ = {$p$ ∈ *Ds* | $\nexists$ $r$ ∈ *Ds*: $r \prec^q_{Ds} p$}. The points in $RSL_q(Ds)$ are called reverse skyline points of *Ds* with regards to *q* ∎

**Definition 8:** Global domination
Given a dataset *Ds* in a *d*-dimensional space *D*, a query point *q* ($q.d_1, q.d_2, ..., q.d_d$) ∈ D and two points p($p.d_1, p.d_2, ...,p.d_d$) ,r($r.d_1, r.d_2, ..., r.d_d$) ∈D, point p will *globally dominate* r with regard to the query point q (denoted as $p \prec^q r$) if ∀i∈{1, . . . , d}: { ($p.d_i$ − $q.d_i$ )($r.d_i$ - $q.d_i$ ) > 0 and |$p.d_i$ − $q.d_i$|≤|$r.d_i$ − $q.d_i$|} and ∃j∈{1, . . . , d}: |p.dj–q.dj| < |r.dj–q.dj|∎

**Definition 9:** Global Skyline
Given a dataset *Ds* in a *d*-dimensional space *D* and a reference point *q* ∈ *D*, The global skyline of a point q, GSL(q), will contains the points which are not globally dominated by another point according to q∎

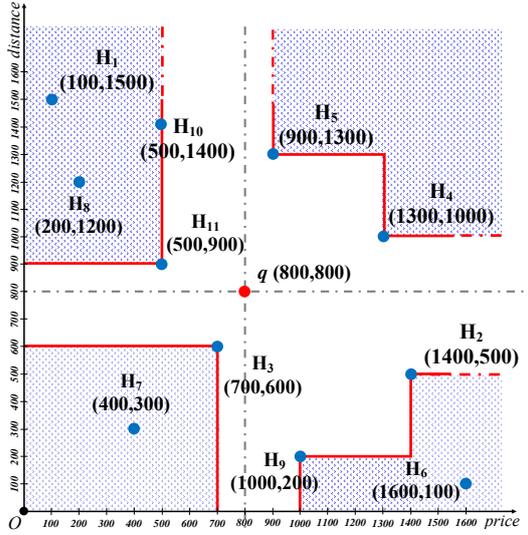
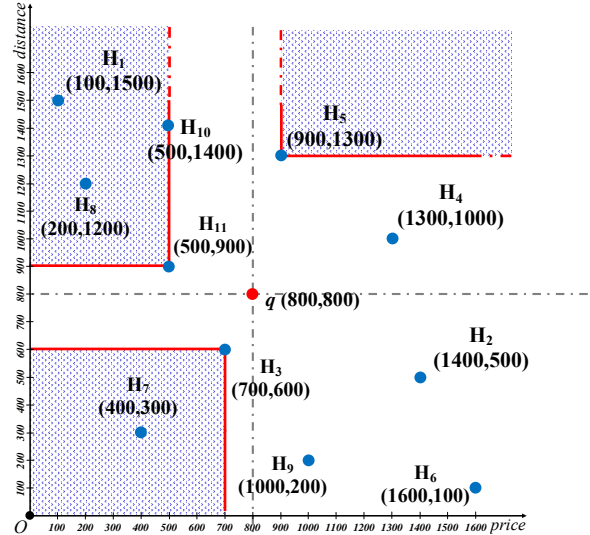

Figure 23: Global skyline and range queries.

Figure 24: Reverse skyline.

In order to compute the reverse skyline (RSL) of the house-metro station example, with regards the query point q (800,800), it is first needed to compute the global skyline GSL(q) as shown in **Figure 23**. As illustrated, the GSL(q) will contain the (reverse skyline candidate) points $H_{11}$, $H_7$, $H_5$, $H_4$, $H_9$ and $H_2$. The resulted reverse skyline **Figure 24** will eventually contain the points $H_5$, $H_7$, $H_{11}$. The rest of the points are discarded because, in order for a reverse skyline candidate p to be a reverse skyline point must not exist any point q in the GSL that is (strictly) better, in terms of distance from q, on all dimension simultaneously.

With the existing introduced algorithms, in order to compute the RSQ of a dataset DS, given a query point q, it is needed to examine all the points of DS by performing a dynamic query (e.g. using BBS) for each point in order to find the points that have q as part of their dynamic skyline. The points that will be retrieved would be the reverse skyline set. A first optimization in this approach would be to stop processing the dynamic skyline of a point when q is already identified as a skyline point since there is no need to compute the entire skyline. To further optimize the identification of Reverse skyline point authors proposed two algorithms *Branch-and-bound algorithm* (BBRS) and *Reversed Skyline Search with Approximation* (RSSA). BBRS is an improved customization of the original BBS algorithm and uses a Multidimensional index (e.g. R-tree). Its goal is to process the reversed skyline of a query point q without applying a space transformation. In order to achieve this, it retrieves the proposed Global Skyline GSL(q) that returns a small subset of the dataset as candidates for RSQ (this subset is still a superset of RSQ), which essentially reduces the search space for the reverse skyline computation. Algorithm RSSA computes the dynamic skyline for each point of the dataset and uses an accurate pre-computed approximation of the skyline in a filter-refinement step to compute the reverse skyline. Along with the RSSA algorithm, authors proposed an optimal algorithm to compute approximations for two-dimensional skylines and a greedy algorithm for higher dimensions. The basic idea of the approximation scheme is to pre-compute the dynamic skyline of each point of the dataset and select a fixed number *k* of it's $K_{max}$ dynamic skyline points ($k \leq K_{max}$).

In the next section is outlined the pseudocode of the BBRS algorithm. The set S represents the candidate reverse skyline points.

| ALGORITHM 8: BBRS (R-tree R, Query point q) [44] |
|---|

**Input:** A dataset R (R-tree).
A query point q.
**Output:** The reverse skyline points.

1. RSL ← {} //set of reverse skyline points
2. insert all entries of the root R in the heap H sorted by distance from q
3. **while** (heap H is not empty) **do**
4.    remove top entry e
5.    **if** (e is globally dominated by some point in S) **then**
6.      discard e
7.    **end if**
8.    **if** (e is an intermediate entry) **then**
9.      **for** (each child $e_i$ of e) **do**
10.        **if** ($e_i$ is not globally dominated by some point in S) **then**
11.          insert $e_i$ into heap H
12.        **end if**
13.      **end for**
14.    **else**
15.      insert the pruning area of $e_i$ into S
16.      execute the window query based on e and q
17.      **if** window query is empty **then**
18.        add e to the result set
19.      **end if**
20.      discard e
21.    **end if**
22.    output RSL
23. **end while**

Some additional applications that the reverse skyline can be applied is the case where is needed to identify customers that would be interested in a particular product by exploring the dominance relationships between other competitor's products, with respect of the user preferences. The reverse skyline can also be applied in situations such as environmental monitoring, where a number of sensors are deployed in order to monitor the area and report data such as temperature and humidity.

In **[83]** authors try to answer the so called why-not questions in reverse skyline queries. In order to answer this type of question need to find why a point does not belong in the reverse skyline and what actions are needed to be performed (to the query point but also to the why-not point) in order to be part of the reverse skyline by incurring only minimum changes to both.

## 3.4. Group-by and Join Skyline Query

In this section are introduced the *Group-by Skyline queries* **[134]** and *skyline queries over joins* **[88].**

### 3.4.1. Group-by Skyline Query

In order to illustrate a Group-by skyline query **[134]** example based on the initial house-metro station dataset, a third attribute is inserted into the original dataset **Table** 1 (without altering any of its values) which represents the number of bedrooms that each house has **(Table 13).** This way a potential buyer can find individual skylines depending on the number of bedrooms. That is to group the houses by the number of their bedrooms and then compute the skyline of each group. In this case the cardinality of distinct values of bedrooms will be equal to the number of individual skylines that will be found. In **Figure 25** are illustrated the individual skylines of each group based on the number of bedrooms.

| House | price (in thousand €) | Distance (m) | No. of bedrooms |
|---|---|---|---|
| H1 | 100 | 1500 | 1 |
| H2 | 1400 | 500 | 3 |
| H3 | 700 | 600 | 2 |
| H4 | 1300 | 1000 | 3 |
| H5 | 900 | 1300 | 2 |
| H6 | 1600 | 100 | 3 |
| H7 | 400 | 300 | 1 |
| H8 | 200 | 1200 | 1 |
| H9 | 1000 | 200 | 2 |
| H10 | 500 | 1400 | 1 |
| H11 | 500 | 900 | 1 |

Table 13: House-metro station dataset with No. of bedrooms.

| House | price (in thousand €) | Distance (m) | No. of bedrooms |
|---|---|---|---|
| H1 | 100 | 1500 | 1 |
| H7 | 400 | 300 | 1 |
| H8 | 200 | 1200 | 1 |
| ~~H10~~ | ~~500~~ | ~~1400~~ | ~~1~~ |
| ~~H11~~ | ~~500~~ | ~~900~~ | ~~1~~ |
| H3 | 700 | 600 | 2 |
| ~~H5~~ | ~~900~~ | ~~1300~~ | ~~2~~ |
| H9 | 1000 | 200 | 2 |
| H2 | 1400 | 500 | 3 |
| H4 | 1300 | 1000 | 3 |
| H6 | 1600 | 100 | 3 |

Table 14: Group-by Skyline.

A *Group-by skyline query* based on the previous logic can be expressed using the SKYLINE OF and the GROUP BY identifier as follows:

SELECT * FROM Houses SKYLINE OF price MIN, distance MIN GROUP BY bedrooms

In order to give a formal definition of the Group-by dominance property and the Group-by skyline it is needed to define the following:

Given a relational table instance *DS* (dataset), in a d-dimension space with equal numeric attributes and a schema *A= (A$_1$, A$_2$,...A$_d$)*, the notation *p[A$_i$]* represents the value of a tuple *p* in the attribute *A$_i$*. Additionally given a set $G \subset A$ of attributes of *DS* that will be used for grouping and an instance *g* of *G* (i.e. one distinct value from the total values of number of bedrooms), *DS(g)* is defined as the set of tuples of *DS* that belong to the group instance of g. That is: $DS(g) = \{p \in D | \forall A_i \in G, p[A_i] = g[A_i]\}$

**Definition 10:** Group-by dominance
Given a set S ⊂ A (S∩G=∅) which this time contains the skyline attributes (that will be checked for dominance), a tuple p dominates another tuple r with respect of *S*, (denoted by p ≻$_s$ r) if and only if ∃ A$_j$∈S such that p[A$_j$]<r[A$_j$] and ∀ A$_i$∈S -{A$_j$}: p[A$_i$]≤r[A$_i$] ∎

**Definition 11:** Group-by Skyline
Eventually the Group-by skyline query will contain all the tuples *p* that are not *Group-by dominated* by any other tuple *r* with respect of *S* and that is: $\Psi(DS, S) = \{p \in DS | \nexists r \in DS, r \succ_S p\}$ ∎

Summarizing a group-by skyline query Q= (G, S), with G representing the grouping attributes and S the skyline attributes (S∩G=∅), computes the skyline result set ψ(DS(g),S) for each group instance g defined on G and the overall query result can be represented as Q (DS).

Based on the dataset *DS* of Table 13, in order to find the group-by skyline with respect to the *No. of bedrooms*, the grouping attributes are defined to be G= {No. of bedrooms} and the skyline attributes S= {Price, Distance} (S∩G=∅). The Table DS is partitioned into groups based on G and then the skyline tuples of each group are computed with respect of S.

A naïve approach to process a Group-by skyline is to create a separate R-tree for each one of the distinct values of bedrooms. Each R-tree will contain the corresponding house entries with their two remaining attributes, depending on the number of bedrooms (grouping attribute), and then an original BBS algorithm on each tree will be invoked. Nevertheless this approach is inefficient since the performance of queries when all attributes are involved is compromised as it may be needed to maximize or minimize the grouping attribute. A more efficient approach which operates on the R-tree that indexes all the attributes is achieved with a variation of BBS **[133].** This variation stores the already found skyline points for every group, in a secondary *(d-1)-dimensional* (in this case) R-tree and maintains a heap with the visited entries. The sorting measure that is used is based only on the *d-1* remaining attributes (without the group-by attribute). The dominance check of a retrieved point, from the original R-tree, is performed on the corresponding by its group R-tree and is inserted in it only if it is not dominated by any of the existing points. Dominance checks for intermediate entries (regions) are more complicated because it is likely to contain hotels of several classes.

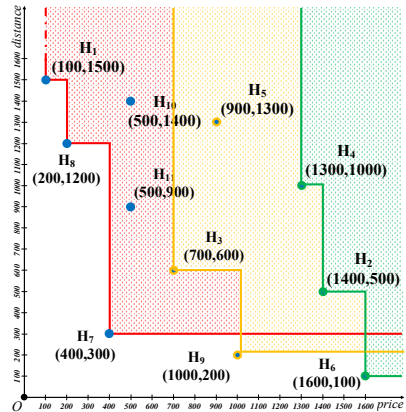

Figure 25: Group-by skyline.

Authors in [126] proposed the aggregate skyline query which combines the skyline and group-by queries. Essentially the aggregate skyline is the set of groups not dominated by other groups. The various groups are defined based on a common property of tuples. In addition authors discussed the differences between the efficiency of the aggregate skyline query processing in relation to the sequential execution of the skyline and group by query.

In [107, 198] authors studied the problem of identifying the k-tuple skyline groups. In this problem authors try to identify groups of k tuples that are not dominated by other, equal sized, groups. In order to compare the groups, each group is associated with an aggregate vector which is computed based on a common aggregate function such as SUM, MIN, MAX. The aggregate values of vectors are computed based on all the attributes of all tuples in a group. A naïve approach to compute the k-tuples skyline groups is to compute the aggregate functions for each k-tuple combination and then invoke a traditional skyline algorithm to identify the skyline groups.

### 3.4.2. Skyline queries over joins

Most of the existing work discusses the computation of skyline queries over data that are stored in a single table. In [88] authors discuss the case where data are stored in multiple tables and thus is required to perform join operations among them, in order to compute the final skyline and propose efficient methods to share the join processing cost with skyline computation cost. More specifically assuming the existence of two (or more) tables (which have one common join attribute) and apply a join operation on them, there is a case that may appear new skyline points that are not in the skyline of the individual tables. Based on this observation a naïve approach would be to compute the join of the two tables and produce a new table that contains the joined records. Afterwards apply a skyline query to the derived table in order to compute the skyline. The problem that may arise in this case is the potential increment of the computational cost of skylines on the joined table due to its increased cardinality and dimensionality. As a solution authors proposed a sort-merge join approach where they group the tuples in three groups according to the values of join attributes and based on whether or not are local skylines in their group and skyline points in the whole table. This involves a first phase of pruning from each table which is achieved with the use of an R-tree. Afterwards each tuple in each group is sorted based on its join attribute value. The next step involves a third (in this case) table which will host the join operation. Each group is inserted individually and merged with the existing tuples by additionally performing a dominance check for each tuple in order to compute final skyline of the join relation.

### 3.5. Top-k Skyline Query

Top-K skyline queries (or ranked skyline queries) were proposed in [133, 134] and return the *K* "most interesting" skyline points of a given dataset, based on a monotone preference function *f*. The user specifies the parameter K, which represents the number of points to be reported and the monotone preference function *f* based on the weighting that wants to apply over the attributes. The query will return the K points of the dataset with the minimum (or maximum) score according to the function *f*. To demonstrate this with an example assume that K=3 and the preference function is f(x)=x+2y (i.e. a lower *distance (y)* is more important than *price (x)* to the user ). The Top-3 points of the house-metro station dataset that will be returned are {(H$_7$, 1000), (H$_9$, 1400), (H$_6$, 1800)}. An SQL query for the above example will have the form:

SELECT * FROM Houses SKYLINE OF price MIN, distance MIN, ORDER BY (price + 2*distance) STOP AFTER 3.

This type of queries can be efficiently solved with BBS algorithm by replacing the *mindist* function with the given preference function. In this case the algorithm will terminate when exactly K points have been retrieved.

## 3.6. Thick Skyline Query

A Thick Skyline [89] extends the conventional skyline (authors use the term *thin skyline*) by returning the conventional skyline points and additionally their nearby non-skyline neighbors that exist within ε–distance, which are similar but not as good as the skyline points. This approach can help the user in cases of nearest neighbor search where the cardinality of the dataset is high and the points of the dataset forms groups. In their work authors extend the concept of *skyline* to *generalized skyline* by adding a user-specific constraint, defined as the *ε-neighbor* of any skyline point, into skyline search space. A *Thick skyline is* composed by a subset of the generalized skyline points.

**Definition 12:** Generalized Skyline.
Given a d-dimensional dataset DS and a set SL={s$_1$, s$_2$, s$_3$,... }, which contains the conventional skyline points of DS, the generalized skyline GL is consisted from the conventional skyline points and additionally the non-skyline points that exist in their vicinity within ε-distance. That is GL=SL∪{p|p∈DS^p∈s$_i$+ε, ∀i, 1≤i≤d}∎

**Definition 13:** Thick Skyline.
Given a d-dimensional dataset DS, a thick skyline is composed by the skyline points of the generalized skyline (named *dense skyline* points), that have in their vicinity (defined by ε) another (strictly) skyline point(s), and additionally the skyline points of the generalized skyline (named *hybrid skyline* points) that have in their vicinity another skyline point(s) and some non-skyline points. Thus, a thick skyline contains all the skyline points of the generalized skyline except of those that do not contain any other point in their vicinity (*outlying skyline* points*) as defined by the* generalized Skyline ∎

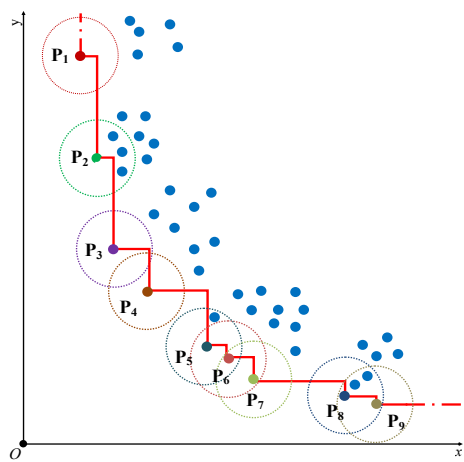

**Figure 26: Dense, hybrid and outlying skyline points.**

**Figure 26** shows the main differences between the *dense*, *hybrid* and *outlying skyline points* in a plane that is consisted from random points, because in the house-metro station dataset this would not be obvious. Point p$_1$, p$_3$, p$_4$ are outline skyline points since are skyline points but they do not contain any point in their vicinity. Point p$_7$ and p$_6$ are dense skyline points since are skyline points and contain another skyline point in their vicinity. Point p$_2$, p$_8$, p$_9$ and p$_5$ are hybrid skyline points since are skyline points and contain other non-skyline points in their vicinity. Additionally point p9 and p8 contain another skyline point in their vicinity.

Authors proposed three algorithms, *Sampling and Pruning*, *Indexing and Estimating* and *Microcluster-based* algorithm for mining thick skylines under three typical scenarios. The first scenario concerns a single file to represent the dataset where the sampling and pruning technique exploits statistics from the database, such as order and quantile in each dimension, in order to identify the thick skyline points by comparing the points of the dataset according to the defined *Strongly Dominating Relationship*.

**Definition 14:** Strongly domination relationship.
A point p strongly dominates a point q (denoted as p▷q), if p+ε dominates q. That is $\forall i \in [1,d]$, $p_i + \varepsilon \leq q_i$ and $\exists j \in [1,d]-\{j\}$, $p_i + \varepsilon < q_i$ ∎

The second scenario concerns a general index structure such as *Index* algorithm **[164]** where points are partitioned in lists ordered by their minimum coordinate and compared in a similar order. The final scenario concerns the partitioning of the dataset into microclusters based on the CF-tree structure **[120].** Then the algorithm follows a similar approach as BBS **[133]** in order to identify the desired points using bounding and pruning techniques. In each case the thick skyline performs approximate selections, as it employs approximate measures and increases the size of the final result set (compared to the conventional skyline) that is returned to users.

Below is the preudocode of the *Index* and *Estimating* method that authors proposed in the second scenario.

| **ALGORITHM 9:** Index & Estimating method **[89]** |
|---|
| **Input:** A B-tree index of $d$ lists. A distance threshold $\varepsilon$. |
| **Output:** The thick skyline $T$. |
| 1. $S = \varnothing$ ; $T = \varnothing$ ; |
| 2. **for** $i = 1$ to $d$ **do**; |
| 3.     $SW_i = \varnothing$ ; $upper_i = |list_i|$; $minC_i = \min\ list_i$; |
| 4. **while** ($Thin - Skyline - Search - Unfinished$) **do**; |
| 5.     Choose the batch with min $minC_1,...,minC_d$, say $minC_k$; |
| 6.     Check each object $p$ in this batch; |
| 7.     **if** $p$ is a skyline object **then** |
| 8.        $S = S \cup \{p\}$; |
| 9.     **if** $(p_j - p_i) < \sqrt{2} \cdot \varepsilon$ **then** |
| 10.        update $upper_j$ to $p_j + \frac{\varepsilon}{\sqrt{2}}$ ; |
| 11.     check $SW_j$ for $\varepsilon$ neighbor; |
| 12.     **if** any $q$ is a $\varepsilon$ neighbor **then** |
| 13.        $T = T \cup \{q\}$; |
| 14.     **else if** $p$ is an $\varepsilon$-neighbor **then** |
| 15.        $T = T \cup \{p\}$; |
| 16.     Move $list_k$ to next batch and update $SW_k$; |
| 17. **while** $list_1 < upper_1 \vee \ldots \vee list_d < upper_d$ **do**; |
| 18.     scan objects to find $\varepsilon$ neighbors and add to $T$; |
| 19. $T = T \cup S$; |
| 20. **Output** thick skyline $T$ |

## 3.7. K-representative and Distance-based representative Skyline Queries

*K-representative skylines points* (top-k RSP) were proposed in **[118]** in order to identify a set of k skyline points that maximize the total number of (distinct) points dominated by one of the *k* skyline points. This type of query was proposed in order to allow users to have a good approximation (returning few but representative skyline points) of the final skyline and let them make a good and quick selection when the skyline is consisted from too many points. Authors also developed an efficient, scalable, index-based randomized algorithm. Authors in their implementation employed the BBS **[133, 134]** algorithm and the *FM probabilistic counting algorithm* **[53].** The FM algorithm is a bitmap based algorithm that can efficiently estimate the number of distinct elements (data points) dominated by a skyline point, overcoming multiple-domination counting.

**Definition 15:** K- Representative Skyline (Top-k RSP).
Given a dataset Ds and an integer k, $\forall p \in Ds$, D({p}) is denoted as the set of points in Ds that are (strictly) dominated by p. For a set S of data points, with $S \subseteq DS$, D({S}) denotes the set of points each of which is strictly dominated by a point s∈S. The set K of the k-representative skyline points will contain k skyline points that Maximizes |D(K)|∎

---

**ALGORITHM 10:**       *Greedy* (k, p) algorithm for top-k RSP [118]

**Input:**    A dataset P.
           An integer parameter k.
**Output:** k skyline points.

1. compute $S_P$
2. $\forall s \in S_P$: compute $D(\{s\})$
3. $S := \emptyset$
4. **while** $|S| < k$ and $S_P - S \neq \emptyset$ **do**
5.      choose $s \in S_P - S$ such that $|D(\{s\} \cup S)|$ is maximized;
6.      $S := \{s\} \cup S$;
7. **return** $S$

---

The previous pseudocode outlines a greedy heuristic for the top-k RSP problem. In this algorithm D(S) denotes the set of points each of which is dominated by at least one point in S and SP is the skyline set of dataset P.

The problem of identifying the *K-representative skyline* is known to be NP-hard [118] in 3 or higher dimensional space. This approach is scale invariant but cannot be considered stable since adding a non-skyline point may alter the final k-representative skyline set. Top-k RSP can be transformed in the *maximum coverage problem* [71] and solved approximately by the author's proposed greedy heuristic.

### 3.7.1. Distance-based Representative skyline

Authors in [165] proposed the *distance-based representative skyline* which is an alternative solution for the problem of *k-representative skyline* points (referred as *max-dominance representative skyline* in this work) where they redefined it. The reasoning was that the set K of k points returned by the *k-representative skyline* can turn out not to be representative, because the produced points may belong to the same cluster or the set K fails to represent the extreme points. From the authors perspective a good representative skyline should have for every non representative skyline point, a nearby representative. Therefore in their work they defined the problem of identifying the *k-representative skyline points* as the set of k points that minimizes the distance between a non-representative skyline point and its nearest representative. The proposed approach it is not a scale invariant, as the previous approach (*k-representative skyline*), since it is based on distances. As opposed, it can be considered to be stable since by adding a non-skyline point in the dataset will not change the final representation (due to the initial algorithm construction).

In this approach it is considered that the data space is normalized in the range [0, 1]. The distance-based representative skyline can be an optimal solution for k-*center problem* [61] of the full skyline. Except from the *distance-based representative skyline* authors introduced the concept of *representation error* of *K*, denoted as *Er*(*K, S*) in order to quantify and check the quality of the representation K of the full skyline S of the dataset Ds, by the k identified representatives. This is achieved by checking the maximum of all distances, between any of the non-representative skyline points in the set *S−K* and their nearest representative in *K*.

**Definition 16:** Distance-based Representative Skyline.
Given a dataset Ds, its skyline set S and an integer value k, the distance-based representative skyline K of Ds is consisted of k-skyline points of S that minimizes the *representation error Er(K,S)*∎

For the 2-dimensional space authors developed a dynamic programming algorithm that optimally finds a solution in polynomial time. For 3-dimensional spaces and higher authors propose a *2-approximate* [72] polynomial algorithm and prove that the problem is still NP-hard [118]. The algorithm can quickly identify the *k* representatives without extracting the entire skyline by utilizing a multidimensional access method (i.e. R-tree). The proposed algorithm is progressive and does not require a specific k value from the user as it continuously returns representatives that are guaranteed to be a *2-approximate* solutions at any moment, until either manually terminated by the user or eventually producing the full skyline.

| **ALGORITHM 11:** | *2D*-opt (S, k) **[165]** |
|---|---|
| **Input:** | The skyline S of a dataset DS. |
| | An integer parameter k. |
| **Output:** | The representative skyline of DS. |

1. **for each** pair of ($i, j$) such that $1 \leq i \leq j \leq m$, derive *radius*($i, j$) and *center*($i, j$).
2. set *opt*($i$, 1) = {*center*(1, $i$)} and *optEr*($i$, 1) = *radius*(1, $i$) **for each** $1 \leq i \leq m$
3. **for** $t = 2$ to $k - 1$ **do**
4.    **for** $i = t$ to $m$ **do**
5.       compute *optEr*($i, t$) by Equation 2 **[165]**
6.       compute *opt*($i, t$) by Equation 3 **[165]**
7. compute *optEr*($k,m$) and *opt*($k,m$) by Equations 2 and 3 **[165]**
8. **return** *opt*($k,m$)

Above is the pseudocode of the k-representative skyline algorithm in a 2-dimensional space. The parameter K represents the number of representatives. The $S_i$ will represent the first i skyline points sorted by their x-coordinates in ascending order. Function *opt(i,t)* represents the optimal size-t skyline of $S_i$ and the *optEr(i,t)* is the representation error of *opt(i,t)* with respect to $S_i$. The function *radius (i,j)* denotes the radius of the smallest circle that covers the sorted skyline point $p_i, p_{i+1}, ..., p_j$ and centers in one of them. The center of the previous mentioned circle is denoted as *center(i,j)*.

The k-representative skyline gives to the user a high-level summary of the entire skyline as it returns only a few points that reflect to the contour of the final skyline and then progressively refines it (contour) by reporting more skyline points. The user may identify interesting representative points and request only the skyline points that are similar to those representatives (i.e. belong to a specific part of the contour).

### 3.8. ε-skyline

The proposed type of query claims that solves the limitations and drawbacks of the original skyline by taking into account that there was no algorithm that can simultaneously control the resulted size of the skyline, has built-in ranking for the points and weighting on dimensions. According to previous considerations authors proposed the ε-skyline **[190]** which allows users to control the number of output skyline points (by increasing or decreasing them depending on an appropriate ε parameter that the user defines), provides a built-in ranking system and integrates weighting factors for each dimension.

The algorithm takes as input a d-dimensional dataset (with its values normalized as on SFS **[35]** in [0, 1] (section 2.3.6), a weight vector W that will contain the weight factors $W_i$, ($i \in [1, d]$) for each dimension (if no weighting is needed all the factors will be equal to 1) and a parameter $\varepsilon \in [-1, 1]$. The dominance property is relaxed according to the ε parameter. In order to manage the dominance relations, authors defined some additional properties such as *irriflexivity*, *loose transitivity* and *loose asymmetry*. The weights that the user inserts are incorporated in the dominance comparisons. For the built-in ranking system to work every point *p* in the dataset has a corresponding ε-max value which represents the largest value of ε, which makes *p* to be a skyline point. Thus the points have a natural order based on ε-max value. This ordering can be used to place top-k ε-skyline queries.

**Definition 17:** *ε-domination.*
Given a d-dimensional dataset *DS*, a weighting vector W={$W_i$ |$i \in [1, d], 0 < w_i < 1$}, a parameter $\varepsilon \in [-1, 1]$ and two points p,r∈DS, p ε-dominates r, denoted with p $\stackrel{\varepsilon}{\prec}$ r if and only if $\exists\ j \in [1,d]$ such that p.dj<r.dj and $\forall\ i \in [1,d]-\{j\}$: p.di $*w_i \leq$ r.di $*w_i+\varepsilon$ ∎

**Definition 18:** *ε-Skyline.*
The *ε*-Skyline of a dataset *DS* contains all the points p∈DS that are not ε-dominated by any other point on the dataset∎

Authors proposed two algorithms, ε-SFS which is progressive and based on the SFS [35] algorithm and IFR (index-based Filter-Refinement) algorithm which uses an index structure such as an R-tree and a filter-refinement framework.

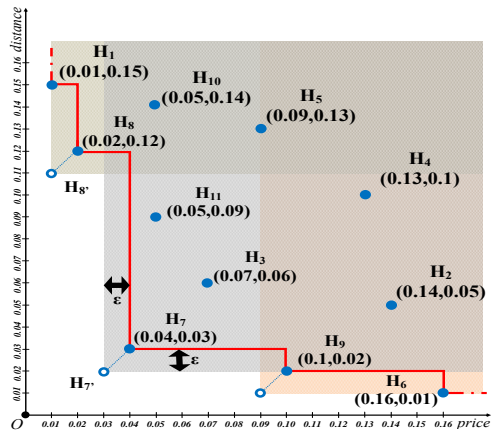
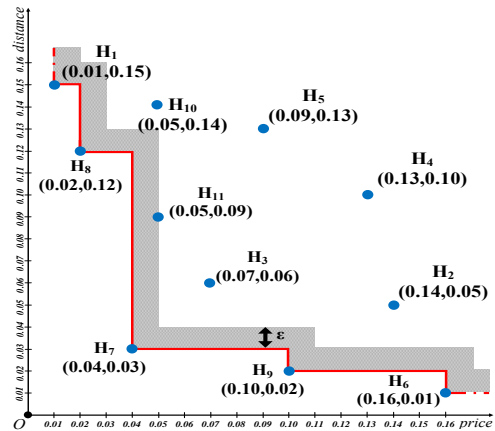

Figure 27: Dominance region of H$_7$' with ε=0.01 .    Figure 28: ε-skyline with ε=-0.01 .

An ε-skyline can monotonically vary from an empty set to the whole dataset depending on the value of ε. An ε-skyline with ε=0 represents the conventional skyline. An ε-skyline with value ε=-1 will return the whole dataset and with value ε=1 the empty set. More generally the case of an ε-skyline, with ε>0 is shown in **Figure 27**. In this case ε=0.01 and as shown the dominance region of (i.e.) H$_7$ will be visualized as the point H$_7$ was moved to the location of H$_7$' in order to fulfill the ε-domination. This way point H$_7$ ε-dominates point H$_9$ which for that reason will not be in the final skyline set. Similar H$_8$ ε-dominates H$_4$ and H$_9$ ε-dominates H$_6$. The final skyline set for ε=0.01 will be S= {H$_7$, H$_8$}. The case of an ε-skyline, with ε<0 is shown in **Figure 28.** In this case the ε-skyline will contain the conventional skyline points and additionally the points that are in the shaded area, as this happens with point H$_{11}$.

### 3.8.1. Approximately dominating representatives

In **[95]** authors introduced the notion of approximately dominating representatives (ADRs). Authors scenario concerns a set of n points in a d-dimensional space and a value ε>0, where it is desired to find the minimum set of points, named ε-ADR, that approximately dominate all the points of the dataset. With this approach they try to minimize the number of (skyline) points to be reported at a small loss of accuracy. The approximation is imposed by a user-defined value ε that extends the dominating region of each point. The data points retrieved by the algorithm when ε=0 (ε-ADR) are guaranteed to be skyline points. In this case the algorithm will return all the existing skyline points. In the cases of ε>0 may exist many different ε-ADRs (for a specific value ε). In this case the points returned are not guaranteed to be skyline points. An example of a case where ε>0 can be considered a dataset that contains a point that approximately dominates all others (i.e. a point very close to the origin of axes if minimization of preferences is desired). In this case the algorithm will return only this point, although it may not be a pure skyline point.

## 3.9. Enumerating and K-dominating Queries

Enumerating queries and K-dominating queries (Top-k dominating queries in general bibliography) were proposed in **[133, 134].** These types of queries do not produce skylines but can work as a measure of "goodness" in various cases.

### 3.9.1. Enumerating queries

An enumerating query **[133, 134]** returns the set of skyline points and additionally the number of points that each skyline point *p* dominates (denoted as *num(p)*). This kind of result could be used to investigate which skyline points are more interesting by means of "number of points that they dominate". To compute the enumerating query the first step is to retrieve the skyline points of the dataset with an existing algorithm (i.e. BBS). The second step performs a query in the R-tree, for every skyline point, in order to count the number of points that exist in their dominance region. In order to avoid multiple node visits with the previous technique (since a node may be dominated by more than one skyline points), a solution is to apply the inverse procedure which is, for each non-skyline point in the dataset, perform a query in the R-tree to find the dominance regions that contains it and accordingly increase the appropriate counters of the skyline points that dominates it. As an example in the house-metro station dataset the enumeration query will return for the house $H_7$, num($H_7$)= 6, for the house $H_{11}$, num($H_8$)= 2 and for the house $H_6$, num($H_6$)=0.

### 3.9.2. K-dominating queries

A variation of the above problem (and also the predecessor of the *k-representative skyline query*) that incorporates the enumerating query (and the constrained skyline queries) is the K-dominating query [133, 134]. This type of query returns the *K* points that dominate the largest number of other points. The points that are returned do not necessary belong to the skyline of the dataset. Below is the pseudocode of the k-dominating queries. The function *num(p)* returns the number of points that fall inside the dominance region of a point p.

| **ALGORITHM 12:** | K-dominating_BBS (R-tree DS, int k) |
|---|---|

**Input:** A dataset DS (R-tree).
An integer parameter K.
**Output:** A set of k-dominated points.

1. compute skyline *S of DS* using BBS
2. for each point in *S* compute the number of dominated points
3. insert the top-*K* points of *S* in *list* sorted on *num(p)*
4. *counter*=0
5. **while** *counter < K* **do**
6.     *p* = remove first entry of *list*
7.     output *p*
8.     *S'* = set of local skyline points in the dominance region of *p*
9.     **if** (*num(p)*-|*S'*|)> *num*(last element of *list*) **then** // *S'* may contain candidate points
10.        **for** each point *p'* in *S'* **do**
11.            find *num(p')* // *perform a window query in data R-tree*
12.            **if** *num(p')* > *num*(last element of *list*) **then**
13.                update list // *remove last element and insert p'*
14.     counter=counter+1;
15. **end while**

The first step to retrieve the K-dominant points is to perform an enumerating query. The query will return the skyline points and the number of points that each one of them dominates. The items that are retrieved are sorted by their descending order of the number of points that they dominate and the first K points are placed in a list. The first point of the list is the first result of the k-dominating query and it is returned to the user, removed from the list and pruned from further computations. Next is applied a local (constrained) skyline with boundaries the (exclusive) dominance region **(Figure 29)** that was defined by the point removed, in order to efficiently find the skyline of the dataset after the removal of the first point and identify potential candidate K-dominant points (that may outnumber points in the list). The second step of the enumeration query is applied on the newly found skyline points (if they exist and are possible candidates) and returns the number of points that they dominate. If any of the points found outnumbers the last point of the list, it replaces it and the list is rearranged. The first point of the list will be the second K-dominating point. The algorithm terminates when it finds the K most dominant points, thus when the new points retrieved from the local skyline cannot outnumber the points in the list. For the House-metro station example a K-dominant query for K=3 will return the points {($H_7$,6),($H_{11}$,3), ($H_8$,2)}. More analytically after the initial enumerating query the list will contain the points {($H_7$,6), ($H_8$,2), ($H_9$,2)}. Point H7 will be the first K-dominant point and returned to the user. After removing point $H_7$, the local skyline point $H_{11}$ is checked and is inserted in the list (the last point of the list is removed resulting in {($H_{11}$,3), ($H_8$,2)}. The local skyline point $H_3$ is also checked, after the removal of $H_7$, but is not inserted in the list because it does not outnumbers any point in it. Thus the second K-dominant point $H_{11}$ is returned to the user but the algorithm terminates **(Figure 30)** since a local skyline, by removing $H_{11}$, has not any candidate that may outnumber the last (and in this case the final) point of the list. So point $H_8$ is also returned to the user.

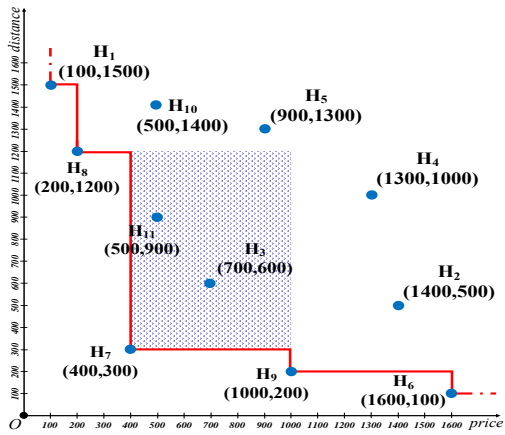

Figure 29: Exclusive dominance region of $H_7$.

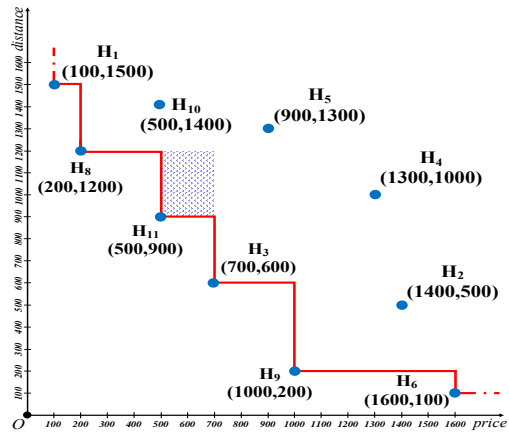

Figure 30: Skyline after removing $H_7$ (final step of algorithm

## 3.10. K-skyband Query

A K-skyband query [134] returns the points that are dominated by at most K points with the case of K=0 representing the original skyline. A K-skyband query follows similar logic with K nearest-neighbor query with K representing the thickness of the skyline.

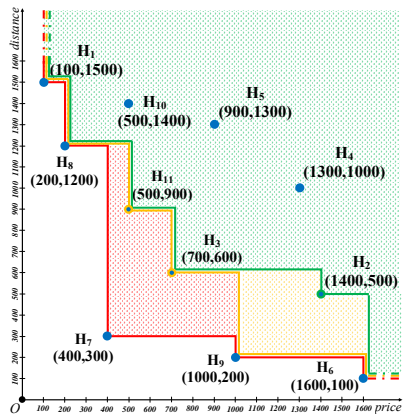

Figure 31: (0, 1, and 2)-skyband query.

In Figure 31 are illustrated a 0-skyband query (red line), a 1-skyband query (yellow line) and a 2-skyband query (green line) of the house-metro station dataset. In detail a 2-skyband query will return points H1, H6, H7, H8, H9 (which are dominated by at most 0 points), H3, H11 (which are dominated by at most 1 points) and H2 (which is dominated by at most 2 points).

A naïve approach to process a k-skyband query is to perform an enclosure (window) query on the R-tree, for every point $p(p.d_1,p.d_2) \in DS$, in order to count the points that exist in the region $[0,p.d_1][0,p.d_2]$. If there exist up to k points in this region then the point p belongs to the skyband. Since this approach is inefficient, because the number of enclosure (window) queries required is equal to the cardinality of the dataset, a more efficient approach involves the processing of k-skyband query with the BBS algorithm. As with the original skyline the algorithm maintains its progressiveness and its only difference is that an entry is rejected only if it is dominated by more than k discovered skyline points.

## 3.11. Summary

The user can apply a dynamic skyline query if in addition with the original skyline computation, wants to apply a space transformation from a (i.e) 3-dimensional space to a 2-dimensional space, (and vice versa) or to find the skyline set based on a given query point. A reverse skyline query can help the user to identify if a given query point is desirable and interesting based on an existing dataset that may represent his/her preferences. A spatial skyline query can be applied when the user wants to find the skyline according to multiple query points such as the case of deploying a number of police cars in order to respond to multiple incidents. The Group-by skyline can help the user to identify the interesting points based on some common attributes (i.e find the best hotels in each 5-star category). A thick skyline can help the user to retrieve not only the skyline points but also some additionally points that may be interested to know even if they are not truly-interesting points but only nearly-interesting points (are very close to a truly interesting point). A top-K skyline query can help the user to retrieve the interesting points of a dataset even if his/her preferences are biased. In example he/she prefers cars with twice as low consumption even if its horsepower is tripled lowered. With a k-representative skyline the user can retrieve a representation of the original skyline which is consisted from a smaller number of points than the original skyline. This representation can be based on dominance or distance from other representatives, depending on the selected query type. This type of query can be useful if the user wants to retrieve a general view of the skyline fast, without retrieving the full skyline. With a ε-skyline the user is able to incorporate the idea of top-k, k-dominating, thick and the k-representative skyline with one algorithm. An enumerating query will help the user to retrieve the skyline points and additionally the number of points that each skyline point dominates while with a k-dominating query can retrieve the k-points that dominate the most points. Finally a k-skyband query will let him to retrieve points based on the number of points that dominate a point which can be useful in cases where the user wants to know the dominance relations.

A performance analysis between BBS [133, 134] and NN [97] based on the application of the various queries types can be found in [134].

| Query type | Specific Algorithms | Based-on | incorporates |
|---|---|---|---|
| Constrained Skyline [133, 134] | Modified BBS or NN | BBS or NN | MBRs |
| Dynamic Skyline [133, 134] | BBS | BBS | *mindist*, distance functions |
| Spatial Skyline [155, 156] | $B^2S^2$ | BBS | Convex hull |
| | $VS^2$ | - | Voronoi diagram / Delaunay graph |
| | $VCS^2$ | $VS^2$ | Voronoi diagram Delaunay graph |
| Reverse Skyline [44] | BBRS | BBS | Global skyline |
| | RSSA | - | Global skyline / Approximation of skyline |
| Group-by Skyline [134] | Modified BBS | BBS | Secondary R-tree / sorting |
| Top-k Skyline [133, 134] | Modified BBS | BBS | *mindist*, distance function |
| Thick skyline [89] | Sampling & Pruning | - | sampling / Strongly Dominating Relationship |
| | Indexing & Estimation | Index [164] | sorting |
| | Microcluster-based | - | microcluster-based index |
| K-representative [118] | Greedy | BBS | sort-merge paradigm |
| | FMGreedy | Greedy - BBS | FM-algorithm[53] / FM sketches |
| | $R^{FM}$-tree | BBS | $R^{FM}$-tree [118] / FM sketches |
| Distance-based K-representative [165] | 2D-opt | - | R-tree / Covering circles |
| | l-greedy | - | R-tree / farthest neighbor search |
| ε-skyline [190] | ε-sfs | SFS | specific monotone function |
| | IFR | - | extra set ToExpand whith MBRs |
| Enumerating query [133, 134] | Modified BBS | BBS | R-tree, find dominance regions |
| K-dominating query [133, 134] | Modified BBS | BBS | Enumerating query, constrained skyline |
| K-skyband query [134] | Modified BBS | BBS | pruning restrictions |

Table 15: Specific algorithms for each query type.

Table 15 outlines the basic algorithms developed for the various types of queries mentioned and notes the fundamental skyline algorithm that is based (if applicable) and the specific structures (geometric) or techniques (approximation) that may incorporate.

| Type | Method | Size of resulted set |
|---|---|---|
| Constraint skyline | Region restrictions | K ≤ S |
| Dynamic Skyline | Space transformation | S |
| Spatial Skyline | Geometric structures | S |
| Reverse Skyline | Space transformation | S |
| Group-by skyline | Grouping attributes | S |
| Thick skyline | Approximate selection | K ≥ S |
| Top-K (ranked) | Point-wise ranking | Exactly k points, ∅⊂ K ⊆Ds |
| K-representative | Exclusive domination | Exactly k points, K < S |
| Distance-based K-representative | Distance aware | Exactly k points, K < S |
| ε-skyline | Multiple methods | ∅⊆ K ⊆Ds |
| K-skyband query | Domination | K ≥ S |
| Enumerating query | Domination | S |
| k-dominating query | Exclusive Domination | Exactly k points |

**Table 16: Skyline queries approaches**

Table 16 illustrates the various skyline related approaches, the general method that is used to retrieve the skyline points of a dataset Ds and the size k of the resulted skyline set in a general case, compared to the size S of conventional skyline query.

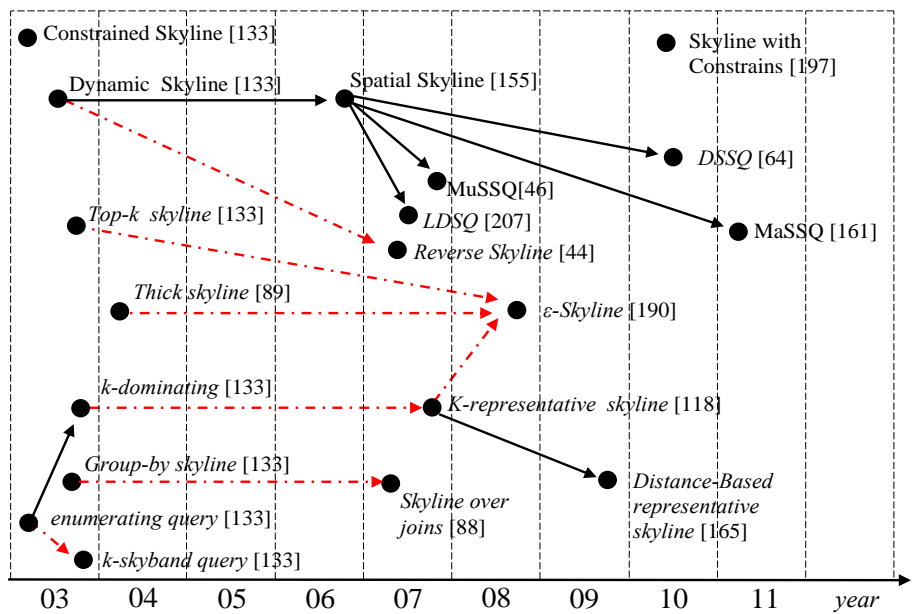

**Figure 32: Chronological order of basic skyline queries.**

Figure 32 illustrates the various queries. The black lines indicate that the algorithm heavily depends or improves a previous algorithm and the red dashed line indicates that the algorithm shares some general main ideas in order to compute the skyline.

## 4. Applications

This section reasons about the various applications where the skyline queries can be applied. With the term "application" is characterized any method or algorithm that reasons about datasets not belonging to the standard space, such as a 2-dimensional or 3-dimensional Euclidean space, but rather on a metric space or even on a subspace of the original space. Additionally as applications are concerned methods and algorithms for a particular environment, such as distributed environments, p2p based architectures, Wireless Sensors Network (WSN) and road networks. Finally datasets that have not numeric or nominal attributes or the datasets that are partial ordered, belong to data streams or have uncertain data are also classified as "applications".

### 4.1. Subspace and Space Partitioning

The fundamental methods for skyline computation are optimized and rely on the fact that the dimensionality of queries is fixed and concerns the full space of the dataset (take into account all the dimensions/attributes of the dataset). Nevertheless different users may be interested about different dimensions/attributes of data and therefore may want to retrieve the skyline by comparing only a specific subset of all dimensions/attributes. Additionally a full space skyline query in high dimensional space may return too many interesting points to the user which will not allow him/her to make an appropriate decision. This problem reveals a different scenario in which a query is placed over fewer dimensions than those of the full space. Formally, given a set of d-dimensional points, a skyline query can be issued on any subset of the *d* dimensions. This subset will be called *subspace* and the corresponding skyline query on those dimensions subspace *skyline query*. Methods that are related with subspace skyline computation can be classified into two categories. The first pre-computes the subspace skylines for all subspaces and organize the results into a structure similar to data cubes in data warehouse environments, called SkyCube, allowing answering any subspace skyline query immediately. The second method computes the subspace skylines on the fly using index structures. In general methods that belong in the first category require larger storage space in order to store all the subspace skylines and are difficult to maintain in dynamic environments.

Essentially for a *d* - dimensional space there will exist $2^d - 1$ subspaces (including the d-dimensional space called *full-space* and without concerning the trivial subspace ∅) and equivalently $2^d - 1$ subspace *skyline queri*es where each one of them has different dimensionality. All these different skyline queries will produce, in general, different results. The strict definition of the subspace dominance is the following:

**Definition 19:** Subspace Dominance
Given two d-dimensional points p, r ∈ Ds with p=($p_1,p_2,...p_d$) and r=($r_1,r_2,...r_d$) and a parameter k that specifies a k-dimensional subspace $DS_k$ of Ds with k≤d, p dominates r in the k-dimensional subspace if and only if ∃ j∈[1,k] such that p.dj<r.dj and ∀ i ∈[1,k]-{j}: p.di ≤ r.di ∎

In order to describe the notion of subspace skyline computation an example is illustrated below based on the original dataset **Table** 1 of the house-metro station example. The dataset is consisted of two dimensions, price and distance which are the two (pure/strict) subspaces that compose the original space. An easy way to find the subspace skylines for this example is to project the points of full space to each of the subspaces. In **Figure 33** all the points are projected on the subspace *Price* and on **Figure 34** are projected on the second subspace *Distance*.

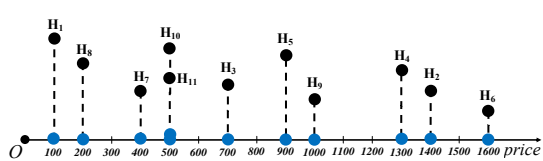

**Figure 33: Subspace *Price* of the full space *Price-Distance*.**

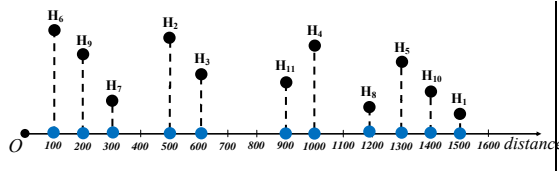

Figure 34: Subspace *Distance* of the full space Price – Distance.

The subspace computation is sufficient to qualify point H1 and H6 of subspace *Price* and *Distance* respectively to be skyline points. On the other hand this is not sufficient for the rest of the skyline points of the *full-space* (H8, H7, H9 as computed in the previous sections), since can only be retrieved if both the subspaces are considered simultaneously. To describe this condition the term "*decisive*" will be used. Thus, subspace *Price* is *decisive* to point H1, subspace *Distance* is *decisive* to point H6 and the subspace {Price, Distance} is *decisive* to points H8, H7, and H9. The *decisive* subspaces of the skyline points and the values of the points in those subspaces help to understand the *semantics* of the points that represent the skyline. That is, not every subspace must contribute in order to qualify a point to be skyline point. This information cannot be captured in the traditional skyline computation.

**Definition 20:** Subspace Skyline
Given a d-dimensional dataset Ds, a data point p ∈ Ds and a parameter k (k≤d) that specifies a k-dimensional subspace $DS_k$ of Ds, point p is said to be a subspace skyline point iff ∄ r∈ Ds such that r≺p in the subspace $DS_k$. ∎

In order to describe more analytically the notion of subspace skyline computation and it's relevant notions through examples, the original house-metro station example based on **Table** 1 is extended by adding two more dimensions/attributes for each house which are the *number of bedrooms* and the *age* of the building, making the dataset as follows:

| House | Price (in thousand €) | Distance (m) | Bedrooms | Age of Building (years) |
|---|---|---|---|---|
| H1 | 100 | 1500 | 1 | 25 |
| H2 | 1400 | 500 | 3 | 5 |
| H3 | 700 | 600 | 2 | 9 |
| H4 | 1300 | 1000 | 3 | 15 |
| H5 | 900 | 1300 | 2 | 17 |
| H6 | 1600 | 100 | 3 | 7 |
| H7 | 400 | 300 | 1 | 20 |
| H8 | 200 | 1200 | 1 | 23 |
| H9 | 1000 | 200 | 2 | 10 |
| H10 | 500 | 1400 | 1 | 6 |
| H11 | 500 | 900 | 1 | 19 |

Table 17: Dataset for the subspace skyline computation.

### 4.1.1. Multiple Subspace Computation

In general, different users may be interested in different dimensions of the dataset, fulfilling different preferences on what attributes want to compare and from which subspaces want to retrieve the skyline. Therefore, the skyline queries can be issued on any subset of the original d-dimensional space. As an example, one's preference may be the subspace *Price* and *Distance* (as previous) and the other one's *Price* and *Age*. Due to the fact that the truly interesting subspaces for each user are unpredictable one approach is to compute all the possible subspace skylines.

Multidimensional subspace skyline computation was proposed simultaneously by two different groups of authors in **[196]** and **[139].** The methodology that they follow is different but the main idea remains the same and is to compute the skylines of all possible subspaces forming a lattice structure similar to the data cube **[1, 62]**. The authors of both groups combined, extended and improved their works in [140]. The initial problem that they state is that none of the existing methods considered skyline computation in subspaces.

*SkyCube*

Authors in [196] proposed the *SkyCube* which pre-computes and stores all the possible subspace skylines, providing minimum response time at query requests. The term SkyCube refers to the set of all possible subspace skyline query results. The result of a subspace skyline query on a subspace U over a set of point S, denoted as SKY$_U$(S), will be called Cuboid U. The SkyCube can be visualized as a lattice structure (similar to that of the data cube), as shown in **Figure 35** for the dataset of **Table 17**. For illustration purposes each dimension is represented by the first letter of their name. However computation a SkyCube is more demanding [196] than computing a *data cube*. When the SkyCube is constructed, all the subspace skyline queries can be efficiently answered with little overhead, since all subspace skylines are pre-computed and stored in corresponding cuboids. In order to answer a subspace skyline query of a subspace *U*, a user can go to the cuboid *U* and return the result points immediately.

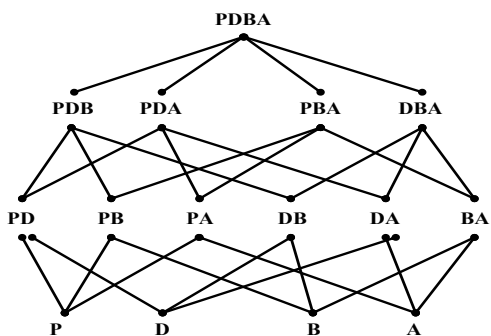

**Figure 35: Lattice Structure of a SkyCube.**

| Cuboid | D | DB | PDB |
|---|---|---|---|
| Skyline | H6 | H6,H7,H9 | H1,H6,H7,H8,H9 |
| Skylist | <D,DB,PDB> = <(H6),(H7,H9),(H1,H8)> | | |

**Table 18: Skylist (path).**

| Subspace (Cuboid) | Skyline |
|---|---|
| PDBA | H1,H2,H3,H6, H7,H8,H9,H10,H11 |
| PDB | H1,H6,H7,H8,H9 |
| PDA | H1,H2,H3,H6, H7,H8,H9,H10,H11 |
| PBA | H1,H2,H7,H8,H10 |
| DBA | H2,H3,H6,H7,H9,H10,H11 |
| PD | H1,H6,H7,H8,H9 |
| PB | H1 |
| PA | H1,H2,H7,H8,H10 |
| DB | H6,H7,H9 |
| DA | H2,H6 |
| BA | H2,H10 |
| P | H1 |
| D | H6 |
| B | H1,H7,H8,H10,H11 |
| A | H2 |

**Table 19: SkyCube/Cuboids of the 4-dimensional dataset.**

Authors proposed the extension of SQL by incorporating the SKYCUBE BY keyword. The *SkyCube query* which results **Table 19,** based on dataset of **Table 17,** can be expressed using this clause as:

SELECT * FROM Houses SKYCUBE BY price MIN, distance MIN, bedrooms MIN, age MIN

The various subspace skyline queries (all the possible combination of all subspaces) can be efficiently computed individually with the existing skyline algorithms. Nevertheless naïve algorithms are not efficient for SkyCube computation since they cannot share the computation of multiple related skyline queries (follow a "share-nothing" approach) and it's designed incurs additional overhead due to certain pre-conditions that have such as sorted or indexed inputs.

The approaches that were proposed are the BUS (bottom-up SkyCube) and the TDS (top-down SkyCube). Their efficiencies mainly come from the computation *sharing strategies* that authors proposed for multiple skyline computations. The methods compute the subspace skylines by traversing the lattice of subspaces either in a top-down or bottom-up manner. Below is the pseudocode of the BUS algorithm. In line 6-7 if a data point is a skyline point of a child cuboid it will be inserted into the skyline, otherwise it will be compared with the current skyline in order to determine if it is a new skyline point using the Evaluate function in line 9. The simplest implementation of the Evaluate function is to perform a dominance test. In order to further optimize their algorithm authors proposed an alternative filter-based approach as shown in algorithm 14, which incorporates the computation of a monotonic function $f_u(p) = \sum_{\forall a_i \in U} p(a_i)$ based on a data point p. With the use of this approach a computation cost reduction is achieved by reducing dominance checks.

**ALGORITHM 13:**      BUS (*S*) [19]

**Input:** A d-dimensional dataset *S*.
**Output:** Every cuboid *V*, SKY<sub>*V*</sub>(*S*).

1. sort S on every dimension $a_i$ (in non-decreasing order) to form d sorted lists $l_{a_i}$ ($1 \leq i \leq d$)
2. **for** each level from bottom to top of the skycube and each cuboid *V* in this level **do**
3.     SKY = the union of all the child cuboids
4.     choose a sorted list $l_{a_i}$ ($a_i \in V$)
5.     **for** each data point q in $l_{a_i}$ **do**
6.       **if** q ∈ SKY **then**
7.         insert q into SKY<sub>*V*</sub>(*S*)
8.       **else**
9.         **Evaluate**(q, SKY<sub>*V*</sub>(*S*))

The bottom-up SkyCube algorithm (BUS), computes each cuboid according to their lattice levels from bottom to the top. Lower level cuboids are merged to form a part of the upper level parent cuboids. The cuboids are computed by a nested loop-based algorithm similar to SFS [35]. The algorithm takes advantage of two sharing techniques, named *sharing result* and *sharing sorting*, which essentially shares the computation and sorting cost among cuboids. Top-Down SkyCube algorithm (TDS) incorporates a novel Shared-Divide-and-Conquer skyline algorithm (SDC) based on the Divide & Conquer skyline algorithm [19] to recursively compute the cuboids from the top to bottom level, by sharing the partitions and merging. TDS also incorporates a novel data structure called *skylist* which stores the skyline points of a given path in the cube lattice, according to the cuboid that they belong, without duplications.

**ALGORITHM 14:**      Evaluate (q,*SP*) [19]

**Input:** A data point q to be evaluated.
       A set *SP* representing the cuboid *U* computed so far.
**Output:** Inserts q into *SP* if it is a skyline point of *U*.

1. **for** each data point p in *SP* **do**
2.     **if** $f_U(q) < f_U(p)$ **then**
3.       insert q into SP; **return**
4.     **else if** $(p \prec q)_U$ **then**
5.       discard q; **return**
6. insert *q* into *SP*

Authors focused on the initial construction of the SkyCube, where the computation cost can be shared over different subspace skylines. However, they did not discuss how a SkyCube can efficiently maintained upon dataset updates and how to balance the query and update costs. Additionally the SkyCube, due to it's construction has redundant information between same level subspaces/cuboids and child-parent cuboids as can be seen in **Table 19**. For example point H1 is stored 9 cuboids and H6 in 7 cuboids out of 15 and additionally H2 in 3 out of 4 same level cuboids.

Materializing the SkyCube enables drill down and roll up analytical queries in order to identify additional interesting features and properties of the SkyCube which will help the user to make additional decisions based on his/her secondary concerns. As an example a skyline query on the 4-th dimensional subspace would return a number of skyline points. Then user can focus on a specific subspace (i.e. 2 dimensional *price - distance*) to see how the skyline points perform concerning only these two dimensions (i.e. are decisive to the full space of the query, belong to a subspace skyline, or are extreme points of a subspace skyline).

*Compressed SkyCube*

Authors in **[188][189]** proposed the *Compressed SkyCube (CSC)* which represents the *complete SkyCube* preserving the essential information of subspace skylines without accessing the whole dataset. The initial reason for this work was that the previously described method SkyCube (or complete SkyCube) did not take into account that the dataset is not always *static* but rather could be *dynamic (not to be confused with the dynamic skyline)*. That is objects of the dataset may change their attribute values over time, or other objects can be added on the dataset. In their work, authors try to balance the query with the update cost, by trading the query's efficiency with better storage and maintenance cost.

One naïve approach (extreme case) to deal with dynamic data is to update the pre-computed SkyCube every time a change occurs. In this case the update cost will be very high since it needs to re-compute a big part of the already pre-computed SkyCube, since every affected cuboid must be recomputed in order to reflect the correct results. In the case of a pre-computed SkyCube the benefit is that the query cost will be almost zero (due to pre-computation). Considerably additional overhead will occur especially when the updates are frequently or the dataset has many objects. Additionally incoming queries will be blocked during the update process.

Another extreme case would be the absence of pre-compute cuboids. In this case the query cost will be expensive since the query may need to access a large part of the dataset. Nevertheless the update cost will be minimized. In order to combine the benefits of the two extreme cases, authors aim to minimize the storage cost of the SkyCube and support more efficient updates, without compromising the overall query efficiency. Thus an efficient update mechanism, which balances the query and update cost would be more efficient than the initial (re)computation.

**Definition 21:** Minimum subspaces
Given a point p and two subspaces U, V, V⊂U of the original space ,the minimum subspaces of p, denoted as mss(p), is the set of all subspaces where ∀U∈*mss(p)*, *p* belongs to the subspace skyline of U and ∀V⊂U, *p* does not belongs to the subspace skyline of V ∎

The *compressed* SkyCube is based on the concept of *minimum subspaces which* makes possible the elimination of many duplicates entries on the complete SkyCube, thus improving its storage cost. This is achieved by storing each skyline point *p* to the cuboids of its minimum subspaces in which p is a skyline and additionally containing only the non-empty cuboids (thus, strictly speaking, cannot be considered that fully pre-materializes all the skylines). An example of minimum subspaces of dataset **Table 17** is shown in **Table 20**. Minimum subspaces can easily derive from **Table 19** looking from bottom to top for each skyline point. Following are the definition of the compressed skycube and its pseudocode.

**Definition 22:** Compressed SkyCube (CSC)
A point *p* is stored in a cuboid *U iff  U∈ mss(p). The compressed skyline cube (CS*C) contains all the non-empty cuboids *that* are created∎

| **ALGORITHM 15:** | QueryCSC($U_q$, $l$) **[19]** |
|---|---|
| **Input:** | Query subspace $U_q$ in level l. |
| **Output:** | The skyline sky($U_q$). |

1. **if** $U_Q$ is full-space, **return** the full-space skyline.
2. SK = ∅. /* sky($U_Q$). */
3. FP = ∅. /* false positives. */
4. For each non-empty level i that i ≤ l
5. For each non-empty cuboid V that V ⊆ $U_Q$,
6. **if** an object in V is in FP, **continue**.
7. **if** an object in V is in SK, **then** push objects dominated by it on $U_Q$ into FP.
8. compare the rest of the objects in V on dimensions $U_Q$, push skyline objects into SK and false positives into FP.
9. **return** SK.

The compressed SkyCube contains less duplicates entries among the different cuboids, in comparison with the complete SkyCube. As shown in Table 21 Compress skycube contains 9 non-empty cuboids in comparison to the complete skycube which contains 15. Additionally compress skycube has a smaller number of duplicates among the cuboids than complete skycube. As an example point H1 is stored in 9 cuboids in complete skycube and only in 2 cuboids in the compressed skycube.

| Point | Minimum Subspace | Point | Minimum Subspace |
|---|---|---|---|
| H1 | P,B | H7 | B, PA, PD |
| H2 | A | H8 | B, PA, PD, |
| H3 | DBA,PDA | H9 | DB, PD, |
| H4 | - | H10 | B, PA, |
| H5 | - | H11 | B, PDA |
| H6 | D | | |

Table 20: Minimum Subspaces.

| Cuboid | Skyline | Cuboid | Skyline |
|---|---|---|---|
| P | H1 | DB | H9 |
| D | H6 | PDA | H3, H11 |
| B | H1, H7, H8,H10,H11 | DBA | H3 |
| A | H2 | | |
| PD | H7, H8, H9 | | |
| PA | H7, H8, H10 | | |

Table 21: Cuboids of the Compressed SkyCube.

Additionally authors, for the case of updates, proposed an object-aware update scheme, in which updates of different objects will trigger different amount of computation, preventing unnecessary disk access and cuboid computations. In each case when a dataset update occurs, the CSC Storage is updated only if it is needed, in order to be up-to-date. Note that the full-space skyline plays a key role in their update scheme, since intuitively as long as the full-space skyline is unchanged, no new cuboid will be added to the compressed skycube.

*Skyey*

Authors in **[139]** discussed primarily the semantics of subspace skyline queries and the importance of the dominance relationships in the subspaces. They studied the *skyline membership query*, which tackles "why and in which subspaces an object belongs to the skyline", by introducing the notions of *skyline group* and using the general idea of *decisive subspaces*. A skyline group (G,U), of a subspace U (with G representing the set of points) contains the skyline points of U that share same attribute values in this subspace. Additionally every point in skyline group (G,U) does not share any value, on any other dimension not belonging to the subspace U, and none of the points that do not belong in G shares same values with points in G (essentially this restrictions forms the conditions of whether or not a subspace is a decisive subspace of a particular skyline group). In general, a skyline group will be formed by a group of points which share some common values in a subspace and the shared values are in the skyline of that subspace. A decisive subspace of a skyline group is a minimal set of attributes/dimensions that qualifies the group in the skyline of some subspaces.

Skyey algorithm works in a top-down manner and computes skylines of all subspaces of a given set of dimensions taking advantage of sharing sorting among subspaces. Essentially, searches all the subspaces for subspace skylines and merges them into skyline groups. Starting from the full space visits all the non-empty subspaces, that are enumerated by an altered version (top-down) of the enumeration tree, identifying the skylines points by initially sorting them in lexicographic order and then creates new skyline groups (forming a skyline group lattice structure with each skyline group forming a node) if some new skyline points are needed to be inserted into an old group. The skyey returns the skyline points of every subspace such as skycube and additionally, the skyline groups in the form of signatures (which contain the points and the decisive subspaces of the skyline group) as the summarization of the skylines in subspaces. However, the skyline group lattice may not be as efficient or scalable as the skycube. This can occurs because a skyline group lattice may have many more skyline groups than the cuboids in skycube, since one subspace may contain many skyline groups especially in high dimensional space.

*Stellar*

In [137] authors improved their work on [139] in order to sufficiently address the efficient skyline group and decisive subspaces computation. In this work also developed the *Stellar* algorithm which computes *skyline groups* and *decisive subspaces* without searching all subspaces for skyline points, by exploiting the skyline groups formed by the full space skyline points. In their performance study, compared *skyey* and stellar and showed that in most cases stellar performs better. *Skyey* computes the SkyCube as a byproduct, and *Stellar* computes the skyline groups directly.

### 4.1.2. Single Subspace Computation

Previously proposed methods, related with the subspace skyline computation computed the skylines of *all* the possible subspaces. This approach was selected because it is not known (unpredictable) in which and how many dimensions a user may want to retrieve the subspace skyline, so it is needed to compute every possible subspace skyline. From another perspective many times most of the users perform queries that are related with a small subset of dimensions (and might also be the same in their majority) with respect the full space.

*SubSky*

Differentiating, authors in [167] studied the computation of the skyline of *one* specified subspace, as opposed to all. This route was selected taking into account that most of the users in a high dimensional space (i.e. 10 dimensions/attributes) will perform a query about the skyline in few dimensions (i.e. in the house-metro station example for the two attributes, price and distance), thus it is not needed to compute all the subspace skylines but rather find a way to efficiently retrieve small-dimensionality subspace skylines. For that reasons, authors proposed the index-based algorithm *SUBSKY* which can efficiently find the skyline of any specified, low dimensionality subspace.

For the case of a uniform distributed dataset, for each point *p* (normalized in [0, 1]) is defined a 1D value $f(p) = max_{i=1}^{d}(1 - p.d_i)$ which is the $L\infty$ distance[1] between point p and the *maximal corner* of the dataspace called anchor point (i.e. the point of the dataset where all dimensions equal to 1). Thus, each of the d-dimensional points are transformed in a 1D value *(f (p))* based on which are indexed in a single B-tree (which will be used for the subspace skyline retrieval) and sorted by their *f* value in a descending order. In such a case the cost of maintaining the tree equals the cost of updating a traditional B-tree. In order to improve the pruning power and the efficiency of the algorithm, when the dataset forms clusters of points, authors proposed to create *m* anchor points and assign the points of the dataset to each one of the anchors. The *f(p) values ($L\infty$* distance) are computed between each point *p* and their corresponding anchor. As previously mentioned, the points are indexed in a single B-tree according to their f(p) value and it's corresponding anchors point. This is achieved by indexing the points with a composite key (*i,f(p)*) consisted of the numeric id *i* of the anchor point to which point p is assigned and it's f(p) value. With this kind of approach the subspace skyline of a specific subspace is retrieved minimizing the computations only to those related with that subspace (due to the construction and computation of *f(p)*) and without scanning the whole dataset (due to the indexing imposed by the B-tree).

| House | Price (in thousand €) | Distance (m) | Hospital (m) |
|---|---|---|---|
| H1 | 100 | 1500 | 1450 |
| H2 | 1400 | 500 | 920 |
| H3 | 700 | 600 | 320 |
| H4 | 1300 | 1000 | 1200 |
| H5 | 900 | 1300 | 230 |
| H6 | 1600 | 100 | 350 |

---

[1] $L\infty$ distance, Chebyshev distance or Maximum metric is a metric where the distance between two vectors is represented by the greatest of their differences along every dimension i.e. for two points on a 2-dimensional space $L\infty=\max(|x_1-x_2|, |y_1-y_2|)$.

| | | | |
|---|---|---|---|
| H7 | 400 | 300 | 1300 |
| H8 | 200 | 1200 | 150 |
| H9 | 1000 | 200 | 1500 |
| H10 | 500 | 1400 | 570 |
| H11 | 500 | 900 | 600 |

Table 22: 3-dimensional dataset for SUBSKY.

In order to illustrate an example consider the 3-dimensional dataset of Table 22. The dataset is based on the house-metro station dataset (Table 1) but has an additional dimension named *Hospital* which represents the distance of the house from the nearest hospital in meters. In this case consider that the 3-dimensional dataset is the full-space. Moreover let a user that wants to identify the subspace skyline of the 2-dimensional space {Distance, Hospital}, named SUB={D,H}. Note that the dataset values must be normalized in [0,1], so all the values are divided with 1600 which is the largest value on the dataset. The normalization of a dataset was also described in section 2.3.6 related with the SFS algorithm. The normalized values of the the 3-dimensional dataset are presented on Table 23. The algorithm first computes the $f_{max}(p)$ values of all points taking into account all the dimensions of full-space and processes the points in ascending order of their values $f_{max}(p)$ as shown below. Additionally computes the *values* $f_{min}(p) = \min_{i \in SUB}(1 - p.d_i)$ *based* only on the values of points in the desired subspace dimensions. Note that the algorithm considers that none of the point remaining can be qualified as skyline point if their $f_{max}(p)$ value is smaller than the largest $f_{min}(p)$ value of all the previously processed points. For that reason it maintains a value U which corresponds to the largest $f_{min}(p)$ value of all points processed so far. Thus the algorithm terminates when $f_{max}(p) < U = f_{min}(p_i)$.

| House | Price (in thousand €) | Distance (m) | Hospital (m) | $f_{max}(p)$ | $f_{min}(p)$ |
|---|---|---|---|---|---|
| H1 | 0,0625 | 0,9375 | 0,90625 | 0,9375 | **0,6250** |
| H6 | 0,875 | 0,3125 | 0,575 | 0,9375 | **0,78125** |
| H8 | 0,4375 | 0,375 | 0,2 | 0,90625 | 0,25 |
| H9 | 0,8125 | 0,625 | 0,75 | 0,875 | 0,0625 |
| H5 | 0,5625 | 0,8125 | 0,14375 | 0,85625 | 0,1875 |
| H7 | 1 | 0,0625 | 0,21875 | 0,8125 | 0,1875 |
| H3 | 0,25 | 0,1875 | 0,8125 | 0,8 | 0,625 |
| ~~H2~~ | 0,125 | 0,75 | 0,09375 | 0,6875 | 0,425 |
| ~~H10~~ | 0,625 | 0,125 | 0,9375 | 0,6875 | 0,125 |
| ~~H11~~ | 0,3125 | 0,875 | 0,35625 | 0,6875 | 0,4375 |
| ~~H4~~ | 0,3125 | 0,5625 | 0,375 | 0,375 | 0,25 |

Table 23: Normalized values for 2-dimensional dataset.

Based on the f(p) values the points will be processed in the order H1, H6, H8, H9, H5, H7, H3, H2, H10, H11, H4. After examining the first point which is H1 is added to the skyline set since is the only point and U is set to 0,0625. Next H6 is added to the skyline set, H1 is pruned since it is dominated by H6 and U is set to 0,78125. Next point to be processed is H8 which is also added to the skyline list but U remains the same. Points H9, H5, H7 are discarded since are dominated by points already in the skyline set. Next point H3 is added to the skyline. Since $f_{max}(H7) < U = f_{min}(H6)$ none of the following points can be a skyline point and the algorithm terminates. Following is presented the pseudocode of the SUBSKY algorithm.

**ALGORITHM 1:** BASIC-SUBSKY(*SUB*) **[168]**

**Input:**   The dimensions $SUB$ that are relevant to the query.
**Output:**   The subspace skyline S$_{sky}$.

1. *DB*=the given set of data points sorted in descending order of their *f*-values.
2. *U*=0; S$_{sky}$=Ø

3.   P=the first point in *DB*.
4.   **while** p≠∅ and U≤f(p) **do**
5.      **if** p is not dominated by any point in *DB*.
6.         **add p to** $S_{sky}$
7.         remove from $S_{sky}$ the points dominated by p.
8.      U=the maximum $\min_{i \in SUB}(1 - p'.d_i)$ of all p'∈ $S_{sky}$
9.      set p to the next point in *DB*.
10.  **return** $S_{sky}$

In [168] extended their work where they discuss about the applicability of existing full-space skyline algorithms in subspace and extend the *SUBSKY* algorithm to compute *k-skyband* [134] and *top-k* queries [133, 134] in subspace.

The problems that arise with the SUBSKY as discussed by authors and outlined in [196] are the following. First, when using single anchor points, the pruning ability of the algorithm deteriorates as the query's dimensionality increases which makes it inefficient for computing the skylines of all spaces. This happens due to the way the stop condition is computed, which results in smaller pruning region than the actual dominance region that a skyline points has. Second, in the case of clustered dataset the anchor points (which their selection affects the pruning power) are never modified since their initial computation, making the algorithm inefficient for dynamic datasets when large updates occurs as verified by authors experiments. Third, it isn't progressive since it cannot determine if a point belongs to the candidate set until the search terminates.

### 4.1.3. Top-k and K-dominant

In this section will be described some skyline ranking algorithms to identify the k most interesting skyline points on a subspace, using some newly proposed ranking metrics.

*Skyline frequency*

To deal with the problem of returning to many interesting points in high dimensional spaces, when different subspaces are considered, authors in [26] focused on ranking skyline objects and introduced a new metric called skyline *frequency*. This metric ranks the interestingness of a skyline point and retrieves the skyline points in a top-k fashion. The ranking is based on how often a point is returned as a skyline point, when different subspaces are considered. The method orders the points in a decreasing order of the number of subspaces in which they belong to a subspace skyline and returns the top-k ranking.

**Definition 23:** Skyline frequency
Given a d-dimensional dataset D and a point p∈D the *skyline frequency* of point *p* denoted as *f(p)*, will be equal to the number of subspaces in which *p* is a skyline point∎

Points with high skyline frequency are more interesting than others, since they can be dominated on fewer subspaces. A naïve subspace-based approach will be to compute the skyline points of each subspace, with an existing algorithm and then compute the skyline frequency of each point *p* by summing the number of subspaces for which *p* is a skyline. This approach requires $2^d-1$ distinct subspace skyline queries (for a d-dimensional space) and essentially enumerates each subspace to compute skylines.

**Definition 24:** Top-k frequent skyline points
The set S of Top-k frequent skyline points of the dataset D will contain the k points that have the highest skyline frequency among all the points in D∎

To avoid the expense of computing all the subspace skylines, authors proposed an approximate algorithm for computing *top-k frequent skylines* which contains the top-k points whose skyline frequencies are the highest. Thus, the problem becomes one of finding top-k frequent skyline points.

Nevertheless the algorithm cannot answer dynamic queries and in some cases points that are intuitively superior may get the same or lower ranking as existing inferior points [25].

### *K-dominant*

As dimensionality increases the chance of one point to dominate another is very low. This leads in the retrieval of numerous skyline points, which cancels any interesting insights on the dataset. An efficient approach is to relax the notion of dominance to k-dominance in order to consider only *k* among *d* dimensions and retrieve only important and meaningful skyline points. For that reason, same authors of previous work ([26]) proposed in [25] the k-dominant skyline query (not to be confused with k-dominating queries). A k-dominant skyline query retrieves a representative subset of skyline points from a high d-dimensional dataset. This type of query can essentially be used to decrease the number of returned skyline points. This will increase the probability of a point to dominate another point, since there exist more points that will be *k*-dominated than *d*-dominated, thus allowing the existence of fewer skyline points. This will also reduce the skyline cardinality. Note that a k-dominant skyline is a subset of the original skyline. In the case of *k*=d, in a d-dimensional space, the k-dominant skyline query becomes the conventional skyline. Authors state that their work maybe useful for *dominant relationship queries* (DRQs) [108] which support microeconomic data mining.

A d-dimensional point p is said to k-dominate (k < d) another point q if p is not worse than q in at least k arbitrary dimensions and p is better than q in at least one dimension. A k-dominant skyline point is a point that is not k-dominated by any other point. The k-dimensions must not be necessarily the same, but this voids the transitivity property as will be discussed after the following formal definitions:

**Definition 25:** K-dominance
Given a d-dimensional space D=$\{d_1,d_2,d_3,...d_d\}$, a subspace S=$\{s_1,s_2,s_3,...s_n\}$, a dataset Ds $\in$D and two points p,r$\in$Ds , point p is said to k-dominate r iff $\exists S\subseteq D$, $|S|=k$, $\exists s_i \in S$ such that $p.d_i<r.d_i$ and $\forall s_j \in S - \{s_i\}:p.d_i \leq r.d_i$ ∎

**Definition 26:** K-dominant skyline
A point p$\in$DS is a k-dominant skyline point iff $\nexists$r$\in$DS –{p}such that r k-dominates p∎

For the *k*-dominant skyline queries the property of transitivity (section 2.1) is no longer valid making the existing algorithms inapplicable. To make it more clear this can happen in the case of any k<d as will be explained. Consider for example thar for points *p,r,z,* holds that point *p* k-dominates point *r, r k'-dominates z* (in k different dimensions than those of *p and r)* and z k''-dominates *p* (also in k different dimensions than those of *r and z),* thus forming a *cyclic dominance*. In this case points *p,r,z* are not considered to be *k-dominant skyline* points. Due to *cyclic dominance* points that are k-dominated cannot be discarded immediately, since they might be used to exclude other points through another k-dominant relationship.

In their work authors proposed three algorithms to solve the k-dominant skyline problem. The proposed algorithms for *k*-dominant skyline queries are the One-Scan-Algorithm (OSA), Two-Scan-Algorithm (TSA) and Sort-Retrieval-Algorithm (SRA). OSA scans the dataset once and identifies the k-dominant skyline points by computing the conventional skyline points and using them to prune away the non *k*-dominate skyline points. TSA algorithm scans the dataset twice. In the first scan the algorithm retrieves a set *R* of candidate dominant skyline points by comparing every point *p* of the dataset *D* with the points that exist in this set. Note that when this step finishes the set *R* may contain false-positives. To eliminate the false positives a second scan eliminates all the non-dominant skyline points (false hits from previous scan) from that set. In this step the algorithm compares all the points of the set R with every point in {D}-{R}. Finally, authors proposed the elimination-based approach SRA, which was inspired from the rank aggregation algorithm proposed by [52], as also happened with *Index* algorithm [164], which pre-sorts points according to their dimensions in separate ranked lists and then "merges" them. Following is the pseudocode of the TSA algorithm.

| ALGORITHM 16: Two-Scan Algorithm(D,S,k) [25] | 5.     if ($p'$ $k$-dominates $p$) then |
|---|---|

```
ALGORITHM 16:   Two-Scan Algorithm(D,S,k) [25]
Input:   A dataset D, a subspace S, and a parameter k.
Output:  The set R of k-dominant skyline points.

1.  initialize set of k-dominant skyline points R = ∅
2.  for every point p ∈ D do
3.      initialize isDominant = true
4.      for every point p' ∈ R do
5.          if (p' k-dominates p) then
6.              isDominant = false
7.          if (p k-dominates p') then
8.              remove p' from R
9.      if (isDominant) then
10.         insert p into R
11. for every point p ∈ D do
12.     for every point p' ∈ R, p' ≠ p do
13.         if (p k-dominates p') then
14.             remove p' from R
15. return R
```

Depending on the value of k, the number of returned k-dominant skyline points can be very large or on the other hand very few and in some cases none, due to cyclic dominance. In order to choose a good k value, that guarantees a small but non-empty set of dominant skyline points authors proposed the top-*δ* dominant skyline query. Its purpose is to find the smallest *k* such that there are more than *δ* *k*-dominant skyline points.

In some cases certain attribute/dimensions of points/space are more important than others. For that reason authors extend the *k-dominant skyline* by incorporating *d* positive weights $w_1,...,w_d$. These weights are assigned to the d-dimensions. A point p is said to be a weighted *w-dominant skyline* point if there does not exist any point *r* that can dominate p on any subset of dimensions whose sum of weights is over *w*.

*Telescope*

Authors in [103, 104] identified that the metrics k-dominant [25] and Skyey [139] cannot adapt their ranking criteria to user specified preferences. For that reasons proposed the personalized skyline ranking algorithm *Telescope* which identifies truly interesting points by dynamically searching over the SkyCube. The identification of the top-k skylines is guided by the user-specified retrieval size and preferences over the available dimensions of the dataset. In order to illustrate the algorithm, a weighted lattice graph is used, which represents the user's weighs, and the relations between subspaces. This graph is then transformed, by pruning multiple links, in order to retrieve the ordering of the subspaces and guarantees that not any node will be visited twice. Additionally to reduce storage overhead authors discuss the adoption of the Compressed SkyCube [188][189].

*SkyRank*

In [177] proposed the skyline ranking algorithm SKYRANK, and defined the interestingness of a skyline point *p* (basic on full space), based on the number of subspace skyline points that dominates, in all the possible subspaces of the original dataspace. SkyRank finds the interesting points by mapping the dominance relationships between skyline objects for different subspaces, in a weighted directed graph, called skyline graph [176]. Initially the algorithm process in the absence of user-defined preference functions but authors shows that can easily be enhanced to support preference skyline queries (personalized top-k skyline queries) by using link-based web ranking technics such us [23].

### 4.1.4. Space partitioning

As with the fundamental algorithms there is an area of research that concentrates on processing the skyline queries by partitioning the dataspace. Most of the partition-based algorithms exploit the property of incomparability and block-based dominance. Especially the property of incomparability is very useful in high dimensional spaces. This section will reason about algorithms that partition the space directly or by incorporating mapping techniques such as specific space filling curves. Most of the algorithms in this section perform full-space skyline computation but are analyzed here because their computations are based on sub-regions or even on subspaces of the original space.

*Z-order based*

The access order of data points has direct impact on the performance of algorithms since the early identifications of dominant and highly-dominating points can eliminate unnecessary domination comparisons. Additionally pairwise point-to-point dominance comparisons have considerably computational and time cost which can be avoided by block based-comparisons.

Taking into account the previous considerations authors in [106] proposed an approach based on the popular dimension reduction technique, Z-order space filling curve (or Z-order curve, in short) [55, 145] which carry many good geometric properties for skyline processing. Z-order curve maps multi-dimensional data points to one-dimensional points, where each point is represented by a bit-string computed by interleaving the bits of it's coordinate values, called *Z-address*. As an example point (1,2) can be represented by the 1-dimensional value 001001, where coordinates (x,y) have a 3-bit representation of (001, 010). Z-order curve, starts from the origin point and passes through all coordinates and data points in space. It's name derives from the access sequence of points which follows an exact (and rotated) 'Z' order. Z-addresses have a hierarchical nature (property) meaning that points located in the same regions have similar Z-addresses. This leads points to be naturally mapped onto Z-order curve segments, which are in turn grouped (clustered) into blocks to facilitate efficient dominance tests and space pruning. A z-order curve provides the properties of *monotonic ordering* and *clustering* (based on their locations in a multidimensional space). These properties perfectly match the *transitivity* and *incomparability* properties of existing skyline problem. Other space filling curves, such as Hilbert and Peano curves are not suitable for skyline query processing since they do not hold the monotonic ordering property [106]. By construction and due to the ordering and transitivity property of Z-address, when points are sorted in non-descending order (of Z-addresses), those points who have large Z-address cannot dominate points with small Z-address. Thus points, that passed the dominance tests against all the previous retrieved skyline points obtained ahead of it, are guaranteed to be skyline points, as also inspired from the sorting-based approaches [35],[59]. The clustering feature has the advantage that if two points or regions are known to be incomparable, dominance tests between them can be avoided. The incomparability inspired from the divide-and-conquer approaches [19] thus facilitating *true block-based dominance* tests

**Figure 36** illustrates an abstract approximation of how a Z-order curve will visit the points of the house-metro station example. Note that the precise curve of this dataset would be hard to illustrate and understood. So for abstraction reasons the curve is visiting only the points in a scale of a hundred in the region [0,0]x[1500,1500].

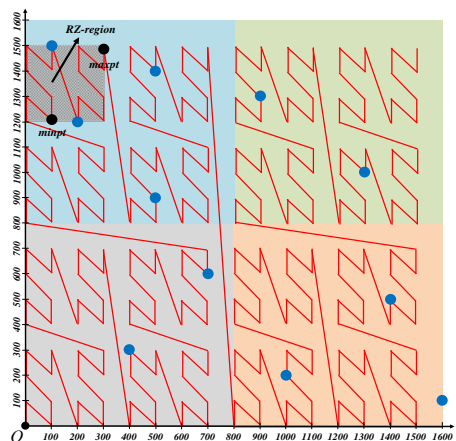

Figure 36: Z-curve access order.

When points are ordered by their Z-addresses will be naturally clustered as regions based on the iterations of Z-curve. Additionally when points are monotonic ordered based on non-ascending order of their Z-addresses, dominating points will be placed before their dominated points. In order to maintain these two properties and allow them to be incorporated in the skyline query processing authors proposed a new index data structure called *ZB-tree*, in order to index the points of the dataset and maintain the set of skyline points. Also developed the algorithms *ZSearch*, *ZUpdate* and *k-ZSearch*, to process skyline queries, skyline result updates, and *k*-dominant skyline queries, respectively.

A ZB-tree is newly proposed index structure which is a variant of B-tree and based on Z-order curve to facilitate searches/updates. It uses the Z-address of the d-dimensional points as key. A ZB-tree divides a Z-order curve into disjoint segments. Each segment represents a region which preserves the clustering property. The space covered by a segment is called *Z-region*. Since each Z-region can be of any size, RZ-regions are used to represents the smallest square region that covers a bounded Z-region. An RZ-region can be specified by two z-addresses such as *minpt* and *maxpt* in **Figure 36.** These bounding regions can be used for region based (block based) dominance test as done by BBS in order to alleviate the point-to-point computational overhead. Following is presented the pseudocode of the algorithm *Dominate* which is invoked by the *Zsearch* algorithm in order to perform dominance tests. The algorithm takes as an input a set *SL* which contains the identified skyline points that are maintained by the *Zsearch* algorithm and the *minpt* and *maxpt* which define the endpoints of an RZ-region. The algorithm checks if a given RZ-region, that belongs to the dataset and is represented by the endpoints, contains potential skyline points by comparing against all the identified skyline points that belong in *SL.* The algorithm traverses the *SL* on a breadth-first way. This way RZ-regions of high level nodes from *SL* will be compared against the RZ-regions defined by the endpoints and drill down *SL* if further examination is needed.

| **ALGORITHM 17:** Dominate(SL,minpt,maxpt) [106] | |
|---|---|
| **Input:** | A ZB-tree SL indexing skyline points. The endpoints *minpt* and *maxpt* of an RZ-region.. |
| **Local:** | A Queue q. |
| **Output:** | *True* if input is dominated and false otherwise. |

1. *q*.enqueue(*SL*'s root);
2. **while** *q* is not empty **do**
3.    var *n*: Node;
4.    *q*.dequeue(*n*);
5.    **if** *n* is a non-leaf node **then**
6.      **forall** children node *c* of *n* **do**
7.        **if** (*c*'s RZ-region's *maxpt* ≺ *minpt* **then**
8.           output TRUE; /* Case 1 of Lemma 1 */
9.        **else if** (*c*'s RZ-region's *mint* ≺ *maxpt*) **then**
10.           *q*.enqueue(*c*); /* Case 2 of Lemma 1 */
11.    **else** /* leaf node */
12.      **forall** children point *c* of *n* **do**
13.        **if** (*c* ≺ *minpt*) **then**
14.           output TRUE;
15.    output FALSE;
16. **END**

Z-search can efficiently identify skyline points in full space and scales well with dimensionality and cardinality. The algorithm computes the skyline set by accessing the points in z-order allowing block-based dominance tests due it's cluster fashion (regions), asserting efficiently if a block of points is dominated. Following is presented the pseudocode of the Zsearch algorithm. The algorithm takes as an input a dataset *SRC* and outputs the set *SL* of skyline points. At each round the RZ-region of each node in *SRC* is compared against *SL* (which at runtime contains the candidate skyline points) by invoking the *Dominate* algorithm that was described previously. If a corresponding RZ-region is not dominated, the node is further explored (line 7-13). Otherwise if a point is not dominated by any candidate skyline point in SL is inserted into it.

| **ALGORITHM 18:** Zsearch(SRC) [106] | |
|---|---|
| **Input:** | A dataset SRC indexed by a ZB-tree. The endpoints *minpt* and *maxpt* of an RZ-region.. |
| **Local:** | A Stack s. |
| **Output:** | A ZB-tree *SL* with the skyline points. |

1. *s*.push(*source*'s root);
2. **while** *s* is not empty **do**
3.    var *n*: Node;
4.    *s*.pop(*n*);
5.    **if** *Dominate*(*SL*, *n*'s *minpt*, *n*'s *maxpt*) **then**
6.      goto line 3.
7.    **if** *n* is a non-leaf node **then**
8.      **forall** children node *c* of *n* **do**
9.        *s*.push(*c*);
10.   **else** /* leaf node */
11.      **forall** children point *c* of *n* **do**
12.        **if** not *Dominate*(*SL*, *c*, *c*) **then**
13.           *SL*.insert(*c*);
14. output *SL*;
15. **END**

Authors additionally proposed ZUpdate which efficiently updates the skyline results if insertions or deletions occur. This algorithm will be described in more detail in the specific section 5.2.3 about skyline maintenance. *k*-ZSearch is used to answers the *k*-dominant skyline queries by utilizing a two-phase filter-and-reexamine approach. This way if a cluster of points is found to be k-dominated, further examination of points belonging to this cluster is prevented. This allows the improvement of search performance. The algorithm can also efficiently handle the cyclic dominance problem which occurs in *k*-dominant skyline queries.

In [105] authors extended their work, proposed new algorithms and studied the problem of ranking and subspace skyline query processing. Particularly proposed (in addition with the previous) the algorithms *ZBand*, *ZRank* and *ZSubspace* to evaluate skyband queries, top-ranked skyline queries and subspace skyline queries (on a specified subset of dimensions), respectively. Along with the previously proposed algorithm *ZUpdate*, introduced the *ZInsert* and *ZDelete* for incrementally updating of the skyline query results. Finally, authors extended the ZB-tree to handle the additional information, in order to facilitate the counting of dominated points, needed by skyband and top-ranked skyline queries.

All the above elements are incorporated in the Z-SKY skyline query processing framework which essentially contains the dataset and a set of skyline candidates placed on a ZB-tree, the algorithm library, and the various dominance relations. Authors extensively compare their suite with the state-of-the-art skyline algorithms SFS [35], SaLSa [9], BBS [133, 134], BBS-Update [134], DeltaSky [186], TSA [25] and SUBSKY [167] showing that their algorithms are at least as good and most of the time outperforms the current algorithms.

*LS algorithm*

The dominance based elimination of points is highly influenced by the dataset distribution. Especially the performance of sorting based algorithms degrades if the distribution of the dataset tends towards anti-correlated. Authors in [130] proposed the Lattice Skyline (LS), to answer skyline queries of low-cardinality domains, which uses a static lattice structure to determine the dominance between the various combinations of distinct attribute values in the dataset. An example of a dataset with low-cardinality is the one which has Boolean-valued attributes in which one attribute might reason if the house has central heating or not which corresponds to True-False values. Note that the attribute value combinations that can derive from a 2-dimensional Boolean-valued dataset are very few (specifically only four). This work defers from the previous ones because, instead of considering a lattice of subspaces, considers the full data space as lattice to evaluate the skylines. Thus it is based on a lattice structure and is independent of the dataset ordering and it's initial distribution. The algorithm is applicable in the case where all attributes derive from low cardinality domains or can be naturally mapped to them. Also is applicable in cases where (only) one attribute derives from a high cardinality domain. When the number or size of the domains is large the algorithm becomes inapplicable [200].

*Incorporating Incomparability*

In order to minimize the computational cost in high dimensional spaces authors in [200] proposed a progressive object-based space partitioning (OSP) algorithm, which recursively divides the d-dimensional space into $2^d$ separate partitions. The algorithm is based on partition-wise dominance, which exploits the incomparability property, and organizes the retrieved skyline points in a search tree, allowing points to be compared only with a small number of skyline points. The partitions are encoded in bitmaps and organized by a left-child/right-sibling (LCRS) tree allowing index space efficiency by indexing only the non-empty partitions and fast bitwise comparisons between them. Additionally allows the partitions and points to be sequentially accessed in a breadth-first fashion. The benefit of this technique is that the number of skyline points and candidate skyline points to be compared decreases as the dimensionality increases. Finally authors showed that their work can be extended to facilitate k- dominant skyline queries, parallel skyline computation, and skyline computation on partially ordered domains.

Authors in **[102]** proposed the BSkyTree which tries to find a cost-optimal strategy for skyline processing by exploiting the properties of both dominance and incomparability. While most of the methods are dominance based, incomparability is very useful in high-dimensional spaces since most pair of points become incomparable. An approach that only optimizes incomparability would be to select a pivot point that partitions evenly the space. In case of dominance optimization would be to select a pivot point than maximizes the dominating region. This method tries to efficiently process the skyline computation by selecting specific pivot points based both on dominance and incomparability.

### 4.1.5. Subspace Skyline Queries & Constraints

In many cases it might be desired to query a specific region of the dataset. These constraints may be also a part of a user's preferences. Taking into account these considerations the following work has been conducted. Note that full space constraint skyline queries have been outlined in section **3.1** based on the work of **[133, 134].**

*Constrained Subspace Skyline Queries*

In **[45]** authors studied the problem of supporting constrained subspace skyline queries which deals with finding subspace skylines within a specific query region. For this problem the data space is vertically partitioned to several low-dimensional subspaces, each of them indexed by an R-tee. Naively, if the queried subspace is entirely covered by an indexed subspace the desired query can be solved with a modified version of the index-based algorithm BBS **[133, 134].** This paper concerns the cases where the queried subspace is not covered entirely by any of the indexed subspaces. In order to compute the constrained subspace skyline in this case authors proposed a *threshold based Skyline (STA)* algorithm which merges synchronously the results of constrained nearest neighbor queries that are placed simultaneously on multiple indexes.

*Skyline queries with Constrains*

In **[197]** authors study the case of supporting skyline queries with constrains on a subspace. They try to deal with the problem of returning to many points to the user, when dimensionality is high and allowing the incorporation of preferences (constrains) in different subspaces. Thus, proposed a progressive algorithm which incorporates a b-tree and compared it with SUBSKY. It is useful to mention that they the problem in a different way than constrained skylines mentioned on **[133, 134]** since the algorithm returns a restricted (by the constraints) set of the retrieved skyline points.

### 4.1.6. Various related topics

In **[193]** authors introduced the concept of group-by skyline cube. Specifically, this work is related with the field of OLAP (On-Line Analytical Processing) which in business environments stands for data access and analysis, based on multidimensional views of business data, in the search of business intelligence. A key element for OLAP techniques are Data cubes. Authors proposed instead of building the data cubes with an aggregate function such as SUM () to use the skyline operator forming the so-called group-by skyline cubes.

### 4.1.7. Summary

In multiple subspace skyline computation the algorithms SkyCube **[196]** and Skyey **[139]** where parallel developed by a different group of authors, reasoning about the same problem. Continuing, their works where individually extended. From previous described algorithms, partition based approaches are the algorithms Z-search, LS, OSP and BskyTree. Partitioned based approaches that are based on incomparability are OSP and BSkyTree. From these approaches Z-search can be extend on subspace, k-dominant and k-skyband. The OSP can be extended to k-dominant skyline. Z-search and LS are based on dominance comparisons while OSP and BSkyTree on incomparability. **Figure 37** illustrates the hierarchy of subspace and space partitioned skyline computation.

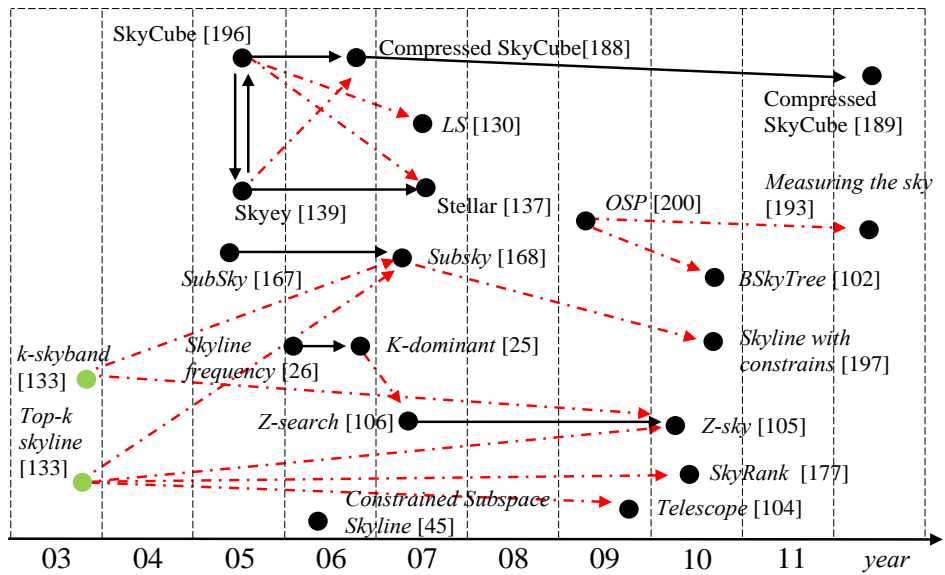

**Figure 37: Subspace skyline queries hierarchy.**

**Table 24** summarizes the state-of-the-art algorithms related with subspace and space partitioned skyline computation. The column *Approach* corresponds to whether the algorithm pre-computes the results, uses indexes or sorting and counting techniques in order to answer a requested query. The *No. of subspaces* corresponds to whether the algorithms make their computations for all the possible subspaces, a single subspace or is generally applicable to full-space computation as with the partitioned based approaches. The column *Dataset* corresponds to weather the algorithm is applicable in dynamic datasets where updates and deletions occur.

| Algorithm | Approach | No. of Subspaces | Dataset | incorporates |
|---|---|---|---|---|
| SkyCube [196] | pre-materializing | all | static | cost sharing strategies |
| Compresed SkyCube [188][189] | pre-materializing | all | dynamic | minimum subspaces |
| Skyey [139] | pre-materializing | all | static | Skyline groups – decisive subspace |
| Stellar [137] | pre-materializing | all | static | skyline groups – decisive subspace |
| Subsky [167] [168] | index | single | dynamic | L∞ distance, anchor points |
| Skyline frequency [26] | counting | single | static | maximal dominating subspace |
| k-dominant [25] | counting/ sorting | single | static | conventional skyline - nested loops |
| **Space Partitioning** | | | | |
| Z-search / Z-sky [106] [105] | sorting - index | full space/single | dynamic | Z-order curve, ZBtree |
| OSP [200] | Index | full space | dynamic | incomparability, LCRS tree, bitmaps |
| LS [130] | Lattice structure exploration | full space | static | lattice structure |
| BSkyTree [102] | Cost-based partitioning | full space | static | pivot points, incomparability |

**Table 24: State of the art subspace skyline algorithms.**

## 4.2. Parallel & Distributed computation

Due to the high skyline query processing cost of previously proposed centralized architectures in section **3**, research has focused in parallel and distributed skyline query processing. However, previously proposed algorithms cannot be directly applied in parallel or distributed environments.[2]

This section will reason at first for parallel and distributed skyline computation without taking into account any relaying infrastructure or any partition technique as happens with ad-hoc networks. Following will be described the methods that involve partitioning in a vertical or horizontal manner. In general the efficiency of many of the proposed methods relies on the use of filtering points, sorting techniques or both.

The main problem that arises with centralized architectures is the increased computation cost, which leads in increased response time, in high cardinality and dimensionality datasets. A first approach to deal with this problem is parallel skyline computation in which servers (processors) communicate in a partially serialized way (avoiding the communication with all server), reporting their local skyline results to other servers. Thought, this leads to an increased waiting time for results from other servers (processors). Additionally many of the transferred results do not contribute to the final skyline. A related approach which concerns random partitioning and parallel skyline computation is **[37]**.

Parallel and distributed skyline computation was proposed in order to deal efficiently with the problem of waiting time and the dependence on the results of other severs. This way it is achieved immediate local skyline computation after a server receives its corresponding dataset. A general approach for this method, in a shared-nothing architecture, involves a central coordinator server, which partitions the dataset (or dataspace) in N partitions equal to the number of servers that are related with him. Then each server computes its local skyline, based on the dataset that has been assigned to him, and reports his results to the coordinator which computes its own local skyline and performs a merging process in order to retrieve the final skyline. This method also described in the literature as *all-local-skyline* (ALS) method **[208]**. Nevertheless, this approach highly depends on the method that is used for the local skyline computation and the partitioning technique followed. Additional the workload isn't evenly distributed to servers since the cardinality of each partition will follow the distribution of the initial dataset. Also the data points that are transferred from the servers to the coordinator contain many points that will not be part of the final skyline resulting in increased communication and merging overhead on the coordinators side. In general servers whose partitions are closer to the axes (in a uniform distribution) or the origin of axes (in a correlated distribution) tend to have more final skyline points and thus contribute more to the final skyline result. For that reason, partitions that lie in the lower corner of the dataset is wisely to be processed first in order to identify as many final skyline points as sooner as possible.

Some early (and the first in this sub-field) proposed work in distributed skyline computation considered the partitioning of the dataset in a vertical manner. Nevertheless this approach isn't adopted by the rest of authors, since the widely adopted partition method in P2P architectures is horizontal partitioning. The main approach that was followed, considering the partitioning technique, is *horizontal partitioning* which will be described in the later **section 4.2.2**. Work related with Wireless Sensor Network (WSN) is outlined in **section 4.9**. The reason for that is the need to discuss further fundamental notions (such as **4.4 Continuous skyline computation**) before investigating this subject.

### 4.2.1. Vertical partitioning

This section will illustrate the (early proposed) work on parallel and distributed skyline computation which considers vertical partitioning of the dataset. This partitioning approach was not adopted by the rest of authors due to the wide adoption of horizontal partitioning on highly distributed environments, such as peer-to-peer networks.

---

[2]The work of **[74]** is the only survey in skyline query processing, which is focused on parallel distributed skyline query processing.

*BDS-IDS/PDS*

Authors in [7] were the first that studied the problem of skyline query processing in distributed environments and especially in a specialized Web setting where each one of the dimensions are stored on a different Web-accessible site (source/database). In their work proposed the algorithms *basic distributed skyline (BDS)* and *improved distributed skyline (IDS)* which are inspired from the TA approach proposed in [52]. As a remainder the same was considered in the Index algorithm [164] in section 2.3.3.

As mentioned previously the d-dimensional dataset is vertically partitioned in d-lists where each list contains the attribute values of points in one of the dimensions. Each site stores one of the lists sorted in ascending order. A centralized site is responsible for probing attribute values from each site in order to calculate the final skyline. In BDS algorithm the retrieval of points is done by using sorted access on each site, in a round-robin fashion through all sites, retrieving one attribute value at each time. The algorithm continuous in the same way until it has finally achieved to retrieve all the attribute values of a (random) point (called the *terminating object*). Points that are not "seen" during this process can be pruned as they must be dominated by the *terminating object*. Then the centralized site retrieves all the attribute values, from all sites, for every point for which at least one dimension was previously retrieved. Finally the algorithm filters out the non-skyline points from this set, retrieving the final skyline. Following Table 25 and Table 26 illustrate the two lists that are stored in the two sites in an ascending order of their values, derived by the 2-dimensional dataset on Table 1. At beginning the algorithm visits *list 1* and retrieves the value of the attribute price of the first point which in this case is house H1. In the next step visits *list 2* and retrieves the attribute value *distance* of the first point of the list which is house H2. In the next round, following the round-robin fashion, algorithm visits again *list 1* and retrieves the attribute value *price* of the next point which is the point H8. Continuing in this way, at the third round and on the second step retrieves the attribute value *distance* of point H7 which is eventually the *terminating object* because the algorithm has retrieved all the attribute values of one point (H7). At this point the algorithm visits all the lists and retrieves all the missing attribute values of all points that had visit till the identification of the *terminating object*. From the set of those points computes and retrieves the final skyline.

| House | H1 | H8 | **H7** | H11 | H10 | H3 | H5 | H9 | H4 | H2 | H6 |
|---|---|---|---|---|---|---|---|---|---|---|---|
| **Price** | 100 | 200 | **400** | 500 | 500 | 700 | 900 | 1000 | 1300 | 1400 | 1600 |

Table 25: List 1- dimension *price*.

| House | H6 | H9 | **H7** | H2 | H3 | H11 | H4 | H8 | H5 | H10 | H1 |
|---|---|---|---|---|---|---|---|---|---|---|---|
| **Distance** | 100 | 200 | **300** | 500 | 600 | 900 | 1000 | 1200 | 1300 | 1400 | 1500 |

Table 26: List 2 - dimension *distance*.

The *IDS* algorithm is an enchased version of *BDS* which does not visit the list in round-robin fashion but rather always accesses the most *promising list,* in terms of earlier *terminating object* identification, in order to reduce the number of visits on the lists. As previously the algorithm accesses *list 1* and retrieves the attribute value of the first point which is H1. Differentiating from *BDS* it performs a random access in *list 2* to identify the position of point H1 which is 11[th]. This indicates that the algorithm must examine 11 points on *list 2* before it can terminate, since at this time this is the terminating object considered. Next the algorithm continuous and retrieves the attribute value of the first point of *list 2* which is H6. Again it randomly accesses *list 1* this time to identify the position of point H6 (retrieved from *list 2*) in *list 1*. Its position in *list 1* is also 11[th] (as is point H1 on *list 2*) and thus none of the two points retrieved so far are more promising than another as an earlier *terminating object*. Thus, the algorithm continuous and retrieves the attribute value of the next point in *list 1* which is H8. The random access on *list 2* indicates that point H8 is in the 8[th] position making it more promising as an early *terminating object*. This happens because the algorithm will need in this case only 7 more sorted accesses (has already done one sorted access on list 2) on list 2, in order to retrieve both the attribute values of H8, rather than 10 more in the case of H1. Thus the algorithm will continue with sorted access retrieval on the *list 2* (even it wasn't its turn). The algorithm retrieves the attribute value of point H9 which is identified to be in the 8[th] position making it again no more promising than H8. Next continues normally its operation retrieving the attribute value of point H7 from *list 1* which is identified to be 3rd in the *list 2* making it more promising from H8 and H9. Thus the algorithm continues with a retrieval in *list 2* and retrieves the second attribute of point H7 which eventually is the *terminating object*. As previously retrieves all the attribute values of points that have been visited and computes the final skyline. However the use of a central site can limit the scalability of the method, and additionally data in p2p systems are typically horizontally partitioned along the peers. Below is outlined the pseudocode of the BDS algorithm. The algorithm in general is consisted from three phases. In the first phase (step one) performs a sorted access until all possible skyline points have been seen. This is guaranteed by the retrieval of the termination point. In the second phase (step 2 and 3) the algorithm accesses all the points with minimum attribute values in the lists and prunes all other points. In third phase (step 4) employees a random access in order to discard all the retrieved points that are dominated by any other point and returns the set of skyline points.

| ALGORITHM 19: | BDS (set P) [7] |
|---|---|

**Input:** A dataset P.
**Output:** The set of skyline points of dataset P.

0. Initialize a data structure P := ∅ containing records with an identifier and n real values indexed by the identifiers, initialize n lists $K_1,…,K_n$ := ∅ containing records with an identifier and a real value, and initialize n real values $p_1,…,p_n$ := 1
1. Initialize counter i := 1.
    1.1. Get the next object $o_{new}$ by sorted access on list $S_i$
    1.2. If $o_{new}$ ∈ P, update its record's $i^{th}$ real value with $s_i(o_{new})$, else create such a record in P
    1.3. Append $o_{new}$ with $s_i(o_{new})$ to list $K_i$
    1.4. Set $p_i := s_i(o_{new})$ and i := (i mod n) +1
    1.5. If all scores $s_j(o_{new})$ (1≤ j ≤ n) are known, proceed with step 2 else with step 1.1.
2. For i = 1 to n do
    2.1. While $p_i = s_i(o_{new})$ do sorted access on list $S_i$ and handle the retrieved objects like in step 1.2 to 1.3
3. If more than one object is entirely known, compare pairwise and remove the dominated objects from P.
4. For i = 1 to n do
    4.1. Do all necessary random accesses for the objects in $K_i$ that are also in P, and discard objects that are not in P
    4.2. Take the objects of $K_i$ and compare them pairwise to the objects in $K_i$. If an object is dominated by another object remove it from $K_i$ and P
5. Output P as the set of all non-dominated objects.

Authors in [121] improved the previous work by proposing the *PDS* (progressive distributed skylining) algorithm. This algorithm returns progressively the skyline points, by determining if a point belongs to the final skyline as soon as it is retrieved from the source by adopting an object-ranking approach. Additionally, based on the ranking approach can efficiently predict the termination object. In order to reduce the number of pairwise comparisons authors proposed the use of R-trees where dominance checks are achieved with the use of containment queries. In order to identify the terminating object of the skyline retrieval process, an object-ranking approach is proposed. Each object, in each site, is characterized by a rank. The rank is defined by numbering the items based on their increasing order of the list. The number of sorted accesses that will be needed to fully retrieve all the attributes of a point from all lists will be equal to the sum of its ranks. Thus a good terminating point will be the one that can be fully retrieve with minimum number of sorted accesses. Additionally, authors discussed the processing of top-k skyline queries and the estimation of completion percentage of skyline points retrieval. The top-k skyline retrieval based on a user defined preference function *f*, is achieved by using a priority queue that keeps the k objects with the highest (or lowest) rank that was computed with the use of the preference function during the skyline point retrieval. The completion percentage is calculated by estimating the total number of sorted access required by the query till it finds the identified terminating point and the number of sorted accesses already performed.

One of the most recent works on vertical data partition is [170]. In this scenario authors assume that the dataset is vertical partitioned in a number of arbitrary servers and each server stores an arbitrary number of dimensions. The computation of skyline is de-centralized and their main goal is to reduce the communication cost between servers taking into account the various updates that may occur. The reduction of the communication is achieved by identifying anchor points in each partition that will prune and minimize the number of points to be transmitted.

### 4.2.2. Horizontal data partitioning

This section outlines the methods that concern horizontal partitioning, in which a portion of the dataset is stored by a peer without taking into account the partitioning imposed on the dataspace. This defers from the approach of horizontal dataspace partitioning, which is described in sections **4.3** and **4.2.4**, where a partitioning scheme is applied to the dataspace and the points of each partition are assigned to peers. In these methods peers (servers) can communicate with their neighboring peers, through a coordinator, or through the use of a backbone structure. Thus this approach does not considers any kind of overlay structure, in which a logical network is built on top of a physical one without considering the physical network structure.

*MANET*

The work on **[77]** is focused in distributed skyline computation on mobile devices that communicate without routing information in shared-nothing environments and especially over ad-hoc networks (MANETs). In this scenario data are stored in a number of light-weighted mobile devices where each device is able to communicate only with its neighbors (devices that are in its communication range) by exchanging messages. The communication between all devices can be achieved via multi-hops. Each mobile device stores a portion of the whole dataset (that can be overlapping) which corresponds to a portion of the data that are related with the geographic area that it covers. The partitioning placed in this case concerns horizontal data partitioning. The scenario studied involves spatial constrains, in order to retrieve only the data related to the geographic area of interest, but they are not related with **3.1** or involved in the skyline operation. The constraints are only used to limit the propagation of the query based on the distance from the (global or local) originator of the query.

One approach to compute the skyline in this scenario would be for an originator to issue a query and send it to all of its neighbor peers in a breadth-first strategy. Each peer computes its local skyline and sends the results back to the originator. Then each of these peers forwards the query to their neighbors. A depth-first approach will include the originator to send the query to only one of its neighbor which will compute its local skyline and send the query to one of its own neighbors that have not received the query yet. When there is no other neighbor to propagate the query or the query has been completed, the results are returned through the same path and merged along the peers. Finally when the results have reported to the originator it will be responsible to compute its local skyline, remove the duplicate tuples that received and compute the final skyline. The processing of the same device twice is prevented by using *tagging* approach incorporated in each query so a device that receives a query can identify if the query has been previously processed by checking its log. Bellow is outlined the local skyline query on a mobile device $M_i$. In authors storage model $R_i$ corresponds to the relations where the data are store and $VDR_i$ corresponds to the dominating region of a point.

---

**ALGORITHM 20:** local skyline($pos_{org}$, $d$, $tp_{flt}$) **[77]**

**Input:** The location $pos_{org}$ of query originator, the distance $d$ of interest and a filtering tuple $tp_{flt}$.
**Output:** The reduced local skyline, and the updated filtering tuple.

// Check if $R_i$'s *MBR* overlaps the query region.
**if** ($mindist(pos_{org}, MBR_i) > d$) **return**;
// Check if $R_i$ is dominated by the filtering tuple
$skip$ = TRUE;
**for each** attribute $j$ of $R_i$
　**if** ($tp_{flt}.p_j > l_j$) $skip$ = FALSE; **break**;
**if** ($skip$) **return**; **else** $SK_i$ = ø;
// Local ID-based SFS processing
**for each** tuple $tp_j$ in $R_i$
　// Too far away from query point
　**if** ($dist(pos_{org}, tp_j) > d$) **continue**;
　$out$ = FALSE;
　**for each** skyline point $sp_k$ in $SK_i$
　　// $sp_k$ dominates $tp_j$
　　**if** ($\forall l > 1$, $sp_k.id_l < tp_j.id_l$) $out$ = TRUE; **break**;
　**if** (!$out$) add $tp_j$ into $SK_i$
// Filtering, and picking up maximum VDR
$idx$ = null; $VDR_m$ = 0;
**for each** skyline point $sp_k$ in $SK_i$
　**if** ($\forall l$, $tp_{flt}.pl < sp_k.pl$) remove $sp_k$ from $SK_i$
　**else if** ($VDR_k > VDR_m$) $idx$ = $k$; $VDR_m$ = $VDR_k$
// Update filtering tuple if necessary
**if** ($VDR_m > VDR_{flt}$) $tp_{flt}$ = $tp_{idx}$

Authors improved the overall query performance by proposing techniques to reduce the communication and computation cost, taking also into account the low storage capabilities of devices. The reduction of communication and computation cost between each peer is achieved by using a filtering-based data reduction technique. In this technique each query request is forwarded in a breadth-first or depth-first manner as mentioned, containing in addition a single, dynamically changing, filter point (computed from a parent node) in order to prune local skyline points, of child nodes, that will not eventually contribute to the final skyline result. The filtering point that will be propagated by the peer is selected based on the maximum *dominating region* (DR) among all the local skyline points computed by the peer. The dominating region that is considered is defined as the volume of the orthogonal region defined by the point selected and the upper boundaries (or lower in case of maximizing the preferences) of the global space. A special case is taken into account when peers lack of information about global boundaries. The dominance region of each point is represented in **Figure 38** by the shaded regions. The region with the maximum volume is the one created by the point H7. Essentially every peer that receives a filtering point knows that any of its points lying in this region will be dominated and can be safely pruned from their local skylines.

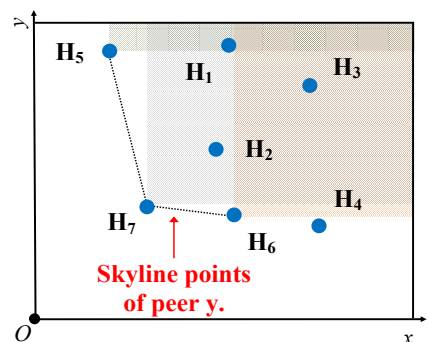

Figure 38: Dominating region.

In **Figure 38** points H1, H2, H3, H4 belong to a peer *x* that has received the query from one of his neighbouring peers *y*, which has also computed its local skyline point set which is consisted from the points H5, H6, H7. As mention previously the point with the largest dominating region among all of the local skyline points of *y,* is imposed by the point H7. For that reason this point is send as a filtering point along with the propagated query to the peer *x*. As seen in **Figure 38** the grey area represents the pruning imposed by the filtering point H7 of the neighboring peer *y* over the local (skyline) points of the peer *x*.

Summarizing, each peer receives along with the query a filtering point. It computes its local skyline and uses the filtering point to prune its local skyline points. Then checks if any of its remaining points has larger dominating region than the one of the filtering point that was received. If so the query that will be propagate will contain as filtering point the point found or in the other case will propagate the filtering point that had received. This essentially limits the data points to be transferred among devices during the breadth-first or depth-first propagation of the query. The reduction of storage cost is achieved by adopting a hybrid storage scheme which stores the spatial coordinate values of points (their location on space) in each device and uses an ID-based storage [154] approach for the rest attribute values that are not related with the coordinate values. However, this approach is not scalable for large scale p2p networks since it must visit all the devices in the network in order to retrieve the skyline.

### *SKYPEER/SKYPEER+*

On [173] authors study the problem of subspace skyline query processing in super-peer networks (*large scale P2P networks*) and proposed the SKYPEER framework where the dataset is horizontally distributed across peers. In these type of networks there exist a number of super-peers among the ordinary peers which have special abilities due to their enhanced features. Super-peers are linked through a backbone and peers are connected to super-peers. Each super-peer maintains information about the peers that have assigned on him in order to achieving efficient routing. In general a query request in this type of network is fist routed among the super-peers backbone and if necessary it is distributed to the peers. Each peer holds only the data that are related with its region (not imposing any space partition) and computes a partial local subspace skyline.

In their approach authors try to find a way to efficient compute a subspace skyline query relying as much possible on super-peers, in order to reduce the communication overhead and avoid a flooding process. Authors approach on this issue is each super-peer to collect from its peers the skyline points of any existing subspace, since an accurate (subspace) skyline computation needs that all data to be taken into account. In order for this to be achieved extended the notion of domination by defining the *extended domination* and *extended skyline*:

**Definition 27:** Extended domination (ext-domination)
For any given d-dimensional dataset *D* and any k-dimensional subspace U, with U⊆D, a point p∈U *ext-dominates* a point r∈U, if $\forall d_i \in U, p.d_i < r.d_i$ ∎

**Definition 28:** Extended skyline (ext-skyline)
The ext-skyline set contains all the points that are not ext-dominated by any other point ∎

Intuitively the *extended skyline set,* will contain all the points that are necessary to efficiently answer a skyline query in any arbitrary subspace. Additionally authors, in order to perform their computations and filtering, proposed the use of the transformations *f(p)* and *dist$_U$(p)* for each point p, inspired by [167]. In general the f(p) values will be used for ordering and *dist$_U$(p)* as a threshold value that will indicate the termination of the subspace skyline retrieval on a super-peer. The *f(p)* value of a d-dimensional point *p* is defined to be f(p)=min(p.d$_i$), thus the minimum attribute value on all dimensions. The value of *dist$_U$(p)* represents the L$_\infty$-distance of a point *p* from the origin of the axes on the subspace *U,* denoted as *dist$_U$(p)*=max$_{d_i \in U}$(p.d$_i$). Note that the f(p) value is computed based on all the dimensions of the dataspace, while *dist$_U$(p)* based on the dimensions of the queried subspace. The main difference of that transformation compared to the one on [167] is that the distances are calculated based on the origin of the axes rather than the right upper corner of the dataset.

The algorithm has a pre-processing step in which each peer computes its local ext-skyline. The results of the local ext-skylines are propagated to the associated super-peers of each peer. Each of the super-peers maintains a list with the ext-skyline points of each peer and computes its own ext-skyline. As a reminder, note that the ext-skyline is a set that can answer a skyline query of any arbitrary subspace. In the case of a query the initiator forwards the skyline query request for a subspace U to a super-peer, which in turn forwards it to his adjacent super-peers. The super-peer is responsible to merge the ext-skyline results of his peers with his results and compute its final local skyline set SKY$_U$ based on the subspace given by the query.

A super-peer can access the stored ext-skyline points in ascending order of their f(p) values. During this retrieval the dominating points are inserted in a local subspace skyline set SKY$_U$. During this process a *threshold* value *T* is defined to be the minimum *dist$_U$(p)* value of all the points. The processing on the super-peer terminates when the *threshold* value *T* is smaller than the f(p) value of the next point retrieved since, based on authors observation, the rest points cannot belong to the skyline set of subspace U.

The results of each super-peer are forwarded through the backbone of super-peers. When a super-peer receives a result set from another super-peers, merges the set with his results and proceeds with forwarding the message. Each result set is sorted based on the f(p) values of points. When the initiator has received all the local subspace skyline results from all super-peers, merges the results and retrieves the final subspace skyline. The access-order on each result set is based on the order imposed. The points that are dominated are pruned and the retrieval stops when the f (p) value of the next point is larger than the threshold value. The algorithm is consisted the following methods. The *local subspace skyline computation* which is performed on each peer upon a request and the *Super-peer merging of subspace skylines* approach which is performed on Supperpeers level in order merge the local skylines of it's peers and retrieve the ext-skyline of space D which is related with those peers. The pseudocode of the *Super-peer merging of subspace skylines* approach is outlined below.

| ALGORITHM 21: Super-peer merging of subspace skylines [167] | |
|---|---|
| **Input:** The super-peer's set of local subspace skyline points | |
| **Output:** The skyline of the subspace U. | |

1. $SKY_U \leftarrow \{\emptyset\}$
2. $threshold \leftarrow MAX\_INT$
3. $SKY_{U1}...SKY_{UNsp}$ the super-peers' set of local subspace skyline points
4. $SKY_{Us} \leftarrow$ the list with the minimum first element
5. $p \leftarrow$ next point based on $SKY_{Us}$
6. **while** $(f(p) < threshold)$ **do**
7.    **if** p is not dominated by any point in $SKY_U$ on subspace U **then**
8.       remove from $SKY_U$ the points dominated by p
9.       $SKY_U \leftarrow SKY_U \cup \{p\}$
10.      $threshold \leftarrow \min_{p_i \in SKY_U}(dist_U(p_i))$
11.    **end if**
12.    $SKY_{Us} \leftarrow$ the list with the minimum first element
13.    $p \leftarrow$ next point based on $SKY_s$
14. **end while**
15. **return** $SKY_U$

The amount of unnecessary transferred data is reduced by the proposed threshold based algorithm which facilitate pruning of dominated data across the peers. Based on this algorithm authors explored the two optimization criteria threshold *propagation* and *merging strategy*, which refines the threshold value of the algorithm and performs progressive merging on local results and results received at each super-peer respectively. The *all-local-skyline (ALS)* algorithm can be considered [208] as a degenerated version of SKYPEER, in the case where the algorithm is used to return the skyline of the full space.

In [174] authors extended their work and proposed the SKYPEER+ algorithm which focusses on efficient routing and improves the performance of skyline queries in the case of non-uniform distributions of data among super-peers. The algorithm employs an appropriate indexing and routing mechanism based on routing indexes which results in reducing the number of super-peers employed in a query.

Another method that does not takes into account the partitioning imposed is [73]. This method reasons about probabilistic skylines and for that reason is outlined in section **4.2.6,** which is related with various topics in distributed skyline computation that are not related with the exact skyline computation.

### 4.2.3. Horizontal data partitioning without overlay networks

This section outlines the work that is based on horizontal data partitioning but do not consider any underling overlay network. That is the query originator can communicate with all the existing peers in order to compute the skyline set.

*PaDSkyline*

Authors in [40, 32] proposed the filter-based PaDSkyline algorithm for parallel constrained skyline query processing in which assume that no overlay exists and any query originator can directly communicate with all servers. Thus, authors do not address the problem of communication among peers but rather deal with wired peers that are connected through the internet. This approach employee and extends the single-point filtering method of *MANET* [77] with the difference that uses multiple filtering points rather than just one, considering that wired connections are faster and more reliable than the wireless.

The data are horizontally partitioned among peers and may overlap. The dataset of each peer is summarized and represented by a Minimum bounding region (MBR). Initially the algorithm collects the MBRs of all peers and partitions them in incomparable groups. In the case that constrains exist, will have also the form of region. Thus the MBRs that will be considered for further computation in this case are the ones that fully lie inside the constrained region. Additionally if the region constrains intersect a MBR it will be considered for further computation only the part of the MBR that belongs in the constrained region. As illustrated in **Figure 39** there exist 3 groups where the first group contains region e1, the second e2, e3 and the third e4, e5, e6, e7. The grouping of regions is based on the property of incomparability where each group is incomparable with the other groups. Thus group 1 contains region e1 which is incomparable with all the other regions. Group 2 contains regions e2 and e3 which are incomparable with all the other regions but they are not mutually incomparable. For that reason both regions belong to the same group. In group 3, regions e4 and e6 are mutual incomparable but belong to the same group because none of them is incomparable with region e5 or e7 and additionally region e7 is not incomparable with region e5. This grouping approach allows the parallel and progressive computation of the query over the groups, since none of the groups depends on the points or the results of another group.

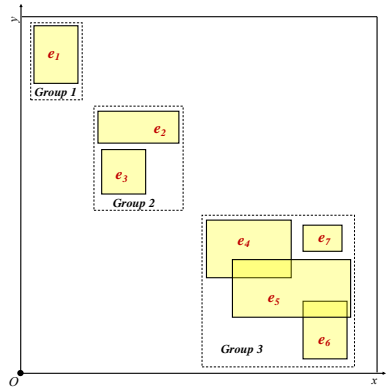

Figure 39: MBR groups.

For each group of MBRs is defined an intra-group query execution order among peers and a group head which will be responsible for the group's local skyline query by merging the results. For the execution order of the query authors proposed two strategies. In the first the query execution order of peers is derived by sorting the MBRs in a non –ascending order of their Euclidean distance from the origin of the data space or the origin of the constrained region if exists. In the second the execution order is defined by a tree which is built during the construction of each group and essentially stores the order on which the MBRs were processed inside the group. Thus, the tree is heavily based on the order that the MBRs were collected.

The communication cost between peers is reduced by employing pruning on each local skyline computation with the use of high-dominating filtering points. The filtering points are selected from the local skyline results. In the case of one filtering point is selected the one whose dominating region has the maximum volume among all others as on MANETS on section **4.2.2**. In the case of selecting $k$ filtering points among the $n$ local skyline points, is selected the set that maximizes the sum of their dominating region's volume or the set in which the distance between the filtering points is maximized. In each peer the filtering points are dynamically updated if it is needed as they compute a potential filter set and compares it with the one that have received.

The PaDSkyline algorithm is consisted from two individual algorithms. These algorithms are the incomparable partitioning algorithm *icmpPartition* and the intra-group skyline computation algorithm named *groupSkyline*. The *icmpPartition* algorithm partitions the space in incomparable partitions while the *groupSkyline* algorithm is responsible for the local skyline of each goup. Next is outlined the general structure of the PaDskyline algorithm which incorporates *icmpPartition* and *groupSkyline* in the form of pseudocode, where $S_{org}$ refers to the originator peer (site) of the query and the $g_i$.plan to the intra-group query execution plan. It is needed to mention that after a group $g_i$'s head receives a query request, operates based on the intra-group skyline algorithm.

| ALGORITHM 22: BaDSkyline (S, C) **[40, 32]** |
|---|
| **Input:** A set S of peers (sites) |
| A set C of constrains |
| **Output:** The constrained Skyline. |
| |
| 1. $\Pi_S$ = icmpPartition($S, C$) |
| // parallel execution |
| 2. **for each** group $g_i \in \Pi_S$ in parallel |
| 3. send {$C, S_{org}, g_i.plan$} to $g_i$'s group head |
| // result merge, triggered by incoming result reply |
| 4. **repeat** |
| 5. receive result reply from a group $g_i$'s head |
| 6. report $g_i.result$ |
| 7. **until** all group heads have replied |

Summarizing, when an originator wants to place a skyline query, it receives and groups all the MBRs from all peers. After the grouping process, sends the query request in parallel to the head of each group. The query will contain the constraints (if applicable) and the execution plan (order). Each head computes its local skyline and the initial filtering points that will be used. Next propagates the query to a peer of its group according to the execution plan. The peer computes its local skyline, prunes its resulted set with the use of the filtering points received and updates them if it is needed. Then sends its refined local skyline to the head of the group and propagates the filtering points and the execution plan (having removed itself from it) to another peer according to the execution plan that has initially received. The head continuous to receive results until all peers of its group have finished the computation. Then merges all the results that has received and reports the local skyline of the group to the originator. When the originator receives a result from a head can directly report it as is incomparable with the results of other groups. The main problem of this method is that the load on peers is unbalanced.

*FDS*

In [208] authors proposed a progressive feedback-based distributed skyline (FDS) algorithm which assumes that a small number of servers are geographically distributed and connected through the internet. Data are partitioned horizontally and assume no overlay structure. FDS is focused on minimizing the transferred data among the network. In this distributed environment, a querying (coordinator) server directly communicates with other servers in order to retrieve the results of their local skylines. In a naïve approach each server first computes its local skyline and reports it to the querying server. However this approach will allow the delivery of many non-skyline points to the querying server, since a server cannot determine if the points of it's set are dominated by the points in a set of another server. For this reason a feedback-based approach can be incorporated allowing the querying server to send a set of feedback points that are derived from the (incomplete) set of skyline points that are collected from other servers till now. The points that are selected as feedback are based on their score derived from a user-specified preference function. This approach worth only if the point that will be send will allow the pruning of pre-defined *n* number of points.

A simple example to demonstrate the general idea of the algorithm, where minimization of preferences is desired, is the following. Assume that the server $S_1$ contains points H1, H5, H8, H10, H11 and server $S_2$ the points H2, H3, H4, H6, H7, H9 of the house-metro station example (Table 1). Assuming that the preference function is defined to be the sum of all coordinates of a point *p*. That is $f(p) = \sum_{i=1}^{d} p.d_i$, where *d* is the number of dimension and *p.di* is the value of the coordinate of *p* on the i-th dimension. When a coordinator wants to compute a skyline query instructs all servers to find their local skyline points. Initially the local skyline computation on each server is done by using an existing centralized algorithm. Additionally each server sorts its resulted set of points in ascending order of their values, derived from the defined preference function, as illustrated in Table 27. Then each server sends to the coordinator (only) the first point of its sorted list. Since the list is sorted in ascending order, this point will be the first point on the list which essentially has the minimum score. This way coordinator will receive points H8(200,1200) and H7(400,300) from server 1 and 2 respectively, which in his turn will sort them based on their score derived from the preference function. At this step the coordinator sends to the servers the first feedback, which will be the maximum score $f_{max}$=1400, derived from the received points since H8 has a value of *f(p)*=1400 and H7 a value *f(p)*=700. In the next step servers will report to the coordinator all of their points that have as an upper score bound the score received from the coordinator. The points that the coordinator have eventually received in this step are illustrated in Table 28. As seen in table Table 27 only point H7 and H9 have as their upper bound the value 1400 that was send by the coordinator. In this step the points of Table 28 are inserted in the final skyline list where a dominance check is performed with the already potential skyline points. Then the newly retrieved points can be reported to the user, since the algorithm is progressive. Continuing the coordinator performs a feedback/refirement phase and identifies from its list with the two received points H8 and H7 the one with the minimum score, which is H7 with $f_{min}$=700 and sends it to $S_1$. That is the server who had the point with the maximum score value. The point is sent only to this server because it would not prune any point on the other server. In its turn the server prunes its local skyline points that are dominated by that point. This completes the first phase of the algorithm and the server's lists at this point are illustrated in Table 29. In the second round, again servers will sent to the coordinator the points of their list which have the minimum score. That is point H8(200,1200) and H6(1600,100) from S1 and S2 respectively. From those two points point H6 (from server S2) has the largest $f_{max}$=1700 value and thus $f_{max}$ is send two servers as a feedback. Servers in their turn return their points who's f(p) value has the upper bound of 1700 as seen on Table 30. Next the coordinator identifies the minimum score among the previously received H8 and H6 which is fmin=1400 and sends it to server S2 in order to prunes his set. In the next round the algorithm continuous in the same way and returns the final skyline point.

| $S_1$ | H8(200,1200) | H11(500,900) | H1(100,1500) |
|---|---|---|---|
| $S_2$ | H7(400,300) | H9(1000,200) | H6(1600,100) |

Table 27: Servers lists.

| $S_1$ | H8(200,1200) | H1(100,1500) |
|---|---|---|
| $S_2$ | H6(1600,100) | |

Table 29: Servers refined lists.

| $S_1$ | ∅ | |
|---|---|---|
| $S_2$ | H7(400,300) | H9(1000,200) |

Table 28: First round - coordinators list.

| $S_1$ | H8(200,1200) | H1(100,1500) |
|---|---|---|
| $S_2$ | ∅ | |

Table 30: Second round - coordinators list.

The algorithm is iterative and can achieve multiple-round feedbacks to the servers by sending different point each time. However in large networks the algorithm may delay to deliver skyline points due to the several round-trips of the feedback. In their work authors additionally sketch an approach to maintain the skyline queries when new points are added to the servers (this type of queries are described on section **4.4**). Finally, point out that their proposed method can also solve subspace skyline **[4.1]**, skycube **[4.1.1]** and k-dominant skylines **[3.9.2].**

#### 4.2.4. Horizontal space partitioning

This section reasons about distributed skyline computation assuming the incorporations of a dataspace partitioning technique. Thus each server will be responsible for a disjoint partition of the data space. This scenario differs from this in the previous section since the partitioning is imposed to the dataspace rather to the dataset itself. All the methods that will be described assume the existence of an overlay network in order to achieve content-based addressing. Methods described in this section can by categorize into two classes. DHT – based (distributed hash table-based), such as CAN **[146]** and balanced-tree based such as BATON **[85]**.

*DSL (CAN)*

Authors in **[187]** proposed the progressive and parallel distributed skyline algorithm (DSL).The algorithm deals with constrained skyline queries, which return the skyline set of a given region, on a shared-nothing architecture. DSL is based on grid-based data space partitioning techniques which horizontally partition the space. The data partitioning is determined by the structured P2P overlay networks CAN **[146]**. CAN is a distributed, decentralized P2P infrastructure, based on a logical d-dimensional Cartesian coordinate space, which incorporates a distributed hash table (DHT) for point and server multi-dimensional indexing. The d-dimensional logical coordinate space is partitioned in n regions according the number of servers that exists on the network. Each point of the dataset is mapped into a region in the logical coordinate space (using a hash function) which is represented by a server. Each server is related only with one region and is the only one that can provide direct access to the data mapped on this server and consequently to this region. Each server in CAN stores along with its points, indexed by a DHT, an additional routing table that contains its neighbors, their locations and their region boundaries. The communication in this type of networks is achieved only through other servers with neighboring zones. Two d-dimensional regions are neighboring only if they abut in (at least) one dimension and their coordinate spans overlap in d-1 dimensions. A node can route a message towards its destination by simply forwarding it to its neighbor with coordinates closest to the coordinates of the desired destination server. In the case of multiple neighboring zones the neighbor with the smallest distance from the server of interest is preferred. The partitioning process work as follows. Every node that is willing to be part of the network, communicates with any existing server. Then picks a random point in space in which wants to be located or a location inside the largest existing region. The nodes in the network are responsible to route him to this server. If its desired location lies inside a region that is represented by another server, the region is partitioned in half and the points stored on that server are distributed among the two. The regions in CANs can have different size, but always have a rectangular shape. After the new server (newly inserted node) placement neighboring servers must update their routing tables about the existence of the new server and the new region boundaries that occurs in each case. Below is outlined the pseudocode of the DSL algorithm. The pseudocode of the algorithm that will be executed in each node is consisted from the two procedures QUERY and COMPLETE and is outlined below.

**ALGORITHM 23:** QUERY($Q_{cd}$, $Q_{ab}$, ID($Q_{cd}$), skyline) **[187]**

**Input:** $Q_{cd}$: current region to evaluate.
$Q_{ab}$: the "parent" region of $Q_{cd}$.
ID($Q_{cd}$): region code for $Q_{cd}$.
*skyline*: skyline results from upstream regions.
$M(Q_{cd})$: master node of $Q_{cd}$.
**Output:** set of local skyline points

1. QUERY($Q_{cd}$, $Q_{ab}$, ID($Q_{cd}$), skyline)
2. **Procedure**
3. calculate predecessor set pred($Q_{cd}$) and successor set succ($Q_{cd}$);
4. **if** all regions in pred($Q_{cd}$) are completed **then**
5.   **if** skyline dominates $Q_{cd}$ **then**
6.     $M(Q_{cd})$.COMPLETE();
7.   **end if**
8.   localresults ← $M(Q_{cd})$.CalculateLocalSkyline(skyline,$Q_{cd}$);
9.   skyline ← skyline ∪ localresults;
10.   **if** $M(Q_{cd})$.zone covers $Q_{cd}$ **then**
11.     $M(Q_{cd})$.COMPLETE();
12.   **else**
13.     $M(Q_{cd})$ partitions $Q_{cd}$ into RS($Q_{cd}$);
14.     **foreach** successor $Q_{gh}$ in RS($Q_{cd}$)
15.       $M(Q_{gh})$.QUERY($Q_{gh}$,$Q_{cd}$,ID($Q_{gh}$),skyline);
16.   **end if**
17. **end if**
18. **End Procedure**
19.
20. COMPLETE()
21. **Procedure**
22. **if** succ($Q_{cd}$) equals to NULL **then**
23.   $M(Q_{ab})$.COMPLETE();
24. **else**
25.   **foreach** successor $Q_{ef}$ in succ($Q_{cd}$)
26.     $M(Q_{ef})$.QUERY($Q_{ef}$, $Q_{ab}$, ID($Q_{ef}$), skyline);
27. **end if**
28. **End Procedure**

In order to achieve a parallel computation authors enforced a partial order hierarchy on query propagation and computation between servers (peers). This order is used to pipeline the participating server's results, since local skyline results in one server (region) may dominate local skyline results on another server (region). This indicates that the local skyline computation on the second server must start after the completion of the skyline computation on the first server. These dependences are called the *skyline precedence relationship*. Note that servers that do not lie inside the constraint region can be safely pruned **Figure 40.** At query time is build a multicast tree **Figure 41** which has as root the lower-left corner server. The points stored on that server are guaranteed to be on the final skyline set. The rest of the servers are placed inside the tree in such a way, so that servers whose data cannot dominate each other, to be queried in parallel. The query is propagated along the tree where each server computes its local skyline. The results are propagated to the root through the same path where intermediate servers prune unnecessary points.

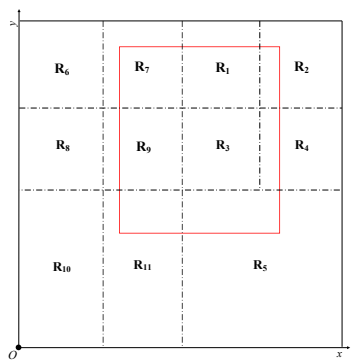

**Figure 40: Partitioned dataspase & constrained query region.**

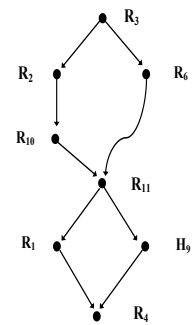

**Figure 41: Multicast tree for partial ordering.**

Authors also reasoned about the load balancing problem that occurs between peers since the algorithm always starts from the bottom-left region of the queried space. The load balance of a node is defined as the number of local skyline queries that has processed excluding the load of disk I/O and control messages which authors assume negligible. In order to distribute the cost evenly authors proposed a *dynamic zone replication* method. In order for this process to be invoked each server generates *n* random points in the d-dimension CAN space. Then asks the servers that represents these points for their query load and compares the samples with its own. If its load is heavier than a threshold *t* of all samples the *zone replication method* is invoked. The result of this method will essentially be to forward the actual local query processing that must be done in a high loaded node to a lightly loaded node using mutual region replication among nodes.

This approach is highly scalable since its performance can be improved by adding new servers. In this case the algorithm will automatically balance the load by distributing data to those servers. However, the parallelism that the algorithm imposes is not always beneficial since succeeding nodes must wait until they receive the results of local skylines of their preceding nodes. Additionally, the algorithm is designed and focused on constrained skyline queries, thus its load balancing mechanism is designed to be efficient in this problem. This causes in many cases a few servers to carry all the computation overhead and the rest to be practically idle. Nevertheless, although the discussion of this paper is based on CAN overlay networks, DSL does not rely on its specific features, such as decentralized routing, which makes it applicable in a more general case of content-based data partitioning problems. It worth to mention that the partitioning scheme adopted in this approach incurs considerably computation and communication overhead in a case of an update on a server dataset, since this affects the computation of the skyline on all servers that were concern in the initial computation phase.

*SSP (BATTON)*

In [180] authors proposed the Skyline Space Partitioning (SSP) approach which is based on BATTON [85]. BATON instead of using a distributed hash table (DHT) as CAN [146] uses a distributed balanced tree for indexing of nodes. As opposed with the previous method SSP processes the unconstrained skyline queries which retrieves skyline points from the whole space.

In a BATTON network each physical point (server) is mapped to a logical node inside the balanced tree. BATON in contradiction with CAN supports one-dimensional indexing schemes. For that reason authors incorporate the z-order curve [55, 145] to map the multidimensional space in one dimensional value. Each region is mapped to a one dimensional value based on that curve as shown in **Figure 42**. Then each node in the balanced tree of BATON is assigned and dedicated to hold a range of the z-order values. Additionally its left child will contain the range of values smaller than his and the right child will contain the range of values higher than his. Each node in the tree holds routing tables with information (links) about his parent, his children and his adjacent nodes. Additionally holds a left and right routing table that contains some of the nodes that exist in the same level. For all nodes mentioned, additional information is hold about the range of values that have been assigned to each one of them. Based on that information, by following the adjacent links we can access the data in ascending order. In the case a node accepts a new node as a child, splits and distributes its range of values among both of them, updates its routing tables and the tables of his children. Additionally updates the information that exists on the neighboring nodes in the same level with him and his children.

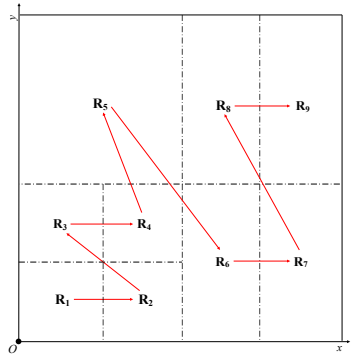

Figure 42: Region ordering based on z-curve.

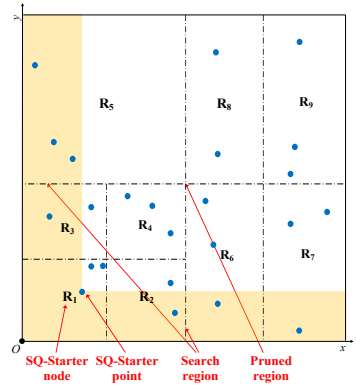

Figure 43: SQ-Starter point and search region.

The SSP algorithm partitions the space into non-overlapping regions. Each region is mapped to a value based on the z-curve and assigned to an existing server (peer) on the network **Figure 42**. This mapping-numbering process is selected so it can be possible to retrieve a region number efficiently and with a good accuracy. The mapped values are organized in a BATON tree represented by nodes. Each node stores, along with the records described previously, additional information that are related with the region's *split histo*ry in order to be possible to estimate the target region number (value) on a query search. The algorithm forwards the query request of a query initiator node to a node (region) (*SQ-Starter* node) whose local skyline results are guaranteed to be in the final skyline. This node will be responsible to start the query process. In the case of a 2-dimensional space, where the minimization of values is desired, this node can be the one that contains as a point the origin of the axes. Thus, the algorithm starts its query process with that node and computes its local skyline points. From the resulted set selects the point with the largest dominating region **[77],** as filtering point in order to minimize the global search space. For the needs of the example this point is named SQ-starter point. As illustrated in **Figure 43** the SQ-starter point belongs in the region R1. The non-shaded area represents the search space that is dominated by that point. Thus regions that completely lie in this area can be skipped, since all of their points are dominated by the SQ-Starter point. The only regions that will be visited are the one that are completely or partial covered from the shaded region. Next the *SQ-Starter* node routes the query to the nodes related with those regions. The filtering point in this algorithm is not adaptive/dynamic. This approach was selected by author in order to further reduce the computation cost on each node, taking into account that this will lead that the pruning power of the filtering point will not be maximized all the time. The order that the nodes (and their children nodes in sub-regions) are visited is based on their order derived from the z-curve. Each node computes its local skyline, refines the search space, forwards the query and returns its local skyline results to the query initiator. When the query initiator has received all the local skyline sets from all the nodes computes and reports the final skyline.

Authors additionally reasoned about *query load-balancing*, where they proposed two approaches, one static and one dynamic. The *query load* is defined to be the sum of the number of local data points retrieved for answering a query and the number of messages that are rooted through a node. In the first static approach load balance is related with the data load (points stored in one node) on nodes and the partitioning strategy followed in a node joining or departure on the tree. In their approach a node joins the tree based on the data load that an existing nodes has rather than simply joining the first qualifying node that does not have enough children. In this way, heavy loaded regions are partitioned resulting smaller regions with distributed data load. The same approach is followed in a node departure where light loaded regions are merged. The second approach is related with sampling of the loads of other nodes during the query process and dynamically balancing the load. The samples of loads that a node takes into account are collected from neighboring or adjacent nodes or by random sampling other not linked nodes in order to reflect, in this case, the global load distribution. In the case of load imbalance detection, data migration is performed. This is achieved by balancing node's load with a neighboring node by repartitioning the dataspace between them or by sharing its load with a node from the ones that randomly received samples.

### SKYFRAME (CAN-BATTON)

Authors in [181] extended their work on [180] and proposed the *SkyFrame framework* which can be applicable on BATON and other structured P2P overlay networks. The *SkyFrame framework* is mainly consisted from the SSP algorithm (renamed as *Greedy Skyline Search (GSS)*), the two previously described load balancing mechanisms, which balances the query workload among peers and a newly proposed algorithm *Relaxed Skyline Search (RSS)*. This sub-section will introduce the RSS algorithm since the rest of the components were introduced in the previous sub-section.

As a reminder, GSS algorithm (which in a previous work was named SSP) must first find the *SQ-Starter* node and its local skyline points, in order to identify the point that will be used for space pruning, before actually executing the skyline query on all the nodes. Related with that, authors observed that most of the time *GSS* involves partitions that abut to the borders of the dataspace and named *SQ-Border* nodes. This happens because nodes that abut to the boundaries of the dataspace or nodes that abut previous nodes are more likely to contain local skyline points that will eventually belong to the final skyline set of points. In **Figure 44** the *SQ-Border* nodes are the nodes associated with regions R1, R3, R5, R7, R10, R11 and R12.

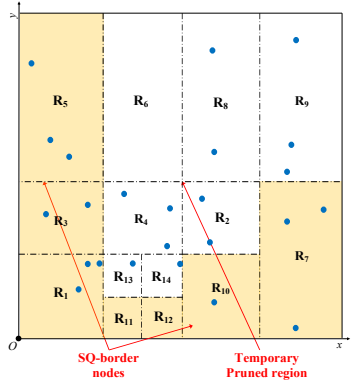 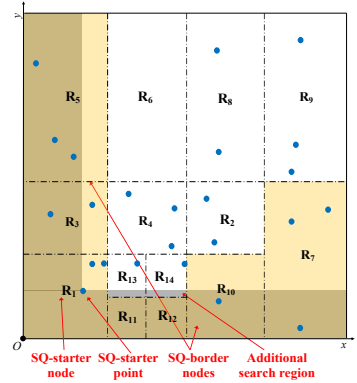

Figure 44: SQ-border nodes and the initial search region.

Figure 45: Search region refinement with SQ-starter node.

Thus, instead of waiting for the SQ-starter node to be identified and compute its local skyline results authors proposed that RSS algorithm will parallel execute the skyline query on all the *SQ-Border* nodes **Figure 44**. This relaxes the boundaries of the *search space*, compared to GSS as shown in **Figure 45**. Although it may return more local skyline points that will not eventually contribute to the final skyline (which will be filtered in a later phase), the process is speed-up by exploiting the time needed from the SQ-starter node to compute its results, to parallelize local skyline computation in other regions. Following is outlined the pseudocode of the GSS algorithm where. $P_{md}$ represents the point with the largest dominating region among a set of points.

| **ALGORITHM 24:** GSS(node n, query q, search_region sr, phase p) [181] | 12. Return local_skyline_points, $p_{md}$, and the region the node is in charge of to the query initiator. |
|---|---|
| **Input:** The routing table RT(n) of a node n. The region Region(n) maintained by a node n. **Output:** Local skyline points <br><br>1. **if** p = 1 **then** <br>2.   **if** n is *SQ-STARTER* of q **then** <br>3.     local_skyline_points = Compute_Skyline_Points <br>4.     pmd = Compute_Pmd <br>5.     SR = Compute_Search_Region <br>6.     Partition *SR* into a disjoint set of subSRs for neighbor nodes in RT(n). <br>7.     **for all** nodes m in RT(n) **do** <br>8.       **if** m is in charge of a subSR **then** <br>9.         GSS(m, q, subSR, 2) <br>10.       **end if** <br>11.     **end for** | 13.  **else** <br>14.    x = a node in RT(n) nearer towards SQ- STARTER <br>15.    GSS(x, q, null, 1) <br>16.  **end if** <br>17. **else** {p = 2} <br>18.    local_skyline_points = Compute_Skyline_Points <br>19.    Return local_skyline_points and the region the node is in charge of to the query initiator <br>20.    **if** Region(n) $\not\supseteq$ sr **then** <br>21.       SR = sr \ Region(n) <br>22.       Partition SR into a disjoint set of subSRs for neighbor nodes in RT(n) <br>23.       **for all** nodes m in RT(n) **do** <br>24.         **if** m is in charge of a subSR **then** <br>25.           GSS(m, q, subSR, 2) <br>26.         **end if** <br>27.       **end for** |

|  28. **end if** |  29. **end if** |
|---|---|

In RSS algorithm the query initiator identifies all the *SQ-Border* nodes and instructs them to compute their local skyline. The results from each node are reported back to the initiator. From the received results finds the point with the maximum dominating region. This point will act in the same way as the SQ-starter point described in SSP, but this time will be used to achieve the correctness of results. In the rest of the description the point with the highest dominating region will be called SQ-starter point. At this phase the query initiator defines the search region based on the SQ-starter point as on SSP **Figure 45**. If the region defined by the SQ-starter point does not contain any other region that was not searched till now, the query initiator computes the final skyline set from the results collected till now and reports the final set of skyline points. If additional regions are identified, that is needed to be searched, query initiator forwards the query to these nodes. Such a region is illustrated in **Figure 45,** were regions R13 and R14 are not abut to the axis of the dataspace but are covered from the search region define by the SQ-starter point. In this case, regions R13, R14 and similar regions must be searched because they may have potential global skyline points.

For that reason the query initiators forwards the query to these regions in order to receive their local skylines. The query forwarding on these nodes can be easily achieved without excessive overhead because each node in the tree stores additional information about his neighboring and adjacent nodes. After the query initiator have received these additional local skylines from nodes, computes and reports the final skyline set.

Concluding, GSS is optimized to prune the search space as much as possible by using high dominant points and RSS is optimized for query response time (in comparison to GSS) by processing skyline queries in parallel at the *SQ-Border* nodes. Compared to CAN, BATTON has smaller search hops in order to reach a target node **[181]**. The skyframe framework is applicable in different structured P2P overlay networks as long as they are capable of representing data regions with nodes, forward the query to the nodes related with the search region and identify the SQ-starter node or *SQ-Border* nodes.

Both *CAN* and *BATTON* methods rely on grid-based partitioning and assume the availability of an overlay network in which each peer stores a disjoint data partition. A major drawback of most these approaches is load balancing on peers, since many data partitions (and equivalently their peers) do not essentially contribute to the final skyline set and the main load is distributed on a small number of peers whose partitions are allocated close to the origin of the axes.

*iSky (BATTON)*

In **[33, 39]** authors proposed the iSky algorithm which is similar to the SSP/skyframe **[180], [181]** and based on the BATON overlay network **[85].** The differences from the SSP/SkyFrame, is the use of the iMinMax **[132, 182]** data transformation, as similarly used in the Index approach **(section 2.3.3)** described in **section 2.3**. As a remainder this approach transforms each d-dimensional point *p* to a one dimensional value *y* according to iMinMax approach and indexes the values with a B-tree. The value y is computed as y=$d_{xmax}$+$x_{max}$ where $x_{max}$ is the maximum of the attribute values of a point x and $d_{max}$ is the dimension that this attribute value exists. Note that the dataset is normalized in [0,1). The only difference between Index algorithm **(section 2.3.3)** and iSky approach related with the indexing of values is that the Index algorithm uses a B-tree to index the values and the isky uses the balanced tree of BATON. The iMinMax transformation was selected because it actually partitions the d-dimensional space in *d* partitions and imposes an ordering on each one of them. Also through this transformation is achieved the progressiveness of the algorithm. Another difference from SSP/SkyFrame is that it uses an adaptive filtering point and a threshold value, in order to further minimize the communication cost. It is noted that the algorithm assumes that maximization of preferences is desired as opposed in previous sections. A practical approach, that is proposed by authors, in order to apply the algorithm in cases where minimization of preferences is desired is to add a negative sign to each value on relevant dimensions. Following is presented the basic pseudocode of the iSky algorithm. $P_{org}$ represents the originator peer of a query q . $\text{SF}_{global}$ and $\text{SF}_{local}$ are the filtering points to be broadcasted and $DR_i$ is the dominance region of a node i. The algorithm peerSkyline **[33, 39]** is called from each peer that receives a query from an originator.

| ALGORITHM 25: | iSky() [33, 39] |
|---|---|

Input:
Output:

```
1. result = ∅; DR_max = 0; idx = 0;
2. foreach (P ∈ P) do send q to P;
3. receive the results from all initial skyline peers;
4. result = merge(received skyline points);
5. // generate skyline filters;
6. foreach (r ∈ result) do
7.     if DRr > DRmax then
8.         DRmax = DRr;
9.         idx = r;
10. SF_local = idx;
11. SF_global = max_{r∈result}(r_min);
12. foreach P in P_org's routing table do
13.     call peerSkyline(q, SF_global, SF_local);
14. result = merge(received skyline points)
```

The Algorithm processes in general in three steps. First identifies a set of peers (named *initial skyline peers*), whose dataset has specific properties based on the iMinMax approach, and retrieves their local skyline sets. From these sets computes the filtering point and the threshold value that will be used. Then forwards the query to the rest of the peers along with the filtering point and the threshold value. After the query initiator has received the local skylines from each peer computes and reports the final skyline.

| $Max_1$ | Dimension 1 | y | $Max_2$ | Dimension 2 | y | $Max_3$ | Dimension 2 | y |
|---|---|---|---|---|---|---|---|---|
| -0.01 | H1 (-0.01,-0.15,-0.145) | -1.01 | -0.01 | H6 (-0.16,-0.01,-0.035) | -2.01 | -0.015 | H8 (-0.02,-0.12,-0.015) | -3.015 |
| -0.05 | H10 (-0.05,-0.14,-0.057) | -1.05 | -0.02 | H9 (-0.1,-0.02,-0.15) | -2.02 | -0.023 | H5 (-0.09,-0.13,-0.023) | -3.023 |
| -0.05 | H11 (-0.05,-0.09,-0.06) | -1.05 | -0.03 | H7 (-0.04,-0.03,-0.13) | -2.03 | -0.032 | H3 (-0.07,-0.06,-0.032) | -3.032 |
| | | | -0.05 | H2 (-0.14,-0.05,-0.092) | -2.05 | | | |
| | | | -0.1 | H4 (-0.13,-0.1,-0.12) | -2.1 | | | |

Table 31: Sorted 2-dimensional dataset.

The Table 31 represents a 3-dimensional version of the house-metro station dataset (Table 22) normalized in the interval [0,1] (as on 2.3.6) and altered with a negative sign in order to indicate that the minimization of preferences is desired, as authors proposed in their work. The points on the Table 31 are sorted based on non-ascending order of their attribute values in each dimension. As seen points H1, H6 and H8/H9 have the maximum attribute value on dimension 1, 2 and 3 respectively. The one-dimensional values derived from the iMinMax approach are the y values. As seen, the original 3-dimensional dataset is mapped into the range (-4,-1) (range of y values). According to BATON protocol each peer will be dedicated to store the points of a non-overlapping subrange. Additionally iMinMax transformation actually separates the dataspace into d partitions where in this case (3-dimensions) are (-2,-1] and (-3,-2] and (4,-3]. As an example based on a BATON tree with three peers, peer $N_1$ is dedicated to store the points that their y value belongs to the range (-2.5, -1.75].

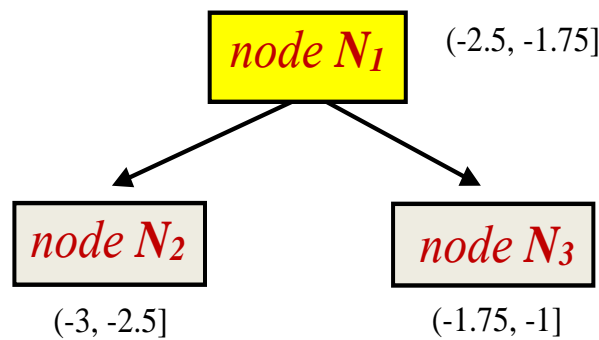

Figure 46: BATON data indexing.

The *initial skyline peers* set is consisted from the peers who store the data points with the maximum value in some dimension. The set that fulfill this requirement is selected due to the high pruning power that their points have. This happen because points that have a maximum attribute value on (at least) one dimension are guaranteed to be in the final skyline set, because they will dominate all others points in those dimensions. This also holds in our case were the negative sign was added as proposed by authors. For **Table 31** one peer that can belong to the *initial skyline peers* set is the one who stores the points H6 (which has the maximum attribute value -0.01 in Dimension 2). These peers can be easily identified by simply checking where their region range overlaps with two adjacent partitions. As observed the sub-region of peer $N_1$ is [-2.5,-1.75) which overlaps with the partitions (-2,-1] and (-3,-2]. The pseudocode of this approach, named *locateISPeers* is outlined bellow.

| **ALGORITHM 26:** | locateISPeers (root) |
|---|---|
| **Input:** | The root of the Baton tree |
| **Output:** | The *id* of peer. |

1. compute $\max_{x \in DPi} (x_{max})$;
2. $\min_i = \min(d_{xmax} + x_{max})$;
3. $\max_i = \max(d_{xmax} + x_{max})$;
4. **for** (k = 1; k ≤ d; k++) **do**
5.    **if** ($\min_i < k + 1 \leq \max_i$) **then**
6.    report *id* to *root*
7.    **break;**

The peers that will first be queried are the ones of the *initial skyline peer* set. The results from those peers are collected by the query initiator. A merging on these sets will construct the *initial skyline set* of points. From this set of points the query initiator determines a filtering point and a threshold value. In order to compute the threshold value, first is retrieved the minimum attribute values of all dimension for all the *initial skyline points*. Then, from those values is selected the maximum and transformed into an iMinMax range value. This value will have the role of the threshold value. With the use of threshold value is achieved that a point can be pruned if its maximum attribute value on all dimensions is smaller or equal to the received threshold. In this example the *initial skyline peer* set will be consisted from node $N_1$ and the *initial skyline points* will be H6, H7 and H9. The minimum attribute values of all the *initial skyline points* are -0.01, -0.02, -0.03. From those values the maximum is -0.01 and thus is selected as a filtering point which is sent along with the query to the rest nodes in order to complete the skyline point retrieval. The filtering point will be the one with maximum dominating region as defined on **[4.2.2 MANET]**. Now the query will be propagated along with the threshold value and the filtering point, to the peers that their stored range of values is larger than the threshold value. The forwarding of the query will be based on the information stored in the BATON tree and concerns the parent nodes, child nodes and neighbor nodes. Each peer that receives the query checks if the threshold value prunes all of its dataset. If that happens the query is forwarded to its neighbors whose range values are larger than the threshold value. Otherwise the peer computes its local skyline and uses the filtering point to prune the unnecessary points that are dominated. Each peer reports its results to the query initiator and after have refined the threshold value and the filtering point forwards the query as defined in the previous case. After the query initiator has received the results from all peers computes and reports the final skyline set.

### 4.2.5. Angle-based partitioning

Previously proposed methods for distributed skyline computation widely adopt grid-base space partition techniques. When a query is parallel processed over a grid partitioned space its processing performance degrades since many of the local results computed do not contribute to the final results, resulting in redundant computation and communication cost. Taking into account these considerations a different partitioning approach was proposed.

*Hyper-spherical coordinates*

In **[172]** authors proposed an angle-based space partitioning approach which uses hyper-spherical coordinates **[67]** of points in order to partition the space in such a way to increase the efficiency of parallel distributed skyline query processing. This method transforms the Cartesian coordinate space in hyper-spherical coordinate space and applies partitioning based on the angular coordinates as shown in **Figure 47.**

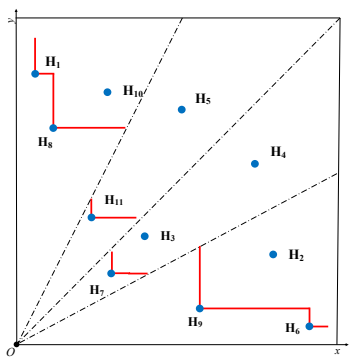
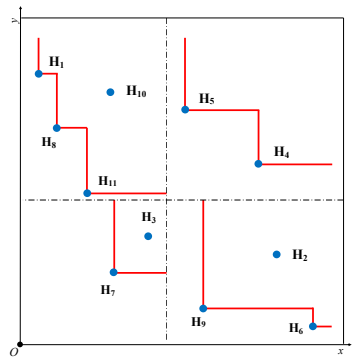

Figure 47: Angle-based partitioned dataset.      Figure 48: Grid-based partitioned dataset.

In **Figure 48** is illustrated the partitioning of the same dataset with a grid-based approach. Both approaches contain four equi-volume partitions resulting in four local skylines. Nevertheless angle-based approach is more efficient from the grid-based as will be explained. First even if the numbers of local skylines are the same, the total number of skyline points that will be reported in the grid-based approach is higher that this on the angle-based approach. As illustrated the grid-based approach will report 8 points but the angle based approach only 6. The global skyline set contains 6 points which mean that the 2 more points returned from the grid-based approach will be pruned and their computation and reporting cost was unnecessary. Additionally the local skyline points reported in the angle-based approach are more likely to be global skyline points in contradiction with those on the grid-based approach. One example is the upper-left partition of the grid-based approach **Figure 48** were none of the points that will report belongs to the final skyline set, in contradiction with angle based approach were each partition essentially contributes to the final skyline set.

In order to determine the partition boundaries authors sketch two algorithms that compute the angular boundaries of partitions for uniform and arbitrary data distributions, namely *equi-volume partitioning* and *dynamic partitioning* respectively. In *equi-volume partitioning,* for uniform data distributions, the key idea is to generate partitions of equal volumes. This will lead to equal number of points in each partition (in a uniform distribution). In the case of a d-dimensional space with volume $V_d$ and *N* number of servers the volume of each partition should be $V_d/N$. In grid-based partition approaches this is trivial and achieved by dividing each dimension in equal parts. For angle based partitioning this approach does not hold. As a counter example imagine one quadrant of the upper hemisphere of Earth (as sphere). The equator is divided into equal parts by the meridians and meridians have the same distance between them. The area of a region near the equator differs from that near the poles of the Earth. The same applies to the corresponding volume adding as a third dimension the radius of the Earth as constant (considered in spherical-earth model). A complex way to achieve equi-volume partitioning is to solve an equation system that satisfies the equality of the partition's volume. To limit the complexity involving this methodology authors proposed a two-step method that involves the division of each dimension of the space in $k = \sqrt[d-1]{N}$ equi-volume partitions using dimension's corresponding angular coordinate, where d is the number of dimension and N the number of partitions. A two dimension space uses one angular coordinate. Intuitively in a 3-dimensional space, that has two angular coordinates, the dataspace is first divided horizontally and then vertically rather than in one step. That is for a 3-dimensional space where we want 9 partitions the value of k will be k=3. First the dataspace must be divided horizontally (or vertically) in 3 equi-volume partitions. The angular coordinate bounds of each partition are computed based on this equality. Then space is partitioned vertically in the same way, computing also the angular coordinate bounds of each partition for the second dimension. Intuitively the partitioning is achieved by dividing each angular coordinate of each dimension of the full space in k not-equal parts. The concept of this inequality in the angle-based approach can be seen in **Figure 47** . As shown partitions that are abut to the axes have larger angular coordinate range than the others in order to be equi-volume.

The *Dynamic space partition* technique can be followed by both grid-based and angle based partitioning methods, in order to achieve balance distribution of points into partitions, even if the dataset is not uniform distributed. The basic idea is to recursively divide the space into partitions trying to maintain an equal number of points at each step of partition splitting by determining the appropriate partition bounds. The partition process can stop when the number of points in a partition reaches the threshold value $n_{max}=2*n/N$ which is defined based on the n points of the dataset and N number of partitions. This idea of a dynamic grid-based partitioning and dynamic angle-based partitioning is illustrated in **Figure 49** and **Figure 50** respectively.

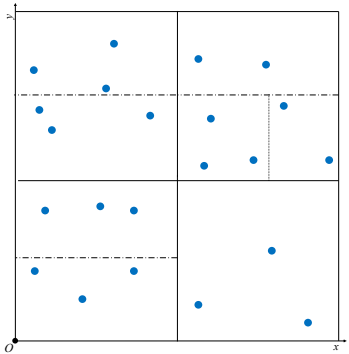
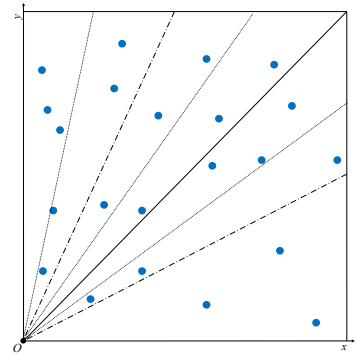

**Figure 49: Dynamic grid-base partitioning.**    **Figure 50: Dynamic angle-based partitioning.**

Finally mapping of points from the Cartesian coordinate space to the hyper-spherical space can be achieved as follows. First, the Cartesian coordinates of each d-dimensional point *p* are transformed to hyper-spherical coordinates. Each d-dimensional hyper-spherical coordinate is consisted by d coordinates where the d-1 represents the angular coordinates. The *d – 1* angular coordinates are compared with the angular boundaries of each partition and the corresponding partition that the point must be placed can be found.

The angle based partitions tends to distribute the space near the origin of the axis more uniformly to the servers, which increases the probability to evenly spread the (final) skyline points among partitions, in contradiction with grid-based partitioning which in general assigns most of them in one or a small number of partitions. This will eventually lead that servers will tend to report small result sets that will (in most cases) considered intact in the final skyline computation. Also in general, points that exist in angle-based partitions achieve higher pruning power than those in grid-based partitions by means that the average volume of the dominating regions of points inside angle-based partitions is higher than this on grid-based partitions **[172]**. However as reasoned in **[171]** the coordinate system that is used cannot easily exploit the skyline properties of points. Additionally cannot answer efficiently queries based on both *min* and *max* preferences due to the specific origin point selection in the partitioning process. Finally it is unclear how this method can efficiently answer other types of skyline variants such as **[44]** since it does not maintain the Cartesian properties that are needed.

### 4.2.6. Various topics

This section will outline various other topics on parallel and distributed skyline computation. This topics concern approaches that are not based on the exact skyline computation, reason about execution order on servers, are case studies or approaching a specific subject.

*Approximate skyline computation*

Authors in **[73]** studied the problem of distributed skyline computation over P2P networks and especially on Peer Data Management Systems (PDMS). It belongs in the horizontal data partitioning category **(4.2.3)** because each peer has its own data schemes and relations. In PDMS systems there is no global knowledge about the location (peer) that data are stored and the only way to retrieve information is by a flooding process. In order to deal with the problems of high communication cost and lack of global knowledge that arises in such type of networks, authors proposed the notion of relaxed skylines and sketched a routing index [38], called Qtree. Routing indexes are helpful in situations where global knowledge about the location (peer) that data information is stored, isn't available. In this scenario it is assumed that each peer is linked with one or more super-peers that incorporate a routing index, as distributed data summaries. The information that contained in one of the routing indexes can be visualized as regions (summaries) representing the existence of other non-local data points, without having the need to know which points and at which precise location exist. This helps to identify in which neighbors must be the query forwarded. This way queries can be forwarded to specific peers that will eventually contribute to the final result rather than to all by a flooding process. In order to further reduce execution costs (at the expense of accuracy) authors proposed to relax the completeness requirement of skyline by introducing relaxed skylines. A relaxed skyline is an approximation of the skyline similar in concept and inspired from thick skylines [89] in subsection 3.6. Recalling Figure 26 of thick skylines, in relaxed skylines a skyline point (center of circle) can be represented by any point within a user predefined distance ε from it (any point inside the circle). The benefit in terms of cost that this method implies is that several skyline points can be represented by one representation. This way is further reduced the numbers of peers to be queried.

In [175] authors studied the problem of distributed skyline computation in bandwidth-constrained environments such as wireless networks. In this approach authors assumed an ad-hoc infrastructure were all wireless devises compute their local skylines based on their assigned dataset and report their results to a central coordinator which deals with the merging process and the computation of the final skyline result. To fulfill the bandwidth constrain each device reports only a part of its local skyline set, resulting in an approximate final skyline set. In order for this to be achieved each device selects the local skyline points that will report based on the highest number of points that each one of them dominates on subspaces based on *skyline frequency* [26] described in subsection 4.1.3.

*Skyplan*

In [148] authors reasoned about the construction of execution plans that will define the order that servers must be visited in distributed environment in order to efficiently compute skyline queries. The execution plans take into account the various dependences that may occur, related with dominance or incomparability between the local datasets of each server. In their approach each server sends to the originator of the query a set of MBRs that describe the points that have in their dataset. Through these MBRs the originator builds the dependences that may occur due to dominance or incomparability of the datasets. These dependences are mapped to a weighted graph called *skyline dependency graph* (SD-graph). The weights on the graph represent the pruning power of each MBR based on the dominance area that imposes normalized by the number of points enclosed in the region, divided with the cardinality of the entire dataset. By comparing each point (representing an MBR) of the SD-graph derives the execution plan.

*Case studies*

The work in [136, 82] are case studies that deal with parallelizing skyline computation in multicore architectures [17]. The algorithms that have been parallelized are BBS [133, 134] and SFS [35]. Parallelization of algorithms was achieved with OpenMP [41] which is a programming environment for parallel computing on shared memory architectures such as multicore architectures. In OpenMP a programmer can annotate parallelizable loops in sequential programs in order to obtain a parallel program. In their work also proposed the algorithms PSkyline [136, 82] and SSkyline [82] in order to perform comparison analysis between the parallelization of existing state of the art algorithms and simple parallelizable techniques such as divide and conqueror (Pskyline) and nested-loop (Sskyline) methods. These algorithms are very simple and do not use any indexing or sorting. PSkyline is a simple D&C approach which does not use any index structure and divides the dataset into smaller group, locally computes skylines and merges the results. SSkyline is a simple nested-loop approach. An additional case study is the work on [56] where authors study the problem of skyline retrieval considering that the multidimensional points are distributed and stored in multiple disks. In their work proposed the two algorithms *Basic Parallel Skyline (BPS)* and *Improved Parallel Skyline (IPS)* algorithms. Both algorithms use parallel R-trees [90] utilizing multiple heaps (based on the number of disks) in order to simultaneously access the stored points on each disk and perform dominance comparisons. The *IPS* algorithm additionally incorporates a pruning strategy based on Dominating Regions.

*Joins*

In **[162]** authors considered the problem of **[88]**.(described on section **3.4.2**), where the dataset is considered to be stored in more than one tables, in distributed environments. Authors reason about joining approaches, in order to efficiently compute the skyline set over the grouped sets. The query operator between the multi-relations is called *skyline-join.* In this approach the tuples of tables are first grouped by the join attribute. Then is invoked a skyline algorithm (such as **2.3.8**) in order to obtain the local skyline points for each group. Then the global skyline points are computed from the local results.

### 4.2.7. Summary

The only survey work that is related with skyline query computation is **[74]** which study the skyline retrieval on parallel and distributed environments. The main classification on this work is based on structured and unstructured P2P systems. Based on the performance analysis conducted by authors and the algorithms outlined previously, isky outperforms skyframe in the case of BATTON overlay network. Additional SSP and Skyframe are more appropriate for skyline computation over CAN. In the case that is possible to have a super-peer structure SKYPEER+ is the appropriate algorithm. In the case that it is not possible to have a super-peer structure and the possibility of nodes failure is high MANET and PDMS are more appropriate solutions. If it is possible to perform a pre-processing in order to construct routing information the most appropriate algorithms are MANET and PaDSkyline. Note that these approaches are not applicable when the dataset is dynamic. For a complete exhaustive performance analysis of all the algorithm related with parallel and distributed skyline computation reader should refer to **[74]**.

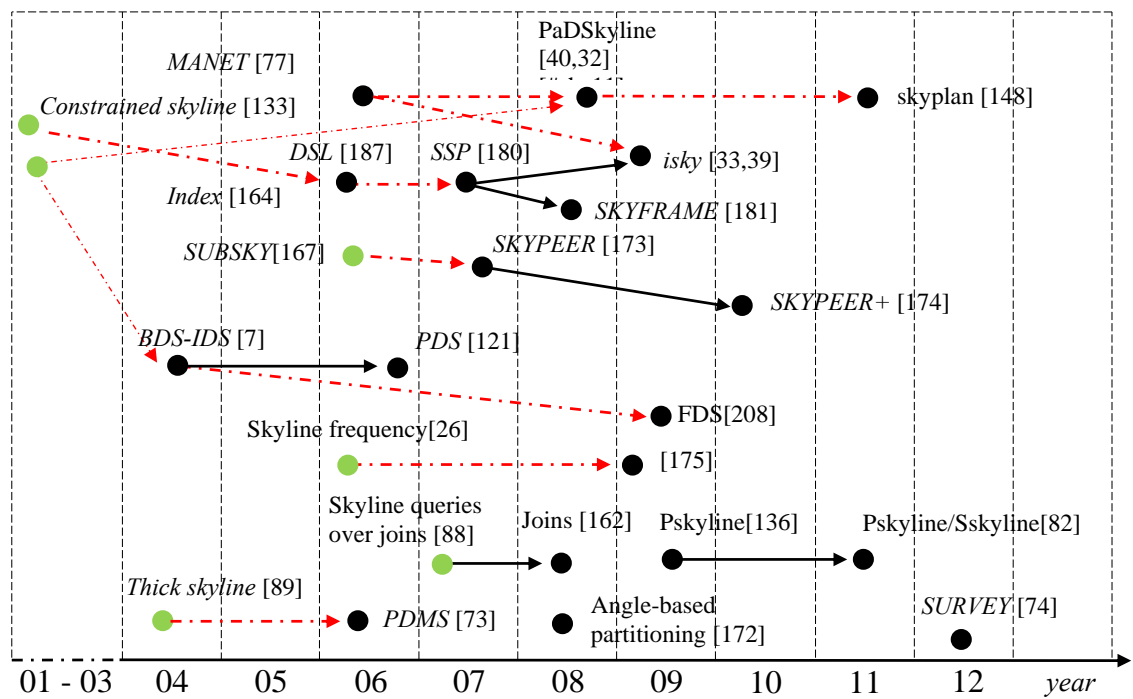

**Figure 51: Parallel and distributed skyline query hierarchy.**

Summarizing our study **Figure 51** illustrates the relation between the previous outlined works. As a remainder the dash red line represents the adoption of a general idea from another work and the black line represents the extension or the heavily dependence on a previous work. The green dot represents work that does not belong to the parallel and distributed skyline computation.

| Algorithm | Incorporates | Routing | Topology | Skyline Query Type | Main drawbacks |
|---|---|---|---|---|---|
| **BDS-IDS / PDS** | Vertical partitioning – sorting | Direct / initiator to peer | Web | | Vertical partitioning |
| **MANET** | Filtering point | Breadth/depth first | Manet | Subspace/ Constrained/ Dynamic | Exhaustive routing |
| **SKYPEER/ SKYPEER+** | Sorting / threshold value Ext-skyline | Super-peers | Super-peers | Subspace | Existence of super-peers |
| **PaDSkyline** | Multiple filtering points & MBRs | Cluster-heads | Clusters | Subspace/ Constrained/ Dynamic | Flooding / Heavy load on cluster-heads |
| **FDS** | Multiple-round filtering /sorting | Direct / initiator to peer | Web | Subspace/ Constrained/ Dynamic | Many rounds on large networks |
| **DSL** | Partial ordering | Local routing table / neighbors | CAN | Constrained | Load balance – High cost on updates |
| **SSP** | Filtering points/Partition ordering (z-order) | Balanced tree adjacent nodes | BATON | Constrained | Load balance |
| **SKYFRAME** | As SSP + Border regions | Balanced tree adjacent nodes | BATON-CAN | Constrained | Load balance |
| **isky** | Sorting/ filtering points/ threshold value / | Balanced tree range search | BATON | | Load balance |
| **Angle-based partitioning** | Hyper-spherical coordinates | - | - | - | Issues on its application |

Table 32: Fundamental algorithms on parallel and distributed skyline computation.

Table 32 outlines the fundamental key points and ideas used, the routing method followed, the topology of the network that an approach takes into account and a general problem that may occur in the application of the algorithm in a general scenario. In the outlined table every algorithm is applicable for the general skyline query but this information is omitted for readability and space efficiency. Note that in different application scenarios the problems that arise with each method may be different.

## 4.3. Attribute & data-specific applications

This section reasons about specialized datasets that wasn't previously considered. Previous methods considered that all attributes of all dimensions are available, for all points. Additionally there is the case of incomplete datasets, where the points miss some of their dimensions/attribute values, partial order datasets, where the ordering of attributes can't be defined or is defined differently by each user and finally uncertain datasets where an object may have different instances that can occur with different probabilities.

### 4.3.1. Partially ordered data

The previous studies focused on total order (TO) domains (dataspace). That is datasets where their attributes have an internal ordering such as numbers. In these domains it is easy to understand which attributes are preferable than others. The lack of ordering or preference among a pair of attributes indicates that a domain is partially ordered (PO). Authors in [24] focused on skyline computation over partial-ordered domains. This type of domains among others can include intervals, hierarchies, domains of set values and preferences.

Initially a naive approach to compute partial-ordered skyline queries (POS-queries) was to incorporate the non-index based BNL [19] algorithm. BNL works for all types of domains but is inferior from the index-based algorithms and lucks of progressiveness. Index-based algorithms such as NN [97] and BBS [133, 134] are the most efficient progressive algorithms for skyline computation in total-ordered domains. Nevertheless, the efficiency of these algorithms may not hold on partially-ordered domains due to their reduced pruning power.

In order to deal with the problem of skyline computation over partial-ordered domains author's basic idea is to transform each partial-order domain into a total-ordered domain in such a way to preserve the partial ordering in the transformed space. Then organize the transformed domain (and subsequently it's values) in an index, in order to compute the skyline with an index-based algorithm. Due to the transformation of space, false-positives may arise that needs to be removed.

Authors in their work proposed the BBS+, SDC (Stratification by Dominance classification) and SDC+ algorithms. BBS+ is an extension of BBS but does not maintain his progressiveness due to the false positives. The SDC and SDC+ exploit properties of domain mapping in order to avoid unnecessary dominance checking. Especially SDC+ processes the data in such a way to handle and prune the false-positives in order to achieve progressiveness.

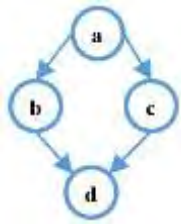

Figure 52: directed acyclic graph of domain D.

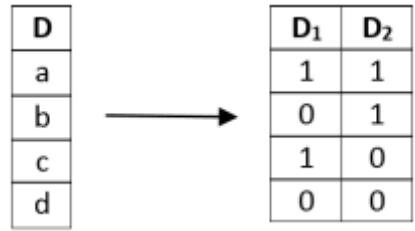

Figure 53: Domain transformation.

The easiest method to achieve an order transformation is to map the partial-ordered domain into boolean valued, total ordered domains. This approach is illustrated in Figure 52 and Figure 53. Initially a partially ordered domain D, which has in it's dataspace four values a,b,c,d, can be represented by a directed acyclic graph (DAG), named Hasse diagram (Figure 52). Each node of the graph corresponds to a value of the domain. In DAG, a direct edge from node x to node y exists only if x<y (value of x is smaller than value of y) and there does not exist any node z for which x<z<y (value of x smaller than value of z, which is smaller from value of y). A value x will be preferred from a value y if there exists a directed path from x to y in the DAG. Based on the previous definition node *a* dominates *b* and *c* and both of them dominate *d*, but there does not exists any dominance relation between *b* and *c*.

The partial ordered domain D can be mapped using two, boolean valued, total ordered domains $D_1$, $D_2$ as shown in Figure 53. Since in a boolean domain we have two distinct values, using two boolean domains we can represent $2^2$ distinct values. This way given two points p,r on the partial ordered domain D, $p_D$ will dominate $r_D$ (in the partial order domain D) if p dominates r in the transformed domain and thus $p_{D1}$ dominates $r_{D1}$ and $p_{D2}$ dominates $r_{D2}$. Nevertheless, this approach suffers from the curse of dimensionality.

$$f(x) = \begin{cases} [0,3], & if(x = a) \\ [0,2], & if(x = b) \\ [1,3], & if(x = c) \\ [1,2], & if(x = d) \end{cases}$$

Figure 54: Mapping function *f*.

In order to use the skyline computation methods implemented for the totally ordered domains and avoid the course of dimensionality authors proposed to use an efficient approximate space transformation. This approach maps the domains by using for each partially ordered value an approximate interval representation in the form of a pair of integer values, as shown in **Figure 54**. Essentially, the encoding scheme that is used is the one on **[2],** but other encoding schemes such as **[57]** can be used. The encoding scheme creates a spanning tree on DAG. The basic idea is to construct a 1-1 mapping function *f* which transforms any value u in the partial ordered domain D, into an interval. Considering the same values on **Figure 52,** function f can be defined as illustrated in **Figure 54**. Using this transformation a value u∈D will dominate u'∈D if f(u) dominates f(u'). That is if u=a and u'=d then *a* dominates *d* since f(d) ⊂ f(a) as [1,2] ⊂ [0,3]. That is if the mapped interval of a node *d* is contained in the mapped interval of another node *a* then the node *a* is preferred from *d*. If two intervals are disjoined then the values retrieved from D are incomparable. Due to the approximate space transformation may exist cases where comparable values (in PO domain) may become incomparable (in TO domain) leading into false positives, by identifying non-skyline point as skyline points. This case occurs when, for two values from the PO domain (even if they are comparable and a dominance relation holds), neither of their transformed interval values contains each other (leading in incomparability). In hierarchical partial order false positives can be avoided with appropriate domain mapping but is inevitable for non-hierarchical partial orders.

Authors in their work used a more restrictive form of dominance named m-dominance. In m-dominane a point p will m-dominate a point r if p is at least as good in all the totally ordered dimensions, its interval value match with those of r for all partial ordered dimensions and exists a totally ordered dimension in which p is strictly better, or a partially ordered dimension where the interval value of r is contained in the interval value of p.

In general, the algorithms work as follows. First, a domain mapping function f(x) is constructed in order to transform each partial-ordered value u∈D in boolean-valued domains. Next an index-based method is used to organize the data and compute the skyline by additionally taking into account the false positives. The selection of the domain mapping function is independent of the choice of the index-base algorithm that will be used to compute the skyline. Considering the BBS+ algorithm the only difference in comparison with BBS is that the two dominance checks (dominance check on a leaf or on an intermediate node) on the BBS are replaced by the m-dominance comparisons. Additionally due to the occurrence of false positives an additional function UpdateSkyline is used. The function detects and removes the false positives by comparing any new point p with all the temporally skyline points maintained in a set S. The SDC algorithm organizes the data in two sets at runtime. The first set will contain the points that are definitely be in the skyline and the other one those that may be false positives. The SDC+ algorithm partitions the data in two or more set where the points on the $i^{th}$ set cannot dominate the points in the $i^{th}$-1 set. This way the points on the $i^{th}$-1 set can be returned before examining the points in set i. The pseudocode of the BBS+ algorithm is presented below.

| **ALGORITHM 27:** UpdateSkylines (e, S) | **ALGORITHM 28:** BBS+ (T, S) |
|---|---|
| **Input:** A data point e in some leaf node of an R-tree. An intermediate set S of skyline points.<br>**Output:** Return an updated set S.<br><br>1. **for** each p ∈ S **do**<br>2.   **if** (e is dominated by p) **then**<br>3.     return S;<br>4.   **else if** (p is dominated by e) **then**<br>5.     Delete p from S;<br>6. Insert p into S;<br>7. **return** S; | **Input:** T is an R-tree.<br>S is an intermediate set of skyline points.<br>**Output:** Set of skyline points.<br><br>1. Initialize heap H to be empty;<br>2. Insert all entries in the root node of T into heap H;<br>3. **while** (H is not empty) **do**<br>4.   Remove top entry e from H;<br>5.   **if** (e is an internal entry) **then**<br>6.     **if** (e is not dominated by any entry in S) **then**<br>7.       **for** each child entry ei of e **do**<br>8.         **if** (ei is not dominated by any entry in S) **then**<br>9.           Insert ei into H;<br>10.     **else**<br>11.       S = UpdateSkylines(e, S);<br>12. **return** S; |

According to authors of **[151]**, previous work **[24]** is only applicable to static skylines and has limited progressiveness and pruning ability. The mapping fails to preserve all the preferences that exist in the original domain. In [151] authors extended the work of [24] and presented a progressive framework named Topologically-sorted Skyline (TSS). In general, TSS topologically sorts the nodes of a DAG and extracts the spanning tree. Then associates with each value in the PO, the ordinal number in the topological sort and multiple intervals determined by the spanning tree. An additional dominance check is performed that is not related with the total ordered domain.

Because TSS does not restrict or modifies the original dominance definition has not any false positives that are needed to be re-examined and thus is progressive returning the skyline points immediately. Authors additionally proposed a novel dominance check that further enhances progressiveness and pruning ability. Also studies the case where the preferences among the partial ordered attributes are not uniquely defined, as users may often have different, even conflicting preferences. For that reason proposed the dynamic skyline over PO domains. In the case of a dynamic skyline in a partial order domain, the method explicitly redefines all the dominance relations, by defining a partial order for each PO domain, every time a query is issued. In contradiction, the previous method would need to re-compute the whole transformed domain which would incur considerably overhead.

The work in [184], [185] considers the case where different users have different preferences and thus the ordering imposed on the dataset changes for each user. Authors assume that they possess a template that contains a partial order for every dimension, which is applicable to every user and is constructed based on previous knowledge of user's preferences or assumptions on the ordering, such as smaller delay is preferable. A user's preference can be modeled as a partial order. Then each user can express his/her preference by refining this template. The method semi-materializes the space, based on this template and efficiently answers any skyline query based on any given preference input.

### 4.3.2. Incomplete data

Another data-related skyline computation approach is the one in [91] which assumes that data are incomplete, meaning that have missing values in some of their dimensions/attributes. Most of the algorithms assume data completeness on all dimensions and transitivity in the dominance relation. However, this is not always the case. On incomplete data, the transitivity does not always holds. A non-transitive dominance property may lead to cycle dominance [section 4.1.3 (K-dominant)] resulting that each point is dominated by at least one other point. Lack of transitivity also affects negatively pruning and indexing techniques. The closest work to this is the one in [25], which also does not assume that transitivity holds.

Each incomplete dimension of a point is denoted with "-". Thus a 3-dimensional point with an unknown value on the third dimension will be denoted as (a,b,-). The rest of the remaining known values are assumed to have a total order. The problem is to find the skyline set of a set of points where there is at least one dimension that is known and there is a non-zero probability for the other dimensions to be unknown. In this type of problem the dominance relation that is applied is the traditional dominance relation considered only over the common, known dimensions. This way for the 4-dimensional points p=(1,-,-,3) and r=(4,3,-,-), the dominance relation will be considered only on the common known dimension which is the first. Since r.d1<p.d1, point r will dominate point p. In the case of points p=(1,-,-,3) and r=(-,4,2,-) there isn't any common dimension and thus the points are considered incomparable. In order to find if two points are comparable authors proposed a bitmap representation. This way for each d-dimensional point a d-bit representation is derived. Given a point p, the ith bit will be zero if the ith attribute value is unknown and in all other cases will be 1. Thus point p=(4,3,-) will have bitmap representation 110, point r=(-,-,1) a bitmap representation of 001 and point s=(2,-,3) a bitmap representation of 101. Based on this representation two points are comparable if the bitwise AND operation of their bitmaps representations has a non-zero value. Points p and r are incomparable because 110&001=000 but points p and s will are comparable because 110&101=100.

Due to incomplete data the previously described dominating relation is non-transitive. The lack of a transitive dominance relation can lead to a cycle dominance. As an example consider the points p1=(4,2,-), p2=(3,-,4), p3=(-3,2). P1 dominates p2, in turn p2 dominates p3 but additionally p3 dominates p1 leading in a cycle dominance as also described in **[section 4.1.3 (K-dominant)]**. In this case none of the points can be considered skyline points.

To solve the overall problem of skyline computation over incomplete data, authors firstly proposed two algorithm namely *replacement* and *bucket* which incorporate the traditional skyline algorithms. These algorithms improve the naïve solution in which an exhaustive pairwise comparison between all points is performed. Next authors proposed the *ISkyline* algorithm which employee's two optimization techniques namely *virtual points* and *shadow skyline* in order to avoid the cycle dominance that may occur due to the non-transitivite dominance relation. The *replacement* algorithm replaces any unknown dimension value ("-") with $+\infty$ (in case of maximization $-\infty$). This way the incomplete dimensions are converted to complete, allowing to apply a traditional skyline algorithm to retrieve the skyline set S of the modified dataset. Then all $-\infty$ values are replaced with "-" as it was in the original dataset and an exhaustive pairwise comparison is performed in order to retrieve the final skyline set S. The *bucket* algorithm divides all the points of the dataset in distinct lists (buckets) in such a way that the points in each list have the same bitmap representation. This way the transitivity will hold inside each list and a traditional skyline algorithm can be applied to retrieve the local skyline of each list by simply ignoring the unknown dimension. The *local* skylines are merged in one list forming the *candidate* skyline points and an exhaustive search is applied to retrieve the final skyline. Nevertheless, the previous proposed methods are not efficient due to the large resulting size of candidate skyline points, due to the union of all the local skylines, and the exhaustive search operation. Additionally the local skyline points derived from a list $l_1$ are not used to filter/prune points in other lists (e.g $l_2$) (and subsequently reduce their size) prior their local skyline computation (e.g prior the local skyline computation on $l_2$). For that reason the ISkyline algorithm was proposed that incorporates two optimization named *virtual points* and *shadow skyline*.

A *virtual point* is used in order to reduce the number of local skyline points in each list and essentially the points in the candidate skyline list. The general idea is to use a (local skyline) point from a list i to reduce the size of a list j with i≠j. A local skyline point retrieved from a list *i* will be transformed into a virtual point, by considering only the common dimensions of it's bitmap representation and those of the bitmap representation of the list that will be placed. As an example, consider the lists *list1* and *list2*, which contain points with bitmaps representations in the form of $l_{1p}$=(4,3,-) and $l_{2p}$=(5,-,3) respectively. The local skyline point from *list1* that will be selected (assuming that $l_{1p}$ is a local skyline point of *list1*), will be transformed in a virtual point for *list2* with the form of $p_v$=(4,-,-). The points of the list that are dominated by a virtual point that was placed in that list are not stored as local skyline points and not propagated to the candidate skyline list. Note that virtual skyline points are also not propagated to the candidate skyline list. The problem that arises with the use of *virtual points* is that we cannot rely on an exhaustive pairwise comparison of the points in the candidate skyline list, in order to retrieve the final skyline set. This happens because points that are not stored in the local skyline list of a list i may dominate points in the candidate skyline list, that came from other lists different from list i. For that reason a point p in the candidate skyline list must be compared with every other point that has comparable bitmap representation regardless if it is candidate skyline or local skyline point because may help to further prune dominating candidate skyline points. Thus a point p that does not belong to the local skyline of a list cannot be discarded since may help in dominating candidate skyline points.

In order to reduce the need of storing and comparing all the data points authors proposed the use of *shadow skyline* that will work among with the virtual skyline points. The shadow skyline of a list *l* is derived only from the points of the list that do not belong in the local skyline of that list. Essentially the shadow skyline will be the skyline of these points. This way only a part of the dataset is kept (that is the shadow skyline for each list), allowing the comparison of the candidate skyline points only with the points in the shadow skyline, rather than with all the points of the dataset.

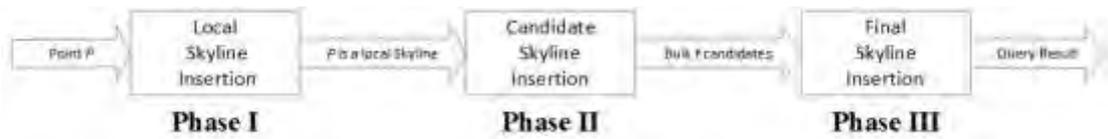

Figure 55: Phases of ISkyline [#105] [91] .

The *ISkyline* algorithm incorporates a parameter *t* based on which uses a number of *t* candidate skyline points each iteration and process them in order to retrieve the final skyline result. Essentially the value of *t* is a tuning parameter that controls the frequency of updating the skyline results. A smaller t value will allow the algorithm to update more frequently the query result. The algorithm stores the *local skyline* lists that may contain real and virtual points, the *shadow skyline* lists that contain only real points and a *updated* flag which indicates if a *shadow skyline* list is modified through the iterations of the algorithm. The flag is an optimization that is used in order to prune the search space by avoiding looking in unmodified lists. Additionally a *candidate* and a *final skyline* list are maintained. The algorithm is consisted by three phases as shown in Figure 55. In the **initialization step** of the algorithm the retrieved points of the dataset are placed in their corresponding list based on their bitmap representation. In the **first phase** every point p in each list N is checked if it is needed to be stored in the *local skyline* list of the list N, in the *shadow skyline* list of the list N or discarded. If the point p is not dominated by any other point in the *local skyline* list is inserted into it the *local skyline* list. If it is dominated by a virtual point, is inserted into the *shadow skyline* list of the list N. In all other cases, is discarded. The algorithm proceeds in the **second phase** considering only the points of the *local skyline* list, where it checks if those points are needed to be stored in the *candidate skyline* list. A point p, from the *local skyline* list, is inserted in the *candidate skyline* list if it is not dominated by any comparable point in the *candidate skyline* list. In this phase are also constructed the virtual points that will be inserted in the N lists, based on the domination comparison of the points with the points already exist in the *candidate list*. When the algorithm has retrieved *t candidate skyline* points proceeds to the **third phase** where the *final skyline* list is updated. This phase is mainly consisted from four steps. In the first step is performed a dominance comparison against the points of the *candidate skyline* list and the current points in the *final skyline* list (that already exist due to previous iterations of the algorithm), in order to remove dominated points from the *final skyline* list. In the second step all the points of the *final skyline* list are compared against all comparable (but not equal) *shadow skyline* lists with an *updated* flag set to true. If a point from the *shadow skyline* list dominates a point in the *final skyline* list, then the point from the *final skyline list* is removed. The third step essentially has the same functionality with the second step with the difference that the comparisons are placed against the points in the *final skyline* list and those in the *candidate skyline* list. Note also that the candidate skyline list has no flag. In the final step the *final skyline* list is formed by combining all the points in the *candidate skyline* list and the points in the *final skyline* list. Additionally all the *updated* flags, of every *shadow skyline* list, are set to false in order to indicate that there is no modification in any of the lists, that was not considered in the computations till now.

In [124] authors studied the skyline queries over crowd-enabled databases. Crowd-enabled databases deal with incomplete data during runtime by requesting missing values or complete tuples from other sources. Authors approach focus on cost optimization of crowd-enabled skyline queries by managing the amount of data that are needed to be retrieved applying *missing data prediction* approaches and *prediction risk* estimation.

### 4.3.3. Uncertain data (probabilistic skyline)

Authors in [138, 87] in order to tackle the problem of skyline computation on uncertain data, proposed a probabilistic skyline model and additionally the p-skyline. Some conditions that may impose uncertainty on data are limitations on receiving and measuring data, missed or delayed data reports and randomness of data. In general, the probability of an object to be in the skyline, is the probability that the object is not dominated by another object. The p-skyline, where p is a user defined threshold probability, will contain all the uncertain points that have probability at least p to be in the skyline, with 0≤p≤1. The uncertainty of data is typically modeled as a *probability density function* (pdf). Any uncertain point u in the dataset D can be represented by a probability density function *f*, where f(u)>0 and for any point u in the dataset D, $\int_{u \in D} f(u)du = 1$.

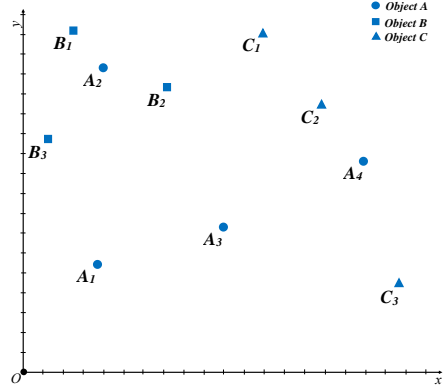

**Figure 56: Dataset of uncertain objects.**

Nevertheless, the pdf function of an uncertain point may not be available. In this case, authors assume that for each uncertain point there are instances/points that each one of them partially represents it. As an example consider **Figure 56,** where exist three uncertain objects A,B,C with A having four instances and B,C three. Those instances are used to approximate the pdf function and can be sampled or retrieved from the dataset. Thus each uncertain point U is model as a set of points in the dataset that will represent its instances, denoted as U={$u_1,u_2,...u_n$}. The number of it's instances is denoted as |U|=n and the uncertain object can have only one instance at a time. The skyline probability Pr($u_i$) of an instance $u_i$ that belongs to an uncertain point U is the probability that this instance exists and is not dominated by any other instance (not related with U) or point.

| Object A | | | | Object B | | | Object C | | |
|---|---|---|---|---|---|---|---|---|---|
| $A_1$ | $A_2$ | $A_3$ | $A_4$ | $B_1$ | $B_2$ | $B_3$ | $C_1$ | $C_2$ | $C_3$ |
| 1 | 0.67 | 1 | 1 | 1 | 0.75 | 1 | 0 | 0.035 | 1 |
| 0.9175 | | | | 0.91 | | | 0.3345 | | |

**Table 33: Skyline probabilities of instances.**

Given two uncertain objects U,V, object U may dominate V, V may dominate U, or they may be incomparable. This approach gives that Pr[U≺V] + Pr[V≺U] ≤ 1. The probability Pr(u), of an instance u of U, that u is not dominated by any other instance/point, and thus u to be in the skyline, is denoted as $\Pr(u) = \prod_{V \neq U}\left(1 - \frac{|\{v \in V | v \prec u\}|}{|V|}\right)$. Since the instances of an object are mutually exclusive, the skyline probability Pr(U) of an uncertain object U={$u_1,u_2,...u_n$} will be equal to the sum of the skyline probabilities of its instances, divided by the number n of its instances and therefore, $\Pr(U) = \frac{1}{n}\sum_{u \in U}\Pr(u)$, represents the skyline probability of U. For a dataset D of uncertain objects, the p-skyline will be the set of uncertain objects U∈D which have probability Pr(U) to be in the skyline, at least p. That is $sky(p) = \{U \in D \mid \Pr(U) \geq p\}$. In **Table 33** is shown the skyline probability of each uncertain instance of every uncertain object and additional the skyline probability of each uncertain object. Instances $A_1, A_3, A_4, B_1, B_3, C_3$ will be always skyline points when they exist. Instance $C_1$ cannot be a skyline point because in any case will be dominated by an instance of B. Instance $A_2$ and $B_2$ can be a skyline point only when instance B3 and A1 respectively does occur. Instance $C_2$ depends almost on all instances of A and B (except instance $A_4$). Folowing is presented the pseudocode of the Bottom-up algorithm. With Pr(U) is denoted the skyline probability of an object U. As Pr$^+$(U) and Pr$^-$(U) is denoted the upper and lower bound of the skyline probability of object U.

**ALGORITHM 29:** *Bottom-up algorithm* **[138, 87]**
**Input:** A set of uncertain objects *S* and the probability threshold p.
**Output:** The p-skyline of S.

1. SKY = ∅;
2. **FOR EACH** object U ∈ S **DO**
3.    Pr$^+$(U) = 1; Pr$^-$(U) = 0;
4.    compute $U_{min}$, the minimum corner of its MBB;
5. **END FOR EACH**
6. build an R-tree to store $U_{min}$ for all U ∈ S;
7. build a heap H on $U_{min}$ for all U ∈ S;
8. **WHILE** H ≠∅ **DO**
9.    let u ∈ U be the top instance in H;
10.    **IF** u is from a non-skyline object **THEN NEXT**;
11.    **IF** u is dominated by another object **THEN**
12.       **GOTO** Line 22; // Pruning Rule 3
13.    **IF** u is the minimum corner of U **THEN**

| | |
|---|---|
| 14.     *find possible dominating objects of U; // Section 3.4.1* | 24.     $Pr^+(U) = Pr^-(U) + U.Pr_{max} \cdot \frac{|\overline{U}|}{|U|}$; |
| 15.     *compute Pr(u); // Section 3.4.2* | 25.     **IF** $\frac{|U-U'|}{|U|} \cdot min u \in U'\{Pr(u)\} < p$ **THEN** |
| 16.     **IF** *Pr(u) ≥ p* **THEN** | 26.         *apply Pruning Rule 4 to prune other objects;* |
| 17.         *partition instances of U to layers; // Section 3.3.2* | 27.     **IF** *Pr⁻(U) ≥ p* **THEN** |
| 18.     **ELSE** *U is pruned; // Pruning Rule 1* | 28.         *SKY = SKY ∪ {U}*; **NEXT**; // *Pruning Rule 2* |
| 19.     **ELSE** | 29.     **IF** *Pr⁺(U) ≥ p* **THEN** |
| 20.         *compute Pr(u); // Section 3.4.2* | 30.         *insert the next instance of U into H;* |
| 21.         $Pr^-(U) = Pr^-(U) + \frac{1}{U} Pr(u)$; | 31.   **END WHILE** |
| 22.         **IF** *u is the last instance at a layer* **THEN** | 32.   **return** *SKY*; |
| 23.             *update U.Prmax;* | |

Based on the previous definitions a bottom-up and a top-down algorithm where proposed. Both algorithms follow a bounding-pruning-refining recursive iteration approach. In general, the bottom-up algorithm computes the skyline probabilities of selected uncertain objects and uses them to prune other instances and uncertain objects. In more detail the bounding stage of the algorithm computes, for an instance $u_i$ of an uncertain object U, an upper and a lower bound of its skyline probability $Pr(u_i)$. Using the equation $\Pr(U) = \frac{1}{n}\sum_{u \epsilon U} \Pr(u)$, is obtained the upper and lower bound of the skyline probability Pr(U) of the uncertain object U, based on the computed skyline probabilities $Pr(u_i)$ of it's instances. In the pruning stage if the lower bound of Pr(U) is larger or equal to the user-defined threshold probability p, then U is in the p-skyline otherwise is not. If p is between the lower and the upper bound an additional iteration is needed in order to refine the results by getting tighter bounds. The iterations continuous until it can be determined for every object if it belongs to the p-skyline or not. Summarizing the algorithm computes and refines the bounds of instances of uncertain objects by selectively computing the skyline probabilities of a small subset of instances. The uncertain objects can be pruned based on the skyline probabilities of their instances or those of other objects. The top-down algorithm represents an uncertain object in the form of a minimum bounding box (MBB) that encloses all of its instances. The MBB is defined based on the minimum and maximum attribute values of all the instances of the uncertain object. This way the minimum and maximum corners of the MBB can be used to bound the skyline probability of an uncertain object. To improve and refine the bounds the algorithm partitions the instances of the uncertain object in subsets, in a recursive way, and prunes subsets and uncertain points. The skyline probability of each subset can be bounded in the same way as previously. Finally, the skyline probability of an uncertain object can be bounded as the weighted mean of the bounds of its subsets.

In [4] authors propose efficient methods to compute the skyline probabilities on all objects. In their work proposed a more general uncertain model where the instances of each uncertain object may have different probabilities to occur and that the probabilities of all instances may sum up to less than 1, meaning that may exist an instance of an object that is not known to us. The last observation is because an instance of the object or the object itself in general may not be present which is known as tuple uncertainty [158]. Additionally authors abandoned the use of a threshold probability value to prune the dataset since they argue that there are cases where low probability results may be useful to the user giving him a more detailed report of all the possible results. The main drawbacks of a threshold based approach that authors (of [4]) outline is that an appropriate threshold may not be easy to be selected leading in too few or too many results. Moreover, the user must be able to identify the interesting points from the skyline according to his/her own utility function. On the other hand, a threshold-based approach may make indirect (implicit) assumptions for the user's utility function. Such an assumption is that user does not want any information about skyline points with low probability, which is not always the case. A specific threshold may prune low probability points that would be highly desirable based on the user's utility function. In their approach, authors compute the skyline probabilities of all instances, from which they compute the skyline probabilities of all objects.

In [5] authors study the same problem with [4] and propose an asymptotically faster algorithm for the worst-case. The algorithm uses the same partitioning technique as in [4] but handles the partitioned sets more efficiently. Additionally authors study the online version of the problem where no query point or instance is known in advance. In [93] authors further studied the problem of [4]. In their

Page 88 of 127

work compute the exact skyline probabilities of all objects in high-dimensional datasets by incorporating a ZB-tree [106]. Additionally, proposed a probabilistic skyline algorithm for uncertain data streams and developed a top-k probabilistic skyline algorithm in order to retrieve the top-k objects with the highest probabilities.

In [110], authors reason about reverse skyline computation over uncertain data in monochromatic and bichromatic cases. In the monochromatic case the point (object) of interest and the query point (object) are of the same type and thus from the same dataset, while in the bichromatic case there exist two different types of points (objects). The monochromatic probabilistic reverse skyline (MPRS) returns the uncertain points whose dynamic skyline contains a given query point with probability greater or equal to a user-defined threshold. The bichromatic probabilistic reverse skyline (BPRS), given two uncertain datasets A and B and a query point q, returns the points that belongs to A whose dynamic skyline on the dataset B contains the query point q. In general, the first phase of the algorithm uses a R-tree to index the uncertain regions of each point. The second phase is a pruning phase which prunes the object using the spatial or probabilistic pruning method in order to reduce the search space as much as possible. In the last phase the algorithm computes the probabilities of regions that could not be pruned previously and returns the final result. This work is partially based on [138][87] which considers the monochromatic case. Considering only the approach related with uncertain data, the difference in the approach of [110] from [138][87] is that the uncertain data that are received, are considered as uncertain regions in the form of hyperrectangles, rather than a precise point. An example of an uncertain range can be considered the price/distance range of a house from a metro station, that a buyer is willing to buy, as the buyer would not specify an exact price or distance. In [111] authors extended their work on [110] and proposed the probabilistic reverse furthest skyline (PRFS) which considers the case where minimization of preferences is desired rather than the maximization. Additionally is proposed a variation of the probabilistic reverse skyline (PRS) query that returns the k objects with the highest probability among all. In addition, authors studied the retrieval of the points with k highest probabilities and a ranking approach on the PRS query results, called top-k reverse skyline that retrieves the k points that have the highest number of dynamic skylines.

Authors in [48, 49] reason about distributed skyline computation over uncertain data. Their scenario is based on the existence of a number of distributed sites that each one of them contains a number of uncertain data and a centralize server that processes the query. In [150], authors reason about contextual skylines taking into account the uncertainty in user's preferences rather the uncertainty of attribute values. The uncertain preferences are based on previously stated preferences for specific contexts. The uncertainty in user's preferences can be defined as the uncertainty imposed due to lack of defining user's preferences for some contexts. In this scenario is assumed that the transitivity property of the dominance relation does not hold. Authors in [199] further studied the problem of [150] without taking into account the independent object dominance assumption that was considered in [150] and in which the object's dominances are considered as mutually independent events. Authors in [203] reason about the top-k skyline computation over uncertain data. In their work proposed an exact method that retrieves the top-k skyline points against skyline probabilities and an approximate algorithm to deal with cases where each uncertain object has a large number of instances or a continuous pdf function.

*Trade-off and Stochastic skylines*

Trade-off skylines were first proposed in [123]. A trade-off is defined as, how much is willing to sacrifice in one dimension in order to gain an improvement in another dimension/attribute. This concept is primary reflected by the top-k retrieval paradigm using weighting over each attribute. This constructs a scoring (or utility) function which is used to compute the overall ranking of an object based an all of it's attributes. Nevertheless a weighting approach, that can be modeled for example as 0,6*price+0.2*bedrooms cannot be easily used by users to express their preferences. This happens because such kind of expression are to restrictive since the user may not want to hold for every case. For that reason authors propose the use of trade-off which is essentially a relation between two attributes. An example of a trade-off $t_1$ between house H1 with price 800.000 and 2 bedrooms and house H2 with price 810.000 and 3 bedrooms will be in the form $t_1:=((810.000,3)\prec(800.000,2))$. That means that a user is willing to trade-off additional 10.000 for an additional bedroom and thus house H1 will be dominated taking into account the trade-off $t_1$. Essentially the trade-off imposes an additional dominance relation over a specific range of values defined by the user. This range of values can be considers as [800.000, 810.000] and [2,3] for dimensions price and bedrooms respectively. In the case of multiple trade-off, where additional dimensions are considered, authors construct a tree-based representation structure in order to materialize all the trade-off chains. Essentially a basic approach to compute the trade-off skyline is to retrieve the skyline set with a conventional skyline algorithm, such as [97] and then apply the trade-off relations in order to refine the set. Note that the trade-off skyline is a subset of the original skyline query.

Stochastic skyline where proposed in [119],[201]. In a stochastic model, the subsequent state of the system can be determined probabilistically based on previous states. As an example, a future stock price will be equal to the current stock price plus an unknown change that can be determined probabilistically. The proposed model uses for each user one scoring function for every dimension. The ranking of a points $u$ is based on the expected value $\mathrm{E}\left[\prod_{i=1}^{d} f_i(u_i)\right]$ which is defined to be the product of all scoring functions $f$ over all the dimensions $i$ that are defined. The algorithm returns the minimum set of candidates for the optimal solutions over all possible monotonic multiplicative utility functions. Given a set F of utility (scoring) functions based on the preferences of all users, an uncertain object u will stochastically dominates an uncertain object v iff E[f(u)]>=E[f(v)] for every preference function that is defined. Given a set U of uncertain objects the stochastic skyline, regarding F will contains all the objects of this set that are not stochastically dominated by any other object in u regarding Authors approach is based on the R-tree which additionally indexes the mean μ and the variance $σ^2$ of objects.

### 4.3.4. Summary

This section outlined the skyline computation on various types of datasets that are different from the previous static/dynamic datasets that were discussed. These datasets in general can be categorized in partially ordered datasets, incomplete datasets and uncertain datasets. Partially ordered datasets represent the case where there does not exist a permanent total order, among the points of the dataset, or the preferred ordering changes depending on user preferences. Incomplete datasets represent the case where specific attribute values, of the points of the dataset, are not known. Uncertain datasets involve a probabilistic approach. The cases that are covered are the case where the points of the dataset have many instances, with different probability each instance to occur and the cases of trade-off and stochastic models, where users may want to bias their preferences based on certain feedback derived from existing data.

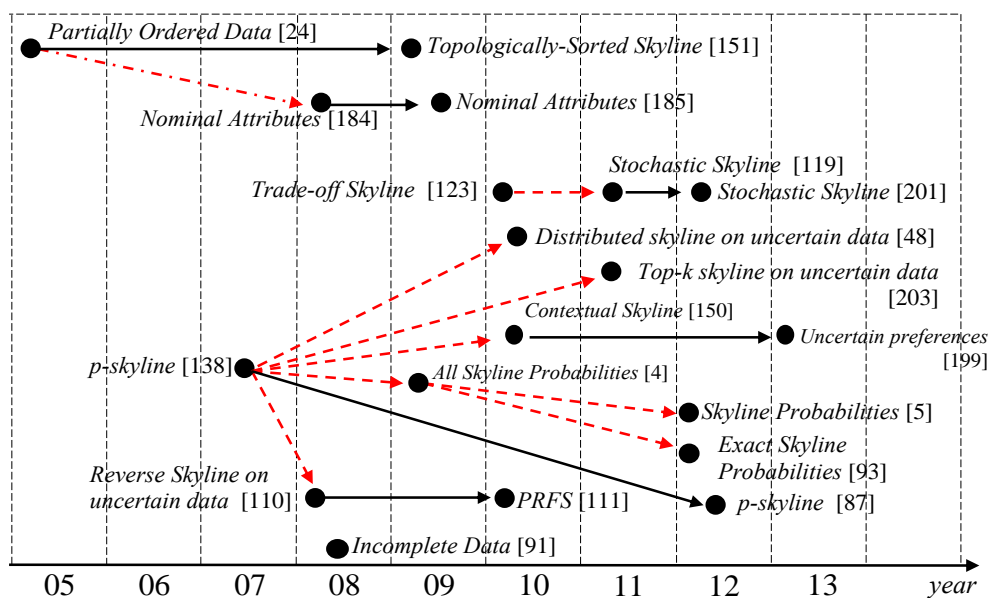

**Figure 57: Attribute & data-specific skyline queries hierarchy.**

Summarizing **Figure 57** illustrates the hierarchy of attribute and data-specific applications of skyline queries. As a remainder the dash red line represents the adoption of a general idea from another work and the black line represents the extension or the heavily dependence on a previous work.

## 4.4. Continuous skyline computation

The existing skyline algorithms are designed to compute skylines over static datasets rather than dynamic, that occurs in streaming environments. Dynamic streaming data can be retrieved with the use of data-streams **[6]** which is a continuous stream of received data point. In the append-only data-streams data points are removed only when they expired. In data-streams the elements are positioned and labeled according to their relative arrival ordering. A similar mechanism to data-stream is a time series **[183]**. A time series continuously reports points at pre-defined intervals. The number of points received in a specific timestamp or time interval is fixed, in contradiction with data-streams. As an example a time series can be related with the data reported by a sensor in pre-defined intervals. A data-stream on the other side can be related with the number of people that disembark when a ship arrives to its destination. Note that in this case the intervals and the number of values that can be reported are not fixed. Nevertheless data-streams can have outdated data that are not needed to be considered in a skyline computation. For that reason the sliding window model **[6]** is used over data-streams in order to evaluate the queries only to the N most recent received points and not to the entire data-stream.

### 4.4.1. Data Streams

Authors in **[117]** reason about on-line computation in the presence of rapid updates of data such as in data-streams. Particularly the scenario concerns append-only data-streams where there is not any deletion of the data elements till they expire and the elements are positioned and labeled according to their relative arrival ordering. Such type of streams are those of wireless sensors networks, where the data collected prior a specific time interval are discarded because they are not representative in comparison with the existing readings of sensors. In their work authors incorporate the sliding time window mechanism. In this case the arrival of an element is always related with the deletion of the expired/oldest item. As indicated by the use of a sliding window author's work try to compute the skyline on the N most recent elements. Additionally due to the various preferences that the user may have, reason about the computation of the skyline over a sample of the n most recent elements of the N elements ($\forall n \leq N$) that were stored from the data-stream. Authors study involves the on-line computation of skyline queries on n-of-N model **[116]** and ($n_1$, $n_2$)-of-N model. Essentially a sliding window model is a special case of n-of-N model where n=N. The ($n_1$, $n_2$)-of-N model is a generalization of the n-of-N model when it is desired to compute the skyline of the elements between the $n_2$-th and $n_1$-th most recent elements ($\forall n_1 \leq n_2 \leq N$). The n-of-N model gives the skyline based on the most recent information and the ($n_1$, $n_2$)-of-N model provides recent "historic" information. This way by combining the results of the two models can be identified a trend change between the most $n_2$ recent elements and the $n_1$ most recent elements.

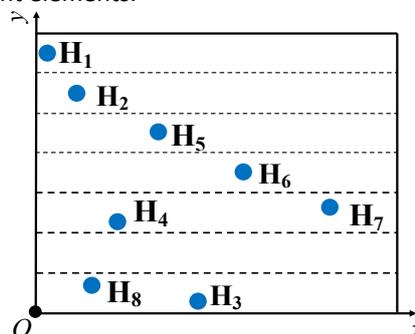

**Figure 58: Data stream.**

An example the continuous skyline computation based on sliding window is illustrated with the help of **Figure 58.** It is assumed that the points have been received in the order imposed from the numbering. The skyline of the whole data-stream will be consisted from the 8 points and the skyline set $S_8$ derived will be the {H1, H2, H4, and H3}. At this point it will be considered that the point H8 has not been received yet. Consider that it is needed to find the skyline of the 5 most recent elements (except H8 as noted). This action will include only the points H3, H4, H5, H6, and H7. The skyline $S_5$ of the five most recent points (except H8) will be consisted from the point H3 and H4. Note that the domination holds as the traditional skyline. If we consider the 3 most recent points H5, H6, H7 the skyline set $S_3$ will contain the points H5, H6 and H7 as skyline points. With the arrival of point H8 the skyline set $S_5$ will be computed over the points H4, H5, H6, H7, H8 and the skyline set will be consisted only from point H8. In the same way the skyline $S_3$ of the three most recent points will be computed over the points H6, H7, H8 and the skyline will be consisted by the points H8. Note that considering the 6 most recent point the skyline $S_6$ will be consisted from the points H3 and H8. Below ia outlined the pseudo code of the basic algorithm. The algorithm continuously receives new items through the data stream and updates the set $R_N$ of non-redundant elements, of the most recent N elements and the interval tree $I_{RN}$. A data element $e$ is redundant with respect to the most recent $N$ if $e$ is expired or is dominated by a newer element. The variable $De_{new}$ denotes the set of redundant elements dominated by a new element $e_{new}$.

| **ALGORITHM 30:** *Maintaining $R_N$ & its Encoding Scheme* **[117]** |
|---|
| **Input:** datastrem elements $e_{new}$. |
| **Output:** Continious Skyline set. |
| 1. **while** new element $e_{new}$ **do** |
| 2.    **if** the oldest $e_{old}$ in $R_N$ is expired **then** |
| 3.       $R_N := R_N - \{e_{old}\}$; |
| 4.       remove $(0, \kappa(e_{old})]$ from $I_{RN}$; |
| 5.       **for** $\forall e_{old} \xrightarrow{c} e$ **do** |
| 6.          update $(\kappa(e_{old}), \kappa(e)]$ in $I_{RN}$ to $(0, \kappa(e)]$ |
| 7.       **end for** |
| 8.    **end if** |
| 9.    find $De_{new} \subseteq R_N$ dominated by $e_{new}$; |
| 10.   **for** $\forall e \in De_{new}$ **do** |
| 11.      remove the intervals in $I_{RN}$ with $\kappa(e)$ as an end |
| 12.   **end for** |
| 13.   $R_N := R_N - De_{new} + \{e_{new}\}$ |
| 14.   determine the critical relation $e \xrightarrow{c} e_{new}$; |
| 15.   add $(\kappa(e), \kappa(e_{new})]$ (or $(0, \kappa(e_{new})]$) to $I_{RN}$ |
| 16. **end while** |

To retrieve the continuous skyline, based on the these two models, it is needed to efficiently maintain and organize the N most recent elements, in order to efficiently compute any n-of-N skyline query. Authors propose a pruning method in order to minimize the number N of elements that are stored in order to answer any n-of-N skyline query, by discarding the redundant points. A redundant point is a point that expired or is dominated by a newly received point that will not allow it to be again a skyline point till its expiration. Moreover authors proposed an efficient encoding scheme and an update technique in order to index and store the data with the use of an R-tree. Additionally they used an interval tree **[43]** to determine if a point belongs in the N most recent elements. Finally proposed a trigger based technique in order to efficiently process continuous n-of-N skyline queries and an extended technique for (n1, n2)-of-N.

Authors in **[166]** also proposed the skyline computation in data stream environments. In this scenario, they take into account only the tuples that arrive in a sliding time window. That is the W most recent timestamps, where W is a parameter that defines the length of the window. As previously, the data-stream considered is "append-only" which means that a tuple is discarded only when it is expired. In their work, authors reasoned about time-based sliding windows but their methods are applicable to count-based sliding windows **[60]**, where a tuple expires after receiving W subsequent tuples.

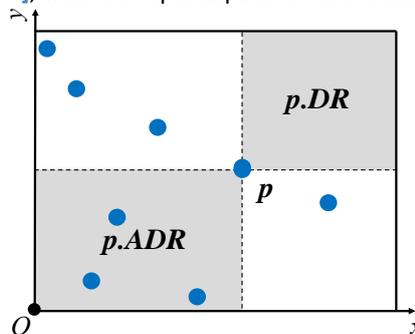

Figure 59: Dominance (DR) and anti-dominance (ADR) regions.

The data in stream environments change frequently and thus it is desired to incremental maintain the skyline query rather re-computing it. Additionally a skyline change occurs when a new tuple arrives or a skyline point expires. For this reason, authors in their work proposed two approaches. The lazy strategy which delays most computations until the expiration of a point and the eager strategy which incorporates a pre-computation phase in order to minimize the storage cost and simultaneously store only the points that may become part of the skyline in the future. In the first strategy these cases are handled by the preprocessing module (L-PM) and the maintenance module (L-MM) respectively. In the first case when a new tuple *r* arrives the algorithm checks if it is dominated by any existing point in the database skyline set, named DBsky, which contains all the skyline points. If it is not dominated is placed in the DBsky set otherwise is maintained in a DBrest set. Note that the point was placed in the DBrest set since it might appear as a skyline point later when another point has expired and thus cannot be completely pruned. When a tuple is placed in the DBsky it might dominate other points that are already in the skyline. These points are pruned if they are not expired yet and additionally their expiration time is prior than this of the points that dominates them. The dominance checks are based on the dominance and anti-dominance regions of points as shown in **Figure 59.** Given a received point *p*, the points that are dominated by the *p* are those that belong to the *p*.DR region and those that dominate *p* are those lying in the *p*.ADR region. The algorithm also stores the earliest expiring time of the current skyline set, which essentially is the time when the oldest skyline point will expire and additionally a pointer to the oldest skyline point.

In [129] authors studied the problem of continuous skyline computation on datastreams where the validity and expiration of points is determined with the use of time intervals. Each point received is associated with an arrival and an expiration time which essentially defines time interval that the point will be valid. In their work introduced *continuous time-interval skylines* and proposed the LookOut algorithm. This approach differentiates from the sliding windows where the skyline is evaluated only on the most recent n points. In contradiction, *continuous time-interval skyline* is a more general approach, which considers and computes the skyline based on all the current valid data points. The data points are continuously added to the dataset (through a data-stream) or removed (expired) from the dataset. Nevertheless, this approach can also be applied to evaluate sliding window queries as well. Differentiating in construction from the previous methods authors proposed the use of a quad-tree [152] rather than R*-tree [11]. Nevertheless, authors implemented their work using both spatial indexes in order to perform a comparison analysis.

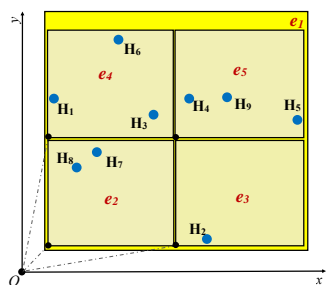
Figure 60: Quad-tree.

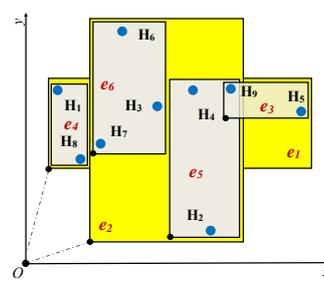
Figure 61: R-tree with overlapping regions.

As shown in **Figure 60** quad-trees use non-overlapping partitions/regions in contradiction with R-trees that the regions may overlap **Figure 61**. Non-overlapping partitions/regions are also supported by the $R^+$-trees but have not been adopted by authors due to performance considerations [129]. Authors have additionally outlined a detailed analysis about the tradeoffs that occur by using the two different spatial indexes.

The R*-tree approach was implemented because it is widely adopted in other methods such as **[NN, BBS, [97, 133].** The quad-tree was selected because its non-overlapping regions helps in a more effective pruning of the regions that are not needed to be traversed. This will speed up the skyline computation and lower the memory consumption, as observed through author's experiments. In general the insertion into a quad-tree is faster than the R*-tree and the traversal of a quad-tree reduces the maximum number elements inserted in the heap.

Each point of the dataset is associated with a time interval for which will be valid. The time interval is consisted by the arrival and expiration time. Thus a continuous skyline can only change if an existing point is expired or a new point is added. Thus the computation is performed over the data that are still valid at query time. In order to give an illustrative example of the algorithm, Table 34 represents a series of data points. The data points are described by the arrival time $t_a$ and its expiration time $t_e$. Additional information about points can be stored in a fully detailed dataset.

| P | H1 | H2 | H3 | H4 | H5 | H6 | H7 | H8 | H9 | H10 | H11 |
|---|----|----|----|----|----|----|----|----|----|-----|-----|
| $t_a$ | 1 | 2 | 3 | 4 | 5 | 6 | 7 | 8 | 9 | 10 | 11 |
| $t_e$ | 15 | 16 | 8 | 20 | 6 | 9 | 22 | 24 | 20 | 29 | 21 |

Table 34: Data stream for continuous skyline.

Figure 62 illustrates the skyline at the time $t_a=7$ based on the received points. Note that point H5 was received, expired and so deleted from the dataset. Figure 63 illustrates the skyline at the time $t_a=11$. From all 11 points received only the 8 of them haven't expired yet.

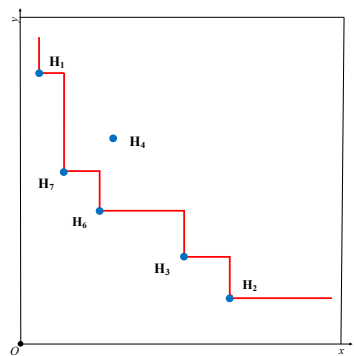
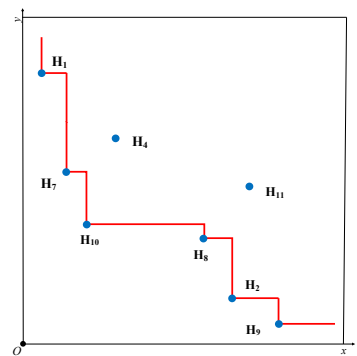

Figure 62: Skyline of the dataset in $t_a=7$ time. | Figure 63: Skyline of the dataset in $t_b=11$ time.

In general when the algorithm receives a point $p$ with arrival time $t_a$ and expiration time $t_e$ adds it to the spatial index (quad-tree or R-tree depending on the implementation). Additionally places it in a heap $H_p$, which keeps all the received (not expired) points sorted, based on their expiration time. Intuitively the spatial index maintains and indexes the points and the heap deals with the efficient identification of the expiration time of the various points. As the time passes the point with the minimum timestamp in the heap $H_p$ is checked if it is expired. If so it is deleted from the heap and the index. Along with the previous described steps, every time a point is received the algorithm checks if it is a skyline point. The skyline points are kept in a list S in order to be efficiently returned in a query request. If $p$ is dominated from other points no action is performed and the point is kept in the spatial index and the heap. Otherwise, if it is a skyline point, the skyline list *S* is updated accordingly, removing also the dominated points by $p$ from the skyline list. Along with the skyline list S an additional heap $H_s$ is kept for the skyline points, sorted by the expiration time, in order to efficiently remove them when the expiration time comes.

In particular, for point $p$ and any other skyline point, when the expiration time $t_e$ arrives is removed from the heap $H_p$, $H_s$ and the spatial index, thus deleted from the dataset. In this case the algorithm must check if there are potential skyline points from the items that were dominated by $p$ and not expired till now. For that reason after a skyline point removal the algorithm finds the skyline of the points that were dominated by the removed skyline point, named *mini-skyline*. Nevertheless, these points may be dominated by other existing skyline points. The points of the mini-skyline are checked if they are definitely skyline points (compared with the other skyline points) and added to the skyline list S and heap Hs if it is needed.

Authors in [34] proposes a method for finding keyword-matched skylines. In their scenario each tuple contains among others a number of textual descriptions that can represent some specific characteristic. The keyword-matched skyline will contain the skyline tuples among the set of tuples whose textual description contains all the query words that where placed.

## 4.4.2. Time Series

In **[86]** authors study the problem of skyline query computation on time series. Time series are useful since they can give information about events that happen in a specific time interval. As an example these events could include the upload bandwidth consumption or the visiting rate of a web page, along the day, moth, year or any other time interval. In their work proposed *interval skyline queries* on time series data.

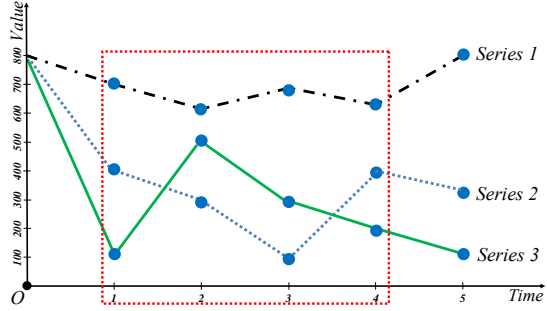

**Figure 64: Time series.**

A time series *S* is consisted of a set of tuples in the form (value, timestamp). The data values are order based on the timestamps. The values of a time series S are denoted as s[i] where i is a specific timestamp. The time series can be written as a sequence of the values that contains, in the form S[1], S[2], S[3] . A time interval [i:j] (i.e.: [1,4] in **Figure 64**) specifies a range in time. This range contains the set of timestamps that exist between timestamp i and timestamp j with i<j. In interval skyline query processing a time series S is interesting in a time interval [i:j] if there does not exists any other time series Q for which Q is better than S on at least one timestamp and is not worse than S on the remaining timestamps of the time interval. Thus the interval skyline query will return the time series that is not dominated by any other time series in a given time interval [i:j]. More formally:

**Definition 29:** Domination in time series
A time series S dominates a time series Q in an interval [i:j], denoted as S$>_{[i:j]}$Q, iff ∃x∈[i,j], S[x]<Q[x] and ∀y∈[i,j]-{x}, S[y]≤Q[y]■

**Definition 30:** Interval Skyline
Given a set T of n time series $s_k$, 1≤k≤n and an interval [i:j] the *interval skyline*, denoted as Sky[i:j], is the set of time series $s_k$∈T, that are not dominated by any other time series $s_m$∈T in [i,j]. That is: Sky[i:j] = {s∈T|∄r∈T, r$<_{[i:j]}$s}■

The previous definition is easy to understand when a specific time series is better in the whole time interval. In order to understand the interval skyline in a more general scenario, where different time series belong to the skyline due to different time intervals the following property is outlined.

**Property:** Given a set T of time series and an interval [i:j], a time series s∈T belongs to the Sky[i:j] if for any number n of timestamps $t_1, t_2, ..., t_n$, with i≤$t_x$≤j, $\frac{\sum_1^n s[t_x]}{n}$ has the minimum value among all the time series in the set T, compared to the specified timestamps, and additionally the tuple with that minimum value is unique. Essentially the unique time series p that belongs to the skyline for a specified interval [$t_1, t_n$] will be derived from the equation $p = arg \min_{s \in S} \left\{ \frac{\sum_1^n s[t_x]}{n} \right\}$, where S represents all the existing time series.

Essentially the interval skyline will contain all time series which achieve the highest average aggregate values on any subsets of timestamps. Based on **Figure 64** both series 2 and series 3 will be in the skyline for the time interval [1,2]. In the skyline for the time interval [2,3] will return only series 2 and for [4,5] series 3. A naïve approach to compute such type of skyline queries would be to compute the query every time it is placed based on the given time interval. However this method computes the skyline from scratch and will not be efficient in on-line/dynamic environments. Additionally in overlapping time intervals there will be no computation sharing. Finally timestamps are considered as dimensions and time series are consider as points on a specific dimension **Figure 64**, with respect of each timestamp. Thus, the dataset will have very high dimensionality. This makes the existing skyline computation methods inapplicable and inefficient. Below is outlined the on-the-fly method that authors proposed. For a time series t, t.max denotes the maximum value of t and t.min[i:j] denotes the minimum value of t in the interval [i,j].

| **ALGORITHM 31:** *on-the-fly query algorithm* [86] |
|---|
| **Input:** A set *S* of time series and an interval [i:j]. |
| **Output:** The skyline in [i:j]. |
| 1. *L = a sorted list of the time series in S in the descending order of their s.max.* |
| 2. *Sky = ∅;* |
| 3. *maxmin = - ∞ ;* |
| 4. *let s be the first time series in L;* |
| 5. **while** *L is not empty and maxmin ≤ s.max* **do** |
| 6.    **if** *no time series in Sky dominates s in [i : j]* **then** |
| 7.       *remove the time series from Sky dominated by s in [i : j];* |
| 8.       *Sky = Sky ∪ {s}* |
| 9.       *maxmin = max$_{q∈Sky}${q.min[i : j]};* |
| 10.    **end if** |
| 11.    *s = the next time series in L;* |
| 12. **end while** |
| 13. **return** Sky |

Authors in their scenario assume that data are collected incrementally and that it is always desired to maintain the w most recent timestamps, that essentially define an interval W. This approach is similar to the sliding window models and based on this consideration proposed two methods. The first one called on-the-fly method and the second one view-materialization. The on-the-fly method maintains the minimum and maximum values for all possible intervals of each time series in radix priority search trees [80] and computes the interval skyline at query time. The view-materialization method maintains a specific data structure D, that has a set of *non-redundant skyline* time series which are the ones that belongs to the skyline in an interval [i,j] and not in the skyline in any subinterval [i':j']⊂[i,j], in order to efficiently answer any *interval skyline* query in any given interval [I:J] ⊆ W based on a set S of time series.

### 4.4.3. Various Topics

This section will outline the work that is based on continuous skyline computations but is applicable in a more tight and specific scope. These works contains continuous skyline computation on categorical data, uncertain data [4.3.3] and approximate skyline computation. Finally additional work is presented which is related with continuous skyline computation on the skyline family variants.

*Approximate continuous skyline computation*

Authors in [204] studied the problem of continuous maintenance of the skyline in the existence of dynamic datasets in client-server architectures. Their methods achieve approximate skyline computation by means that the reported points may not be at the exact reported location. Nevertheless all the existing skyline points are reported. In their scenario a server continuously receives records and updates from the various clients connected on him. Updated information about a point must be transferred from a client to the server when at least one attribute of the point has changed. Considering this, authors proposed a *Filter* method in order to reduce the number of updates transmitted and additionally maintain the approximate skyline. For that reason server defines regions, for every record, that bounds their attribute values and sends the boundary regions to the clients. The client will only need to report the updates where the attribute values of points exceed these boundaries. Additionally with this method authors proposed the *frequent skyline query over sliding windows* (FSQW), which reports only the skyline points that consistently appear in the skyline over several timestamps, in order to avoid the continuous change of the skyline. To further reduce the communication cost authors proposed a sampling method which retrieves an approximation of FSQW by instructing the client to report updates only in a specific rate. Finally, they combined the *Filter* method and *FSQW* method in one *hybrid* method.

*Moving objects*

Authors in **[78]** study the problem of continuous skyline query processing for moving objects by exploiting the spatiotemporal coherence of the problem. Spatiotemporal coherence is a method used in computers graphics **[68]** in order to relate various properties of one part of a scene with another part in order to reduce the processing cost, such as in area filling. The moving objects that exist in a 2-dimensional space can be represented by their starting position ($x_p,y_p$), their velocities ($u_{px},u_{py}$) and some non-spatial attributes $at_i$ as ($x_p,y_p,u_{px},u_{qy},at_1,at_{2,...}$). The velocities of stationary points are set to zero. The location of points can be computed through their velocities. The skyline computation is performed only once at the beginning of the algorithm. In the rest of the time the skyline is updated and not computed from the scratch at each time. In order to maintain the skyline authors use kinetic data structures (KDS) **[10]**. On these structure it easy to maintain the validity of a relation between points in a period of time t, with the use of *certificates* that consists from algebraic conditions. As an example the expiration of a certificate can trigger an update process in order to ensure that a point is not dominated after a specific time. In this scenario the distance between a moving query point q and a moving data point p at time t can be defined by the function $dist(p(t),q(t)) = \sqrt{at^2+bt+c}$, where a,b,c are constants that are determined by their starting positions and velocities. For simplicity authors use $f_p(t) = at^2+bt+c$ to denote the square distance when the data point p is static and the query point is moving. A change in the skyline can be identified with the use of this function by locating the intersections of the distance function curves $f_p(t)$, for all points p in the dataset. Nevertheless, an intersection does not quaranties a change in the skyline.

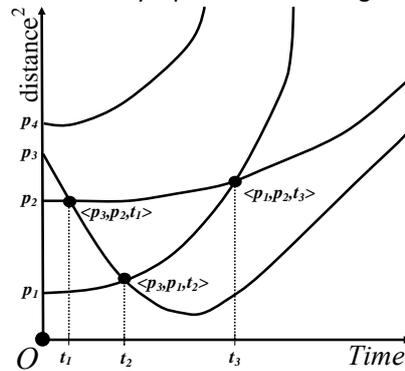

**Figure 65: Distance function curves.**

**Figure 65** illustrates the square distance between several data points $p_i$ and a moving query point. The distance changes as the query point is moving. An intersection between the distance functions with respect point $p_i$ and $p_j$ at time $t_x$ is represented by the <$p_i,p_j,t_x$>. In **Figure 65** the intersection <$p_1,p_2,t_3$> indicates that point $p_1$ is getting closer to the query point, than the point $p_2$, after time $t_x$. This indicates a potential change in the skyline and thus the skyline may be needed to be updated.

In **[153]** authors studied the problem of skyline computation based on partial ordered, categorical data, over streaming environments. Essentially this work extends the work of **[24]** in order to be applied in categorical streaming environments. Authors imply a topological sorting in the streaming data. Based on the topological sorting they construct a grid-based index structure in order to obtain and maintain the skyline set. Authors in **[92]** reason about uncertain data that are in the form of continuous ranges. Their objective is to identify and return the points that have probability *p* to be in the skyline within a given tolerance threshold *δ*. Their methods gradually bound the probability of each point in order to identify if it will be in the skyline result. In **[202]** authors studied the problem of continuous skyline computation based on uncertain data streams and given probability thresholds. Additionally studied the scenarios were multiple probability thresholds are available and the retrieval of top-k skyline objects. The general problem that try to solve is the retrieval of the skyline points that exist in the N most recent elements which have skyline probability not smaller than a given threshold t with 0<t≤1. The computed probabilities derive from past received data which have a single instance in the database. In order to deal with dominance relations and store information related with the probabilities authors use aggregate-Rtrees. In **[50]** authors extended the previous work and assumed that each object has multiple instances. In general the proposed method continuous monitors the points that exists over the sliding window and reports the points whose skyline probability in the current timestamp is greater than a given threshold probability t. The dominance relations are computed with the use of dominance and anti-dominance regions [4.4.1]. In order to efficient maintain the skyline authors use a *candidate list* approach which contains the candidate points that might belong to the skyline in a future timestamp. Additionally research has been conducted on skyline variants over datastreams. Such work is [96] which study the k-dominant skyline queries [25] over data-streams. Subspace skyline computation (section 4.1) over sliding windows is studied in [76]. Finally distributed skyline computation (section 4.2) over data-streams was studied in [163]. The streams in this scenario are derived from horizontally partitioned distributed Sources.

In [114] author reason about range-based skyline query computation and take into account that the query locations may not be an exact point or a line segment. In their work they proposed the index-based I-SKY algorithm and the non-index-based N-SKY algorithm. It is worth to mention that range-based skyline computation problem can also be solved by BBS [133, 134] with the use of R-trees.

### 4.4.4. Summary

As on all previous sections **Figure 66** illustrates the relation that exists between the outlined works. The black line represents the extension or the heavily dependence on a previous work. The dash red line represents the adoption of a general idea from another work. The green dot represents work that isn't related with continuous skyline queries.

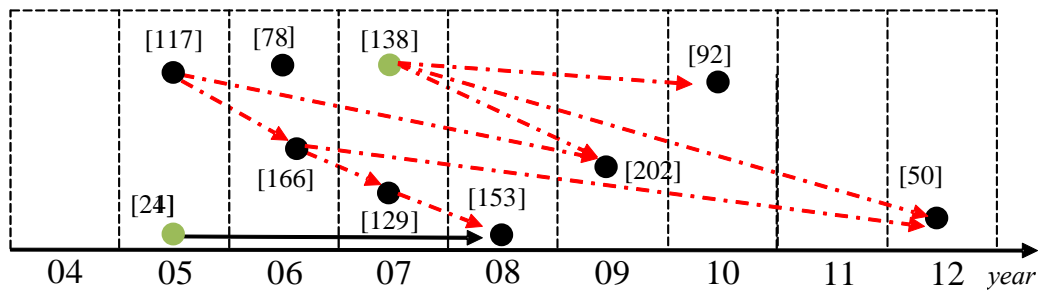

Figure 66: Continuous skyline queries hierarchy.

**Table 35** outlines the fundamental key points used by the state of the art algorithms, the environment that is considered, the indexing mechanism that is used and the number of points over which the skyline is computed.

| Algorithm | Incorporates | Environment | Indexing method | Skyline on # points |
|---|---|---|---|---|
| [177] | Stabbing queries [177] | Sliding-windows | R-tree | n-of-N & (n1,n2)-of-N |
| [166] | Dominance & anti-dominance regions | Sliding-windows | R-tree | N most recent |
| [129] | continuous time-interval skyline | Time-intervals | R-tree / Quad-tree | All valid received points |
| [86] | interval skyline | Time series | Radix priority tree | #Time-series* #timestamps |

Table 35: Fundamental algorithms on continuous skyline retrieval.

## 4.5. Route skylines queries and road networks

This section will reason about in-route skyline algorithms that are related with the identification of efficient routes or detours on road networks and efficient locations that satisfy the desired minimization criteria among several user-defined points. In general route planning methods that are based on road networks usually find the shortest route between two points with the use of Dijkstra's algorithm [47] and additionally the A*-algorithm [100] which directs the pruning of the search space. Same security related work such as [113],[28] that reason about location based skyline queries will be discussed in specific security section [4.6].

### 4.5.1. Route skyline queries

In [75] authors reason about skyline computation on road networks and particularly in-route skyline queries and in-route k[th]-order skyline queries which concern normal domination and the points that are dominated by less than k other points respectively. The problem that study is related with location-based queries. In general user movement on Location-based services (LBSs) can be categorized based on three scenarios. The first is unconstrained movement in physical space. The second is constrained movement, where constrains derive from the various physical obstacles of space. Third is network constrained movement where users are constrained in a transportation network and which authors are based. The distance computation in this scenario is based on travel distance (network distance) rather than Euclidian distance.

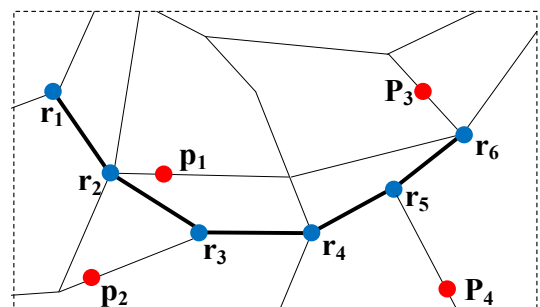

**Figure 67: Data model.**

The scenario concerns a user that moves on a pre-defined route and the algorithm tries to find the minimum detour in order to visit a point of interest, before reaching the destination. Points of interest, such as grocery shops illustrated as points P1, P2, etc. with a red dot in **Figure 67,** are located within the road network. For example a user may will to visit the nearest grocery shop to his route during his way back from the beach to his hotel. Thus, a user issues a query in order to find and visit points of interest while moving along his pre-defined root. The user's location represents the query point. In this scenario user's location, user's destination and the route that will be followed are known and retrieved from navigation system or from past behavior [21]. The algorithm does not use any index but instead the points are organized in a list structure. The data model used in order to represent the road network is a labeled graph G=(V,E), where V is the set of vertices and E the set of edges. Vertices can essentially represent the start or the end of a road segment. An edge e∈E has the form e=(u,v,l) where u,v∈V are vertices and l is the length (or a weight) that describes the edge (road). The user's query point (user's location) and the points of interest are called *data points*. A *data point* has the form p=(e,pos$_u$) where e∈E is the edge on which the point is located and pos$_u$ represents the distance of the point from the edge's vertex u. The distance is computed from the length l of the edge, minus the point's position pos$_u$. That is, l-pos$_u$. A route on this data model is given by a set of neighboring vertices {r1, r2, r3, r4, r5, r6} as illustrated in **Figure 67**. The distance function used in order to compute the distance between to vertices is defined as the sum of lengths of the edges, along the shortest path, that connects them. To identify and compute the distance of the shortest path between two points can be achieved with Dijkstra's shortest path algorithm [47].

The scenario followed in this work assumes that a user follows a predefined route in order to reach its destination and visits points of interest by leaving its route if this is necessary. This is done based on the three cases, Traverse case, General Case, Best case. The traverse case identifies the nearest point of interest along the route, which can be visited from any vertex of the route. Then the user leaves the route at the vertex which has the smallest distance from the point of interest and return back to the route from the same path. Having the route defined in **Figure 67** point P1 is the closest point of interest in the route, thus user will leave its route at vertex r2, will visit the point of interest and return back to the vertex r2 in order to continuous on its predefined route. The traverse case is similar to [157] with the difference that this approach uses Euclidian distance. The General case considers that the user leaves its route at the vertex ri and return again in a vertex rj which is prior to the destination. For example user visits point P1, leaving vertex r2 and returns on vertex r4 in order to reduce the total travel distance. Finally in the best case, user leaves the route and returns directly towards its destination. Again in **Figure 67** user visits point of interest p1 and then continuous directly to reach the destination point r6 without revisiting the route in order to achieve the lower traveled distance. In every case, in order to refine the results, the last step to be performed is the k-th order skyline query (K-skyband query in section **3.10**), which returns the points that are dominated by less than k other points. Below is outlined the pseudocode of the algorithm for the traverse case. As D(c,r,R) is denoted the distance from the location of the query point c, to a vertex r along a route R. The variable T and P, represent two queues that are used to store the result data points of the NN (nearest neighbor query) and the candidate points for the skyline query. The variable dis is related with distance values from a specific destination and the variable det is related with the distance of a detour. As previously mentioned the algorithm returns the points of interest from the set P filtered by a k-skyband operator comparing the values dis and det for each tuple in P.

| **ALGORITHM 32:** $TraverseSQ(k, c, R)$ **[75]** |
|---|
| **Input:** An order k, the query point location c and a route $R=\{r_0,r_1,\ldots,r_l\}$. |
| **Output:** filtered-set of points of interest. |
| 1.  $P \leftarrow \varnothing$ |
| 2.  $T \leftarrow NNQ(k, r_1)$ |
| 3.  **for each** $t \in T$ |
| 4.      $dis \leftarrow \mathbf{D}(c, r_1, R) + D(r_1, t)$ |
| 5.      $det \leftarrow 2D(r_1, t)$ |
| 6.      $P \leftarrow P \cup \{t, dis, det\}$ |
| 7.  $d \leftarrow$ distance from $r_1$ to its $k$th neighbor |
| 8.  **for each** $r_i$, $i = 2, \ldots, l$ |
| 9.      $T \leftarrow RNNQ(k, r_i, d)$ |
| 10.     **if** $T$ not empty |
| 11.         **for each** $t \in T$ |
| 12.             $dis \leftarrow \mathbf{D}(c, r_i, R) + D(r_i, t)$ |
| 13.             $det \leftarrow 2D(r_i, t)$ |
| 14.             $P \leftarrow P \cup \{t, dis, det\}$ |
| 15.         $d_1 \leftarrow$ distance from $r_i$ to its $k$th neighbor |
| 16.         **if** $d_1 < d$ |
| 17.             $d \leftarrow d_1$ |
| 18. **return** $(SKYLINE(k, P, \{dis, det\}))$ |

In **[98]** authors consider route skylines in road networks with multiple preferences as opposed with the previous method. As an example the preferences can be related with the distance, the speed limit or the number of traffic lights of a road segment. Each route can be represented by a d-dimensional point, where d will be equal to the number of preferences defined as attributes/weights. The route domination comparisons are achieved by comparing the cost of each path, which derives from the sum of costs of each road segment. The cost of a road segment is defined based on the preference selected each time. A route skyline will return a subset of all possible paths, between a start and a destination point, that are optimal for any arbitrary user preference that has been defined. Then the user can select the path that fits best his needs. In this work authors model the road network as a multiple-attribute network graph (MAG). The multiple attribute graph is a directed network graph G(V,E,W), were V represents a set of vertices, E a set of Edges and W a set of weight vectors that describe the attributes of each road segment. Authors proposed the algorithms BRSC (Basic Route Skyline Computation) and ARSC (Advanced Route Skyline Computing). The BRSC algorithm uses the forward cost estimation of the A*-search **[100]** as a global pruning criterion for routes. The ARSC algorithm uses an additional local pruning criterion based on which any sub route of a skyline route must be a pareto optimal path.

### 4.5.2. Multi-source skyline queries

In **[46]** authors reasoned about multi-source skyline queries where multiple query points are considered at the same time, in constrained space and especially on road networks. The multiple query points represent the locations or the points of interest from which a user wants to minimize its distance. As an example a user will select the hotel(s) (data point(s)) that minimizes the distance from a metro-station, a grocery shop, a coffee shop and the beach (query points). The problem of multi-source skyline queries on Euclidian space has been studied in **[155].** However this work is based on Euclidian distance and thus cannot be applied on network distances. Authors in this work reason about *relative skyline queries* where the minimization is based on user given data points and not on the static attribute values of the data-space. *Relative skyline que*ries are also known as dynamic skyline queries in **section 3.2**.

As previously a road network can be represented by a graph G=(E,V) where E is a set of non-directional edges that can represent the roads and V is a set of nodes that represent the road junctions. Two nodes p,r∈V are adjacent if they are linked by an edge in E. Adj(p) denotes all the adjacent nodes of node p∈V. The distance between two nodes p,r∈V is denoted by d(p,r) and is equal to the total length of edges that must be traveled in order to go from the one point to the other. Authors denoted as $d_N(u,v)$ the distance of the network shortest path that exist in the network and $d_E(u,v)$ as the Euclidian distance of points p,r∈V. The shortest path between two nodes can be identified and computed on-the-fly with the use of Dijkstra's algorithm **[47]** or the A* algorithm **[100]**. In order to avoid on-the-fly distance computation authors use a middle layer that partially stores the network distances $d_N$ For a point p that exists on a network edge e∈E, between two adjacent nodes u,v∈V, the algorithm computes the distances of p from both nodes and stores them in the middle layer along with the id of the point. This middle layer is indexed by a b-tree based on the ids of points.

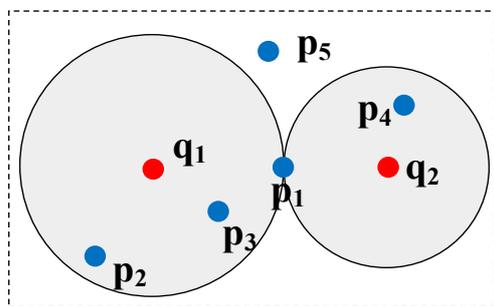

Figure 68: Termination of first phase.

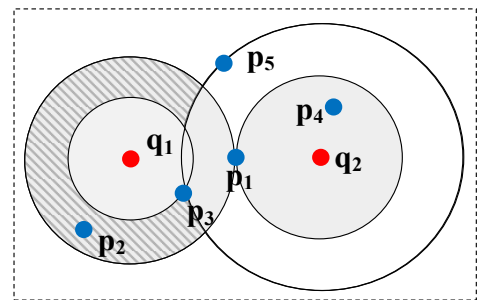

Figure 69: Second object visited.

In their work authors proposed the three algorithms, Collaborative Expansion (CE), Euclidean Distance Constraint (EDC) and Lower Bound Constraint (LBC). In the first phase the CE algorithm identifies the nearest neighbor of a query point, using the Euclidian distance by defining an incrementally expandable circle around each query point. This way the points around the query points are visited in ascending order of their network distances from the query point. When a data point p is identified as a nearest neighbor of a query point q, is said that p is visited by q. The first phase of the algorithm terminates when the first data point p has been visited by all the query points. In this case the circle expansion is terminated and the point p is definitely a skyline point. All the data points that have been visited till this step by the query points are kept in a candidate set C. This can be illustrated with the use of **Figure 68**. The point p1 is the first point that is visited by all the query points and thus is the first skyline point. Points p2, p3, p4, p5 that were visited before point $p_1$ are placed in the candidate set C. Point p5 is not visited by any query point and cannot be a skyline point since $d_N(p_1,q) \leq d_N(p_4,q)$ and thus it is pruned. Essentially at this point the points that are not placed in C are dominated by p. The shaded area on **Figure 68** represents the candidate search area in order to identify potential skyline points. After the first phase a similar second phase is performed in order to identify the rest of skyline points that exist in the set C. In this phase point $p_3$ is found as the first point that is visited by all the query points as illustrated in **Figure 69**, in which the shaded area represents the area that is pruned by $p_3$. Point $p_3$ is compared with the existing skyline points. If it is not dominated by any other existing skyline point is reported as the next skyline point. The points in the candidate set C that are dominated by the point $p_3$ are pruned. In this case this is the point $p_2$. The algorithm follows in the same way until all skyline points have been discovered. In this case the final skyline point is $p_5$.

The main problem with the CE approach is that the search is expanded in all directions and may lead to unnecessary distance computations. This is solved by the EDC algorithm which first identifies the multi-source skyline points on Euclidean space. Note that the Euclidian skyline points may not be network skyline points. Then the algorithm computes the network distance between the Euclidian space skyline points and each query point by using the directional expansion heuristic of the A* algorithm. Again EDC algorithm, as CE, can lead in many network distance computations which will incur considerably distance computation cost. For that reason authors proposed the LBC algorithm which is based on the network nearest neighbor.

The algorithm stores the network skyline points in a set S. For any of the query points the algorithm performs the following steps. First finds the (next) network nearest neighbor of a query point q. Assuming that this point is r (the identification of this point will be described in the next paragraph). The network distance of r from all the query points is computed. If r is not dominated by any skyline point s' of S (when $d_N(r,q) < d_N(s',q)$ holds) then point r is added to S.

The identification of the next nearest neighbor that was mentioned previously is performed as follows. First the algorithm identifies the Euclidian nearest neighbor p of the query point q, which exists in the area that is not dominated by any of the points of S. The algorithm terminates when cannot be identified any such point for any of the query points. Next the algorithm finds the next network nearest neighbor. The network distance between q and p is computed with the use of the A* algorithm and p is added in a candidate set C along with the values $d_N(p,q)$. If there is a point $c \in C$ for which $d_N(c,q) \leq d_E(p,q)$, then the point in C with the minimum network distance to q is returned as the next nearest neighbor. Otherwise the algorithm continuous again from the beginning and finds the next Euclidian nearest neighbor is found.

### 4.5.3. Summary

**Figure 70** illustrates the hierarchy and the chronological order of the algorithms related with the in-route skyline computation and the skyline computation on road networks. The dash red line represents the adoption of a general idea from another work. The green dot represents work that is not directly related with this section.

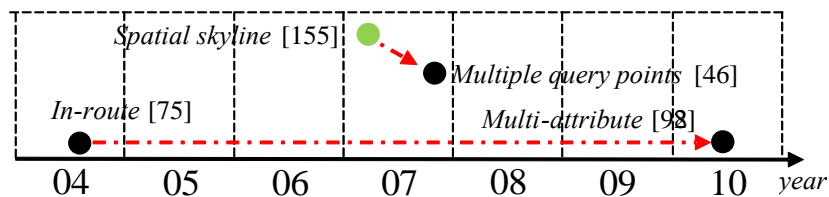

Figure 70: In-route & road networks algorithm hierarchy.

**Table 36** outlines some key point ideas used in the previous described algorithms. The *skyline of* column represents the type of object that the skyline notion will be applied. This involves the identification of the most efficient detour, the dynamic skyline query of a set of points or the identification of the best candidate routes based on various preferences. The Data model column represents the underlying data model that is used. The *Incorporates* column outlines the various key point methods that are used. The attribute column represents the optimization criterion that the algorithm will be based in order to find the skyline. The query points represent the points over which the query will be considered. That is user's current location, the various desired optimization points and the start and destination points of a route.

| Algorithm | Skyline of | Data model | Incorporates | Attributes | Query points |
|---|---|---|---|---|---|
| **[75]** | detours | graph | NN-search / network distance | Single (length) | single |
| **[46]** | distance from data points | graph | NN-search / Euclidian-network distance & Dynamic/spatial skyline [133, 134] / [155, 156] | Single (length) | Multiple |
| **[98]** | preference-based routes | multiple-attribute graph | Pareto optimality | Multiple attributes | start / destination |

Table 36: Fundamental algorithm in In-route and road network skyline computation.

## 4.6. Security

This section will reason about security related approaches that are based or use skyline queries. The first topic that will be discussed concerns the location-based skyline queries. In many cases a dataset is maintained by an outsourced Location based service (LBS) which is responsible to store the dataset and deal with the query processing. Users will issue queries and get the results from the LBS and thus must be able to authenticate the results that receive. Another topic that will be discussed is user's privacy under the existence of multiple anonymized datasets and an adversary with external knowledge

### 4.6.1. Authentication

Authors in [113] study the problem of authentication of location-based skyline queries (section 4.5.1). The scenario that is followed assumes that the spatial data are stored in a spatial database and are outsourced to a location-based service provider (LBS) which will handle the queries issued by users. This scenario is a widely known scenario in the real life time since the database owner (DO) does not need to maintain hardware or software resources. The users will query the LBS server and will get the results. Thus in such scenarios the LBS is not the real owner of the data. For that reason users may want to authenticate the soundness and completeness of the results that will receive. This leads to the problem of *authenticated query processing* [109].

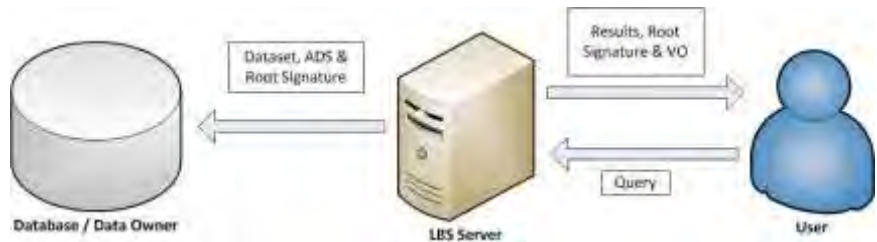

Figure 71: Authenticated query processing.

The general authentication framework that is followed is illustrated in **Figure 71**. In this scenario when the data owner (DO) wants to provide a spatial database to an LBS it also constructs and provides an authenticated data structure (ADS) for his dataset. The ADS is often a tree-based index structure where the root is signed by the DO with his own private key. Thus the LBS has the spatial dataset, the ADS and the signature constructed by the DO. When a user issues a query, the LBS server returns to the user the results. Additionally returns the root's signature and an additional verification object (VO) that is constructed based on the query results. This way the user can verify the results using the returned VO, the signature of the root and the public key of the DO. The whole process between the user and the LBS server involves the result set, the ADS of the dataset, the construction of the VO for each LSQ query placed on the LBS server and finally the result verification by the user. The key problem in this framework is the construction of the VO in order the user to be able to verify the results.

In their work authors proposed two authentication methods. The one is based on the MR-tree [191] which is a generic ADS that indexes the (spatial) data and is based on the MB-tree [109] and the R-tree. The other one is based on the MR-Sky-tree which is a newly proposed ADS that indexes the solution space in the form of *skyline scopes* of each spatial object which is defined as the area in which the point will contribute to the final skyline result. This way the computation cost of VO and its size are reduced. The MR-Sky-tree approach has larger construction cost than the MR-tree but its runtime performance is good for static or infrequent updated datasets in contradiction to the MR-tree that fits best for dynamic datasets.

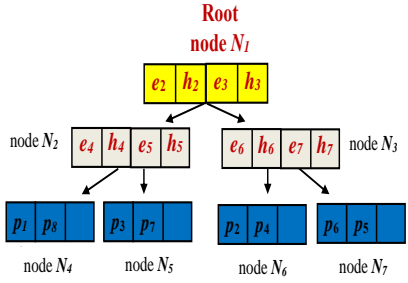

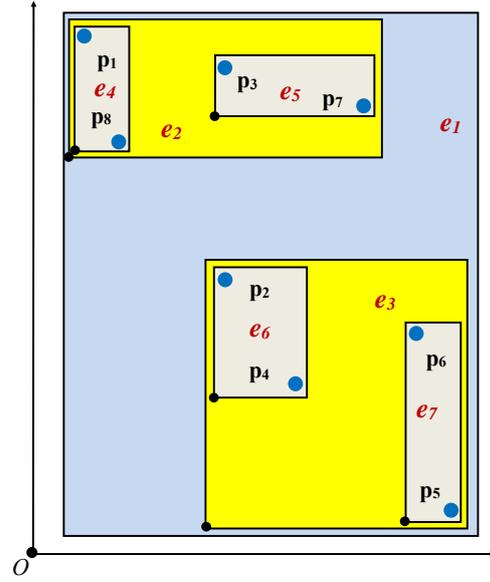

| |
|---|
| h₁ = hash(e1\|h1\|e2\|h2) |
| h₂ = hash(e4\|h4\|e5\|h5) |
| h₃ = hash(e6\|h6\|e7\|h7) |
| h₄ = hash(p1\|p8) |
| h₅ = hash(p3\|p7) |
| h₆ = hash(p1\|p2) |
| h7=hash(p6\|p5) |

Table 37: Digest values.

Figure 72: MR-Tree.

Figure 73: Graphical representation of the MR-Tree.

Reviewing the first method, due to its applicability in dynamic datasets, an MR-Tree is similar to the R-tree. Specifically the leaf nodes of the MR-tree are identical to those of the R-tree. Each internal node will contain triplets of the form <P, M, H> where P is a pointer to a child node, M is the MBR that covers all the child's MBRs (as in the conventional R-tree) and H is the hashed digest of this child. Essentially the hash digest of each internal node will represent the MBRs and the digests of all its children nodes. Finally the hash digest of each leaf node is constructed by hashing the concatenation of the binary representation of all points in the node. The hash digest of an internal node summarizes the MBRs and the digests nodes. An illustration of an MR-tree is presented in **Figure 72** and **Figure 73**.

In the previous outlined authentication scenario the MR-tree can be used as an ADS. A naïve approach to authenticate a skyline query result would be to return the whole MR-tree excluding the digests. The signature of the root will be in the VO. This way the user can authenticate the results that will receive. Nevertheless in this approach the size of the VO can be too large. Authors proposed to prune some index nodes from the MR-tree in order to reduce the size of the VO (R-tree + signature) without affecting the verification process. This can be achieved due to the digests. The pseudocode of the basic method is outlined bellow.

| **ALGORITHM 33:** | VO Construction in the Basic Method [113] |
|---|---|
| **Input:** | Root *mrRoot* of the MR-tree and a query point *q*. |
| **Output:** | Skyline Set *S* and the VOTree *voTree*. |

1. S ← ∅
2. initialize voTree with *mrRoot* (excluding the digest)
3. insert *mrRoot* into a min-heap H ordered by mindist
4. **while** H is not empty **do**
5.    get the top element e from H
6.    **if** e is an index node **do**
7.       **if** e is dominated by some object in S **do**
8.          prune e and keep its digest in the parent node
9.       **else**
10.         insert e's children into H and *voTree*
11.    **if** e is an object **do**
12.       **if** e is not dominated by any object in S **do**
13.          S ← S ∪ {e}

Essentially the algorithm constructs the VO as a tree structure (*voTree*) and keeps a set S that will contain the skyline points. An additional heap H is kept in order to process the nodes and the points of the dataset based on their mindist [2.3.5]. The processing of the dataset is performed in the same way with the R-tee. Initially the VO and H will contain the root of the MR-tree (without the digest) and the set S will be empty. The algorithm each time retrieves the first element e of the heap H. If e is an intermediate node and is not dominated by any skyline point in S then all of its children nodes will be inserted in the heap H and in the VO set (as with the R-tree). In the other case if it is dominated by a skyline point in S, then the node is pruned from further processing and is kept only its digest in the *voTree*. Finally, in the case where e is a point (leaf) and is not dominated by any point in S is added to S and kept in the voTree. In order to verify the results the user can compare the root that is signed by the DO and the digest that is dynamically computed from the voTree and the digests of all the nodes that were received. Since the voTree will contain all the skyline points and the digests of all regions that were pruned the user can safely determine the soundness and the correctness of the results. Authors extended their work in [115] where they considered the problem of authenticating location-based skyline queries in arbitrary subspaces.

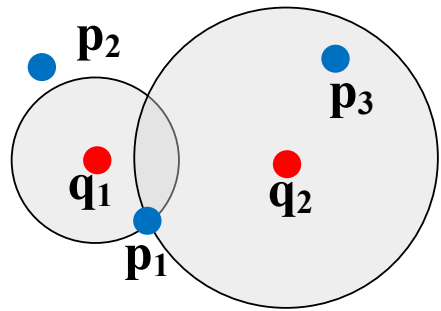

**Figure 74: Area to be authenticated.**

Authors in [122] proposed an additional method to generate VOs for spatial skyline queries. Their goal is to reduce the communication cost by reducing the number of digests to be reported and thus the size of the VO that is sent to the user. In their work observed that only a part of the search space is needed in order to authenticate the results. The area that needs to be authenticated is determined by the relative locations of the query point and the data points. This space is irregular and is consisted by a set of intersecting circles that represents the domination area of points relative with the query points. As shown in [Figure 74], with respect the query points q1 and q2, any point outside the shaded area is dominated by $p_1$. Thus the VO needs to authenticate only the shaded regions.

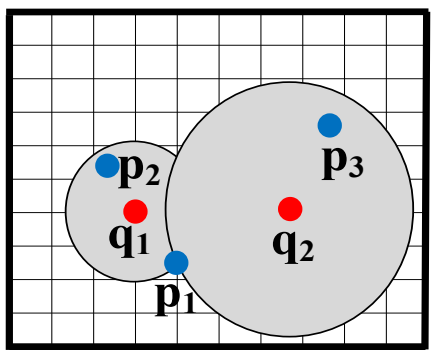

**Figure 75: Single square VO covering all the area.**

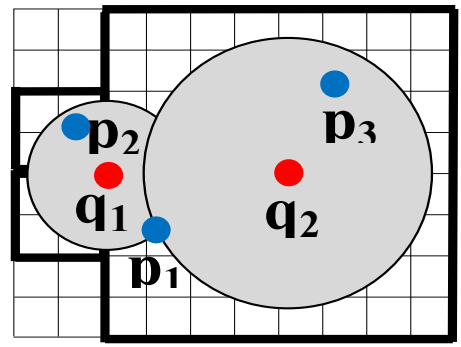

**Figure 76: Three square VOs to cover the same area.**

In their approach authors use a grid to divide the space, in order to create multiple VOs. Each VO covers a square region of the space defined by the grid. The size of each VO can vary and thus determines the number of VOs that will be needed in order to cover a specific region. In general smaller VOs will represent the area more precisely [Figure 76] but will be needed a larger number of VOs rather if we had one large VO [Figure 75]. Intuitively the problem is modeled as a painting problem where the space that is needed to be authenticated is covered by as few VOs as possible. The quality of the solution and the time that will be needed to compute the solution depends on the search method that will be employed.

### 4.6.2. Privacy

Authors in [28] reason about privacy in the presence of external knowledge. Their work builds upon and extends the work of [127] and quantifies the adversary's external knowledge in order to identify privacy threats and enforce privacy requirements. Authors assume that the data owner has a data table D that contains among others, sensitive attributes. By anonymizing the dataset, a table D* is produced which hides the direct relation between the tuple and the respective sensitive attributes. Thus the dataset satisfies specific privacy criterions that were placed. In a general scenario the adversary tries to predict if a specific tuple (an individual t) has a target sensitive value s with the use of external knowledge that has. The privacy criterion should place an upper bound to the confidence of the adversary about the prediction of the existence of the sensitive attribute s on an individual t. For that reason authors describe a theoretical framework for computing the *breach probability* where the adversary can retrieve more information than those given by the anonymized dataset. Additionally, authors quantify the adversary's external knowledge based on [127] and developed algorithms to generate and check the safety of the released candidate datasets.

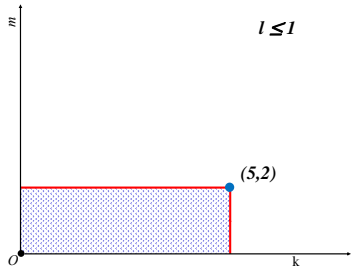

Figure 77: Basic privacy criterion.

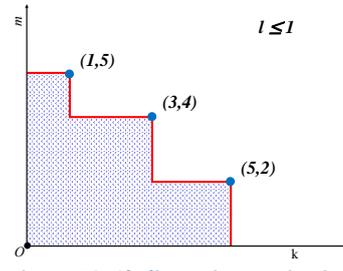

Figure 78: Skyline privacy criterion.

In order to achieve this they break down the initial quantification of the global external knowledge in meaningful components based on three different types of knowledge. Considering an adversary who wants to identify if a target individual t (i.e a person) has a target sensitive value σ (i.e a specific disease), the three types of knowledge can be defined as the knowledge about a target individual t, named $K_{\sigma|t}$, the knowledge about individuals ($u_1,...u_k$) other than t, named $K_{\sigma|u}$ and the knowledge about the relation between t and other individuals ($v_1,...,v_m$), named as $K_{\sigma|v,t}$. In order to quantify the three types of knowledge, the triplet (l,k,m) is used, where l denotes the l sensitive values that a target individual does not have, the sensitive values of k other individuals and m the m members (i.e. a group of people) who tend to have the same sensitive values with t. Intuitively the last definition indicates that if one of the m individuals has a disease σ then t also has σ. Based on this the adversaries knowledge can be expressed as $\lambda_{t,\sigma}(l,k,m) = K_{\sigma|t}(l) \wedge K_{\sigma|u}(k) \wedge K_{\sigma|v,t}(m)$, where t,σ denotes that the target individual is t and the target sensitive value is σ. The adversaries confidence, given a release candidate D*, that an individual t has a sensitive value σ can be defined as $\Pr(\sigma \in t[S]|\lambda_{t,\sigma}(l,k,m), D^*)$. The breach probability, where an adversary can retrieve additional knowledge from the dataset, can be defined as $\max \{ \Pr(\sigma \in t[S]|\lambda_{t,\sigma}(l,k,m), D^*)\}$. Thus a basic privacy criterion should guarantee that the adversaries confidence should not exceed a given threshold value c, which is define as $\max \{ \Pr(\sigma \in t[S]|\lambda_{t,\sigma}(l,k,m), D^*)\} < c$. Based on this authors defined the skyline privacy criterion where given a skyline consisted from triplets {($l_1,k_1,m_1$),...,($l_r,k_r,m_r$)}, the release candidate D* is safe for a sensitive value σ, if $\max \{ \Pr(\sigma \in t[S]|\lambda_{t,\sigma}(l_i,k_i,m_i), D^*)\} < c$, with 1≤i≤r. An example is illustrated with the help of **Figure 77** and **Figure 78**. Consider that the data owner specifies a triplet (l,k,m)=(1,5,2) and c=50% for a sensitive value (disease) σ. The basic privacy criterion guaranties that the adversary cannot predict that any individual t has a disease σ with confidence ≥ 50%, if knows l≤1 sensitive values that the target individual t has, knows the sensitive values of k≤5 other individuals, and additionally knows m≤2 members who have the same sensitive value. If D* is safe under (1,5,2) then it safe under any triplet (l,k,m) with l≤1, k≤5 and m≤2 that is represented by the shaded area in **Figure 77**. Nevertheless this privacy criterion cannot sufficient express all the data owners desired levels of privacy since an additional desired triplet (1,3,4) does not provide any protection guaranty according to the basic privacy criterion, since it is not in the shaded region of **Figure 77**. In that point the skyline privacy criterion was crated which with the help of skyline, lets the data owner to specify the set of incomparable points that will define the skyline that will be used.

Based on **Figure 78** the data owner defined the skyline to be consisted from triplets {(1,1,5), (1,3,4), (1,5,2)}. This way the release candidate will be safe with confidence ≤ 50% given any adversaries knowledge with amount beneath the shaded area of skyline in **Figure 78**. The privacy criterion should provide a worst-case guarantee where the adversaries confidence should not exceed a given threshold value c and thus the release candidate to be safe for a specific sensitive value. Additionally authors extended the basic privacy criterion to allow the data owner to specify and use incomparable points, the set of which is called skyline. The values of these points are based on the quantified types of knowledge and defined with the use of the privacy criterion. This way it can be determined with the use of the skyline if a released candidate is safe, given the adversary a specific amount of knowledge in different conditions.

Authors in **[20]** reason about skyline queries over encrypted data. Their approach uses the SFS algorithm **[35]** in order to identify the skyline and a set of invertible matrices as the key of their encryption scheme.

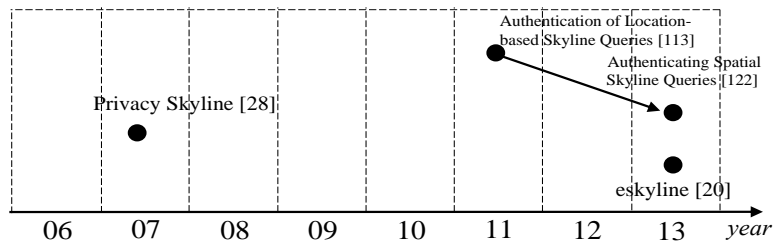

**Figure 79: Hierarchy of security related skyline queries.**

## 4.7. Quality of (web) services

In **[3]** authors study the problem of efficient selection of web services. The identification of the best web service among several similar in function services is a multi-criterion decision problem. Many times a web service that the user needs is composed from multiple sub services. In **Figure 80** illustrates a web service composition of a RealEstate web service which suggests combinations of houses, loan and insurance offer for every retrieved result. The selected sub services that will compose the final service should optimize the overall required QoS level of the application. For that reason a Service Level Agreement (SLA) is used between the user and the service provider in order to define the expected overall QoS level. The QoS-based service composition aims to find the best combination of sub services in order to satisfy the end-to-end QoS constrains that where placed with a given SLA.

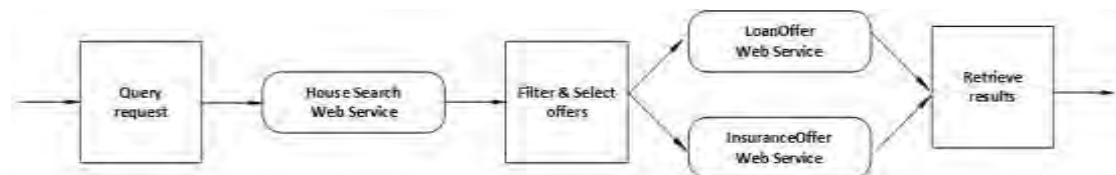

**Figure 80: Service Composition for a RealEstate Web Service.**

One naïve approach to find the best combination of services, that would fulfill an SLA, would be to exhaustive search all the possible combinations. This approach cannot be applied since is very time consuming and has high computation cost even for a small number of candidate services. In order to deal with this problem authors incorporate the notion of skyline in order to reduce the number of candidate services and efficiently select the services that will compose the final application. In general, authors considered dominance relations between the QoS attributes of each service. Typically, the various services are grouped in a set S of service classes. Each service class Si contains all the services that have the same functionality. Each service contains a set of generic QoS attributes such as availability, quality, price, response time, reputation and throughput. Some of those attributes need to minimized and others need to maximized. The various attributes can be computed with the use of the various QoS aggregation functions. The typical aggregation functions are summation, multiplication and minimum relation. The evaluation and quantification of the various services is achieved with the use of utility functions which involves the scaling of the various QoS values to allow a uniform measurement without taking into account the units or ranges and a weighting process in order to represent the preferences of the user. The overall QoS value of a composite service is determined by the QoS values of the component sub services and the composition structure. In this work authors use the sequential composition model **[84]**. The QoS service composition is an optimization problem where it is desired to maximize or minimize the overall utility value, while satisfying the global QoS constrains. One approach for the service composition would be to select from each class the service which maximizes or minimizes the overall utility function. In many cases this may not be the best case since the various criteria placed by the user might be violated. For that reason authors proposed the use of the skyline. First they perform a skyline query on the services of each class in order to identify the potential candidate services forming the *skyline services* of each class. The skyline service computation can be conducted offline and not at request time since it does not depend on the users query.

Additionally authors reason about the case that the resulting skyline sets are large. This can happen when the datasets are anti-correlated which, in this scenario is the most common case. To solve this problem authors proposed to identify a set of representative skyline services. To achieve this they proposed a hierarchical clustering method, where the services are clustered in k clusters and one representative is selected from each cluster. The representative service would be the one with the highest utility value.

In **[194]** authors additionally considered the problem where the quality of the various services and service providers change over time. In many cases the aggregate QoS values may not perfectly reflect the actual performance of a service that is given by a service provider. Additionally a service provider may not deliver the services according to the quality that declares. For that reason authors investigated the problem of *computing services skylines* from *uncertain QoS*. In order to solve the previous problem proposed the *p-dominant service skyline* which is based on p-skyline **[138, 87]**. Specifically a service provider S will belong to the *p-dominant service skyline* if the possibility of S to be dominated by any other service provider is less than p which is a probability threshold. This way the users can efficiently select the service providers that constantly have good QoS values.

In **[179]** authors reason about the cloud-based web service composition. In their work, use the skyline operator in order to prune unqualified services and reduce the related search space. In the next step they perform a Particle Swarm Optimization (PSO) **[143]** in order to find the optimal services based on the user's QoS constrains. In **[195]** authors follow a similar path to **[3]** but instead of using the skyline operator in order to select the set of interesting services that will potentially be part of the composition they focus on finding the set of the most interesting compositions.

In **Figure 81** is illustrated the hierarchy of the QoS skyline queries. With a red line is represented the adoption of a general idea from another work and with a black line an extension or a heavily dependence on a previous work.

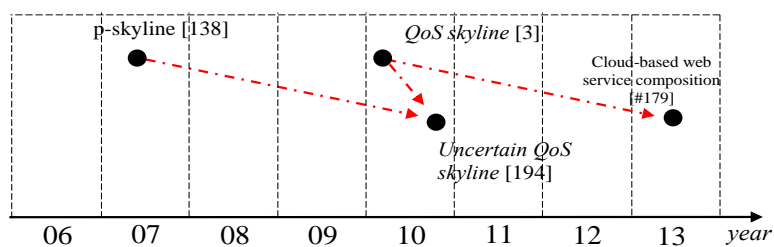

Figure 81: Hierarchy of QoS skyline queries.

## 4.8. Metric space

Skyline computation in metric space **[131]** was first proposed in **[30]**. In this work the dataset belongs to a metric space rather than a multi-dimensional space (i.e. Euclidean) as opposed in previous works. The most familiar metric space is the Euclidean space of a dimension n, which is denoted by $R^n$. In general a topological space contains among others all the metric spaces, where one of the metric spaces is the Euclidean space. The reason that metric spaces involved to the skyline computation is because in many cases the dataset cannot be represented as vectors, something that is fundamental for the *Euclidean space*. An example in bioinformatics is the *DNA searching problem* where the *DNA sequences* are usually modeled and represented by strings and is desired to find the strongest sequence similarity. An additional example where the metric skyline can be incorporated is *image retrieval* [159],[63] where it is desired to identify pictures that have the highest similarity, since an exact match may not exist.

In a metric space we cannot incorporate any geometric information to guide the pruning process during the skyline computation. The only information that can be obtained is the distance between each pair of data objects. However, previous works assume that the data can be represented as vectors in order to utilize the geometric properties of the *Euclidean space* and prune the dataset efficiently, something that makes them inapplicable in metric spaces. In order to describe the related work is needed to give the following important definitions:

**Definition 31:** Metric Space.
A metric space is an ordered pair $(X,d_f)$ which is consisted of a set X of points and a distance (metric) function $d_f$ where $d_f:X \times X \rightarrow \mathbb{R}$, which takes two points belonging to set X and finds their distance $d_f$ as a numeric real value ∎

In the previous definition the set X of points could be the space $\mathbb{R}$ or a finite set of points such as $X=\{k_1,k_2,k_3,k_n\}$. The function $d_f$ could be of the form $d(x,y)=|x-y|$, where x,y are two points of the set X. The distance function must satisfies the following conditions $\forall p,r,s \in X$ :

      **(1) Positivity:** $d_f(p,r) \geq 0$
      **(2) Identity:** $d_f(p,r)=0 \Leftrightarrow p=r$
      **(3) Symmetry:** $d_f(p,r)=d_f(q,r)$
      **(4) Triangle Inequality:** $d_f(p,r)+d_f(r,s) \geq d_f(p,s)$ ∎

The dominance relation in metric spaces is similar to the already described. A full definition considering the metric space will be the following:

**Definition 32:** Metric Space Dominance.
Given a metric space $(X,d_f)$ and a set of query points $Q=\{q_1,q_2,...q_n\}$, $\forall p,r \in X$, p dominates r with regard all the qi *iff* $\exists q_j \in Q$ such that $d_f(p,q_i)<d_f(r,q_i)$ and $\forall q_i \in Q-\{q_j\}$, $d_f(p,q_i) \leq d_f(r,q_i)$ ∎

**Definition 33:** Metric Space Skyline query.
The metric space skyline of a metric space $(X,d_f)$ with a set of query points $Q=\{q_1,q_2,...q_n\}$, is the set of points in X which are not dominated by any other point in X, with regard to Q ∎

In [30] authors proposed a *triangle-based pruning* method that incorporates the *triangle inequality property* in order to safely and efficiently prune the dataset (since distance computation can be very expensive [22] in metric spaces). Additionally they proposed an efficient Metric Skyline Query (*MSQ*) procedure that incorporates the M-tree [36] metric index structure, in order to retrieve the metric skyline points without scanning the whole dataset and without making any particular assumption about the data format and the metric distance function. Metric trees exploit properties of metric spaces such as the triangle inequality to make accesses to the data more efficient. The most related problems to this work are the dynamic skyline [133, 134], the spatial skyline [155, 156] and the multi-source skyline on road networks [46]. However dynamic skyline queries consider only the Euclidean distance as dimension function. The spatial skyline queries require the dataset to be on the Euclidean space, in order to apply the geometric properties of the dataset for the pruning process and the multi-source skylines queries utilizes the geometric information (shortest distance) of data objects during the pruning process, which limits it's application to road networks. This limitation makes these types of skyline queries inapplicable in the metric space.

In [31] authors extended their work on [30] by constructing an optimized metric index structure in order to minimize the cost of the metric skyline retrieval. In [54] authors try to improve [30] by proposing the *dynamic indexing* and the *k-dispersion* techniques in order to minimize the number of dominance tests distance computations respectively. Essentially the *k dispersion* points are analogously to the vertices of the convex hull of the query points in Euclidean space. Also proposed the algorithms N2RS (Nearest-Neighbor-Range-Skyline) which uses nearest neighbor (NN) and range queries requiring two scans of the whole dataset and B2MS2 (Branch-and-Bound Metric Space Skyline), which returns progressively the Metric Space Skyline (referred as MSS) points by scanning the database only once. In **Figure 82** is illustrated the hierarchy of metric space skyline queries. The dash red line represents the adoption of a general idea from another work and the black line represents the extension or the heavily dependence on a previous work. The green dot represents a work not directly related with this section.

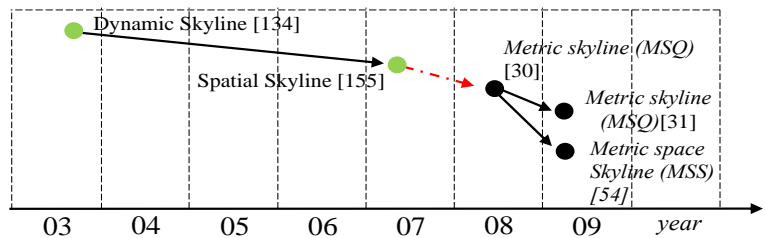

**Figure 82: Metric space skyline queries hierarchy.**

## 4.9. Various topics

This section will outline various topics that do not belong to any of the previously outlined sections. The related work concerns variations of the traditional dominance relation and the wireless sensors network (WSN).

In **[205]** authors propose a new type of dominance relation, named Cone Dominance, in order to deal with the uncontrollable size of the resulting skyline sets. In this type of dominance the dominance region of a point p can be define by a cone region, rather than a rectangle as in traditional dominance as illustrated in **Figure 83**. In this way authors can specify the output result size by appropriately defining the size of the cone. Additionally authors summarize the types of dominance relations that exist and study the problem where the dataset contains tuples who have some of their attribute values missing.

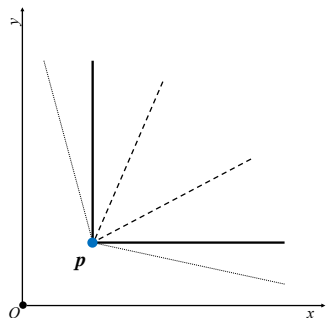

**Figure 83: Cone dominance.**

A WSN network **[192]** is consisted from a number of physical sensors, that can collect information about current conditions such as temperature, humidity, pollution related data and other information depending on their application and construction. The sensors are placed in key locations in space in order to record and report information from that position. Sensors can communicate and report values to a server based on specific rooting algorithms that could involve (but not restricted to) ad-hoc communication or communication based on clusterheads. A simple example of WSN network would be the one that reports atmospheric pollution related values to a server. Consider the scenario where sensors are placed near populated areas and each sensor reports to a server the pollution in that location. Knowing the location of each sensor and the pollution values reported, a user can identify the populated areas that are in great risk. A skyline query on this dataset can reveal the locations that are in high risk by having high pollution but also the locations where the pollutions is small but is close to a populated area.

Based on the previous example authors in [29] reason about skyline computation in WSN networks. The points of the dataset continuously changes and thus the problem that is addressed concerns continuous skyline in sensor networks. For that reason authors distinguish their work from distributed skyline computation. In their work proposed a naïve approach where all nodes report their points to the root node which computes the skyline and a threshold-based approach that tries to minimize the transmitting data over the network in order to maximize the network lifetime. Authors assume that the skyline points, between two rounds of the continuous skyline query, will not deviate much. Thus consider that the current skyline points will have high probability to be skyline points in a next round. Thus authors impose a hierarchical threshold-based approach that defines which points will be reported by each node. Authors in [112] also try to maximize the network lifetime of WSN networks. In their work proposed a distributed filter-based skyline computation approach. The filtering is based on local skyline points that are propagated to other nodes. The algorithm refines the results in a multi-round approach and retrieves the final skyline set.

Authors in [178] studied the problem of reverse skyline computation in WSN. In their work they propose an energy efficient approach which is based on the reduction of the unnecessary transmissions. Additionally authors reasoned about range reverse skyline queries.

In [66] authors proposed the SSPL algorithm which computes the skyline on Big Data. The algorithm utilizes pre-constructed data structures which require low space overhead in order to reduce I/O cost and speed-up the skyline retrieval in such environments. In the first phase the algorithm prebuilds a sorted index list for each attribute and in the second phase sequentially scans and computes the skyline result.

In [142, 141] authors used the skyline operator in order to identify the important members from social networks since the importance of members is a multi-objective issue. This was performed in order to identify the non-skyline member that could be promoted into skyline members at a minimum cost, thus solving the problem of member promotion in social networks. In [51] authors further research the area of social networks and skyline queries. In their work proposed the *geo-social skyline* queries in which they take into account the physical location of persons and their social connectivity in a social network in order to identify persons that are both socially and spatially close to each other. Their research considered the wide adaption of GPS-based mobile devises which support social network check-in capabilities.

## 5. Cardinality estimation & updates maintenance

This section will reason about efficient estimation and computation of cardinality of a skyline set of a given dataset. Additionally are outlined efficient methods for updating skyline. These methods are part or extensions of previously proposed methods. In general skyline cardinality computation tackles the problem of "curse of dimensionality" of skyline computation in high dimensional spaces. The estimation of the cardinality of the dataset can help to perform specific queries in order to reduce the size of the skyline set that is returned. Techniques that try to reduce the number of skyline points that are returned are the *skyline frequency* (section 4.1.3) and *k-dominant* (section 4.1.3).

### 5.1. Cardinality estimation

The cardinality of a maximal set of vectors, which is a similar problem to the skyline operator, was first studied in [21]. Authors consider distinct numeric value conditions and that the dimensions are independent (luck of any correlation or anti-correlation on the dataset). In their work proved that the estimated skyline cardinality m of a d-dimensional set consisted of n points is bounded by the (d-1)-th order harmonic of n thus m= $H_{d-1,n}$. In general the harmonics are defined to be $H_{0,n}=1$ for n>0 and $H_{j,n} = \sum_{i=1}^{n} \frac{H_{j-1,i}}{i}, 1 \leq j \leq d-1$. Finally authors conclude that a loose bound of the estimated skyline cardinality is $O((\ln n)^{d-1})$. Further studies, such as work on [21], [58] proves that the skyline cardinality can be defined as $\Theta((\ln n)^{d-1}/(d-1)!)$, where n is the cardinality of the dataset and d its dimensionality. This indicates that the skyline cardinality increases with the dimensionality. Specifically in [58] authors try to estimate the cardinality of the skyline query results based on the initial dataset, without relying on a distinct value condition. In their model assume that each dimension (attribute) of the input dataset has no duplicate values and that each dimension is statistical independent.

A naive approach to estimate the skyline cardinality would be to retrieve a sample using random sampling without replacement, identify the set of skyline points and compute its cardinality. Then scale the sample's skyline cardinality in order to estimate the cardinality of the whole dataset. Nevertheless, as mentioned in [27] the major problem of this approach is that the cardinality of the skyline set in not a linear function of the cardinality of the dataset. Thus, further study on this subject was performed.

### 5.1.1. Log Sampling

Authors in [27] extend the work of [58] in order to handle numerical and categorical attributes and additionally different distributions without assuming the existence of a distinct value condition. Also authors proposed cost estimation approaches related with skyline computation based on BNL [19] and BNL with presorting [35] algorithms. Their general method assumes that dimensions are independent, thus data distribution is not significant correlated or anti-correlated. In order to deal with the various correlations reasoned about the use of sampling techniques and histograms. Their proposed approach for skyline cardinality estimation is a parametric technique called Log Sampling (LS). The term parametric describes that authors suppose that exists a relation between the cardinality of the dataset and the cardinality of the skyline set. Authors generalized the formula of [58] and followed a hypothetical model that assumes that the estimated cardinality of a skyline s follows the model |SKYs|=A(log(n))$^B$, where n is the cardinality of the dataset. The values A and B are constant parameters that are calculated with the use of small sample of the initial dataset. The sampled set is collected with uniform random sampling without replacement. In order to compute the two parameters A and B the collected sample is split in two parts s1, s2 that have different cardinalities, thus |s1|≠|s2|. Then the skylines SKYs1 and SKYs2, of the sets s1 and s2, are computed in order to find their cardinalities |SKYs1| and |SKYs2|. With the use of the previous computed skyline cardinalities it is possible to compute the A, B values of the model that is used since it is known that |SKYs1|=A(log(|s1|))$^B$ and |SKYs2|=A(log(|s2|))$^B$. Finally the cardinality of the skyline of the whole dataset is estimated with the use of the computed A, B values and the same model. That is the cardinality of the skyline of the whole dataset will be $|SKY_{ds}| = A(\log(n)^B)$ where n is the cardinality of the dataset.

### 5.1.2. Kernel-based

The drawback of the previous proposed algorithm is that is based in a hypothetical empirical model. Additionally in most cases the dataset can be significantly correlated or anti-correlated. Authors in [206] extended the work on [27] and proposed the kernel-based (KB) skyline cardinality estimation approach which is heavily based on kernels [79]. KB approach is nonparametric thus it takes no assumption between the relation of skyline cardinality and the dataset cardinality. Additionally it can handle clustered dataset where the LS method fails [206]. Authors also studied the cardinality estimation of k-dominant skyline [133, 134], which is particular useful in high-dimensional datasets, by extending the LS and KB methods.

The KB approach uses kernels in order to estimate the probability density function (PDF) [101] at an arbitrary position q of the dataset with the use of a random sample s without replacement. The PDF of an arbitrary point q of a d-dimensional space, with the use of a sample s can be computed as $PDF(q) = \sum_{S_i \in S} \left( \frac{1}{|s|h^d \sqrt{det(\Sigma)}} K\left(\frac{dist_\Sigma(q,s_i)}{h}\right) \right)$, where K is the Gaussian kernel [79] function defined as $K(x) = \frac{1}{\sqrt{2\pi}} e^{-\frac{x^2}{2}}$, h is the kernel bandwidth, Σ is a dxd matrix which defines the kernel orientation, $det(\Sigma)$ is the determinant of Σ and $dist_\Sigma(q, s_i)$ is the Mahalanobis distance [79] between q and the sampled point $s_i$. The Mahalonobis distance differences from Euclidian distance as it takes into account the correlation of the dataset.

Below is presented the pseudocode of the KB algorithms. It is noted that authors proposed additional algorithms to compute the kernel bandwidth h.

**ALGORITHM 34:**       KB [206]

**Input:**     A Dataset DS.
           A random sample S of DS and the skyline SKYs of the set S.
           A fixed kernel bandwidth h.
**Output:**  Estimated global skyline cardinality m.

1. Organize all sample points in a grid index of cell length $2h$
2. For each point $p$ in $SKYS$
3. Initialize $Sp$ to empty set
4. For each cell $C$ that intersects with, or is adjacent to $IDR(p)$
5. For each sample point $s \in C$
6. If $MinDist(s, IDR(p)) \leq 2h$, add $s$ to $Sp$
7. Compute $\Omega p$ (using Equation (13) )
8. Compute $m$ (using Equations (6) and (8) )
9. return $m$

Initially a random sample s is retrieved from the dataset. Then the set of local skyline points SKYs of the sample s is retrieved. For each skyline point p is computed the probability $\Omega_p$ that a random point of the dataset falls in the inverse dominance region (IDR) of p. IDR region is equivalent to anti-dominance region in **Figure 59**. The probability incorporates the PDF function over the whole region of IDR and can be computed as $\Omega_p = \int_{IDR(p)} PDF(q)dq$. The estimation of the skyline cardinality of the whole dataset ds can be retrieved from the equation $|SKY_{ds}| = |ds| \times \frac{1}{|s|} \sum_{p \in SKY_i}(1 - \Omega_p)^{|ds|-|s|}$ .

### 5.1.3. PS

The drawback that the KB approach has is that it needs to perform complex computations. Additionally the integration of PDF function over IDR regions suffers from the curse of dimensionality. For that reason authors in [125] proposed the purely sampling-based (PS) approach and compared their method with LS and KS method. As KS, PS is not parametric allowing its application in any dataset distribution. Additionally it can handle high dimensional datasets. Authors approach is consisted from three steps. First retrieve a random sample s of size m, without replacement, form the initial dataset ds. Identify its local skyline points SKYs and compute its cardinality |SKYs|. Finally estimate the cardinality |SKYds| of the whole dataset, using the skyline cardinality |SKYs| of the sampled set s. An unbiased estimation of the cardinality of the skyline set of the whole dataset can be computed through $|SKY_{ds}| = \left(\frac{|SKY_s|}{m}\right) \times n$, where n is the cardinality of the dataset and m is the number of sampled points.

An important aspect of the previously described method is the computation of the skyline set. A naïve approach to compute the skyline points of a sample s is to compare all the points with the points of the whole dataset. Nevertheless this will incur considerably cost. The approach that authors suggest is consisted from three steps. First identify the local skyline points SKYs of the sample s. Then identify the rest local skyline points SKYds-s of the remaining points of the dataset defined as ds-s. Finally check if a point p in SKYs is dominated by any points in the SKYds-s. If no point dominates the point p then this point is a final skyline point.

**ALGORITHM 35:**       checkDominance [125]

**Input:**     A point p, a dataset DS, its
           dimensionality d and it's size n.
**Output:** True or False depending on the
           dominating relation.

1. **for** (int i=0; i<n; i++) **do**
2.   **if** (ds[i]!=p) **then**
3.     int lt=0, eq=0;
4.     **for** (int j=0; j<d; j++) **do**
5.       **if** (ds[i][j]<p[j]) **then**
6.         Lt++;
7.       **end**
8.       **else if** (ds[i][j]==p[j]) **then**
9.         eq++;
10.       **end**
11.     **end**
12.     **if** (lt +eq==d && lt>0) **then**
13.       **return** False;
14.     **end**
15.   **end**
16. **end**
17. **return** True;
18. **end**

In order to reduce the computational cost of dominance comparison authors implemented the checkdominance function which return true if p is not dominated by any point and false otherwise.

| Algorithm | Estimation | Dataset applicability | General approach |
|---|---|---|---|
| **Log Sampling (LS) [27]** | $|SKY_{ds}| = A(\log(n))^B$ | Not on clustered | Two samples |
| **Kernel-based (KB) [206]** | $|SKY_{ds}| = |ds| \times \frac{1}{|s|} \sum_{p \in SKY_s} (1-\Omega_p)^{|ds|-|s|}$ | ANY | Samples based on Kernels |
| **PS [125]** | $|SKY_{ds}| = \left(\frac{|SKY_s|}{m}\right) \times n$ | ANY | Single sample |

Table 38: Fundamental algorithms in skyline cardinality estimation.

**Table 38** outlines the fundamental key aspects of the previous proposed methods. Additionally as in other sections **Figure 84** illustrates the hierarch of the skyline cardinality estimation related methods.

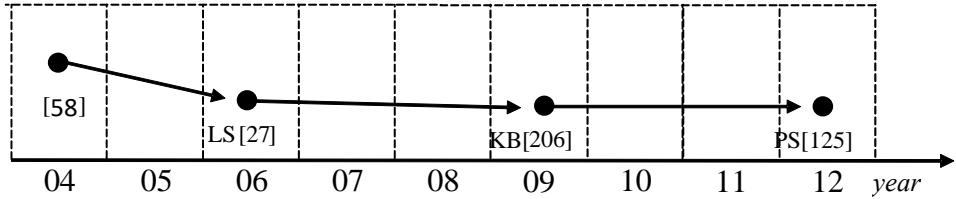

Figure 84: Skyline cardinality estimation methods hierarchy.

## 5.2. Skyline updates & maintenance

Authors research focused on skyline maintenance in order to efficient maintain the skyline when updates or deletions occur on the dataset. The reason of this research interest is that the update of an already computed skyline will have less computation cost than the recomputation of the skyline from scratch. The work in this section is similar with the work on **section 4.4.** The difference is that the work in this section does not assume the existence of data-streams or time series but rather considers that the updates (insertions or deletions) of points are placed over the existing stored dataset.

### 5.2.1. BBS-Update

Authors in **[134]** where the first that studied the problem of incremental skyline maintenance when occur updates over the stored dataset. Their approach in the literature is named BBS-update. This approach was proposed by the authors that proposed the BBS algorithm and essentially discuss how BBS algorithm can efficiently maintain the skyline when various insertions or deletions occur. The simplest approach to maintain the skyline is to recompute the skyline from scratch. Nevertheless this approach is highly inefficient.

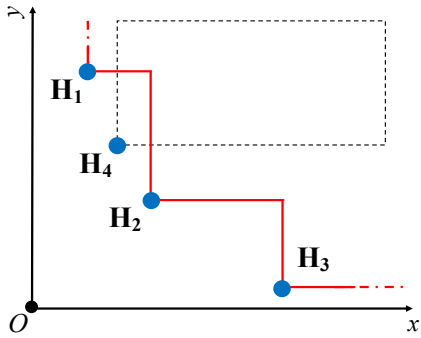

Figure 85: Dominance region without domination.

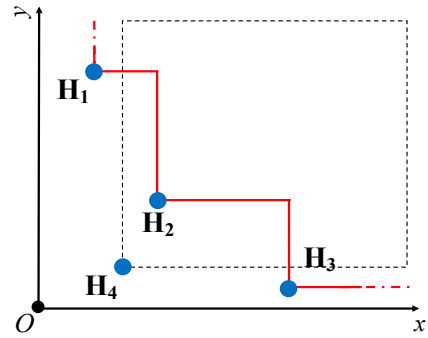

Figure 86: Dominance region with domination.

In BBS-update approach when a new point is retrieved the algorithm checks if that point is dominated by any existing point on the dataset. If it is dominated it is discarded, otherwise the algorithm performs a window query in order to retrieve the points that the new point dominates and thus that can be removed. As illustrated in **Figure 85** the window query on the newly added point H4 does not dominate any point and thus the query won't return any point. The skyline in this case will contain all four points of the dataset. Alternative in **Figure 86** the window query on H4 will return point H2 which will eventually be removed. The final skyline in this case will be the set H1, H3, H4.

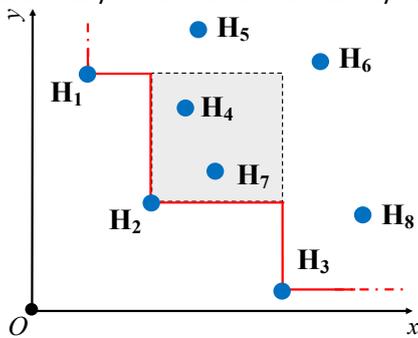

Figure 87: Exclusive dominating region.

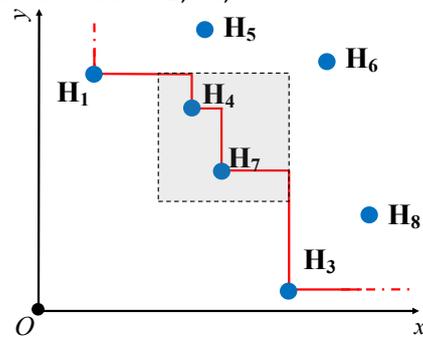

Figure 88: Skyline after deletion of $H_2$.

In the case of a deletion, if a point does not belong to the skyline it can be safely removed. This can be easily checked with using the main-memory R-tree and the point coordinates. In the other case the skyline must be partially reconstructed. Assume that point H2 in **Figure 85** is deleted. The algorithm needs to find the partial skyline based on the exclusive dominating region (EDR) of point H2. The EDR is the region that is dominated by only one specific skyline point. For the skyline point H2 the EDR is represented by the shaded area in **Figure 85**. After the deletion of point H2 a new potential skyline point can only be found inside this region, since everywhere else there will be an existing skyline point that will dominate any other point. In the case of **Figure 85** for any point outside the shaded region, these skyline points could be H1 or H3. Thus the local constrained skyline on the EDR is computed. The two new skyline points will be the points H4 and H7. Authors also proposed another "lazy" method that does not compute the constrained skyline immediately after a point deletion, in order to reduce the number of disk accesses. This approach is based in the observation that when a point is deleted the probability that this will eventually result in the deletion of a skyline point will be very low in large datasets, because the cardinality of skyline points will be relatively small compared to the cardinality of the dataset. The "lazy" method places the various deleted points and the newly inserted points (after the initial skyline computation) in a buffer, in order to efficiently determine, when it is needed, the changes that will occur in the skyline considering only the points that essentially change and their influence on the skyline. Specifically first places the deleted skyline point H2 in a buffer. Then if a new point p is inserted in the dataset is also placed in the buffer. If point p dominates the deleted point H2 (that is placed in the buffer) then H2 is removed from the buffer. This will indicate that it is not needed to perform any other action regarded with the deleted point (i.e a partial skyline computation on the dominance region of H2), since the points that where dominated by the deleted skyline point H2 will also be dominated by the inserted point p. The buffer maintains the updates that occur and prunes the dominated points on the buffer. When there are no more updates or the user issues a query, a local constrained skyline query is placed over the union of EDRs of points that are inside the buffer in order to compute a partial skyline and then merge it with the rest of the skyline points.

### 5.2.2. DeltaSky

Authors in [186] proposed Deltasky that extends the BBS-Update [134] and reasons only about deletions since the insertion process was sufficient studied by the author that proposed BBS-Update. The deletion maintenance on BBS-Update is achieved by issuing constrained skyline queries over the EDRs. Although, EDR regions in high dimensional space (d>2) can have irregular and complex shape making quite expensive their computation. Additionally the computation of EDR regions beyond 2-dimensional spaces is not fully developed in [134].

Again the efficient maintenance of the skyline is based on accessing only the EDRs of the deleted skyline points. Thus the basic problem is to efficient compute the EDRs. As previously mentioned, the computation of the d-dimensional EDR regions in high dimensional spaces is quite complex and expensive. For that reason authors try to reduce the computation cost by representing a EDR as a union of $O(m^d)$ disjoined hyper-rectangles where m is the skyline result set. In the naïve approach the first step involves the computation of the hyper-rectangles that construct the EDR region. Next, this complex EDR region is placed as a constrained region in the algorithm that will be used (in this case BBS). The algorithm will return true when an R-tree region intersects the set of hyper-rectangles and thus additionally actions should performed in order to maintain the skyline.

Additionally the authors proposed the I/O optimal Deltasky algorithm which efficiently determines if an MBR intersects an EDR without calculating the EDR itself. In their work they use a Boolean function to represent the EDR region of a deleted skyline point. The Boolean's function value essentially can indicate if a given point in the dataset belongs in the EDR of a deleted skyline point.

### 5.2.3. Z-Update

The drawback of DeltaSky is that it must scan all the sorted lists. In addition if the skyline is issued on a high dimensional dataset the sorted lists will be large. ZUpdate and ZInsert + *ZDelete* [106, 105] efficiently update the skyline results if insertions or deletions occur by utilizing the sorting property of Z-order curve. This way it can efficiently identify the blocks of skyline points that are not dominated by a newly inserted point and blocks of skyline points that is needed to perform dominance tests. ZUpdate incorporates ZInsert and *ZDelete* in order to efficient handle multiple simultaneous insertions and deletions in order to reduce the computation cost.

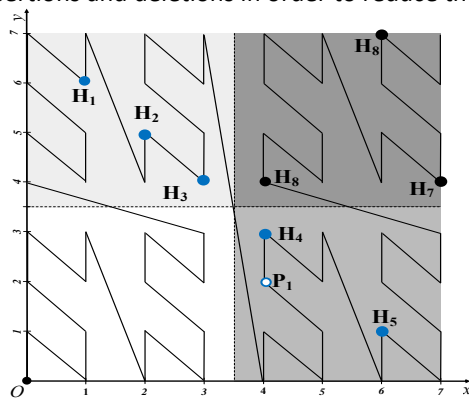 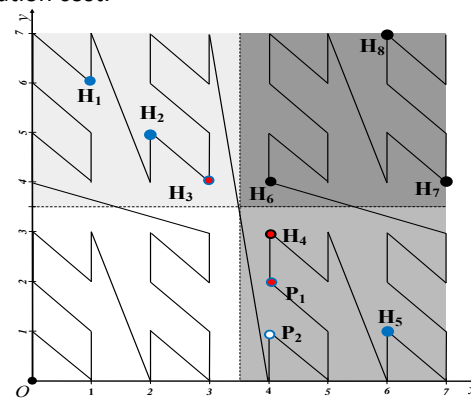

Figure 89: Z-update insertion.      Figure 90: Z-update deletion.

The ZInsert algorithm can be described with the help of **Figure 89** which illustrates an existing dataset and the newly inserted point *p*. The existing skyline set of the dataset is consisted from points H1, H2, H3, H4 and H5. Its location on the z-order curve is between the points H3 and H4. Due to the ordering imposed in the dataset and the skyline set the newly inserted point is needed to be checked if it is dominated only with the skyline points that have smaller Z-addresses. In this case these points are the H1, H2 and H3. If it passes the dominance test is inserted in the skyline set. Furthermore it is needed to check if the newly inserted point dominates other points in the skyline set. This procedure is called candidate reexamination and the points that will be examined are those whose Z-addresses are larger than this of the inserted point p1. During this reexamination the points that will be checked are the H4 and H5 where point H5 will be retained as skyline point and point H4 will be pruned since it is dominated by p1.

The Zdelete algorithm can be described with the help of **Figure 90** which shows the dataset of **Figure 89** after the insertion of point p1. In this case points H3, H4 and p1 are deleted. Additionally a new point p2 is inserted that will be considered in a later phase to demonstrate the Zupdate algorithm. A skyline point deletion may promote other points, which have been pruned to be skyline points. These points will exist only in the dominating region of the deleted point and will be those that have larger z-addresses than the delete point. The algorithm searches for promoted skyline points based on the ordering imposed inside the dominating region of the deleted point. Thus point H3 will consider the region that contains points H6, H7 and H8. Point P1 will consider the points H6, H7 and H8. For point H4 it is not necessary to take any other action except its pruning since it is not a skyline point. This computation will promote point H6 to be a skyline point and the skyline set will be consisted from the points H1, H2, H6, H5. Note that point p2 is not considered at this phase. The search for promoted skyline points is not necessary to be performed at the time a point is deleted but rather can be performed after multiple point deletion and in the occurrence of a point insertion or a skyline query request.

| **ALGORITHM 36:** | ZUpdate (*SRC, SL, P_{UPD}*) [106, 105] |
|---|---|
| **Input:** | A ZBtree *SRC* for source datasets. |
| | A ZBtree *SL* for skyline points. |
| | A set $P_{upd}$ of inserted or deleted data points. |
| **Local:** | A set $P_{ins}$ of inserted data points. |
| | A set $p_{del}$ of deleted skyline points. |

10. $P_{ins} \leftarrow \{p \in P_{upd} \mid p$ is an inserted data point$\}$;
11. *ZInsert(SL, P_{ins})*;           /*ZInsert algorithm [105] */
12. $P_{del} \leftarrow \{p \in P_{upd} \mid p$ is a deleted data point and $p \in SL\}$;
13. *ZDelete(SRC, SL, P_{del})*;      /*ZDelete algorithm [105] */

The Zupdate algorithm incorporates both *Zinsert* and *Zdelete* algorithm in order to group the update operations in insertions and deletions. The algorithm performs all the insertions prior the deletions, since an inserted point may prune some of the points to be deleted, avoiding this way the increased cost of a deletion process. As an example further computation due to the deletion of the skyline point p1 can be eliminate since it is pruned by the inserted point p2. This way the performance of the algorithm is improved. In general, based on the transitivity property, the points that will be dominated by a deleted point will also be dominated by an inserted point that dominates the deleted point. In their work authors compere BBS-update, DeltaSky and Z-update. Since DeltaSky concerns only deletions, in order to be compared with BBS-Update and Z-update authors in adapted with the insertion algorithm of BBS-Update.

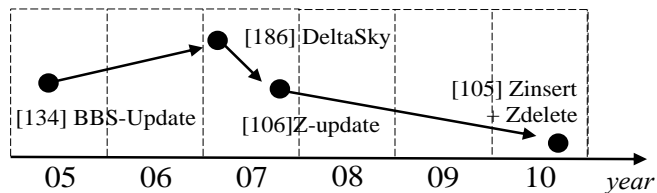

**Figure 91: Skyline update methods hierarchy.**

**Figure 91** shows the hierarchy of the previous proposed methods.

# 6. Future work

Through the previous analysis on the topic of skyline processing we can identify future directions for research. The major research interest these days is concentrated in distributed skyline computation (section **4.2**). Distributed skyline computation attracts the research interests since it is related with the evolution of technology which introduced multicore architectures and distributed informational systems. Cloud-based skyline related applications and skyline computation using mapreduce is also an emerging research field. Researchers trying to cope with this introduced algorithms for distributed skyline computation taking into account various types of datasets. Distributed skyline computation is the biggest research area. Recalling its distributed nature cloud related research has been emerge based especially on optimal service selection and composition. Skyline computation on wireless sensor networks has a little but constant interest mainly focused on energy efficiency.

Another promising research area which keeps the attention of researchers is skyline computation over uncertain data. Uncertainty of data has a major impact in data analysis since it can play a significant role in decision making. As the technology evolves and the need of information analysis is becoming more and more important the need to extract specific analytics even in cases where the data sources or data values are not certain is increased.

Furthermore location-based and spatial skyline queries as long as skyline queries over road networks is always an active research topic. Skyline query processing on subspaces is also an active topic of research but the research interest seems to degrade over time. Specific types of queries such as reverse skyline and range skyline queries are open issues. Since the field of skyline computation is big, it can support surveys on sub-topic level of skyline computation such as a survey on a specific issue in distributed skyline computation. A nice work that could boost research would be a survey on *subspace skyline computation* (section 4.1) and in *attribute and data specific applications* (section 4.3). The sub-field of Continuous and route/road skyline computation may not need a survey at these time since it has not attracted much attention till now.

In more detail based on author's declaration about their future work we will present the major proposed work for future development based on the most recent articles. Authors of [124] among others consider to adapt their approach on skyline queries in crowd-enabled databases in the domain of top-k retrieval on incomplete data. In [143] authors would extend their Cloud-based web service composition approach to a Cloud composition system. In [114] authors are willing to extend their range-based skyline approach to road networks. In [20] an effort will be done to extend their work by validating it with very large datasets and will consider the case of distributed data. In [93] authors are willing to extend their probabilistic and top-k probabilistic skyline algorithm in distributed architectures. In [178] they will explore in more depth the uncertainty of WSN data in skyline computation. In [199] will extend their work of probabilistic skyline computation over uncertain preferences by applying the top-k evaluation framework for uncertain databases that is proposed in [147]. Authors of [34] which investigated skyline queries on keyword match data are willing to proceed with keyword-matched skyline in streams and subspaces but also and the probability of keyword-matched skyline. Authors in [195] that researched the use of skyline queries in service composition are heading to exploit data incomparability in order to further improve the performance of their method. Authors in [201] mention a wide area of future research. At first they will extend their work on stochastic skyline computation by considering sampling approaches on their pdf function (probabilistic density function) that includes stochastic skylines, subspace skyline computation and skyline computation over uncertain objects.

## 7. Conclusions

As the technology evolves the amount and rate of retrieved data that is increased and in many cases derives from multiple sources. Sometimes the data retrieval rate exceeds the processing capabilities and additionally the data quality and validness varies. Skyline processing was developed in order to cope with all these facts and to provide analytic insights over the retrieved data making it a very useful tool for assisting multi-criterion decision-making applications. Skyline query processing does not apply only in computer science but can assist in every area where a decision like "which of my options are truly better than the others" is important such as in business analytics, medical, location-based and GIS (Geographical Positioning System) applications.

This work summarizes and highlights the important work that has been conducted in the field of skyline processing. This survey is essentially consisted from two parts. The first part analyzes the skyline query and its family of related queries along with the state-of-the-art related algorithms. The second part analyzes the various applications of skyline queries along with the state-of-the-art algorithms for each sub-field of research. The selection of articles was based their significance in the research community.

Skyline query processing is a research field that constantly evolves. There are many problems that needed to be solved and more awaiting to have new efficient solutions.

# 8. References

[1] Sameet Agarwal, Rakesh Agrawal, Prasad Deshpande, Ashish Gupta, Jeffrey F. Naughton, Raghu Ramakrishnan, and Sunita Sarawagi. On the computation of multidimensional aggregates. In *VLDB*, pages 506–521, 1996.
[2] Rakesh Agrawal, Alexander Borgida, and H. V. Jagadish. Efficient management of transitive relationships in large data and knowledge bases. In *SIGMOD Conference*, pages 253–262, 1989.
[3] Mohammad Alrifai, Dimitrios Skoutas, and Thomas Risse. Selecting skyline services for qos-based web service composition. In *WWW*, pages 11–20, 2010.
[4] Mikhail J. Atallah and Yinian Qi. Computing all skyline probabilities for uncertain data. In *PODS*, pages 279–287, 2009.
[5] Mikhail J. Atallah, Yinian Qi, and Hao Yuan. Asymptotically efficient algorithms for skyline probabilities of uncertain data. *ACM Trans. Database Syst.*, 36(2):12, 2011.
[6] Brian Babcock, Shivnath Babu, Mayur Datar, Rajeev Motwani, and Jennifer Widom. Models and issues in data stream systems. In *PODS*, pages 1–16, 2002.
[7] Wolf-Tilo Balke, Ulrich Güntzer, and Jason Xin Zheng. Efficient distributed skylining for web information systems. In *EDBT*, pages 256–273, 2004.
[8] O. Barndor-nielsen and M. Sobel. On the distribution of the number of admissible points in a vector random sample. 1966.
[9] Ilaria Bartolini, Paolo Ciaccia, and Marco Patella. Salsa: computing the skyline without scanning the whole sky. In *CIKM*, pages 405–414, 2006.
[10] Julien Basch, Leonidas J. Guibas, and John Hershberger. Data structures for mobile data. *J. Algorithms*, 31(1):1–28, 1999.
[11] Norbert Beckmann, Hans-Peter Kriegel, Ralf Schneider, and Bernhard Seeger. The R*-tree: an efficient and robust access method for points and rectangles. *SIGMOD Rec.*, 19(2):322–331, May 1990.
[12] R.E. Bellman. *Adaptive control processes: a guided tour*. Rand Corporation Research studies. Princeton University Press, 1961.
[13] Jon Louis Bentley. Multidimensional divide-and-conquer. *Commun. ACM*, 23(4):214–229, 1980.
[14] Jon Louis Bentley, Kenneth L. Clarkson, and David B. Levine. Fast linear expected-time algorithms for computing maxima and convex hulls. *Algorithmica*, 9(2):168–183, 1993.
[15] Jon Louis Bentley, H. T. Kung, Mario Schkolnick, and Clark D. Thompson. On the average number of maxima in a set of vectors and applications. *J. ACM*, 25(4):536–543, 1978.
[16] Kevin S. Beyer, Jonathan Goldstein, Raghu Ramakrishnan, and Uri Shaft. When is "nearest neighbor" meaningful? In *ICDT*, pages 217–235, 1999.
[17] Geoffrey Blake, Ronald G Dreslinski, and Trevor Mudge. A survey of multicore processors. *Signal Processing Magazine, IEEE*, 26(6):26–37, 2009.
[18] Christian Bohm and Hans-Peter Kriegel. Determining the convex hull in large multidimensional databases. In *Proceedings of the Third International Conference on Data Warehousing and Knowledge Discovery*, DaWaK '01, pages 294–306, London, UK, UK, 2001. Springer-Verlag.
[19] Stephan Börzsönyi, Donald Kossmann, and Konrad Stocker. The skyline operator. In *Proceedings of the 17th International Conference on Data Engineering*, pages 421–430, Washington, DC, USA, 2001. IEEE Computer Society.
[20] Suvarna Bothe, Panagiotis Karras, and Akrivi Vlachou. eskyline: processing skyline queries over encrypted data. *Proceedings of the VLDB Endowment*, 6(12):1338–1341, 2013.

[21] Agne Brilingaite, Christian S. Jensen, and Nora Zokaite. Enabling routes as context in mobile services. In *GIS*, pages 127–136, 2004.
[22] Sergey Brin. Near neighbor search in large metric spaces. In *VLDB*, pages 574–584, 1995.
[23] Sergey Brin and Lawrence Page. The anatomy of a large-scale hypertextual web search engine. *Computer networks and ISDN systems*, 30(1):107–117, 1998.
[24] Chee Yong Chan, Pin-Kwang Eng, and Kian-Lee Tan. Stratified computation of skylines with partially-ordered domains. In *SIGMOD Conference*, pages 203–214, 2005.
[25] Chee Yong Chan, H. V. Jagadish, Kian-Lee Tan, Anthony K. H. Tung, and Zhenjie Zhang. Finding k-dominant skylines in high dimensional space. In *SIGMOD Conference*, pages 503–514, 2006.
[26] Chee Yong Chan, H. V. Jagadish, Kian-Lee Tan, Anthony K. H. Tung, and Zhenjie Zhang. On high dimensional skylines. In *EDBT*, pages 478–495, 2006.
[27] Surajit Chaudhuri, Nilesh N. Dalvi, and Raghav Kaushik. Robust cardinality and cost estimation for skyline operator. In *ICDE*, page 64, 2006.
[28] Bee-Chung Chen, Raghu Ramakrishnan, and Kristen LeFevre. Privacy skyline: Privacy with multidimensional adversarial knowledge. In *VLDB*, pages 770–781, 2007.
[29] Hekang Chen, Shuigeng Zhou, and Jihong Guan. Towards energy-efficient skyline monitoring in wireless sensor networks. In *EWSN*, pages 101–116, 2007.
[30] Lei Chen and Xiang Lian. Dynamic skyline queries in metric spaces. In *Proceedings of the 11th international conference on Extending database technology: Advances in database technology*, pages 333–343. ACM, 2008.
[31] Lei Chen and Xiang Lian. Efficient processing of metric skyline queries. *Knowledge and Data Engineering, IEEE Transactions on*, 21(3):351–365, 2009.
[32] Lijiang Chen, Bin Cui, and Hua Lu. Constrained skyline query processing against distributed data sites. *IEEE Trans. Knowl. Data Eng.*, 23(2):204–217, 2011.
[33] Lijiang Chen, Bin Cui, Hua Lu, Linhao Xu, and Quanqing Xu. isky: Efficient and progressive skyline computing in a structured p2p network. In *ICDCS*, pages 160–167, 2008.
[34] Hyunsik Choi, HaRim Jung, Ki Yong Lee, and Yon Dohn Chung. Skyline queries on keyword-matched data. *Inf. Sci.*, 232:449–463, 2013.
[35] Jan Chomicki, Parke Godfrey, Jarek Gryz, and Dongming Liang. Skyline with presorting. In *ICDE*, pages 717–719, 2003.
[36] Paolo Ciaccia, Marco Patella, and Pavel Zezula. M-tree: An efficient access method for similarity search in metric spaces. In *VLDB*, pages 426–435, 1997.
[37] Adan Cosgaya-Lozano, Andrew Rau-Chaplin, and Norbert Zeh. Parallel computation of skyline queries. In *HPCS*, page 12, 2007.
[38] Arturo Crespo and Hector Garcia-Molina. Routing indices for peer-to-peer systems. In *ICDCS*, pages 23–, 2002.
[39] Bin Cui, Lijiang Chen, Linhao Xu, Hua Lu, Guojie Song, and Quanqing Xu. Efficient skyline computation in structured peer-to-peer systems. *IEEE Trans. Knowl. Data Eng.*, 21(7):1059–1072, 2009.
[40] Bin Cui, Hua Lu, Quanqing Xu, Lijiang Chen, Yafei Dai, and Yongluan Zhou. Parallel distributed processing of constrained skyline queries by filtering. In *ICDE*, pages 546–555, 2008.
[41] Leonardo Dagum and Ramesh Menon. Openmp: an industry standard api for shared-memory programming. *Computational Science & Engineering, IEEE*, 5(1):46–55, 1998.
[42] Mark De Berg, Otfried Cheong, Marc Van Kreveld, and Mark Overmars. *Computational geometry: algorithms and applications*. Springer, 2008.
[43] Mark De Berg, Marc Van Kreveld, Mark Overmars, and Otfried Cheong Schwarzkopf. *Computational geometry*. Springer, 2000.
[44] Evangelos Dellis and Bernhard Seeger. Efficient computation of reverse skyline queries. In *VLDB*, pages 291–302, 2007.
[45] Evangelos Dellis, Akrivi Vlachou, Ilya Vladimirskiy, Bernhard Seeger, and Yannis Theodoridis. Constrained subspace skyline computation. In *CIKM*, pages 415–424, 2006.
[46] Ke Deng, Xiaofang Zhou, and Heng Tao Shen. Multi-source skyline query processing in road networks. In *ICDE*, pages 796–805, 2007.
[47] EW Dijkastra. A note on two problems in connexion with graphs. *Numerische Mathematic*, pages 269–271, 1959.

[48]     Xiaofeng Ding and Hai Jin. Efficient and progressive algorithms for distributed skyline queries over uncertain data. In *ICDCS*, pages 149–158, 2010.
[49]     Xiaofeng Ding and Hai Jin. Efficient and progressive algorithms for distributed skyline queries over uncertain data. *IEEE Trans. Knowl. Data Eng.*, 24(8):1448–1462, 2012.
[50]     Xiaofeng Ding, Xiang Lian, Lei Chen, and Hai Jin. Continuous monitoring of skylines over uncertain data streams. *Inf. Sci.*, 184(1):196–214, 2012.
[51]     Tobias Emrich, Maximilian Franzke, Nikos Mamoulis, Matthias Renz, and Andreas Züfle. Geo-social skyline queries. In *Database Systems for Advanced Applications - 19th International Conference, DASFAA 2014, Bali, Indonesia, April 21-24, 2014. Proceedings, Part II*, pages 77–91, 2014.
[52]     Ronald Fagin, Amnon Lotem, and Moni Naor. Optimal aggregation algorithms for middleware. *J. Comput. Syst. Sci.*, 66(4):614–656, 2003.
[53]     Philippe Flajolet and G Nigel Martin. Probabilistic counting algorithms for data base applications. *Journal of computer and system sciences*, 31(2):182–209, 1985.
[54]     David Fuhry, Ruoming Jin, and Donghui Zhang. Efficient skyline computation in metric space. In *EDBT*, pages 1042–1051, 2009.
[55]     Volker Gaede and Oliver Günther. Multidimensional access methods. *ACM Comput. Surv.*, 30(2):170–231, 1998.
[56]     Yunjun Gao, Gencai Chen, Ling Chen, and Chun Chen. Parallelizing progressive computation for skyline queries in multi-disk environment. In *DEXA*, pages 697–706, 2006.
[57]     Joseph Gil and Yoav Zibin. Efficient subtyping tests with pq-encoding. *ACM Trans. Program. Lang. Syst.*, 27(5):819–856, 2005.
[58]     Parke Godfrey. Skyline cardinality for relational processing. In *FoIKS*, pages 78–97, 2004.
[59]     Parke Godfrey, Ryan Shipley, and Jarek Gryz. Maximal vector computation in large data sets. In *VLDB*, pages 229–240, 2005.
[60]     Lukasz Golab and M. Tamer Özsu. Processing sliding window multi-joins in continuous queries over data streams. In *VLDB*, pages 500–511, 2003.
[61]     Teofilo F. Gonzalez. Clustering to minimize the maximum intercluster distance. *Theoretical Computer Science*, 38(0):293 – 306, 1985.
[62]     Jim Gray, Surajit Chaudhuri, Adam Bosworth, Andrew Layman, Don Reichart, Murali Venkatrao, Frank Pellow, and Hamid Pirahesh. Data cube: A relational aggregation operator generalizing group-by, cross-tab, and sub totals. *Data Min. Knowl. Discov.*, 1(1):29–53, 1997.
[63]     Venkat N. Gudivada and Vijay V. Raghavan. Design and evaluation of algorithms for image retrieval by spatial similarity. *ACM Trans. Inf. Syst.*, 13(2):115–144, 1995.
[64]     Xi Guo, Yoshiharu Ishikawa, and Yunjun Gao. Direction-based spatial skylines. In *Proceedings of the Ninth ACM International Workshop on Data Engineering for Wireless and Mobile Access*, pages 73–80. ACM, 2010.
[65]     Antonin Guttman. *R-trees: a dynamic index structure for spatial searching*, volume 14. ACM, 1984.
[66]     Xixian Han, Jianzhong Li, Donghua Yang, and Jinbao Wang. Efficient skyline computation on big data. *IEEE Trans. Knowl. Data Eng.*, 25(11):2521–2535, 2013.
[67]     Michiel Hazewinkel. *Encyclopaedia of Mathematics*, volume 9. Springer, 1993.
[68]     Donald Heare and M Pauline Baker. Computer graphics (c version), 1998.
[69]     J.M. Hellerstein, R. Avnur, A. Chou, C. Hidber, C. Olston, V. Raman, T. Roth, and P.J. Haas. Interactive data analysis: the control project. *Computer*, 32(8):51 –59, aug 1999.
[70]     Gsli R. Hjaltason and Hanan Samet. Distance browsing in spatial databases. *ÁCM Trans. Database Syst.*, 24(2):265–318, June 1999.
[71]     Dorit S Hochbaum. Approximation algorithms for the set covering and vertex cover problems. *SIAM Journal on Computing*, 11(3):555–556, 1982.
[72]     Dorit S Hochbaum. *Approximation algorithms for NP-hard problems*. PWS Publishing Co., 1996.
[73]     Katja Hose, Christian Lemke, and Kai-Uwe Sattler. Processing relaxed skylines in pdms using distributed data summaries. In *CIKM*, pages 425–434, 2006.
[74]     Katja Hose and Akrivi Vlachou. A survey of skyline processing in highly distributed environments. *VLDB J.*, 21(3):359–384, 2012.

[75] Xuegang Huang and Christian S. Jensen. In-route skyline querying for location-based services. In *Proceedings of the 4th international conference on Web and Wireless Geographical Information Systems*, W2GIS'04, pages 120–135. Springer-Verlag, 2005.

[76] Zhenhua Huang, Sheng-Li Sun, and Wei Wang. Efficient mining of skyline objects in subspaces over data streams. *Knowl. Inf. Syst.*, 22(2):159–183, 2010.

[77] Zhiyong Huang, Christian S. Jensen, Hua Lu, and Beng Chin Ooi. Skyline queries against mobile lightweight devices in manets. In *ICDE*, page 66, 2006.

[78] Zhiyong Huang, Hua Lu, Beng Chin Ooi, and Anthony K. H. Tung. Continuous skyline queries for moving objects. *IEEE Trans. Knowl. Data Eng.*, 18(12):1645–1658, 2006.

[79] Jenq-Neng Hwang, Shyh-Rong Lay, and Alan Lippman. Nonparametric multivariate density estimation: a comparative study. *IEEE Transactions on Signal Processing*, 42(10):2795–2810, 1994.

[80] Christian Icking, Rolf Klein, and Thomas Ottmann. Priority search trees in secondary memory (extended abstract). In *WG*, pages 84–93, 1987.

[81] Ihab F. Ilyas, George Beskales, and Mohamed A. Soliman. A survey of top-k query processing techniques in relational database systems. *ACM Comput. Surv.*, 40(4):11:1–11:58, October 2008.

[82] Hyeonseung Im, Jonghyun Park, and Sungwoo Park. Parallel skyline computation on multicore architectures. *Inf. Syst.*, 36(4):808–823, 2011.

[83] Md. Saiful Islam, Rui Zhou, and Chengfei Liu. On answering why-not questions in reverse skyline queries. In *29th IEEE International Conference on Data Engineering, ICDE 2013, Brisbane, Australia, April 8-12, 2013*, pages 973–984, 2013.

[84] Michael C. Jaeger, Gregor Rojec-Goldmann, and Gero Mühl. Qos aggregation for web service composition using workflow patterns. In *EDOC*, pages 149–159, 2004.

[85] H. V. Jagadish, Beng Chin Ooi, and Quang Hieu Vu. Baton: A balanced tree structure for peer-to-peer networks. In *VLDB*, pages 661–672, 2005.

[86] Bin Jiang and Jian Pei. Online interval skyline queries on time series. In *ICDE*, pages 1036–1047, 2009.

[87] Bin Jiang, Jian Pei, Xuemin Lin, and Yidong Yuan. Probabilistic skylines on uncertain data: model and bounding-pruning-refining methods. *J. Intell. Inf. Syst.*, 38(1):1–39, 2012.

[88] Wen Jin, Martin Ester, Zengjian Hu, and Jiawei Han. The multi-relational skyline operator. In *ICDE*, pages 1276–1280, 2007.

[89] Wen Jin, Jiawei Han, and Martin Ester. Mining thick skylines over large databases. In *PKDD*, pages 255–266, 2004.

[90] Ibrahim Kamel and Christos Faloutsos. Parallel r-trees. In *SIGMOD Conference*, pages 195–204, 1992.

[91] Mohamed E. Khalefa, Mohamed F. Mokbel, and Justin J. Levandoski. Skyline query processing for incomplete data. In *ICDE*, pages 556–565, 2008.

[92] Mohamed E. Khalefa, Mohamed F. Mokbel, and Justin J. Levandoski. Skyline query processing for uncertain data. In *CIKM*, pages 1293–1296, 2010.

[93] Dongwon Kim, Hyeonseung Im, and Sungwoo Park. Computing exact skyline probabilities for uncertain databases. *IEEE Trans. Knowl. Data Eng.*, 24(12):2113–2126, 2012.

[94] Kazuki Kodama, Yuichi Iijima, Xi Guo, and Yoshiharu Ishikawa. Skyline queries based on user locations and preferences for making location-based recommendations. In *GIS-LBSN*, pages 9–16, 2009.

[95] Vladlen Koltun and Christos H. Papadimitriou. Approximately dominating representatives. *Theor. Comput. Sci.*, 371(3):148–154, 2007.

[96] Maria Kontaki, Apostolos N. Papadopoulos, and Yannis Manolopoulos. Continuous k-dominant skyline computation on multidimensional data streams. In *SAC*, pages 956–960, 2008.

[97] Donald Kossmann, Frank Ramsak, and Steffen Rost. Shooting stars in the sky: an online algorithm for skyline queries. In *Proceedings of the 28th international conference on Very Large Data Bases*, VLDB '02, pages 275–286. VLDB Endowment, 2002.

[98] Hans-Peter Kriegel, Matthias Renz, and Matthias Schubert. Route skyline queries: A multi-preference path planning approach. In *ICDE*, pages 261–272, 2010.

[99] H. T. Kung, Fabrizio Luccio, and Franco P. Preparata. On finding the maxima of a set of vectors. *J. ACM*, 22(4):469–476, 1975.

[100]     Ru-Mei Kung, Eric N. Hanson, Yannis E. Ioannidis, Timos K. Sellis, Leonard D. Shapiro, and Michael Stonebraker. Heuristic search in data base systems. In *Expert Database Workshop*, pages 537–548, 1984.

[101]     R. Larson. *Calculus: An Applied Approach, 9th ed.: An Applied Approach*. Textbooks Available with Cengage Youbook. BROOKS COLE Publishing Company, 2011.

[102]     Jongwuk Lee and Seung won Hwang. Bskytree: scalable skyline computation using a balanced pivot selection. In *EDBT*, pages 195–206, 2010.

[103]     Jongwuk Lee, Gae won You, and Seung won Hwang. Telescope: Zooming to interesting skylines. In *DASFAA*, pages 539–550, 2007.

[104]     Jongwuk Lee, Gae won You, and Seung won Hwang. Personalized top-k skyline queries in high-dimensional space. *Inf. Syst.*, 34(1):45–61, 2009.

[105]     Ken C. K. Lee, Wang-Chien Lee, Baihua Zheng, Huajing Li, and Yuan Tian. Z-sky: an efficient skyline query processing framework based on z-order. *VLDB J.*, 19(3):333–362, 2010.

[106]     Ken C. K. Lee, Baihua Zheng, Huajing Li, and Wang-Chien Lee. Approaching the skyline in z order. In *VLDB*, pages 279–290, 2007.

[107]     Chengkai Li, Nan Zhang, Naeemul Hassan, Sundaresan Rajasekaran, and Gautam Das. On skyline groups. In *CIKM*, pages 2119–2123, 2012.

[108]     Cuiping Li, Beng Chin Ooi, Anthony K. H. Tung, and Shan Wang. Dada: a data cube for dominant relationship analysis. In *SIGMOD Conference*, pages 659–670, 2006.

[109]     Feifei Li, Marios Hadjieleftheriou, George Kollios, and Leonid Reyzin. Dynamic authenticated index structures for outsourced databases. In *SIGMOD Conference*, pages 121–132, 2006.

[110]     Xiang Lian and Lei Chen. Monochromatic and bichromatic reverse skyline search over uncertain databases. In *SIGMOD Conference*, pages 213–226, 2008.

[111]     Xiang Lian and Lei Chen. Reverse skyline search in uncertain databases. *ACM Trans. Database Syst.*, 35(1), 2010.

[112]     Weifa Liang, Baichen Chen, and Jeffrey Xu Yu. Energy-efficient skyline query processing and maintenance in sensor networks. In *CIKM*, pages 1471–1472, 2008.

[113]     Xin Lin, Jianliang Xu, and Haibo Hu. Authentication of location-based skyline queries. In *CIKM*, pages 1583–1588, 2011.

[114]     Xin Lin, Jianliang Xu, and Haibo Hu. Range-based skyline queries in mobile environments. *IEEE Trans. Knowl. Data Eng.*, 25(4):835–849, 2013.

[115]     Xin Lin, Jianliang Xu, Haibo Hu, and Wang-Chien Lee. Authenticating location-based skyline queries in arbitrary subspaces. *IEEE Trans. Knowl. Data Eng.*, 26(6):1479–1493, 2014.

[116]     Xuemin Lin, Hongjun Lu, Jian Xu, and Jeffrey Xu Yu. Continuously maintaining quantile summaries of the most recent n elements over a data stream. In *ICDE*, pages 362–373, 2004.

[117]     Xuemin Lin, Yidong Yuan, Wei Wang, and Hongjun Lu. Stabbing the sky: Efficient skyline computation over sliding windows. In *ICDE*, pages 502–513, 2005.

[118]     Xuemin Lin, Yidong Yuan, Qing Zhang, and Ying Zhang. Selecting stars: The k most representative skyline operator. In *ICDE*, pages 86–95, 2007.

[119]     Xuemin Lin, Ying Zhang, Wenjie Zhang, and Muhammad Aamir Cheema. Stochastic skyline operator. In *ICDE*, pages 721–732, 2011.

[120]     Tian Zhang Raghu Ramakrishnan Miron Livny. Birch: an efficient data clustering method for very large databases. In *ACM SIGMOD international Conference on Management of Data*, volume 1, pages 103–114, 1996.

[121]     Eric Lo, Kevin Y. Yip, King-Ip Lin, and David W. Cheung. Progressive skylining over web-accessible databases. *Data Knowl. Eng.*, 57(2):122–147, 2006.

[122]     Hans Lo and Gabriel Ghinita. Authenticating spatial skyline queries with low communication overhead. In *CODASPY*, pages 177–180, 2013.

[123]     Christoph Lofi, Ulrich Güntzer, and Wolf-Tilo Balke. Efficient computation of trade-off skylines. In *EDBT*, pages 597–608, 2010.

[124]     Christoph Lofi, Kinda El Maarry, and Wolf-Tilo Balke. Skyline queries in crowd-enabled databases. In *Joint 2013 EDBT/ICDT Conferences, EDBT '13 Proceedings, Genoa, Italy, March 18-22, 2013*, pages 465–476, 2013.

[125]     Cheng Luo, Zhewei Jiang, Wen-Chi Hou, Shan He, and Qiang Zhu. A sampling approach for skyline query cardinality estimation. *Knowl. Inf. Syst.*, 32(2):281–301, 2012.

[126]     Matteo Magnani and Ira Assent. From stars to galaxies: skyline queries on aggregate data. In *EDBT*, pages 477–488, 2013.

[127]     David J. Martin, Daniel Kifer, Ashwin Machanavajjhala, Johannes Gehrke, and Joseph Y. Halpern. Worst-case background knowledge for privacy-preserving data publishing. In *ICDE*, pages 126–135, 2007.

[128]     Priti Mishra and Margaret H. Eich. Join processing in relational databases. *ACM Comput. Surv.*, 24(1):63–113, March 1992.

[129]     Michael D. Morse, Jignesh M. Patel, and William I. Grosky. Efficient continuous skyline computation. *Inf. Sci.*, 177(17):3411–3437, 2007.

[130]     Michael D. Morse, Jignesh M. Patel, and H. V. Jagadish. Efficient skyline computation over low-cardinality domains. In *VLDB*, pages 267–278, 2007.

[131]     James R Munkres. Topology, 2000.

[132]     Beng Chin Ooi, Kian-Lee Tan, Cui Yu, and Stéphane Bressan. Indexing the edges - a simple and yet efficient approach to high-dimensional indexing. In *PODS*, pages 166–174, 2000.

[133]     Dimitris Papadias, Yufei Tao, Greg Fu, and Bernhard Seeger. An optimal and progressive algorithm for skyline queries. In *Proceedings of the 2003 ACM SIGMOD international conference on Management of data*, SIGMOD '03, pages 467–478, New York, NY, USA, 2003. ACM.

[134]     Dimitris Papadias, Yufei Tao, Greg Fu, and Bernhard Seeger. Progressive skyline computation in database systems. *ACM Trans. Database Syst.*, 30(1):41–82, March 2005.

[135]     A.N. Papadopoulos. *Nearest Neighbor Search: A Database Perspective*. Series in Computer Science. Springer, 2004.

[136]     Sungwoo Park, Taekyung Kim, Jonghyun Park, Jinha Kim, and Hyeonseung Im. Parallel skyline computation on multicore architectures. In *ICDE*, pages 760–771, 2009.

[137]     Jian Pei, Ada Wai-Chee Fu, Xuemin Lin, and Haixun Wang. Computing compressed multidimensional skyline cubes efficiently. In *ICDE*, pages 96–105, 2007.

[138]     Jian Pei, Bin Jiang, Xuemin Lin, and Yidong Yuan. Probabilistic skylines on uncertain data. In *VLDB*, pages 15–26, 2007.

[139]     Jian Pei, Wen Jin, Martin Ester, and Yufei Tao. Catching the best views of skyline: A semantic approach based on decisive subspaces. In *VLDB*, pages 253–264, 2005.

[140]     Jian Pei, Yidong Yuan, Xuemin Lin, Wen Jin, Martin Ester, Qing Liu, Wei Wang, Yufei Tao, Jeffrey Xu Yu, and Qing Zhang. Towards multidimensional subspace skyline analysis. *ACM Trans. Database Syst.*, 31(4):1335–1381, 2006.

[141]     Zhuo Peng and Chaokun Wang. Member promotion in social networks via skyline. *World Wide Web*, 17(4):457–492, 2014.

[142]     Zhuo Peng, Chaokun Wang, Lu Han, Jingchao Hao, and Xiaoping Ou. Discovering the most potential stars in social networks with infra-skyline queries. In *Web Technologies and Applications - 14th Asia-Pacific Web Conference, APWeb 2012, Kunming, China, April 11-13, 2012. Proceedings*, pages 134–145, 2012.

[143]     Riccardo Poli, James Kennedy, and Tim Blackwell. Particle swarm optimization. *Swarm intelligence*, 1(1):33–57, 2007.

[144]     Franco P. Preparata and Michael Ian Shamos. *Computational Geometry - An Introduction*. Springer, 1985.

[145]     Frank Ramsak, Volker Markl, Robert Fenk, Martin Zirkel, Klaus Elhardt, and Rudolf Bayer. Integrating the ub-tree into a database system kernel. In *VLDB*, pages 263–272, 2000.

[146]     Sylvia Ratnasamy, Paul Francis, Mark Handley, Richard M. Karp, and Scott Shenker. A scalable content-addressable network. In *SIGCOMM*, pages 161–172, 2001.

[147]     Christopher Re, Nilesh N. Dalvi, and Dan Suciu. Efficient top-k query evaluation on probabilistic data. In *Proceedings of the 23rd International Conference on Data Engineering, ICDE 2007, The Marmara Hotel, Istanbul, Turkey, April 15-20, 2007*, pages 886–895, 2007.

[148]     João B. Rocha-Junior, Akrivi Vlachou, Christos Doulkeridis, and Kjetil Nørvåg. Efficient execution plans for distributed skyline query processing. In *EDBT*, pages 271–282, 2011.

[149]     Nick Roussopoulos, Stephen Kelley, and Frédéric Vincent. Nearest neighbor queries. *SIGMOD Rec.*, 24(2):71–79, May 1995.

[150]     Dimitris Sacharidis, Anastasios Arvanitis, and Timos K. Sellis. Probabilistic contextual skylines. In *ICDE*, pages 273–284, 2010.

[151]    Dimitris Sacharidis, Stavros Papadopoulos, and Dimitris Papadias. Topologically sorted skylines for partially ordered domains. In *ICDE*, pages 1072–1083, 2009.

[152]    Hanan Samet. The quadtree and related hierarchical data structures. *ACM Comput. Surv.*, 16(2):187–260, 1984.

[153]    Nikos Sarkas, Gautam Das, Nick Koudas, and Anthony K. H. Tung. Categorical skylines for streaming data. In *SIGMOD Conference*, pages 239–250, 2008.

[154]    Rajkumar Sen and Krithi Ramamritham. Efficient data management on lightweight computing device. In *ICDE*, pages 419–420, 2005.

[155]    Mehdi Sharifzadeh and Cyrus Shahabi. The spatial skyline queries. In *VLDB*, pages 751–762, 2006.

[156]    Mehdi Sharifzadeh, Cyrus Shahabi, and Leyla Kazemi. Processing spatial skyline queries in both vector spaces and spatial network databases. *ACM Transactions on Database Systems (TODS)*, 34(3):14, 2009.

[157]    Shashi Shekhar and Jin Soung Yoo. Processing in-route nearest neighbor queries: a comparison of alternative approaches. In *GIS*, pages 9–16, 2003.

[158]    Sarvjeet Singh, Chris Mayfield, Rahul Shah, Sunil Prabhakar, Susanne E. Hambrusch, Jennifer Neville, and Reynold Cheng. Database support for probabilistic attributes and tuples. In *ICDE*, pages 1053–1061, 2008.

[159]    Tomáš Skopal and J Lokoc. Answering metric skyline queries by pm-tree. In *Proceedings of the Dateso 2010 Workshop*, volume 567, pages 22–37, 2010.

[160]    Wanbin Son, Seung-won Hwang, and Hee-Kap Ahn. MSSQ: manhattan spatial skyline queries. *Inf. Syst.*, 40:67–83, 2014.

[161]    Wanbin Son, Seung won Hwang, and Hee-Kap Ahn. Mssq: Manhattan spatial skyline queries. In *SSTD*, pages 313–329, 2011.

[162]    Dalie Sun, Sai Wu, Jianzhong Li, and Anthony K. H. Tung. Skyline-join in distributed databases. In *ICDE Workshops*, pages 176–181, 2008.

[163]    Shengli Sun, Zhenghua Huang, Hao Zhong, Dongbo Dai, Hongbin Liu, and Jinjiu Li. Efficient monitoring of skyline queries over distributed data streams. *Knowl. Inf. Syst.*, 25(3):575–606, 2010.

[164]    Kian-Lee Tan, Pin-Kwang Eng, and Beng Chin Ooi. Efficient progressive skyline computation. In *Proceedings of the 27th International Conference on Very Large Data Bases*, VLDB '01, pages 301–310, San Francisco, CA, USA, 2001. Morgan Kaufmann Publishers Inc.

[165]    Yufei Tao, Ling Ding, Xuemin Lin, and Jian Pei. Distance-based representative skyline. In *ICDE*, pages 892–903, 2009.

[166]    Yufei Tao and Dimitris Papadias. Maintaining sliding window skylines on data streams. *IEEE Trans. Knowl. Data Eng.*, 18(2):377–391, 2006.

[167]    Yufei Tao, Xiaokui Xiao, and Jian Pei. Subsky: Efficient computation of skylines in subspaces. In *ICDE*, page 65, 2006.

[168]    Yufei Tao, Xiaokui Xiao, and Jian Pei. Efficient skyline and top-k retrieval in subspaces. *IEEE Trans. Knowl. Data Eng.*, 19(8):1072–1088, 2007.

[169]    Eleftherios Tiakas, Apostolos N Papadopoulos, and Yannis Manolopoulos. Skyline queries: An introduction. In *Information, Intelligence, Systems and Applications (IISA), 2015 6th International Conference on*, pages 1–6. IEEE, 2015.

[170]    George Trimponias, Ilaria Bartolini, Dimitris Papadias, and Yin Yang. Skyline processing on distributed vertical decompositions. *IEEE Trans. Knowl. Data Eng.*, 25(4):850–862, 2013.

[171]    George Valkanas and Apostolos N. Papadopoulos. Efficient and adaptive distributed skyline computation. In *SSDBM*, pages 24–41, 2010.

[172]    Akrivi Vlachou, Christos Doulkeridis, and Yannis Kotidis. Angle-based space partitioning for efficient parallel skyline computation. In *SIGMOD Conference*, pages 227–238, 2008.

[173]    Akrivi Vlachou, Christos Doulkeridis, Yannis Kotidis, and Michalis Vazirgiannis. Skypeer: Efficient subspace skyline computation over distributed data. In *ICDE*, pages 416–425, 2007.

[174]    Akrivi Vlachou, Christos Doulkeridis, Yannis Kotidis, and Michalis Vazirgiannis. Efficient routing of subspace skyline queries over highly distributed data. *IEEE Trans. Knowl. Data Eng.*, 22(12):1694–1708, 2010.

[175]    Akrivi Vlachou and Kjetil Nørvåg. Bandwidth-constrained distributed skyline computation. In *MobiDE*, pages 17–24, 2009.

[176] Akrivi Vlachou and Michalis Vazirgiannis. Link-based ranking of skyline result sets. In *Proceedings of the 3rd Multidisciplinary Workshop on Advances in Preference Handling, M-Pref*, 2007.

[177] Akrivi Vlachou and Michalis Vazirgiannis. Ranking the sky: Discovering the importance of skyline points through subspace dominance relationships. *Data Knowl. Eng.*, 69(9):943–964, 2010.

[178] Guoren Wang, Junchang Xin, Lei Chen, and Yunhao Liu. Energy-efficient reverse skyline query processing over wireless sensor networks. *IEEE Trans. Knowl. Data Eng.*, 24(7):1259–1275, 2012.

[179] Shangguang Wang, Qibo Sun, Hua Zou, and Fangchun Yang. Particle swarm optimization with skyline operator for fast cloud-based web service composition. *MONET*, 18(1):116–121, 2013.

[180] Shiyuan Wang, Beng Chin Ooi, Anthony K. H. Tung, and Lizhen Xu. Efficient skyline query processing on peer-to-peer networks. In *ICDE*, pages 1126–1135, 2007.

[181] Shiyuan Wang, Quang Hieu Vu, Beng Chin Ooi, Anthony K. H. Tung, and Lizhen Xu. Skyframe: a framework for skyline query processing in peer-to-peer systems. *VLDB J.*, 18(1):345–362, 2009.

[182] Shuguang Wang, Cui Yu, and Beng Chin Ooi. Compressing the index - a simple and yet efficient approximation approach to high-dimensional indexing. In *WAIM*, pages 291–304, 2001.

[183] William Wu-Shyong Wei. *Time series analysis*. Addison-Wesley Redwood City, California, 1994.

[184] Raymond Chi-Wing Wong, Ada Wai-Chee Fu, Jian Pei, Yip Sing Ho, Tai Wong, and Yubao Liu. Efficient skyline querying with variable user preferences on nominal attributes. *PVLDB*, 1(1):1032–1043, 2008.

[185] Raymond Chi-Wing Wong, Jian Pei, Ada Wai-Chee Fu, and Ke Wang. Online skyline analysis with dynamic preferences on nominal attributes. *IEEE Trans. Knowl. Data Eng.*, 21(1):35–49, 2009.

[186] Ping Wu, Divyakant Agrawal, Ömer Egecioglu, and Amr El Abbadi. Deltasky: Optimal maintenance of skyline deletions without exclusive dominance region generation. In *ICDE*, pages 486–495, 2007.

[187] Ping Wu, Caijie Zhang, Ying Feng, Ben Y. Zhao, Divyakant Agrawal, and Amr El Abbadi. Parallelizing skyline queries for scalable distribution. In *EDBT*, pages 112–130, 2006.

[188] Tian Xia and Donghui Zhang. Refreshing the sky: the compressed skycube with efficient support for frequent updates. In *SIGMOD Conference*, pages 491–502, 2006.

[189] Tian Xia, Donghui Zhang, Zheng Fang, Cindy X. Chen, and Jie Wang. Online subspace skyline query processing using the compressed skycube. *ACM Trans. Database Syst.*, 37(2):15, 2012.

[190] Tian Xia, Donghui Zhang, and Yufei Tao. On skylining with flexible dominance relation. In *ICDE*, pages 1397–1399, 2008.

[191] Yin Yang, Stavros Papadopoulos, Dimitris Papadias, and George Kollios. Authenticated indexing for outsourced spatial databases. *VLDB J.*, 18(3):631–648, 2009.

[192] Jennifer Yick, Biswanath Mukherjee, and Dipak Ghosal. Wireless sensor network survey. *Computer Networks*, 52(12):2292–2330, 2008.

[193] Man Lung Yiu, Eric Lo, and Duncan Yung. Measuring the sky: On computing data cubes via skylining the measures. *IEEE Trans. Knowl. Data Eng.*, 24(3):492–505, 2012.

[194] Qi Yu and Athman Bouguettaya. Computing service skyline from uncertain qows. *IEEE T. Services Computing*, 3(1):16–29, 2010.

[195] Qi Yu and Athman Bouguettaya. Efficient service skyline computation for composite service selection. *IEEE Trans. Knowl. Data Eng.*, 25(4):776–789, 2013.

[196] Yidong Yuan, Xuemin Lin, Qing Liu, Wei Wang, Jeffrey Xu Yu, and Qing Zhang. Efficient computation of the skyline cube. In *VLDB*, pages 241–252, 2005.

[197] Ming Zhang and Reda Alhajj. Skyline queries with constraints: Integrating skyline and traditional query operators. *Data & Knowledge Engineering*, 69(1):153 – 168, 2010.

[198] Nan Zhang, Chengkai Li, Naeemul Hassan, Sundaresan Rajasekaran, and Gautam Das. On skyline groups. *IEEE Trans. Knowl. Data Eng.*, 26(4):942–956, 2014.

[199] Qing Zhang, Pengjie Ye, Xuemin Lin, and Ying Zhang. Skyline probability over uncertain preferences. In *EDBT*, pages 395–405, 2013.

[200] Shiming Zhang, Nikos Mamoulis, and David W. Cheung. Scalable skyline computation using object-based space partitioning. In *SIGMOD Conference*, pages 483–494, 2009.

[201] Wenjie Zhang, Xuemin Lin, Ying Zhang, Muhammad Aamir Cheema, and Qing Zhang. Stochastic skylines. *ACM Trans. Database Syst.*, 37(2):14, 2012.

[202] Wenjie Zhang, Xuemin Lin, Ying Zhang, Wei Wang, and Jeffrey Xu Yu. Probabilistic skyline operator over sliding windows. In *ICDE*, pages 1060–1071, 2009.

[203] Ying Zhang, Wenjie Zhang, Xuemin Lin, Bin Jiang, and Jian Pei. Ranking uncertain sky: The probabilistic top-k skyline operator. *Inf. Syst.*, 36(5):898–915, 2011.

[204] Zhenjie Zhang, Reynold Cheng, Dimitris Papadias, and Anthony K. H. Tung. Minimizing the communication cost for continuous skyline maintenance. In *SIGMOD Conference*, pages 495–508, 2009.

[205] Zhenjie Zhang, Hua Lu, Beng Chin Ooi, and Anthony K. H. Tung. Understanding the meaning of a shifted sky: a general framework on extending skyline query. *VLDB J.*, 19(2):181–201, 2010.

[206] Zhenjie Zhang, Yin Yang, Ruichu Cai, Dimitris Papadias, and Anthony K. H. Tung. Kernel-based skyline cardinality estimation. In *SIGMOD Conference*, pages 509–522, 2009.

[207] Baihua Zheng, Ken CK Lee, and Wang-Chien Lee. Location-dependent skyline query. In *Mobile Data Management, 2008. MDM'08. 9th International Conference on*, pages 148–155. IEEE, 2008.

[208] Lin Zhu, Yufei Tao, and Shuigeng Zhou. Distributed skyline retrieval with low bandwidth consumption. *IEEE Trans. Knowl. Data Eng.*, 21(3):384–400, 2009.